universität ulm

Fakultät für Naturwissenschaften

# Optimization Techniques in Quantum Information

Dissertation zur Erlangung des Doktorgrades Dr. rer. nat.
der Fakultät für Naturwissenschaften der Universität Ulm

**vorgelegt von**

Benjamin Desef

aus Memmingen

**Amtierender Dekan:**
Prof. Dr. Kay-E. Gottschalk

**Erstgutachter:**
Prof. Dr. Martin B. Plenio
**Zweitgutachter:**
Prof. Dr. Koenraad Audenaert

**Tag der Promotion:**
30.10.2025

# Eigenständigkeitserklärung

Ich versichere hiermit, dass ich

- die Arbeit selbständig angefertigt habe und keine anderen als die angegebenen Quellen und Hilfsmittel benutzt,
- die wörtlich oder inhaltlich übernommenen Stellen als solche kenntlich gemacht,
- die Satzung der Universität Ulm zur Sicherung guter wissenschaftlicher Praxis in der jeweils gültigen Fassung beachtet,
- keine KI-basierten Hilfsmittel zur Erstellung herangezogen und
- diese Arbeit noch nicht an einer anderen Stelle zur Prüfung vorgelegt habe.

# Author's declaration

I hereby affirm that I

- created this Thesis independently and did not use any but the acknowledged sources and tools,
- marked all literal or substantive passages in a clearly recognizable way,
- obeyed the statutes of the University of Ulm for good scientific practice,
- did not use any AI-based tools, and
- have not submitted this Thesis in whole or part for a degree anywhere else.

Benjamin Desef

# List of scientific publications

## Published papers and preprints

- Díaz, M. G. et al.: Accessible coherence in open quantum system dynamics. In: *Quantum* 4 (2020), page 249. ISSN: 2521-327X. doi:10.22331/q-2020-04-02-249
- Desef, B.: *yquant: Typesetting quantum circuits in a human-readable language*. 2021. arXiv: 2007.12931 [quant-ph]
- Desef, B.; Plenio, M. B.: Optimizing quantum codes with an application to the loss channel with partial erasure information. In: *Quantum* 6 (2022), page 667. ISSN: 2521-327X. doi:10.22331/q-2022-03-11-667

## Scientific software (selected)

- `yquant` (2020) Typesetting quantum circuits in a human-readable language
  `https://github.com/projekter/yquant`
- `NamedTensor.m` (2020) Support for named indices in Mathematica tensors
  `https://github.com/projekter/NamedTensor.m`
- `ChannelLoss` (2021) Source code for [DP22b] and the extension in section 5.3.2
  `https://github.com/qubit-ulm/ChannelLoss`
- `PolynomialOptimization.jl` (2023) Julia package for polynomial optimization
  `https://github.com/projekter/PolynomialOptimization.jl`
- `StandardPacked.jl` (2023) Julia package for packed matrix storage
  `https://github.com/projekter/StandardPacked.jl`

# Acknowledgments

It has been a long road to finish this Thesis, and many people supported me along the way. The open, friendly, and stimulating atmosphere at the Institute of Theoretical Physics at Ulm University has allowed me to persue the topics I found interesting, finding support in highly motivated and competent people, and broadening the knowledge in many different aspects of physics. A big thanks therefore goes to Martin Plenio, who made this possible and provided input, guidance, and security during this PhD.

The helpful comments and intuition of Koenraad Audenaert convinced me, after many failed attempts, that—and how—my idea for fast interpolant bases laid out in section 4.9.3.3 can *still* work. Even if it now ended up being quite different from what we discussed, thank you for the essential input!

My family had a lot of absence despite being present to endure during the writing, and only their continuous support made it possible to invest this amount of time at all. I will do better in the future! A special thanks goes to my father Thorsten for proofreading the whole Thesis, making it a much smoother reading experience. I hope that none of my corrections introduced new weird sentences or words, else I am of course entirely to blame.

*Benjamin Desef*
*May 2025*

The work during the PhD was partly funded by the Federal Ministry of Education and Research (BMBF) grants number 16KIS0875 and 16KISQ006. Larger numerical data was generated on the JUSTUS 2 cluster, funded by the state of Baden-Württemberg through bwHPC and the German Research Foundation (DFG) through grant no INST 40/575-1 FUGG.

# Contents

# Summary

Many quantities of interest in quantum information are naturally formulated as optimization problems. This applies both to measures characterizing for example states, operations, or rates, but also concrete instances of protocols to perform certain tasks. While the extremal value of these problems might be a distance to some optimal protocol, often a fidelity, the point(s) where it is attained reveals how such a protocol can be realized.

The often straightforward formulation of the optimization problem rarely leads to easily-solvable problems. In recent years, the toolset employed by convex, and in particular semidefinite, programming has proven very fruitful. Although many interesting quantum information problems may not be immediately expressible in terms of linear and semidefinite constraints only, oftentimes, researchers have found ingenious reformulations to cast them in such a way. The allure of semidefinite programming lies in the fact that within a time polynomial in the number of variables and the precision, a global solution to the original problem can be found numerically.

This thesis focuses on the intersection of mathematical and computational optimization and quantum information. A good grasp on which kinds of formulations are relevant in quantum information—for readers from mathematical programming—as well as a thorough understanding of the numerical techniques underlying both nonconvex and convex algorithms—for readers from the physical community—is required to successfully bridge these two fields, and this thesis aims to provide both.

Main contributions come in the form of open-source software code: Since semidefinite reformulations are hard to obtain, if possible at all, a hybrid approach mixing "traditional" nonconvex and convex methods can make difficult problems more accessible. A demonstration of how to efficiently implement such a hybrid algorithm, combining state-of-the-art solvers while avoiding bottlenecks at the interface is provided, finding optimal protocols to establish entanglement through a lossy channel.

The central software package developed during this thesis addresses polynomial optimization problems. While a semidefinite formulation might be difficult to get, many problems naturally involve only a polynomial objective and constraint polynomials. Using relaxations developed in the early 2000s, such problems can *automatically* be cast into semidefinite programs that provide a hierarchy of outer approximations, approaching a minimum from below—as opposed to nonconvex algorithms, which usually work within the feasible set and approach the minimum from above. Constructing the problems is automatic, and several software frameworks exist to relieve the user of this task. The resulting problems are often so large and scale so unfavorably with respect to the variable number and degree involved that the boundary of the doable is reached quickly. However, technical progress both in hardware as well as in algorithms has pushed this boundary further and further—but software frameworks for polynomial optimization have not followed in the same manner, often now making *them* the bottleneck that before was the solver. The package `PolynomialOptimization.jl` developed during this thesis aims to fill the gap and provide a very resource-efficient intermediate layer together with a wide number of algorithms to reduce the problem size, and naturally supporting complex numbers and semidefinite constraints ubiquitous in quantum information problems. Its application on an entanglement distribution problem is demonstrated, validating that the solver again becomes the bottleneck, and showing that even relaxations with semidefinite matrices of three- and four-digit size can be solved using a convenient user interface.

To complement these practical advances, a new way to calculate interior-point barriers for the cone of sums-of-squares matrices—the central object in polynomial optimization—in a nearly time-optimal way (up to a logarithmic factor) is developed in this thesis. Its efficient implementation for a nonsymmetric interior-point has the potential of further reducing resource consumption in solvers.

# Notation

**Sets and spaces**

| | |
|---|---|
| $\mathcal{A}$, $\mathcal{B}$, ... | generic set or space (curly letters) |
| $\mathcal{H}$ | Hilbert space |
| $\mathcal{C}$ | cone |
| $\mathcal{C}^*$ | dual cone |
| $\mathcal{V}^{\circledast}$ | dual vector space |
| $\partial\mathcal{A}$ | boundary of a subset $\mathcal{A}$ in a topological space |
| $\mathring{\mathcal{A}}$ | interior of $\mathcal{A}$ |

*Specific sets*

| | |
|---|---|
| $\mathbb{N}$ | natural numbers, $\{1, 2, \dots\}$ |
| $\mathbb{N}_0$ | natural numbers from zero, $\{0, 1, 2, \dots\}$ |
| $\mathbb{Z}$ | integers, $\{0, \pm 1, \pm 2, \dots\}$ |
| $\mathbb{R}$ | reals |
| $\mathbb{C}$ | complexes |
| $\mathbb{S}^d$ | real-symmetric or complex-Hermitian matrices of side dimension $d$ |
| $\mathbb{F}[\boldsymbol{x}]$ | polynomials in the variables $\boldsymbol{x}$ with coefficients from the field $\mathbb{F}$ |
| $P[\boldsymbol{x}]$ | nonnegative polynomials in the variables $\boldsymbol{x}$ |
| $\Sigma[\boldsymbol{x}]$ | sums-of-squares polynomials in the variables $\boldsymbol{x}$ |
| $\mathcal{L}(\mathcal{B} \leftarrow \mathcal{A})$ | linear maps from $\mathcal{A}$ to $\mathcal{B}$ |
| $\mathcal{T}(\mathcal{H})$ | trace-class operators on $\mathcal{H}$ |

*Landau symbols* for an implicit critical point $x_{\mathrm{c}}$ (typically, $x \in \{0, \infty\}$)

| | |
|---|---|
| $o(\bullet)$ | of a smaller order than: $f \in o(g) \Leftrightarrow \lim_{x \to x_{\mathrm{c}}} \left\lvert \frac{f(x)}{g(x)} \right\rvert = 0$ |
| $\mathcal{O}(\bullet)$ | of the order of: $f \in \mathcal{O}(g) \Leftrightarrow \limsup_{x \to x_{\mathrm{c}}} \left\lvert \frac{f(x)}{g(x)} \right\rvert < \infty$ |
| $\Omega(\bullet)$ | of an order not smaller than: $f \in \Omega(g) \Leftrightarrow \limsup_{x \to x_{\mathrm{c}}} \left\lvert \frac{f(x)}{g(x)} \right\rvert > 0$ |

## Operators

| | |
|---|---|
| $A$, $B$, ... | generic linear operator |
| $\hat{A}$, $\hat{B}$, ... | self-adjoint linear operator (hat) |
| $\mathsf{A}$, $\mathsf{B}$ | superoperator (sans serif) |
| $\mathbb{P}_{\hat{M}:\lambda}$ | projection onto the eigenspace of $\hat{M}$ associated with eigenvalue $\lambda$ |
| $\mathbb{P}_{\mathcal{A}}$ | projection onto the set $\mathcal{A}$ |
| $\mathbb{P}[\lvert x\rangle]$ | projection operator $\lvert x\rangle\langle x\rvert$ |
| $\mathbb{1}$ | identity operator/matrix |
| tr | trace of a linear operator |
| rk | rank of a linear operator |
| spec | spectrum of a self-adjoint linear operator |
| $\nabla f$ | gradient of the function $f$ |
| $\nabla^2 f$ | Hessian of the function $f$ |
| diag$(A)$ | diagonal of a matrix $A$ |
| Diag$(\boldsymbol{x})$ | diagonal matrix made up of the vector $\boldsymbol{x}$ |

## Vectors and Hilbert spaces

| | |
|---|---|
| $\boldsymbol{a}$, $\boldsymbol{b}$, ... | generic vector |
| $\lvert a\rangle$, $\lvert b\rangle$, ... | vector in a Hilbert space |
| $\langle \boldsymbol{a}, \boldsymbol{b}\rangle$ | scalar product between two vectors $\boldsymbol{a}$ and $\boldsymbol{b}$, sesquilinear in the first and linear in the second argument |
| $\langle a\vert b\rangle$ | scalar product between two vectors $\lvert a\rangle$ and $\lvert b\rangle$ in a Hilbert space, sesquilinear in the first and linear in the second argument |
| $^\top$ | transposition of a vector |
| $^\dagger$ | adjoint of a vector |
| $\lVert\bullet\rVert$ | $\ell_2$-norm of a vector or Frobenius norm of a matrix |
| $\otimes$ | tensor product |

**Miscellaneous**

| | |
|---|---|
| $\text{prob}(A)$ | probability of event $A$ |
| $\text{prob}(A \mid B)$ | conditional probability of event $A$ given $B$ |
| $\circ$ | composition of functions |
| $a \succeq_{\mathcal{C}} b$ | conic membership $(a - b) \in \mathcal{C}$ |
| $\boldsymbol{a} \succeq \boldsymbol{b}$ | conic membership w.r.t. the nonnegative cone (elementwise inequality) |
| $A \succeq B$ | conic membership w.r.t. the positive semidefinite cone |
| $A \succ B$ | conic membership w.r.t. the positive definite cone |
| $\bar{\bullet}$ | conjugation |
| $\lceil x \rceil$ | ceiling (smallest integer upper bounding $x$) |
| $\lfloor x \rfloor$ | floor (largest integer lower bounding $x$) |
| $\subset$ | subset relation (not necessarily proper) |
| $\delta_{i,j}$ | Kronecker Delta: $0$ for $i \neq j$, $1$ for $i = j$ |

# 1 Introduction

The field of Quantum Information Theory began as curiosity-driven research asking questions about whether and how well-known quantities from classical Shannon information theory [Sha48] can be translated into a realm in which the laws of quantum mechanics become relevant for the information carriers themselves. While this has been limited to theory only for years, the expertise in experiments and nowadays even commercial manufacturing has increased to a point where these questions actually become relevant for something that *can be realized in practice*.

Though the manipulation, transmission, and storage of quantum states continuously, or from time to time even dramatically, improves on the hardware side, it is still quite clear that a carefree execution of algorithms or protocols will suffer under too many errors to make them practically useful. In quantum computing, this is simply due to the fact that there are still orders of magnitude that separate the error range of classical devices—which define what would be considered "useful" today, after decades of evolution—from that of quantum computers. While quantum communication is easier than computing in that it does not require support for universal operations, it is at the same time harder, as the communication line between the two parties is more difficult to control than a computer, which is in one laboratory or room.

The culprit lies in the *carefree* execution: indeed, by adding suitable encodings and error correction procedures, these issues can be tackled. While the underlying concept is the same as in classical information theory—namely, exploiting redundancy to combat noise—the laws of quantum mechanics, explicitly spelled out in CHAPTER 2 of this thesis, now dictate which operations are feasible. This in particular means that the paradigmatic example of redundancy, copying a message multiple times, is no longer possible due to the no-cloning theorem[1] [WZ82]. Quantum error correction comes to the rescue—by encoding the logical states in a very particular manner into entangled physical states, information can be made more robust against a set of errors. Provided the amount of errors is below a certain threshold, scaling up the code (which of course requires a code *family* that is parameterized by a size) will suppress them to any desired level, in complete analogy with the classical case.

chapter 2

1 This of course only holds true if the quantum nature of the information actually is to be exploited. If one only wishes to use quantum carriers is a "classical" manner, i.e., with operations diagonal in one fixed basis and not building up any entanglement, their quantum nature does not prevent cloning.

Quantum error correction was invented in 1995 by Peter Shor [Sho95; KL97]; and soon after that, Daniel Gottesman developed the stabilizer formalism [Got96; Got97] that allows to conveniently describe a large class of error correction codes[2]. Recent years have seen the emergence of a number of classes of error correction codes that are inspired, e.g., by the geometry of the physical information carriers (surface codes [Fow+12] for a 2D array), the ground state of certain Hamiltonians (topological codes [LP17]), requirements on the abstract (Calderbank-Shor-Steane codes [CS96; Ste96]) or local (quantum low-density parity check codes [BE21]) structure of their stabilizers[3]. Despite this vast zoo of codes that are available by now[4], a fundamental question remains: given a certain set of interesting operations and a number of physical constraints due to the system in use, including more or less well-characterized errors, how can the operations be carried out in the best possible way?

While answering this question can be possible in certain scenarios—for example, if information-theoretical considerations lead to a capacity and one can find an encoding that matches this capacity—it would certainly be helpful if such an answer could be found in a systematic, even automatic, way, without the need to employ clever tricks and investing months of research. Given the vague specification of the problem, the most promising avenue to this is probably to rely on numerics. Numerical methods can cope with vastly different problems as long as they can be formulated in a concise mathematical language and are sufficiently regular, where the meaning of "regular" is dependent on the algorithm—but it refers to some mathematical properties such as continuity or differentiability which are likely to be fulfilled among a vast number of physical problems. However, while the question above can be formulated as an optimization problem and numerical algorithms exist to "solve" such problems, there are important considerations to be taken into account. One is related to resource consumption: how long and how much memory does it take? The second is about reliability: does the returned solution actually satisfy the constraints—this is usually fulfilled up to some error that can be customized by trading runtime—and is the solution in fact globally optimal, or did the algorithm get stuck in a local optimum? To answer these questions, it is important to know about the mathematics of optimization, not only from a purely theoretical perspective, but also how it is applied in popular algorithms.
chapter 3 Chapter 3 therefore presents an introduction to numerical optimization that is much more deep than what a physicist is usually exposed to, although it does not try to mimic the rigor of a mathematical lecture, deferring proofs to the literature. The knowledge imparted in this chapter should be helpful both for the selection of appropriate algorithms as well as in

2 In fact, the connection to error correction is omnipresent, but not necessary: the stabilizer formalism simply allows to describe certain states and their evolution under quantum operations, which is the basis for fast stabilizer-based Clifford simulators such as `STIM` [Gid21].

3 Note that these examples are not disjoint: surface codes are in all of the other three classes.

4 See `errorcorrectionzoo.org` for a curated and quite exhaustive list.

understanding the effects of tunable parameters, hopefully leading to better solutions in shorter time.

While optimization algorithms can usually only yield local solutions, this is not always true. In fact, in recent years, the use of numerical methods has gained a lot of popularity even in “hardcore” analytical proofs for information-theoretical properties of quantum systems. The reason can be found in the success of semidefinite programming in numerical optimization. Often, problems can be formulated in such a way that a linear objective is to be extremized under a set of linear constraints and membership in the cone of positive semidefinite matrices. Such a description is *convex*, with the implication that any local minimum is also global; and while in some cases, writing down the Lagrangian allows to find a solution even analytically, numerical methods exist that can find solutions to a user-defined and certifiable precision, and, importantly, requiring time and memory that are polynomial in both the number of variables and the precision. Therefore, it does not matter how intricate the constraints may look like—as long as the problem can be written down as a semidefinite program, it is essentially solved. As always, the use of “essentially” should raise some red flags; fortunately, here, it does not mean that the algorithms are galactic with unbearably high constants. However, it *does* mean that the scaling of popular algorithm classes with respect to the matrix side dimension is not quite as favorable as one would like—see Remark 3.41. Combined with the exponential growth of the state space in quantum mechanics, this often leads to practical issues—annoying on the time side and roadblocks as regards to memory—once the realm of toy models is left behind.

The other issue is the ability to cast the question of interest into a semidefinite programming problem. While with a lot of ingenuity many measures in quantum information have been written in this form, this is often quite difficult for practical tasks. A *task* naturally consists of multiple steps; at least one preparation of something, one operation, and a gathering of results, where all parts are to be determined by the optimization. This multi-stage process involves the multiplication of the objects that are involved[5], which in turn introduces a nonconvexity into the problem that prevents semidefinite programming methods from being applied directly. The structure of the problem is now *polynomial*—and it will remain so, regardless of how many operations are carried out after the other. With some notable exceptions, the field of polynomial optimization methods has not found much attention in quantum information until recently; only the last two or three years have seen a considerable increase in publications that try to apply these methods to the quantum domain. This is hardly surprising, considering that in solving polynomial optimization problems, semidefinite relaxations of potentially huge sizes are constructed, giving the impression that the rather spectacular hopes of solving a large and very relevant class of nonconvex

5 While it is formally correct to say that the steps can be combined into a single object, then, additional constraints have to be imposed on this object that in the end only shift the complexity, but usually do not alleviate it.

optimization problems to global optimality are practically unachievable. However, developments both on the side of numerical solvers as well as increasing sophistication in the tool kit of the polynomial optimization community shed doubt on this pessimism. Therefore, chapter 4 CHAPTER 4 shines a light on polynomial optimization and shows both well-known and recent developments. Important for practical applications, a list of current software is included.

## Contributions in this thesis

The interplay between the "solver" and the "solution method" aspect, i.e., viewing chapter 4 in the light of chapter 3, is particularly evident in section 4.9.3, which considers the development of a solver for polynomial optimization problems that is *inspired* by the semidefinite relaxations, but avoids their direct construction. The theoretical groundwork for this has been known for many years [Nes00; LP04] and developed into a fully-fledged solver in [PY19]. As original research in this thesis, I enhance these ideas further by asking which representation of the problem will actually lead to the most efficient numerical operations. This results in Theorem 4.72, giving a close-to-optimal performance; i.e., the overhead introduced in solving a relaxation by a general non-approximative second-order solver is reduced to almost the bare minimum of operations required, up to a logarithmic factor.

chapter 5 In CHAPTER 5, based in parts on my publication [DP22b], I apply the numerical optimization methods to the problem of state transmission through a lossy channel, which is the paramount source of errors in quantum communication scenarios. The methods used there are a mixture of convex and nonconvex approaches, relying on the theory of chapter 3. As a consequence, global optimality is not guaranteed, but by employing various methods, confidence can be built. Going beyond the original publication, the polynomial optimization methods introduced in chapter 4 are applied to indeed provide a certificate—up to a well-defined precision—for global optimality of smallish problem instances.

As a further addition, the symmetry assumption made in [DP22b] to reduce the size of the search space is lifted, and the resulting nonlinear and polynomial optimizations are performed. The bounds obtained by the latter again often allow to verify optimality of the data point (describing preparation and recovery) found by nonlinear optimization methods; however, sometimes, the bounds do not match these results. Examples are provided where in some cases, better initial points can be generated from the solutions underlying the higher bounds, helping the nonlinear optimizer to attain the global optimum. In other cases, extracting such solutions fails, and several characteristic examples round off this thesis, demonstrating both the promise, caveats and pitfalls offered by polynomial optimization methods.

Note that when trying to apply polynomial optimization methods even to the smallest of such problem instances, the resulting relaxations grow very quickly. Solving such large semidefinite programs is hard; but it turned out that even harder is feeding the problem to the solver! The way in which existing software frameworks for polynomial optimization were created meant that the task of building the relaxation from the original problem and telling the solver what to do was already too demanding in terms of memory[6]. As a main contribution of this thesis, I developed a new framework for commutative polynomial optimization, incorporating many recently published suggestions to reduce relaxation sizes—though due to their sheer magnitude only a subset—as well as providing efficient open-source implementations for new suggested solver concepts. Quantum information problems with complex values or semidefinite constraints can easily be entered. The main focus of this software were efficiency and usability, including extensive documentation; comprehensive unit testing ensures correctness in the light of the complexity of some algorithms. The considerable efforts made to streamline the algorithms indeed proves its worth: where other frameworks take minutes or hours, if they do not cease functioning due to memory issues, `PolynomialOptimization.jl` often takes only seconds.

6 This is the reason why some packages might, in new versions, see a completely redesigned internal representations, as has happened for example in `SOSTools` [Pap+21]. The bottleneck of problem formulation has also been realized by others during the writing of this thesis [GA24], though the software developed there has a different focus from mine, see section 4.10.1.

# 2 Basics in quantum information

This chapter will give a very brief introduction to the mathematical formalism of quantum mechanics (QM) and important definitions from quantum information (QI), focused on the essentials relevant for formulating and understanding the optimization tasks in the succeeding chapters. In particular, instead of following the classic textbook approach in which the formalism starts with idealized pure states or wave functions, the axiomatization will directly build on mixed states. The underlying concepts from linear algebra will be used without further reference; for an introduction, see [MS23; Ner70; Hog13].

The focus of this thesis will be on the finite-dimensional case, although the language in this chapter is chosen such that it is also suitable for infinite-dimensional QM; however, see [Thi81; Fac15] for a rigorous mathematical background or [Sch16b] for a light exposition.

The quantities in this chapter are introduced in a purely formal way, as this is all that is necessary to obtain computable and useful predictions; in this respect, whether they have an ontic nature or not is irrelevant. For an overview of interpretations of QM, see chapter 4 in [Sch16b] or the review [Dru19].

## 2.1 Axioms of quantum mechanics

**Axiom 2.1** (Quantum states)**.** The state $\hat{\varrho}$ of a *quantum mechanical system* is described by a positive semidefinite trace-class operator on some Hilbert space $\mathcal{H}$.

**Axiom 2.2** (Closed-system time evolution)**.** The time evolution of a quantum state $\hat{\varrho}$ (in natural units) is given by the Liouville–von Neumann equation:

$$\frac{\partial \hat{\varrho}}{\partial t} = -\mathrm{i}[\hat{H}, \hat{\varrho}], \tag{2.1}$$

where $\hat{H}$ is the *Hamiltonian*, the energy observable (see Axiom 2.5) of the closed quantum system. It is a self-adjoint operator and bounded from below.

**Theorem 2.3** (Unitary evolution)**.** The initial value problem ruled by the Liouville–von Neumann equation (2.1) has a unique solution given by conjugating the initial state

$\hat{\varrho}(0)$ with the element $U(t)$ of a strongly continuous one-parameter unitary group:

$$\hat{\varrho}(t) = U(t)\hat{\varrho}(0)U^\dagger(t). \tag{2.2}$$

$U$ is also called the *time-evolution operator* of the system.

*Proof.* This is Stone's theorem [Sto30; Neu96; Fac15]. □

**Remark 2.4** (Closed quantum system). The validity of equation (2.1) and the unitary time evolution hinges on the fact that the system be "closed," i.e., that there is no interaction with anything else.
Remarkably, the Liouville–von Neumann equation still remains valid even in a semiclassical treatment, where external (classical) fields impose some kind of outside influence. These may even be changing in time, which would then make the Hamiltonian time-dependent.

Apart from this semiclassical exception—which is already troublesome computationally—the preceding postulates do not allow for the convenient (and for all practical purposes unavoidable) separation between an *observer* (experimentalist) and the system of study. In fact, for the former to even know about the latter, the observer has to be *part* of the system: forbidding interactions between the system and anything else precludes this "anything else" to even know about the existence of the system. To bypass this problem, a third axiom allows to break the closed-system time evolution and introduces an interaction with an outside observer who can now obtain actual falsifiable information out of the purely theoretical concept of a "state."

**Axiom 2.5** (Observables, projective measurements, and the Born rule). A quantum mechanical observable is represented by a self-adjoint operator $\hat{M}$ on $\mathcal{H}$.
The possible outcomes of a (projective) measurement of this observable can be labeled by its eigenvalues $\lambda \in \operatorname{spec} \hat{M}$; the post-measurement state is given by conjugating the initial state $\hat{\varrho}$ with the projection operator $\mathbb{P}_{\hat{M}:\lambda}$ on the associated eigenspace:

$$\hat{\varrho} \mapsto \mathbb{P}_{\hat{M}:\lambda}\hat{\varrho}\,\mathbb{P}_{\hat{M}:\lambda}. \tag{2.3}$$

The probability that, upon measuring the state $\hat{\varrho}$ with the observable $\hat{M}$, the outcome $\lambda$ is obtained, is given by the *Born rule*,

$$\operatorname{prob}(\lambda \mid \hat{\varrho}) = \frac{\operatorname{tr}(\mathbb{P}_{\hat{M}:\lambda}\hat{\varrho})}{\operatorname{tr}\hat{\varrho}}. \tag{2.4}$$

**Remark 2.6** (Normalization)**.** By "measuring everything", i.e., $\hat{M} = \mathbb{1}$, the outcome probability is always $1$. The notation in equation (2.4) suggests to view the right-hand side as a conditional probability according to Bayes's rule, so that $\operatorname{tr}\hat{\varrho}$ is the probability for $\hat{\varrho}$ to "be there."
So on the one hand, it is often convenient to restrict the loose definition of quantum states according to Axiom 2.1 to normalized operators with $\operatorname{tr}\hat{\varrho} = 1$. This avoids the normalization in the Born rule, but requires renormalization of the post-measurement state in equation (2.3). On the other hand, only demanding that the *initial* state be normalized without changing the axioms allows to conveniently obtain the total probability after a series of measurements by just looking at the trace of the output state vector. Due to the following Proposition, this remains valid even when the system is subjected to intermediate time-evolution, and it is the convention that will be used in this thesis.

**Proposition 2.7** (Trace preservation)**.** The Liouville–von Neumann equation implies that the trace of the state $\operatorname{tr}\hat{\varrho}$ is preserved.

*Proof.* This is a direct consequence of Theorem 2.3:

$$\operatorname{tr}\hat{\varrho}(t) = \operatorname{tr}\big[U(t)\hat{\varrho}(0)U^\dagger(t)\big] = \operatorname{tr}\big[U^\dagger(t)U(t)\hat{\varrho}(0)\big] = \operatorname{tr}\hat{\varrho}(0), \tag{2.5}$$

as unitary operators are bounded, the product of a bounded and a trace-class operator is again trace-class, and therefore all traces exist.

Note that in the case of bounded Hamiltonian, this argument would have allowed to derive trace preservation even without the use of Stone's theorem, only exploiting the Liouville–von Neumann equation:

$$\frac{\partial}{\partial t}\operatorname{tr}\hat{\varrho} = \operatorname{tr}\frac{\partial}{\partial t}\hat{\varrho} = -\mathrm{i}\operatorname{tr}[\hat{H}, \hat{\varrho}] = 0, \tag{2.6}$$

as now the trace of the commutator exists and therefore vanishes. □

## 2.2 Open quantum systems

Although equation (2.1) allows for some degree of external control by time-dependent Hamiltonians, this is entirely classical. Usually, it is not possible to isolate—up to measurements—a quantum system well enough that the no-interaction policy mentioned in Remark 2.4 is obeyed; sometimes, such an isolation is not even desirable.

Clearly, the total system can always be chosen large enough that the criterion is satisfied, potentially at the expense of describing the whole universe. This total system follows a corresponding total Hamiltonian according to the Liouville–von Neumann equation.

It is then possible to single out a part of this total system, and perform measurements on the part alone. To describe such a situation, one more axiom is required.

**Axiom 2.8** (Tensor product decomposition)**.** The Hilbert space $\mathcal{H}$ of any quantum system can be partitioned into Hilbert spaces of countably many subsystems according to a tensor product structure:

$$\mathcal{H} = \mathcal{H}_1 \otimes \mathcal{H}_2 \otimes \dots . \tag{2.7}$$

Operators that only act on a subsystem are implicitly extended to the total system by taking tensor products with identity operators.

**Remark 2.9.** If one is willing to accept preparation independence—i.e., that it is possible to prepare either one of two quantum systems in a state independent of the other—then Axiom 2.8 is actually a consequence of the state postulate (Axiom 2.1) and the Born rule (Axiom 2.5) [CMA21]. This is quite a natural assumption, as it only means that it must be possible to "have system 1" and to "have system 2" at the same time, regardless of what those systems actually are.

**Definition 2.10** (Partial trace)**.** Let $\mathcal{H} = \mathcal{H}_1 \otimes \mathcal{H}_2$ be a tensor product of two Hilbert spaces. The partial trace with respect to the subsystem $j \in \{1, 2\}$ is given by

$$\operatorname{tr}_j \colon \mathcal{H} \to \mathcal{H}_{3-j}, \quad \operatorname{tr}_j \bullet := \sum_{i=1}^{\dim \mathcal{H}_j} {}_{\mathcal{H}_j}\langle i|\bullet|i\rangle_{\mathcal{H}_j}. \tag{2.8}$$

Here, $\{|i\rangle_{\mathcal{H}_j}\}_i$ is a (generalized[1]) basis of $\mathcal{H}_j$; according to Axiom 2.8, an element $|i_1, i_2\rangle_{\mathcal{H}}$ of the total basis of $\mathcal{H}$ can be multi-indexed and decomposed as

$$|i_1, i_2\rangle_{\mathcal{H}} = |i_1\rangle_{\mathcal{H}_1} \otimes |i_2\rangle_{\mathcal{H}_2} \ \forall i_j \in \{1, \dots, \dim \mathcal{H}_j\}. \tag{2.9}$$

As spaces in this thesis are finite-dimensional, the discrete notation will always be used, though $i_j$ may also be a continuous Hamel basis index in general.
The partial trace can be inductively extended to general multipartite partitions.

**Remark 2.11.** If $\mathcal{H} = \mathcal{H}_1 \otimes \mathcal{H}_2$, then naturally

$$\operatorname{tr} \equiv \operatorname{tr}_{1,2} \equiv \operatorname{tr}_1 \circ \operatorname{tr}_2 \equiv \operatorname{tr}_2 \circ \operatorname{tr}_1 \equiv (\operatorname{tr}_1 \otimes \operatorname{tr}_2) \tag{2.10}$$

and the partial trace is invariant with respect to cyclic permutation within the traced-out subspace only.

1 For infinite-dimensional spaces, the basis may also be chosen from the Fréchet embedding of the Hilbert space.

**Definition 2.12** (Reduced quantum states)**.** Let $\hat{\varrho}$ be a quantum system on $\mathcal{H} = \mathcal{H}_1 \otimes \mathcal{H}_2$. The (reduced) quantum state on $\mathcal{H}_1$ is defined as

$$\hat{\varrho}_1 = \operatorname{tr}_2 \hat{\varrho} \tag{2.11a}$$

and similarly,

$$\hat{\varrho}_2 = \operatorname{tr}_1 \hat{\varrho}. \tag{2.11b}$$

Indeed, Definition 2.12 is a meaningful replacement of Axiom 2.1 in the multipartite case due to the following Theorem, which takes the place of Axiom 2.5.

**Theorem 2.13** (Projective measurements and the Born rule for reduced states)**.** Let $\mathcal{H} = \mathcal{H}_1 \otimes \mathcal{H}_2$ and let, without loss of generality, $\hat{\varrho}_1$ be the reduced quantum state of $\hat{\varrho}$ on $\mathcal{H}_1$. Finally, let $\hat{M}_1$ be an observable on $\mathcal{H}_1$.
The possible outcomes of a (projective) measurement of this observable can be labeled by its eigenvalues $\{\lambda\}$; the reduced post-measurement state is given by conjugating the initial reduced state $\hat{\varrho}_1$ with the projection operator $\mathbb{P}_{\hat{M}_1:\lambda}$ on the associated eigenspace:

$$\hat{\varrho}_1 \mapsto \mathbb{P}_{\hat{M}_1:\lambda} \hat{\varrho}_1 \mathbb{P}_{\hat{M}_1:\lambda}. \tag{2.12}$$

The probability that, upon measuring the reduced state $\hat{\varrho}_1$ with the observable $\hat{M}_1$, the outcome $\lambda$ is obtained, is given by

$$\operatorname{prob}(\lambda \mid \hat{\varrho}_1) = \frac{\operatorname{tr}(\mathbb{P}_{\hat{M}_1:\lambda} \hat{\varrho}_1)}{\operatorname{tr} \hat{\varrho}_1}. \tag{2.13}$$

*Proof.* 1. According to Axiom 2.8, the reduced observable $\hat{M}_1$ on $\mathcal{H}_1$ corresponds to the full observable $\hat{M} = \hat{M}_1 \otimes \mathbb{1}_2$ on the total system. But taking a tensor product with the identity operator leaves the spectra unchanged; the eigenvalue multiplicities are merely multiplied by the dimension of $\mathcal{H}_2$. Therefore, the outcomes of the (projective) measurement of $\hat{M}_1$ can indeed be labeled by the eigenvalues of $\hat{M}_1$.

2. The partial post-measurement state can be obtained by applying the partial trace to the total post-measurement state according to equation (2.3):

$$\hat{\varrho}_1 \mapsto \operatorname{tr}_2(\mathbb{P}_{\hat{M}:\lambda} \hat{\varrho} \mathbb{P}_{\hat{M}:\lambda}) = \operatorname{tr}_2\big((\mathbb{P}_{\hat{M}_1:\lambda} \otimes \mathbb{1}_2) \hat{\varrho} (\mathbb{P}_{\hat{M}_1:\lambda} \otimes \mathbb{1}_2)\big) = \mathbb{P}_{\hat{M}_1:\lambda} \hat{\varrho}_1 \mathbb{P}_{\hat{M}_1:\lambda}. \tag{2.14}$$

3. To obtain the outcome probabilities, apply the Born rule, equation (2.4), on the total state and use Remark 2.11 and linearity:

$$\operatorname{prob}(\lambda \mid \hat{\varrho}) = \frac{\operatorname{tr}\big((\hat{M}_1 \otimes \mathbb{1}_2)\hat{\varrho}\big)}{\operatorname{tr} \hat{\varrho}} = \frac{(\operatorname{tr}_1 \otimes \operatorname{tr}_2)\big((\hat{M}_1 \otimes \mathbb{1}_2)\hat{\varrho}\big)}{(\operatorname{tr}_1 \otimes \operatorname{tr}_2)(\hat{\varrho})} = \frac{\operatorname{tr}(\hat{M}_1 \hat{\varrho}_1)}{\operatorname{tr} \hat{\varrho}_1}. \tag{2.15}$$

□

While Axioms 2.1 and 2.5 were easy to translate to the reduced point of view, the Liouville–von Neumann has no simple equivalent. If there were a time evolution operator that acted on the reduced state alone, it would have to fulfill certain properties:

**Theorem 2.14** (Properties of a quantum time evolution operator)**.** Let $\hat{\varrho}$ be a quantum state on $\mathcal{H} = \mathcal{H}_1 \otimes \mathcal{H}_2$. A valid time evolution operator $\Lambda(t)[\hat{\varrho}_1]$ should satisfy the following properties for any valid quantum state $\hat{\varrho}_1(0)$ and any $t$:

1. $\Lambda(t)$ must be linear.
2. $\operatorname{tr} \Lambda(t)[\hat{\varrho}_1(0)] = \operatorname{tr} \hat{\varrho}_1(0)$. *"trace-preserving"*
3. $\Lambda(t)[\hat{\varrho}_1(0)]$ must be self-adjoint. *"hermiticity-preserving"*
4. $\Lambda(t)[\hat{\varrho}_1(0)]$ must be positive semidefinite. *"positivity-preserving"*
5. $(\Lambda(t) \otimes \mathbb{1}_{\text{aux}})[\hat{\sigma}]$ must be positive semidefinite for all quantum states $\hat{\sigma}$ on any Hilbert spaces $\mathcal{H}_1 \otimes \mathcal{H}_{\text{aux}}$ with $\dim \mathcal{H}_{\text{aux}} \leq \dim \mathcal{H}_1$. *"completely positive"*

Any operator that satisfies these properties is called a *quantum map*, with the special attribute *dynamical map* typically reserved for the map that describes the time evolution.

*Proof.* For the first and second property, note that the time evolution on a subsystem necessarily arises as a part of the time evolution on the total system, Theorem 2.3:

$$\Lambda(t)[\hat{\varrho}_1(0)] = \operatorname{tr}_2(U(t)\hat{\varrho}(0)U^\dagger(t));$$

so linearity follows from linearity of unitary conjugation and the partial trace; and trace preservation follows because unitaries and the partial trace leave the total trace invariant.

The third and fourth property are immediately evident from the definition of a quantum state in Axiom 2.1.

For the fifth property, a physical argument rather than a mathematical one is required. Let $\mathcal{H}$ be the whole universe, which is always the ultimate choice for a closed quantum system; even if $\mathcal{H}_2$ were chosen in a smaller way, it can be seamlessly embedded. Clearly, there are various possible partitions $\mathcal{H} = (\mathcal{H}_1 \otimes \mathcal{H}_{\text{aux}}) \otimes \mathcal{H}_{\text{rest}}$. Now considering $\mathcal{H}_1 \otimes \mathcal{H}_{\text{aux}}$ as a single open system, the fourth property must hold for this system, i.e., the reduced state on the system itself must be positive semidefinite at all times. This must hold for all possible dynamical maps on the open system that can arise from the universe's time evolution by taking all the different bisections between $\mathcal{H}_{\text{aux}}$ and $\mathcal{H}_{\text{rest}}$. Ockham's razor certainly advocates that these can be anything; and if this is the case, then Choi showed [Cho75] that it is indeed enough to only check for the identity map acting on systems of a dimension smaller or equal to $\dim \mathcal{H}_1$. □

**Remark 2.15** (Complete positivity). Trying to simplify the long-winding argument for complete positivity, it may colloquially be phrased like this: Whether or not another system $\mathcal{H}_2$ is present somewhere else should not affect the validity of the description of $\mathcal{H}_1$, in particular, if nothing is ever done with $\mathcal{H}_2$; it could as well be on the other side of the universe[2].
Complete positivity is physically quite meaningful, but more importantly, it simplifies the mathematical theory of quantum dynamical maps considerably (see Theorem 2.18), and therefore is also much more convenient than "just" positivity-preservation.

Note that in the sense of a phenomenological system description, one might decide to dispense with any of the criteria as long as they allow to fit a time evolution operator "mostly" better to experimental data, at the expense of having an unphysical description where it "does not hurt too much." However, quantifying this is very hard; therefore, the theoretically sound first four criteria and the convenient and convincing criterion five are usually required of a proper quantum map.

## 2.3 Quantum operations

Theorem 2.14 already hinted at the presence of quantum maps that do not necessarily describe the dynamics of a certain system. Strictly speaking, a quantum map ought to change a physical state into another physical state—so if this is supposed to actually happen, there has to be a dynamics that effectively describes exactly this map. In this sense, there is no quantum "non-dynamical" map.

However, conceptually, it is very convenient to separate the action of what something is *effectively doing* from the dynamical process of *actually doing* the task. In this input/output formalism, the quantum map simply means shining a flashlight on the evolution of a state only at certain points in time. Everything that happens between these points is then a quantum operation; time no longer plays a role. In fact, the very same quantum operation may be realized on a multitude of different physical systems with totally different Hamiltonians, which still *effectively* all give rise to the same operation; and conversely, if the desired action can be shown not to yield a valid quantum operation, then no Hamiltonian will ever be able to produce it.

The subject of quantum information asks what can happen to information (in the classical Shannon sense [Sha48]) that is encoded in some quantum systems on which operations can be carried out subsequently. As with classical information, where the physical bit is abstracted away, the same can be done now: the physical quantum system itself or the classical information that was used to produce the system can be disregarded, yielding

2 While this sounds very plausible, a superdeterministic mindset might reject that the rest of the universe can be anything.

the *qubit* as the fundamental unit of quantum information. The measurement process at the end that closes the cycle (preparation according to classical information, quantum manipulation, obtaining classical information by measurement) can be disregarded similarly to the preparation—for fundamental considerations, there is no fixed measurement, but only the possibility to carry out any one. For a detailed account on quantum information, see [Wil19].

This makes the quantum map—an object that satisfies the properties in Theorem 2.14, but may also map between different spaces by incorporating other systems or tracing them out—the focus of interest, as all manipulations of information in the end are realized by means of quantum maps. Complete positivity then allows for a very concise mathematical description of quantum maps, for which one additional tool is required.

**Definition 2.16** (Choi-Jamiołkowski isomorphism [Cho72; Jam72]). Given a linear map $M\colon \mathcal{L}(\mathcal{H}_1) \to \mathcal{L}(\mathcal{H}_2)$, its associated Choi state $\mathscr{C}[M] \in \mathcal{L}(\mathcal{H}_2 \otimes \mathcal{H}_1)$ is given by

$$\mathscr{C}[M] := \sum_{i,j=1}^{\dim \mathcal{H}_1} M[|i\rangle_{\mathcal{H}_1}\langle j|] \otimes |i\rangle_{\mathcal{H}_1}\langle j|. \tag{2.16}$$

**Remark 2.17** (Choi state indices). The map $M$ can be interpreted as a rank-four tensor with the indices `output row`, `output column`, `input row`, and `input column`. The Choi state $\mathscr{C}[M]$ is typically seen as the square matrix that arises from $M$ by grouping together the `output row` and `input row` into one row index and the `output column` and `input column` into one column index.

**Theorem 2.18** (Representations of completely positive maps). For any linear, hermiticity-preserving operator $\Lambda\colon \mathcal{T}(\mathcal{H}_1) \to \mathcal{T}(\mathcal{H}_2)$, the following statements are equivalent:

1. $\Lambda$ is completely positive.
2. $\Lambda$ admits a Kraus decomposition of the form

$$\Lambda[\bullet] = \sum_i K_i \bullet K_i^\dagger, \tag{2.17}$$

   where $K_i \in \mathcal{L}(\mathcal{H}_1 \to \mathcal{H}_2)$. $\Lambda$ is trace-preserving if and only if

$$\sum_i K_i^\dagger K_i = \mathbb{1}_{\mathcal{H}_1}. \tag{2.18}$$

3. There exists a Stinespring dilation of $\Lambda$, i.e., a Hilbert space $\mathcal{H} \simeq \mathcal{H}_1 \otimes \mathcal{H}_3 \simeq \mathcal{H}_2 \otimes \mathcal{H}_4$ (where the subdivisions may be partially overlapping), a subunitary $U$ on $\mathcal{H}$ and an initial quantum state $\hat{\sigma} \in \mathcal{T}(\mathcal{H}_3)$ such that for any quantum state $\hat{\varrho}$,

$$\Lambda[\hat{\varrho}] = \operatorname{tr}_4\bigl(U(\hat{\varrho} \otimes \hat{\sigma})U^\dagger\bigr). \tag{2.19}$$

$\Lambda$ is trace-preserving if and only if $U$ is unitary.

4. The Choi state of $\Lambda$ is positive semidefinite. $\Lambda$ is trace-preserving if and only if tracing over the output system in the tensor interpretation of $\Lambda$ gives the identity on $\mathscr{H}_1$.

*Proof.* $1 \Leftrightarrow 2$ was established by Kraus in [Kra83]; $1 \Leftrightarrow 3$ is Stinespring's theorem [Sti55], and $1 \Leftrightarrow 4$ is Choi's theorem [Cho75]. The trace-preserving property follows from immediate calculations. □

## 2.4 Optimization problems in quantum information

To conclude this introduction, this section will motivate why it is worth developing sophisticated numerical optimization algorithms and software for QI problems. Naturally, this only resembles a tiny fraction of the power of good optimization.

### 2.4.1 Resource theories and measures

Quantum information theory wants to answer questions about quantum state transformations and the transfer of classical and quantum correlations within systems and between parties to achieve certain tasks. The framework of resource theories [Gou24; CFS16; HO13] provides a very pragmatic, yet accurate view: Given a well-defined set of states and operations that are easy to produce ("free") and others that are resourceful, this allows to define a "currency" of the resource in question—for example, this may be entanglement [Hor+09; PV07], coherence [SAP17], athermality [Bra+13], or asymmetry/frameness [GS08; Vac+08; GMS09]. It is then possible to quantify via measures how costly or valuable a certain operation is, how much of the given currency can be obtained by sacrificing a certain number of states (distillation), or how much of the currency has to be invested to produce a certain state (cost).

While resource theories and measures are not the focus of this thesis, by writing down their general formulation, one fact becomes evident. Let $\hat{\sigma}$ be the currency state of the resource theory—e.g., for entanglement theory, this would be the maximally entangled two-qubit state $\hat{\sigma} = \frac{1}{2}\mathbb{P}[|00\rangle + |11\rangle]$—then [PV07]

$$m_{\text{dist}}(\hat{\varrho} \to \hat{\sigma}) = \sup_r \Bigl\{ r : \lim_{n\to\infty} \inf_{\Lambda \text{ free}} \bigl\| \Lambda[\hat{\varrho}^{\otimes n}] - \hat{\sigma}^{\otimes \lfloor rn \rfloor} \bigr\| = 0 \Bigr\} \tag{2.20a}$$

$$m_{\text{cost}}(\hat{\sigma} \to \hat{\varrho}) = \inf_r \Bigl\{ r : \lim_{n\to\infty} \inf_{\Lambda \text{ free}} \bigl\| \Lambda[\hat{\sigma}^{\otimes \lfloor rn \rfloor}] - \hat{\varrho}^{\otimes n} \bigr\| = 0 \Bigr\}. \tag{2.20b}$$

Both measures—and these are only particularly relevant examples of resource-theoretic measures—can be stated and understood relatively easily in terms of optimization problems. *Performing* these optimizations is certainly far from trivial: they involve the limit of infinite copies (the *regularization*), which complicates calculations a lot unless the measure itself is additive, they contain a nested optimization over the set of all free operations, which may be very hard to describe, and they require some norm with an operational meaning in the resource theory.

Even numerically, these challenges cannot be overcome easily. For sure, for non-additive measures, the limit of infinite copies must be dealt with somehow, either[3] by deriving an analytical bound with a simple scaling in the number of copies together with a remainder that is still to be optimized; or by studying the behavior for fixed values of $n$. The set of free operations may or may not be amenable to some computable characterization; it may, if necessary, also be approximated by a smaller (giving a bound) or larger (giving an intuition) set. Finally, the norm itself might have an analytic expression or be the result of yet another optimization problem. Once a parameterization has been found that expresses membership in the set of free operations in terms of a number of equalities and inequalities, the calculation of the measure, an approximation or bound, is up to the numerical optimizer. Depending on the underlying properties of the parameterization, the optimization might work very well and even be able to give a verifiably correct result; or it might be completely stuck, abort due to stalling, or give some allegedly optimal result without any quantifiers of its quality.

Semidefinite programming [VB96; BN01] has established itself in the last decade as an extremely powerful tool to solve QI problems [SC23]. If it is possible to write the optimization in this particular format, a globally and certifiably optimal[4] result can be obtained in a time that scales only polynomially with the number of variables. Due to their global nature, semidefinite programs can be employed even in theoretical proofs of fundamental results.

However, few problems naturally come up in semidefinite form; some can be converted, others relaxed. Still, these conversions or relaxations often come with a significant increase in the number of variables—and even if "polynomial scaling" is often equated with "good enough," optimizers still hit bounds on computational feasibility. Developing and improving global optimization algorithms therefore is of paramount importance to the field of QI.

3 A very recent result [LMR25] was even able to reduce the specific case of the entanglement cost with respect to free operations that preserve the positive partial transpose of a state—an important subclass of local operations and classical communication—to a hierarchy of semidefinite programs with exponentially fast convergence, showing for the first time that the curse of regularization can be tamed in some cases.

4 Given that this is a numerical method, "certifiably optimal" always means that there are precise bounds available on the error; naturally, the optimization must also scale well with respect to the desired error, typically $\mathcal{O}(\log(\varepsilon))$ or at most $\mathcal{O}(\mathrm{poly}(\varepsilon^{-1}))$ if $\varepsilon$ is the maximum difference allowed between the reported and the actual optimal value.

### 2.4.2 Quantum computation and communication

When doing computations in classical computers, the question of whether the hardware will actually do what it is supposed to do rarely arises. While there certainly are known bugs and defects in CPUs, RAMs, or other hardware (see, e.g., [Goo24]), they are the result of errors in the design or production process, and very rarely individual faulty devices. Overall, classical devices have an extremely low error rate: in 2011, the logic error rate was reported to be $10^{-27}$ per gate operation [Szk+11], and in recent years, the International Technology Roadmap for Devices and Systems [IRD23] does not even focus on error rates any more. In fact, cosmic rays, solar storms, and background radiation from earth sources are major sources of errors—giving rise to the so-called "soft error rate"—in these devices [Aut+12; ITR11; Zie+96].

While miniaturization raises this error rate again [MV16], requiring active error correction for the memory in critical applications or data centers [Mez18], the world of classical computing is still blessed.

The same holds true for classical communication: Here, photons are the carriers of information; however, situations in which one photon carries one bit of information are highly experimental proofs-of-concept, which suffer under the trade-off between sensitivity and bandwidth [Cap07] that limits them to megabits-per-second rates [Bor+14; Hop+06; Gre+15; Rob+06]—although recently, some progress has been made using phase-sensitive optical amplifiers [KSA20]. To accurately process photons, usually at least two- to three-digit numbers of photons per bit are required; and to transmit these over larger distances, the signal has to be amplified in order to make sure that enough photons still arrive. Signal amplification is not a problem in classical physics—the state that is encoded by the photon is known exactly, either it encodes a zero or a one, so more photons of this type can be produced easily. Remaining errors can be dealt with using error correction.

The situation for computation and communication changes dramatically when going to the quantum regime. There are no longer just two distinct logical states that are represented by continuous physical quantities; instead, quantum computation and communication relies on manipulating precisely the continuous range of physical states. Cloning these states to build in redundancy is either physically impossible in case they are unknown [WZ82] or impractical if they are fully known, but heavily entangled, precluding naive error correction.

However, cloning is just a very simplistic instance of error correction. When wanting to account for classical bit errors or losses, the information simply has to be spread out among multiple carriers in such a way that knowledge of an incomplete set is enough to restore the full information. The same is possible for quantum states [Sho95]: a (physical) quantum state is a ray in a linear space. If a *logical* state is now no longer associated with a single ray, but a set of them—due to linearity, a whole subspace—then physical errors that convert one physical state into another *in the same subspace* do not correspond logical errors—the state was protected. This simple argument shows that it is indeed possible to perform operations on quantum states in a way that tolerates errors; but how to find a way to actually do so?

Given a certain task that is to be implemented in some way on a quantum hardware—this corresponds to a quantum map $M$—and a certain physical system that should be used to carry out the task, it is important to have a detailed knowledge about this system: which errors will occur, and how likely are they? Next, the physical system will impose some restrictions in that only a certain set of operations are allowed at various steps in the process—these are experimental, budgetary, or even fundamental physical constraints. Given one particular chain of operations that is to be carried out in a certain way, the knowledge about errors can be effectively captured by a stochastic modification of the intermediate states according to the list of errors—this will result in the application of quantum (error) maps. Finally, an end result occurs, which could be a quantum state, a classical measurement outcome, or both. Now taking a distance measure between the actual and the desired outcome tells how well the chosen chain of operations realized the particular task; naturally, this distance should be as small as possible.

The description in the previous paragraph can be formalized in the following way:

$$\inf_{\substack{0\preceq\hat{\varrho}_{\mathrm{in}}\in\mathcal{T},\\ \mathrm{tr}\,\hat{\varrho}_{\mathrm{in}}=1,\\ \Lambda_1\times\cdots\times\cdots\Lambda_n\in\mathcal{O}}} \left\|(E_n(\Lambda_n)\circ\Lambda_n\circ\cdots\circ E_1(\Lambda_1)\circ\Lambda_1\circ E_0(\hat{\varrho}_{\mathrm{in}}))[\hat{\varrho}_{\mathrm{in}}]-\hat{\varrho}_{\mathrm{des}}\right\|. \tag{2.21}$$

This optimization program finds an optimal input state $\hat{\varrho}_{\mathrm{in}}$ together with a sequence of $n$ quantum operations out of a set $\mathcal{O}$ of permissible operations dictated by experimental constraints, and minimizes the distance to the desirable state $\hat{\varrho}_{\mathrm{des}}$ after applying all those quantum operations—always followed by an error map $E$ (that might potentially also depend on the choice of the last applied map).

When reducing the scenario to a simple scheme with a preparation (encoding), one error channel, and a single processing (recovery) step—both maps only restricted to be valid quantum maps—this is a standard scheme in quantum computation and communication; but even then, viewing this as a numerical optimization problem for both encoding and recovery is rarely done (see [RW05] for one such approach, and [FSW07] for the, with current knowledge, rather trivial optimization over the recovery channel alone). Note that a very recent result [Zhe+24] suggests that the single-stage case can be approximated quite well by an easily-computable quantity, the near-optimal channel fidelity, making the problem more tractable. Including multiple stages of errors and operations into the process resembles a quantum comb [CDP08], which provides a nice visualization, but is of no real help when carrying out the optimization. Additionally, the comb itself—the joint superoperator formed by all the error operators—should not depend on its arguments.

So even when looking at quantum errors from an optimization perspective, equation (2.21) is a much more detailed and realistic description than is usually done. This is even more true for the "orthodox" way of constructing quantum error correction codes, independent of any task that is to be achieved. Biased Pauli noise models are the closest as regards to hand-tailoring them to a particular situation—and while there are quite a number of error correcting codes that are inspired by geometry, they are then often applied to situations in which the original geometry is only marginally present. This generality is advantageous in that solving something akin to equation (2.21) is extremely challenging. However, it comes with a cost, even more, with an unknown cost: it is completely unclear how far from optimal in terms of resource consumption the given code is, and how much more performance could be teased out of a given hardware platform.

In fact, if it were possible to solve equation (2.21), it is not a huge step to next ask the question about sensitivity with respect to the error operators and find the part whose improvement (i.e., tuning the hardware) can give the most gain.

Again, this example shows that solving intricate optimization problems would greatly benefit the development of protocols to implement quantum algorithms or to exchange quantum information.

# 3 Numerical optimization

Optimization problems are ubiquitous in QI; section 2.4 only gave an insight into a small number of examples. In this chapter, the foundations of optimization problems and numerical approaches to finding a solution are laid out. It will start with a general overview of important characteristics that typical optimization problems in QI have. Subsequent sections will introduce methods to locally solve unconstrained (section 3.3) and constrained (section 3.4) problems, followed by a discussion of relevant global approaches (section 3.5). The special class of polynomial optimization problems plays a major role in this thesis and will be the subject of chapter 4. However, it will fundamentally depend on the theory presented in this chapter.

## 3.1 Problem and solution requirements

Typical QI optimization problems seek to minimize or maximize one objective that depends on numerous decision variables and a number of constraints. Objective and constraints are smooth, and typically even infinitely differentiable. The feasible set, i.e., the set of points that satisfies the constraints, is compact, but cannot be parameterized easily. In traditional problems, the objective and constraints are linear or semidefinite, but in more complicated cases, they are nonlinear. The decision variables are usually real- or complex-valued; but problems such as circuit optimization may also involve integer variables. It is often useful to also keep external parameters (such as error rates) in the problem.

single objective
multivariate
constrained
differentiable
compact
nonlinear/partially convex
mixed-integer
parametric

To be usable in fundamental results, the solution should be global with exactly verifiable error bounds. For real-world applications, it needs to be robust against small variations in the problem parameters. The solution should be found at most in polynomial resources (time, memory) with small exponents and prefactors. The gradients or Hessians can be provided, even though they are often expensive to calculate.

global
robust
$\mathcal{P}$-time/-memory
second-order possible

For practicality, given a theoretical formulation of the problem, it should not require a lot of effort to obtain the solution without detailed knowledge in optimization or specific case-by-case analysis. Furthermore, if multiple global solutions exist—probably infinitely many—it would be helpful to get a particularly “simple” instance of such a solution. This might be defined in terms of sparsity, low rank, absence of imaginary parts, or some other criteria—which is too vague to make it a fixed criterion, but keeping this in mind may inspire solver algorithms that aim at providing such solutions.

easy to use
simple solution

## 3.2 General form of optimization problems

In order to apply the machinery of numerical optimization to physical problems, they first have to be described formally. In this thesis, the most generic form that will be considered is given by

$$\inf_{\boldsymbol{x}\in\mathcal{X}} \{f(\boldsymbol{x}) : g_i(\boldsymbol{x}) \leq 0 \ \forall i \in \mathcal{I}, h_j(\boldsymbol{x}) = 0 \ \forall j \in \mathcal{E}\} \tag{3.1}$$

with some index sets $\mathcal{I}$ and $\mathcal{E}$. The objective $f\colon \mathcal{X} \to [-\infty, \infty]$ is—without loss of generality—to be minimized over some set $\mathcal{X} \subset \mathbb{F}^n$, $\mathbb{F} \in \{\mathbb{R}, \mathbb{C}\}$. Additionally, some constraint functions $\{g_i\}_{i\in\mathcal{I}}$ have to be nonpositive, while other functions $\{h_j\}_{j\in\mathcal{E}}$ must be exactly zero. All points in $\mathcal{X}$ that also satisfy the explicit constraints comprise the *feasible set*[1].

The task of finding a solution is carried out by *solvers*, which are very specialized programs that require a lot of additional restrictions on all the quantities involved in equation (3.1). As addressing solvers directly is often cumbersome and error-prone, optimization *frameworks* allow for much more flexibility in the problem formulation and perform the task of rewriting everything so that it fits whatever format the solver expects.

While it is in some sense unnecessary to have both an explicit set $\mathcal{X}$ different from the reals[2] *and* constraint functions[3], it is quite natural to allow for both. Typically, solvers are designed to work particularly well with constraints of a certain type. These should then be specified in the set $\mathcal{X}$ in such a way that the solver understands the *nature* of the constraint:

**Example 3.1** (Equivalent descriptions). Set

$$\mathcal{X} = \left\{\boldsymbol{x} \in \mathbb{R}^4 : x_1 \geq 0, x_2 \leq 0, x_1 + 3x_3 + 5x_4 = 7, \begin{pmatrix} x_1 & x_2 & x_3 \\ x_2 & x_4 & x_2 \\ x_3 & x_2 & x_1 \end{pmatrix} \succeq 0\right\}. \tag{3.2}$$

This is an intersection of half-spaces with a linear and a positive semidefinite constraint—a *semidefinite program (SDP)*—for which very specialized algorithms exist. Using the characterization of the positive semidefinite cone in terms of alternating signs of the coefficients of the characteristic polynomial [HJ12, corollary 7.2.4], all constraints could

1 In equation (3.1), inf was written instead of min. The subtleties of whether the feasible set is open or closed are beyond the grasp of inexact numerics; however it is quite relevant to find out whether the problem is unbounded or infeasible, and this is something that numerical algorithms can do in important cases. There might even be cases in which the objective will be constant; then, equation (3.1) is a *feasibility problem*, checking only whether the feasible set is empty.

2 Complex numbers are a necessity for QI problems, but solvers rarely support them directly, so they have to be emulated in some way. Real numbers are not supported either, but usually approximated in floating-point arithmetic. Special algebraic solvers might also work with rationals or algebraic numbers.

3 The technicality that restrictions in the range of definition of the functions should be enforced via $\mathcal{X}$ is not important in practice. The objective may always be rewritten such that it is infinite where it was not defined previously; analogous rewritings can be done for the constraints. Without loss of generality, all involved functions can therefore be assumed to be defined everywhere.

also be formulated by instead setting $\mathcal{X} = \mathbb{R}^4$ and defining

$$\begin{aligned} g_1(\boldsymbol{x}) &= -x_1; & g_4(\boldsymbol{x}) &= -x_1^2 + 2x_2^2 + x_3^2 - 2x_1x_4; \\ g_2(\boldsymbol{x}) &= x_2; & g_5(\boldsymbol{x}) &= -(x_1 - x_3)((x_1 + x_3)x_4 - 2x_2^2); \\ g_3(\boldsymbol{x}) &= -2x_1 - x_4; & h_1(\boldsymbol{x}) &= x_1 + 3x_3 + 5x_4 - 7. \end{aligned} \tag{3.3}$$

While this problem is an exact translation of the set $\mathcal{X}$ to explicit constraints, from a numerical perspective, it is completely different. The functions $g$ and $h$ in equation (3.2) are typically specified as black-box callbacks that calculate the output given $\boldsymbol{x}$. There is no information about particular properties that the constraints induce; and indeed, $g_3$, $g_4$, and $g_5$ are polynomials, all of them unbounded, partly with complicated individual solution sets. However, taken together, they contain hidden convexity and actually allow to make global statements.

In order to solve equation (3.1), a typical numerical solver will start with an initial point $\boldsymbol{x}_0$ and iteratively try to improve it. Solvers usually require an interior initial point, i.e., $\boldsymbol{x}_0 \in \mathring{\mathcal{X}}$, or even an interior feasible point, i.e., one that additionally satisfies the explicit equality and inequality constraints strictly. To find an interior feasible point when only an interior point is known is not harder than solving the original problem: this is called a *Phase-I* method; for more details, see [BV04, section 11.4.1].

## 3.3 Unconstrained local optimization

Even though equation (3.1) contains most interesting problems, it is too general to be solved directly. In this section, only optimization problems with a twice continuously differentiable objective and with no constraints—i.e., $\mathcal{X} = \mathbb{F}^n$, $\mathcal{I} = \mathcal{E} = \emptyset$—will be considered. While in particular the latter restriction makes the object of study far less interesting, solvers for constrained problems will internally convert the desired full problem into a sequence of unconstrained ones and solve these; so understanding unconstrained optimization is vital for everything that is of more direct interest. More background information and proofs can be found in [NW06].

The minimization can now be rewritten as

$$\text{find } \boldsymbol{x}^\star \text{ such that } f(\boldsymbol{x}^\star) \leq f(\boldsymbol{x}) \ \forall \boldsymbol{x} \in \mathbb{F}^n, \tag{3.4}$$

which is the formal definition of a global minimum. Without particular knowledge about the problem, solvers can often only find a local minimum:

$$\text{find } \boldsymbol{x}^\star \text{ such that } f(\boldsymbol{x}^\star) \leq f(\boldsymbol{x}) \ \forall \boldsymbol{x} \in \mathcal{N}(\boldsymbol{x}^\star), \tag{3.5}$$

where $\mathcal{N}$ is a neighborhood. Local minima of differentiable functions are much easier to find because they have to satisfy criteria that can be directly verified using derivative information.

**Theorem 3.2** (Necessary criteria for local minimizers)**.** Any local minimum $\boldsymbol{x}^\star$ of $f$ has to satisfy $\boldsymbol{\nabla} f(\boldsymbol{x}^\star) = \boldsymbol{0}$ and $\boldsymbol{\nabla}^2 f(\boldsymbol{x}^\star) \succeq 0$.

**Theorem 3.3** (Sufficient criterion for local minimizers)**.** If $\boldsymbol{\nabla} f(\boldsymbol{x}^\star) = \boldsymbol{0}$ and additionally $\boldsymbol{\nabla}^2 f(\boldsymbol{x}^\star) \succ 0$, then $\boldsymbol{x}^\star$ is a local minimum of $f$.

However, there is a particular class of functions that make global optimization no harder than local optimization:

**Definition 3.4** (Convex sets and functions)**.** A set $\mathcal{S}$ is called *convex* if any line segment connecting two points in $\mathcal{S}$ lies entirely in $\mathcal{S}$:

$$\{\alpha\boldsymbol{x} + (1-\alpha)\boldsymbol{y} : \boldsymbol{x}, \boldsymbol{y} \in \mathcal{S}, \alpha \in [0,1]\} \subset \mathcal{S}. \tag{3.6}$$

Equivalently, $\mathcal{S}$ is the intersection of a (possibly infinite) number of half-spaces; or, every tangent to $\mathcal{S}$ never intersects $\mathring{\mathcal{S}}$.
A function $f\colon \mathcal{S} \to \mathbb{R}$ is called *convex* if its domain $\mathcal{S}$ is a convex set and every secant of $f$ stays above the function graph, i.e.,

$$f(\alpha\boldsymbol{x} + (1-\alpha)\boldsymbol{y}) \le \alpha f(\boldsymbol{x}) + (1-\alpha) f(\boldsymbol{y}) \ \forall \boldsymbol{x}, \boldsymbol{y} \in \mathcal{S} \ \forall \alpha \in [0,1]. \tag{3.7}$$

Equivalently, $f$ is the point-wise maximum of a (possibly infinite) number of linear functions; or, every tangent to $f$ stays below the function graph.

**Theorem 3.5** (Convex functions in optimization)**.** For convex functions, every local minimum is a global minimum.

In order to solve equation (3.5), there are two main methods: *line searches* (section 3.3.1) and *trust regions* (section 3.3.2). They both use derivative information; but there are also methods that do not require the evaluation of derivatives (sections 3.3.3 and 3.3.4).

### 3.3.1 Line search methods

In the $k^{\text{th}}$ iteration of a line search method, the solver first picks a search direction $\boldsymbol{p}_k$ and then has to carry out only a simple univariate optimization that finds the step length:

$$\boldsymbol{x}_{k+1} := \operatorname*{arg\,min}_{\alpha>0} f(\boldsymbol{x}_k + \alpha \boldsymbol{p}_k). \tag{3.8}$$

Line search methods can be summarized colloquially as “direction, then distance.”

#### 3.3.1.1 Direction

There are various methods for finding a suitable descent direction, i.e., one that reduces the objective.

The most obvious choice is *steepest descent* (also called *gradient descent*):

$$\boldsymbol{p}_k = -\frac{\boldsymbol{\nabla} f(\boldsymbol{x}_k)}{\|\boldsymbol{\nabla} f(\boldsymbol{x}_k)\|}. \tag{3.9}$$

While this choice gives rise to a *first order method*—as it only requires the evaluation of a gradient—and is indeed globally convergent, it usually converges very slowly [LY08].

**Example 3.6** (Steepest descent zigzagging)**.** Consider the Rosenbrock function [Ros60]

$$f(x, y) = 100(y - x^2)^2 + (1 - x)^2 \tag{3.10}$$

which has a global minimum at $x = y = 1$ with value $0$. The following visualization shows that the current direction of steepest descent continually changes orthogonally; after a long first step, the progress is reduced to almost nothing. In fact, even after one thousand iterations, the distance to the optimal point is still larger than $0.18$!

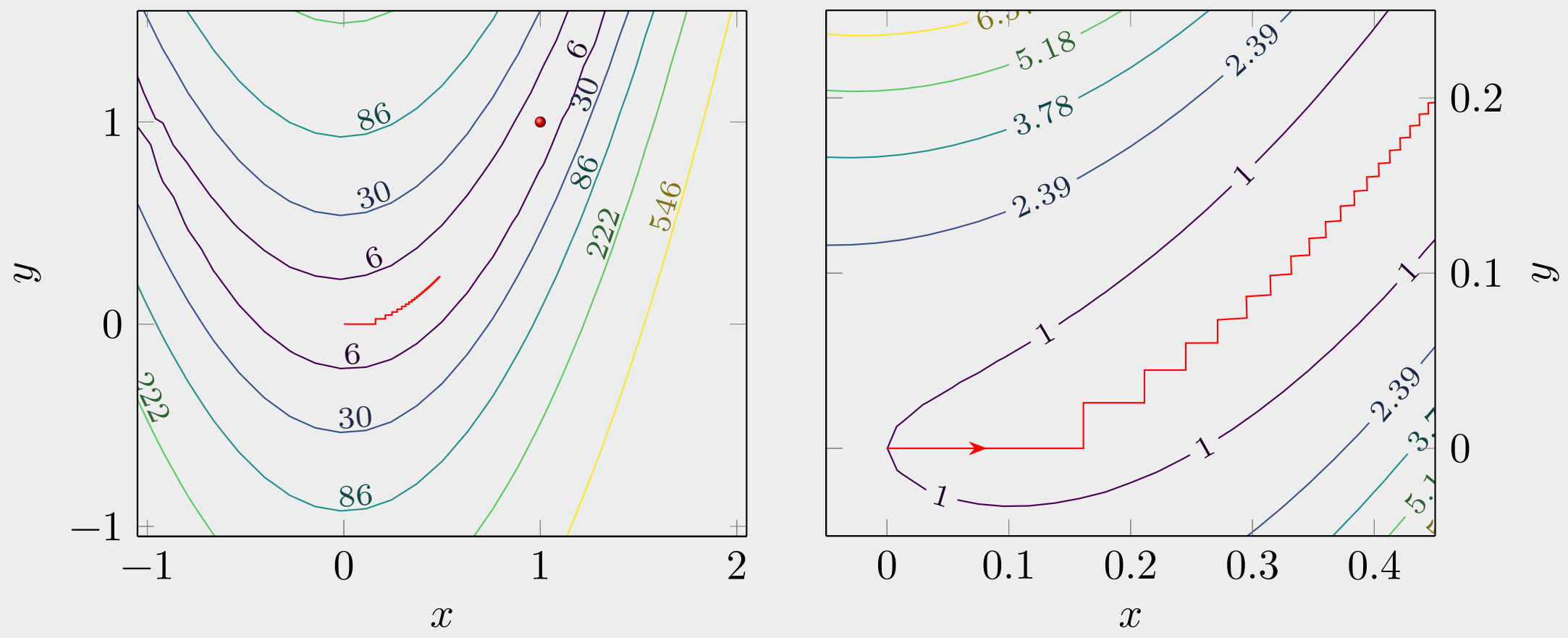

**a)** Overview (optimal point •) **b)** Zoom

**Figure 3.1.** Contours of the Rosenbrock function and first $50$ steps of steepest descent with optimal step length starting at the origin

This raises the question whether for the optimization of nonlinear functions the simple search along a straight line is a good idea at all; this has indeed motivated the development of *arc search* methods [Yan09; Yan11; Zha+19; YIY22], which provide a better point for the next iteration—at the cost of a much more difficult calculation of this point via an arc instead of a straight line, though there are ways to keep this cost under control [Yan25].

However, even a simple line search can do much better than steepest descent simply by choosing a good direction. The best possible generic choice is given by the *Newton direction*:

$$\boldsymbol{p}_k = -(\boldsymbol{\nabla}^2 f(\boldsymbol{x}_k))^{-1} \boldsymbol{\nabla} f(\boldsymbol{x}_k). \tag{3.11}$$

Similar to arc search, the calculation of the direction is now much more difficult, as the Hessian, gradient, and the solution of a linear system have to be calculated; this is a *second order method*. Note that $\boldsymbol{\nabla}^2 f \succ 0$ is not guaranteed, so equation (3.11) might not even give a descent direction. Still, the Newton direction comes with fast quadratic convergence guarantees[4] in a region around the optimum; and additionally, a step length of $1$ is typically a very good choice, largely eliminating the need to even search along the line.

**Example 3.7** (Rosenbrock revisited)**.** Applying the Newton direction with fixed step length $1$ to the Rosenbock function in equation (3.10) is much more successful; in fact, so successful that it is not even worth a plot. Starting from the origin, the next step yields $(1, 0)$, and already the second step is exactly at the position of the optimum, $(1, 1)$. Even for points that are far from the optimum, no more than four or five steps are necessary to yield the exact optimal point. For example, starting at $(-50, -75)$ jumps to $(-50, 2500) \to (1, -2600) \to (1, 1)$ in four steps. While this was rounded liberally with two decimals, the fifth step step is exact up to $10^{-8}$ and the sixth step is indistinguishable from the optimum in double precision. This is despite the Hessian at $(-50, -75)$ having condition number $44\,570$.

To avoid the need for heavy computations, various *quasi-Newton direction* methods have been developed, in which either the Hessian or the inverse Hessian are approximated and updated after each step. To save memory, these may also be low-rank estimators. If the procedure works such that the approximations *along the search direction* become better and better, quasi-Newton methods provide superlinear convergence. There are various quasi-Newton methods, the most prominent one is certainly BFGS [Bro70; Fle70; Gol70; Sha70].

#### 3.3.1.2 Step length

A particular benefit of the Newton direction is that due to its scale invariance, a constant step length is already quite a good choice. While this approach can also be pragmatically carried over to other directions (for example, a complete worst-case convergence analysis for steepest descent with constant step length was recently conducted in [RGP24]), this is usually not a good idea. Instead, an optimal step length should be chosen, which is now

4 If descent directions are not assured, this is fixed by heuristics that invalidate most convergence proofs, but that practically work well.

"just" a univariate optimization problem—but the evaluation of the objective may still be very expensive. Trying to find the global optimum even with respect to just one parameter is neither recommendable nor necessary for a good overall algorithm[5]. In fact, it is enough if the step length is chosen such that it leads to a "sufficient" decrease. To avoid extremely short steps when the function has a steep slope, an additional curvature condition should be imposed. Sufficient decrease and curvature are known as the *Wolfe conditions* [Wol69] (note that the first equation is also called *Armijo condition* [Arm66]):

$$f(\boldsymbol{x}_k + \alpha \boldsymbol{p}_k) \leq f(\boldsymbol{x}_k) + c_1 \alpha \langle \boldsymbol{\nabla} f(\boldsymbol{x}_k), \boldsymbol{p}_k \rangle \qquad \text{with } c_1 \in (0,1) \tag{3.12a}$$

$$\langle \boldsymbol{\nabla} f(\boldsymbol{x}_k + \alpha_k \boldsymbol{p}_k), \boldsymbol{p}_k \rangle \geq c_2 \langle \boldsymbol{\nabla} f(\boldsymbol{x}_k), \boldsymbol{p}_k \rangle \qquad \text{with } c_2 \in (c_1,1) \tag{3.12b}$$

where the constants are algorithm-dependent. There are always intervals of $\alpha$ that satisfy the Wolfe conditions; and as soon as one value is found that satisfies them, the iteration successfully found a new candidate.

## 3.3.2 Trust-region methods

In the $k^{\text{th}}$ iteration of a trust region method, a model function $m_k$ is chosen that closely approximates $f$ within some region around the iterate, so that the problem then reads

$$\inf_{\boldsymbol{p}} m_k(\boldsymbol{x}_k + \boldsymbol{p}) \text{ such that } \boldsymbol{x}_k + \boldsymbol{p} \in \text{trust region.} \tag{3.13}$$

Usually, the trust region is given by the ball $\|\boldsymbol{p}\| \leq \delta_k$. Trust region methods can be summarized colloquially as "distance, then direction."

### 3.3.2.1 Model

The choice of model function is quite clear: it should be a simple, yet accurate function, such that the minimization is easy. This makes a Taylor expansion around the current iterate the natural model candidate, and the degree cutoff is the only open question. When the expansion is cut off after the linear term, the minimization within the ball can be calculated exactly and will give $-\delta_k \frac{\boldsymbol{\nabla} f(\boldsymbol{x}_k)}{\|\boldsymbol{\nabla} f(\boldsymbol{x}_k)\|}$. Note that this is precisely the steepest descent direction with a fixed distance $\delta_k$, and therefore not a good choice.

A quadratic expansion again requires the Hessian—either exactly, then the function is approximated up to $\mathcal{O}(\|\boldsymbol{p}\|^3)$, or an estimate, which gives the approximation up to $\mathcal{O}(\|\boldsymbol{p}\|^2)$. While this sounds similar to the Newton method, now a Hessian that is not strictly positive definite no longer leads to issues in the proof. However, there is no analytic expression for the minimization of the model within the trust-region, requiring more effort here.

5 However, also see [SS25] for a recent different perspective on this; [GSW24] for optimized steepest-descent step sizes; and [SRW25], which modifies the direction so that unit step length becomes a good choice.

#### 3.3.2.2 Region size

A good choice of the trust region size is essential for quick convergence. Large regions might allow to make big progress in a single step; however, the farther away from the point of expansion, the more unreliable the model becomes. Small regions, in turn, provide very accurate models, but also imply small progress. Algorithms should choose the trust region size based on experience in previous iterations, increasing if the actual reduction in the objective closely matches the prediction by the model, and decreasing (and repeating the iteration) if the objective did not decrease.

### 3.3.3 Derivative-free methods

Derivative-free or *zeroth order* methods do not use the information from the gradient and Hessian to solve the optimization problem—since their calculation and storage might be quite expensive, this can be very relevant. Finding a direction for the line search or obtaining an accurate model is therefore not possible in the way that was presented before. Theoretically, the gradient might be constructed using finite differences, but the amount of evaluations necessary to obtain suitable accuracy does not suggest this approach.

However, the objective itself can still be evaluated; and in principle, if it is evaluated at enough points, a good model can be constructed via interpolation or some other means. A search direction for line search methods might also be constructed via some other means, e.g., conjugate directions [NW06], as it is done in Powell's method [Pow64].

The most famous derivative-free method Nelder–Mead [NM65] uses a completely different approach: it keeps track of a number of points of interest in the search space and at each iteration replaces the one with the worst function value by a better one that is constructed from the set of all points using a properly defined strategy. In practice, this works extremely well, but convergence analysis for such a procedure is quite hard.

### 3.3.4 Stochastic methods

All methods presented before are in principle deterministic, and when run on the same hardware with the exact same inputs, they will indeed give the same results in every run. Stochastic methods sacrifice this property in some way. This may be realized through sampled gradients instead of exactly calculated ones, calculating model functions via sampling, or by making the choice of the next step itself inherently random.

There are many stochastic algorithms, but they are not the focus of this thesis and will not be of relevance later on.

## 3.4 Constrained local optimization

Ultimately, quantum information optimization problems have constraints; the definition of a local minimum in equation (3.5) must be adapted such that the neighborhood is intersected with the feasible set. The idea of how to handle this more complicated feasible set goes back to Lagrange [Lag88].

### 3.4.1 The Lagrangian

Formally, a constrained optimization problem as in equation (3.1) is completely equivalent to the following unconstrained problem:

$$\inf_{\boldsymbol{x}\in\mathcal{X}} f(\boldsymbol{x}) + \begin{cases} 0 & \text{if } g_i(\boldsymbol{x}) \leq 0 \ \forall i \in \mathcal{I} \wedge h_j(\boldsymbol{x}) = 0 \ \forall j \in \mathcal{E} \\ \infty & \text{else.} \end{cases} \tag{3.14}$$

Of course, this formal equivalence does not help in solving the problem, as the objective is now no longer differentiable, let alone continuous; and if a test point violates the constraints, there is no information about a direction that points to the feasible set. This does not change if infinity is replaced by a numerically more adequate large value.

Instead of using a step function, a more suitable *penalty function* is required that still gives some direction information. The simplest possible choice is a linear function, which is illustrated in figure 3.2. For nonpositive constraints, a penalty function with positive slope is chosen. Figure 3.2*a* reveals that indeed, any violation of the constraint will increase the objective, and the more the constraint is violated, the larger this penalty. However, it also shows that making the constraint function more negative will *decrease* the objective,

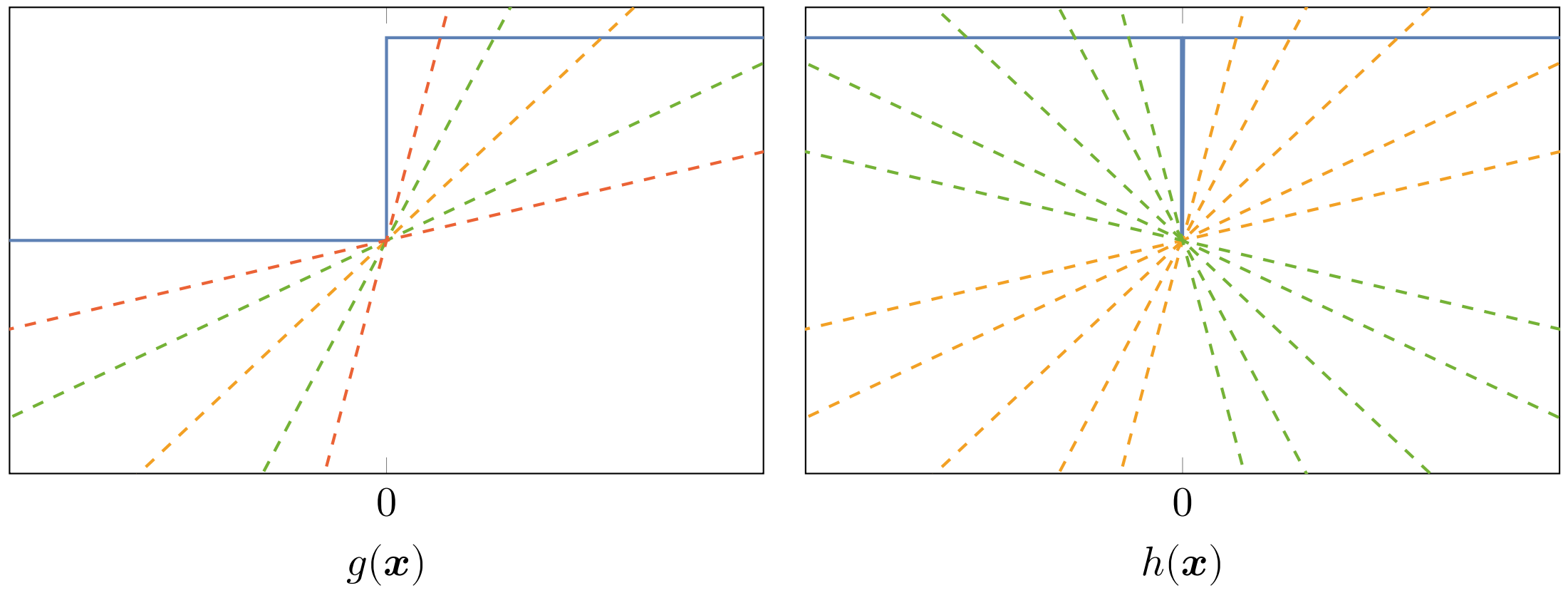

**a)** Nonpositive constraint: positive slopes only **b)** Zero constraint: all slopes invalidating the result. But while a single linear penalty function is of course unable to reproduce a step function behavior, figure 3.2*a* makes clear that the *pointwise supremum* of all linear penalty functions with positive slopes exactly gives the step function!

**Figure 3.2.** Illustration of linear penalty functions with various slopes. The blue curve represents a discontinuous penalty function with finite step size.

Similarly, for zero constraints, figure 3.2*b* shows that depending on the sign of the slope, non-zero function values are either penalized or encouraged; but the maximum of all linear penalty functions with all possible slopes exactly reproduces the step function.

Therefore, the original problem in equation (3.1) is also exactly equivalent to the nested optimization

$$\inf_{\boldsymbol{x}\in\mathcal{X}}\left[f(\boldsymbol{x})+\sum_{i\in\mathcal{I}}\sup_{\lambda_i\geq 0}\lambda_i g_i(\boldsymbol{x})+\sum_{j\in\mathcal{E}}\sup_{\mu_j\in\mathbb{R}}\mu_j h_j(\boldsymbol{x})\right]. \tag{3.15}$$

Note that here, only the explicit equality and inequality constraints were moved to the objective; for the moment, the optimization over $\mathcal{X}$ is assumed to be doable.

This motivates the definition of the Lagrangian.

**Definition 3.8** (Lagrangian)**.** Given a constrained optimization problem of the form equation (3.1), its associated *Lagrangian* is defined as

$$\mathscr{L}(\boldsymbol{x},\boldsymbol{\lambda},\boldsymbol{\mu}) := f(\boldsymbol{x})+\sum_{i\in\mathcal{I}}\lambda_i g_i(\boldsymbol{x})+\sum_{j\in\mathcal{E}}\mu_j h_j(\boldsymbol{x}). \tag{3.16}$$

For the purpose of the optimization, the *decision variables* $\boldsymbol{x}$ are now augmented by the *Lagrangian multipliers* or *dual variables* $\boldsymbol{\lambda}$ and $\boldsymbol{\mu}$ (the latter name will be justified in Lemma 3.10).

**Theorem 3.9** (KKT conditions)**.** Any local minimum $(\boldsymbol{x}^\star,\boldsymbol{\lambda}^\star,\boldsymbol{\mu}^\star)$ of $f$ has to satisfy the *Karush–Kuhn–Tucker conditions* [Kar39; KT51]:

$$\nabla_{\boldsymbol{x}}\mathscr{L}(\boldsymbol{x}^\star,\boldsymbol{\lambda}^\star,\boldsymbol{\mu}^\star)=\mathbf{0} \tag{3.17a}$$

$$g_i(\boldsymbol{x}^\star)\leq 0 \qquad \forall i\in\mathcal{I} \tag{3.17b}$$

$$h_j(\boldsymbol{x}^\star)=0 \qquad \forall j\in\mathcal{E} \tag{3.17c}$$

$$\lambda_i^\star\geq 0 \qquad \forall i\in\mathcal{I} \tag{3.17d}$$

$$\lambda_i^\star g_i(\boldsymbol{x}^\star)=0 \qquad \forall i\in\mathcal{I} \tag{3.17e}$$

Note that the last condition is known as *complementary slackness*.
Additionally, if the gradients of all *active* constraints—that is, all constraints that are satisfied with equality—are linearly independent (*linear independence constraint qualification, LICQ*), or if the constraints are all linear, there is only one possible choice for the Lagrangian multipliers at a local minimum.

The KKT conditions form the cornerstone of constrained optimization, similar to the first-order necessary condition in the unconstrained case (Theorem 3.2); they must be solved in every iteration of the solver. For example, in a trust-region method where the model function is a quadratic approximation of the Lagrangian, setting the gradient with respect to $\boldsymbol{x}$ to zero amounts to solving a linear system of equations. While second-order conditions also exist, they are much more intricate.

As nonlinear optimization problems are composed of just a few number of main ingredients, e.g., the unconstrained solver and the solution of the KKT conditions, a recent work was able to describe a plethora of existing algorithms in terms of a unifying framework [VL24].

### 3.4.2 Duality

According to equation (3.15), the solution of the original problem is obtained by minimizing over $\boldsymbol{x}$, and *within* this optimization by further maximizing with respect to the Lagrangian multipliers. It is trivial to see that for any function $p$ and sets $\mathcal{Q}$ and $\mathcal{R}$,

$$\sup_{\boldsymbol{q}\in\mathcal{Q}} \inf_{\boldsymbol{r}\in\mathcal{R}} p(\boldsymbol{q},\boldsymbol{r}) \leq \inf_{\boldsymbol{r}\in\mathcal{R}} \sup_{\boldsymbol{q}\in\mathcal{Q}} p(\boldsymbol{q},\boldsymbol{r}). \tag{3.18}$$

Therefore, now interchanging the order of optimization, a lower bound can be obtained.

**Lemma 3.10** (Weak duality)**.**

$$\sup_{\substack{\boldsymbol{\lambda}\in[0,\infty)^{|\mathcal{I}|}\\ \boldsymbol{\mu}\in\mathbb{R}^{|\mathcal{E}|}}} \inf_{\boldsymbol{x}\in\mathcal{X}} \mathscr{L}(\boldsymbol{x},\boldsymbol{\lambda},\boldsymbol{\mu}) \leq \inf_{\boldsymbol{x}\in\mathcal{X}} \sup_{\substack{\boldsymbol{\lambda}\in[0,\infty)^{|\mathcal{I}|}\\ \boldsymbol{\mu}\in\mathbb{R}^{|\mathcal{E}|}}} \mathscr{L}(\boldsymbol{x},\boldsymbol{\lambda},\boldsymbol{\mu}). \tag{3.19}$$

The new problem on the left-hand side is referred to as the *dual problem*; the original problem on the right-hand side is the *primal problem*.

It is *a priori* not clear whether the inequality in equation (3.19) can be saturated; this depends on the objective and feasible set. If it is satisfied, this is called *strong duality*. Note the fundamental importance of duality: if a solver is able to report a (local) solution for both the primal and the dual problem, the difference between them is an upper bound to how far the reported solution is from the true global minimum; and if they happen to coincide to a satisfactory tolerance, indeed the optimization was *globally* successful!

Weak duality therefore is a useful tool to get an upper bound to the global suboptimality of a potential solution—though the lower bound to the minimum might well be $-\infty$, in which case it does not help. For an important class of problems, strong duality actually allows to give guarantees; a sufficient criterion for membership in this class is given by Slater's condition.

**Theorem 3.11** (Slater's constraint qualification [Roc70, Theorem 28.2])**.** If the optimization problem in equation (3.1) only involves

- a convex objective,
- convex inequality constraints,
- affine equality constraints,
- a convex cone $\mathcal{X}$ (see Definition 3.18),
- and there exists a feasible point in the relative interior of $\mathcal{X}$ that satisfies all non-affine inequality constraints with strict inequality, then strong duality holds.

## 3.5 Global optimization of structured problems

In the previous section, the particular form of the constraint set $\mathcal{X}$ was left unspecified; it was supposed to restrict the feasible set in a very "structured" way, so that this information can be exploited in optimization strategies.

This section introduces the two most important kinds of structure that directly lead to huge benefits in the optimization: linear and convex programs. After defining the types of problems, important solution algorithms will be discussed that are employed in this thesis; these comprise interior-point, first-order, and conic bundle methods.

### 3.5.1 Linear programming

An optimization problem is called a *linear program* if both its objective and all of its constraints are linear (or rather, affine).

**Definition 3.12** (Linear program)**.** The standard form for *linear programs* is

$$\inf_{\boldsymbol{x} \in [0,\infty)^n} \{\langle \boldsymbol{c}, \boldsymbol{x} \rangle : A\boldsymbol{x} = \boldsymbol{b}\}. \tag{3.20}$$

where $A \in \mathbb{R}^{m \times n}$, $\boldsymbol{b} \in \mathbb{R}^m$, $\boldsymbol{c} \in \mathbb{R}^n$.

**Lemma 3.13** (Dual linear program)**.** The dual problem for a linear program in the form of equation (3.20) is given by

$$\sup_{\substack{\boldsymbol{y} \in \mathbb{R}^m \\ \boldsymbol{s} \in [0,\infty)^n}} \{\langle \boldsymbol{b}, \boldsymbol{y} \rangle : A^\top \boldsymbol{y} + \boldsymbol{s} = \boldsymbol{c}\} \tag{3.21}$$

*Proof.* Writing $\boldsymbol{x} \in [0,\infty)^n \wedge A\boldsymbol{x} = \boldsymbol{b}$ wholly in terms of explicit constraints $\boldsymbol{g}(\boldsymbol{x}) = -\boldsymbol{x}$, $\boldsymbol{h}(\boldsymbol{x}) = \boldsymbol{b} - A\boldsymbol{x}$, the Lagrangian for equation (3.20) follows from equation (3.16) and reads

$$\begin{aligned}
\mathscr{L}(\boldsymbol{x},\boldsymbol{\lambda},\boldsymbol{\mu}) &= \langle \boldsymbol{c},\boldsymbol{x}\rangle - \sum_{i=1}^{n} \lambda_i x_i + \sum_{j=1}^{m} \mu_j \Big( b_j - \sum_{i=1}^{n} A_{j,i} x_i \Big) \\
&= \sum_{i=1}^{n} \Big( c_i - \lambda_i - \sum_{j=1}^{m} A_{j,i}\mu_j \Big) x_i + \sum_{j=1}^{m} b_j \mu_j \\
&= \langle \boldsymbol{c} - \boldsymbol{\lambda} - A^\top \boldsymbol{\mu}, \boldsymbol{x}\rangle + \langle \boldsymbol{b},\boldsymbol{\mu}\rangle. \qquad (3.22)
\end{aligned}$$

Therefore, the dual problem according to equation (3.19) is given by

$$\sup_{\substack{\boldsymbol{\lambda}\in[0,\infty)^n\\ \boldsymbol{\mu}\in\mathbb{R}^m}} \inf_{\boldsymbol{x}\in\mathbb{R}^n} \mathscr{L}(\boldsymbol{x},\boldsymbol{\lambda},\boldsymbol{\mu}) = \sup_{\substack{\boldsymbol{\lambda}\in[0,\infty)^n\\ \boldsymbol{\mu}\in\mathbb{R}^m}} \begin{cases} \langle \boldsymbol{b},\boldsymbol{\mu}\rangle & \text{if } \boldsymbol{c} - \boldsymbol{\lambda} - A^\top\boldsymbol{\mu} = \boldsymbol{0} \\ -\infty & \text{else} \end{cases}$$

and relabeling $\boldsymbol{\mu} \mapsto \boldsymbol{y}$, $\boldsymbol{\lambda} \mapsto \boldsymbol{s}$, the statement follows. □

**Lemma 3.14** (Primal–dual correspondence)**.** The optimal Lagrangian multipliers for a primal linear problem are the optimal decision variables in the dual problem and vice versa.

*Proof.* Every optimal point needs to satisfy the KKT conditions, which read

$$A^\top \boldsymbol{\mu}^\star + \boldsymbol{\lambda}^\star = \boldsymbol{c}; \quad x_i^\star \geq 0\ \forall i; \quad A\boldsymbol{x}^\star = \boldsymbol{b}; \quad \lambda_i^\star \geq 0\ \forall i; \quad \lambda_i^\star x_i^\star = 0\ \forall i \qquad (3.23)$$

for the primal problem. Note how the first KKT condition was already crucial in arriving at the dual problem. The Lagrangian of the dual problem (which is a maximization) reads

$$\begin{aligned}
\tilde{\mathscr{L}}((\boldsymbol{y},\boldsymbol{s}),\tilde{\boldsymbol{\lambda}},\tilde{\boldsymbol{\mu}}) &= -\langle \boldsymbol{b},\boldsymbol{y}\rangle - \sum_{i=1}^{n} \tilde{\lambda}_i s_i + \sum_{j=1}^{n} \tilde{\mu}_j \Big( c_j - \sum_{i=1}^{m} A_{i,j} y_i - s_j \Big) \\
&= \langle -\boldsymbol{b} - A\tilde{\boldsymbol{\mu}}, \boldsymbol{y}\rangle - \langle \tilde{\boldsymbol{\lambda}} + \tilde{\boldsymbol{\mu}}, \boldsymbol{s}\rangle + \langle \boldsymbol{c}, \tilde{\boldsymbol{\mu}}\rangle, \qquad (3.24)
\end{aligned}$$

so that the KKT conditions of the dual problem are

$$A\tilde{\boldsymbol{\mu}}^\star = -\boldsymbol{b}; \quad \tilde{\boldsymbol{\lambda}}^\star = -\tilde{\boldsymbol{\mu}}^\star; \quad s_i^\star \geq 0\ \forall i; \quad A^\top \boldsymbol{y}^\star + \boldsymbol{s}^\star = \boldsymbol{c}; \quad \tilde{\lambda}_i^\star \geq 0\ \forall i; \quad \tilde{\lambda}_i^\star s_i^\star = 0\ \forall i.$$

Eliminating $\tilde{\boldsymbol{\mu}}^\star$,

$$A\tilde{\boldsymbol{\lambda}}^\star = \boldsymbol{b}; \quad s_i^\star \geq 0\ \forall i; \quad A^\top \boldsymbol{y}^\star + \boldsymbol{s}^\star = \boldsymbol{c}; \quad \tilde{\lambda}_i^\star \geq 0\ \forall i; \quad \tilde{\lambda}_i^\star s_i^\star = 0\ \forall i, \qquad (3.25)$$

which are the primal KKT conditions with $\boldsymbol{\lambda}^\star = \boldsymbol{s}^\star$, $\boldsymbol{\mu}^\star = \boldsymbol{y}^\star$, and $\tilde{\boldsymbol{\lambda}}^\star = \boldsymbol{x}^\star$. □

**Lemma 3.15** (KKT sufficiency)**.** The KKT conditions of a primal or dual linear program are sufficient for global optimality.

*Proof.* For the primal problem, let the KKT conditions in equation (3.23) be fulfilled for $\boldsymbol{x}^\star$. Let now $\bar{\boldsymbol{x}}$ be any other feasible point (i.e., $A\bar{\boldsymbol{x}} = \boldsymbol{b}$ and $\bar{x}_i \geq 0 \ \forall i$), then, exploiting all KKT conditions,

$$\begin{aligned} \langle \boldsymbol{c}, \bar{\boldsymbol{x}} \rangle &= \langle A^\top \boldsymbol{\mu}^\star + \boldsymbol{\lambda}^\star, \bar{\boldsymbol{x}} \rangle = \langle \boldsymbol{\mu}^\star, \boldsymbol{b} \rangle + \langle \boldsymbol{\lambda}^\star, \bar{\boldsymbol{x}} \rangle \\ &\geq \langle \boldsymbol{\mu}^\star, \boldsymbol{b} \rangle = \langle \boldsymbol{\mu}^\star, A\boldsymbol{x}^\star \rangle = \langle A^\top \boldsymbol{\mu}^\star, \boldsymbol{x}^\star \rangle \\ &= \langle A^\top \boldsymbol{\mu}^\star + \boldsymbol{\lambda}^\star, \boldsymbol{x}^\star \rangle = \langle \boldsymbol{c}, \boldsymbol{x}^\star \rangle, \end{aligned} \tag{3.26}$$

so no feasible point can have a lower objective value than $\boldsymbol{x}^\star$.

For the dual problem, let the KKT conditions in equation (3.25) be fulfilled for $(\boldsymbol{y}^\star, \boldsymbol{s}^\star)$. Let now $(\bar{\boldsymbol{y}}, \bar{\boldsymbol{s}})$ be any other feasible point (i.e., $A^\top \bar{\boldsymbol{y}} + \bar{\boldsymbol{s}} = \boldsymbol{c}$ and $\bar{s}_i \geq 0 \ \forall i$), then

$$\begin{aligned} \langle \boldsymbol{b}, \bar{\boldsymbol{y}} \rangle &= \langle A\tilde{\boldsymbol{\lambda}}^\star, \bar{\boldsymbol{y}} \rangle = \langle \tilde{\boldsymbol{\lambda}}^\star, A^\top \bar{\boldsymbol{y}} \rangle = \langle \tilde{\boldsymbol{\lambda}}^\star, \boldsymbol{c} - \bar{\boldsymbol{s}} \rangle \\ &\leq \langle \tilde{\boldsymbol{\lambda}}^\star, \boldsymbol{c} \rangle = \langle \tilde{\boldsymbol{\lambda}}^\star, A^\top \boldsymbol{y}^\star + \boldsymbol{s}^\star \rangle \\ &= \langle A\tilde{\boldsymbol{\lambda}}^\star, \boldsymbol{y}^\star \rangle + \langle \tilde{\boldsymbol{\lambda}}^\star, \boldsymbol{s}^\star \rangle = \langle \boldsymbol{b}, \boldsymbol{y}^\star \rangle, \end{aligned} \tag{3.27}$$

so no feasible point can have a higher objective value that $(\boldsymbol{y}^\star, \boldsymbol{s}^\star)$. □

**Corollary 3.16** (Strong duality)**.** If either the primal or dual linear program has a finite solution, then so does the other, and the objectives are equal.

*Proof.* Solving either problem gives optimal decision variables and multipliers that by Lemma 3.14 directly correspond to multipliers and decision variables of the other problem that satisfy its KKT conditions. By Lemma 3.15, this is already sufficient for optimality. Finally, using either equation (3.26) with $\boldsymbol{\mu}^\star = \boldsymbol{y}^\star$ or equation (3.27) with $\tilde{\boldsymbol{\lambda}}^\star = \boldsymbol{x}^\star$ shows that $\langle \boldsymbol{y}^\star, \boldsymbol{b} \rangle = \langle \boldsymbol{x}^\star, \boldsymbol{c} \rangle$, so the objectives are indeed the same. □

Two cases are still uncovered: the (primal or dual) problem might be unbounded in the direction of optimization; then, there is no point where the objective will attain this infinite value, so the previous proof is not directly applicable. The same problem holds if the feasible set is empty. However, there is an important theorem of alternatives that still allows to relate these corner cases:

**Lemma 3.17** (Farkas [Far02])**.** If and only if either problem is unbounded, the other is infeasible.

The previous results show that linear programs possess very convenient duality properties that allow a simple verification of their optimality, for which the KKT conditions are indeed necessary *and* sufficient.

To then actually find a solution, there are two main approaches: the simplex method and interior-point methods. The simplex method, developed in 1947 by George Dantzig [Dan90] was the first seemingly efficient method for solving large-scale optimization problems; "seemingly," because in the worst case, the algorithm has exponential run time[6]. However, its average runtime is indeed polynomial [Bor82a; Sch98; Bor82b], as it is when worst-case input is slightly perturbed [ST01; DH20]. This means that practically, the simplex method is efficient—it even benefits from warm-starting, i.e., similar problems can be optimized faster if previous solutions are known—and variants of it are still in active use today.

However, given that the simplex method is tailored specifically for linear programming problems and cannot be extended to more general cases, it will not be described further. Instead, the focus will be on the more versatile—and, as regards worst-case complexity analysis, much better—interior point methods in section 3.5.3.

### 3.5.2 Convex conic programming

While useful, the scope of linear programming is quite limited—but it is possible to extend the types of allowed functions and constraints in such a way that most of the helpful properties of linear programs are preserved. Equality constraints still have to be linear; however, now, the set $\mathcal{X}$, which before was given by the nonnegative real numbers, is extended to *pointed convex cones*; the resulting problems are then called *convex conic programs*.

**Definition 3.18** (Pointed convex cone [BN01]). A set $\mathcal{X}$ is a *pointed convex cone* if it satisfies the following conditions:

1. $\mathcal{X}$ is nonempty and closed under addition:
$$\boldsymbol{x}, \boldsymbol{y} \in \mathcal{X} \Rightarrow \boldsymbol{x} + \boldsymbol{y} \in \mathcal{X}. \tag{3.28a}$$
2. $\mathcal{X}$ is a cone:
$$\boldsymbol{x} \in \mathcal{X}, \alpha \geq 0 \Rightarrow \alpha \boldsymbol{x} \in \mathcal{X}. \tag{3.28b}$$
3. $\mathcal{X}$ is pointed:
$$\boldsymbol{x} \in \mathcal{X} \wedge -\boldsymbol{x} \in \mathcal{X} \Rightarrow \boldsymbol{x} = \mathbf{0}. \tag{3.28c}$$

6 The first method with a proven polynomial runtime is the ellipsoid method [Kha79; BGT81]. However, it serves as an example that "polynomial" does not equal "practical:" In practice, the ellipsoid method turns out to perform quite badly. In particular, the number of iterations depends quadratically on the number of variables—which is polynomial, but a huge deterrent for even medium-sized problems.

**Definition 3.19** (Convex conic program)**.** The standard form for *convex conic programs* is given by

$$\inf_{\boldsymbol{x}\in\mathcal{C}} \{\langle \boldsymbol{c}, \boldsymbol{x}\rangle : A\boldsymbol{x} = \boldsymbol{b}\} \tag{3.29}$$

where $\mathcal{C} \subset \mathbb{R}^n$ is a pointed convex cone, $A \in \mathbb{R}^{m\times n}$, $\boldsymbol{b} \in \mathbb{R}^m$, $\boldsymbol{c} \in \mathbb{R}^n$.

When constraints were previously introduced into the problem, they were immediately moved to the objective. This was first done via an infinite step function in equation (3.15), but figure 3.2 then quickly showed that there is an equivalent formulation using just a linear penalty function together with nested optimizations, which eventually led to the definition of the Lagrangian in Definition 3.8. The dual problem defined by the optimization of this Lagrangian in reversed order was always a lower bound to the true problem (Lemma 3.10).

Everything remains valid for convex conic problems, as it was defined and derived irrespective of the problem type. But as consequences of the linear programming formulation, convenient properties such as primal–dual correspondence (Lemma 3.14), KKT sufficiency (Lemma 3.15), and strong duality (Corollary 3.16) arose; and they can be mostly carried over to the convex conic case if the definition of the penalty function is slightly modified.

Basically, the picture in figure 3.2 must be extended so that the abscissa no longer hides the black-box function $g(\boldsymbol{x})$ that reduces all the decision variables to a single number. If the abscissa morphs into a high-dimensional space, the cut between "inside" and "outside" has a much more complicated geometry; this is illustrated in figure 3.3 for two variables. Still, the basic idea remains the same: instead of a function with a discontinuous infinite step (indicated in green), the "slopes" in various directions might now be varied. If this is done properly (e.g., previously, for nonpositive constraints only positive slopes were allowed), it stands to reason that the maximum over all those "slopes" effectively corresponds to the true step function.

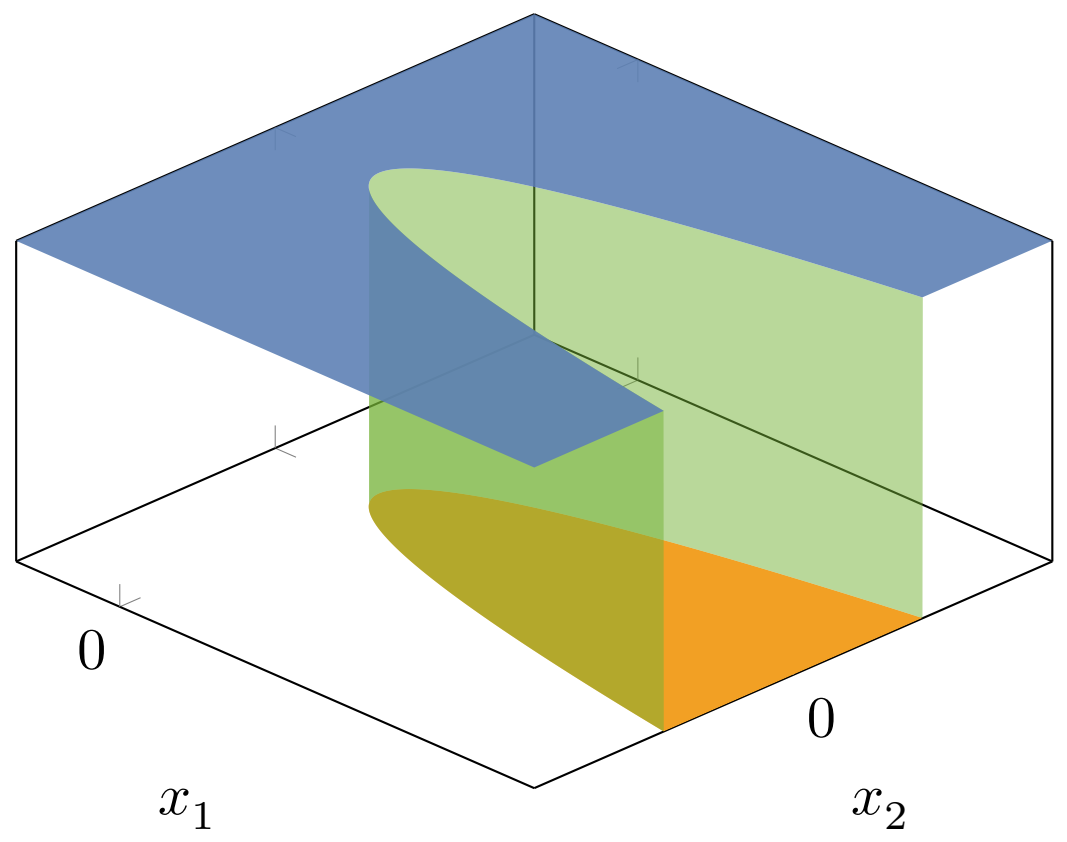

**Figure 3.3.** Illustration of a step penalty function for the constraint $x_1 \geq x_2^2$ indicated in orange.

This idea can be formalized by introducing the dual cone.

**Definition 3.20** (Dual cone)**.** Given any set $\mathcal{X} \subset \mathbb{R}^n$, its *dual cone* is defined as

$$\mathcal{X}^* := \{\boldsymbol{\lambda} \in \mathbb{R}^n : \langle \boldsymbol{\lambda}, \boldsymbol{x} \rangle \geq 0 \ \forall \boldsymbol{x} \in \mathcal{X}\}. \tag{3.30}$$

Here, every point $\boldsymbol{\lambda}$ from the dual cone $\mathcal{X}^*$ defines a hyperplane with normal vector $-\boldsymbol{\lambda}$ that supports the set $\mathcal{X}$ at the origin.

**Remark 3.21** (Properties of the dual cone)**.** The following properties hold:

1. For any set $\mathcal{X}$, $\mathcal{X}^*$ is a closed convex cone.
2. For any two cones $\mathcal{C}_1 \subset \mathcal{C}_2$, $\mathcal{C}_2^* \subset \mathcal{C}_1^*$.
3. For closed convex cones $\mathcal{C}_1$ and $\mathcal{C}_2$, their Cartesian product is a closed convex cone, and its dual is the Cartesian product of the individual duals.
4. For a closed convex cone $\mathcal{C}$, the dual of $\mathcal{C}^*$ is the cone: $(\mathcal{C}^*)^* = \mathcal{C}$.
5. For a closed convex cone $\mathcal{C}$, $-\mathcal{C}^* = \{\boldsymbol{\lambda} \in \mathbb{R}^n : \mathbb{P}_{\mathcal{C}}\boldsymbol{\lambda} = \mathbf{0}\}$, where $\mathbb{P}_{\mathcal{C}}$ is the Euclidean projection onto $\mathcal{C}$.

Many more properties are given in [Dat19, section 2.13.1.2].

**Corollary 3.22** (Conjugation)**.** Let $\mathcal{C}$ be a closed convex cone, then

$$\boldsymbol{x} \in \mathcal{C} \Leftrightarrow \langle \boldsymbol{\lambda}, \boldsymbol{x} \rangle \geq 0 \ \forall \boldsymbol{\lambda} \in \mathcal{C}^*. \tag{3.31}$$

*Proof.* By closedness, the dual of the dual is the cone itself, so $\boldsymbol{x} \in \mathcal{C} \Leftrightarrow \boldsymbol{x} \in (\mathcal{C}^*)^*$; the equation then merely expresses the definition of the dual cone. □

The dual cone now allows to incorporate the set membership $\boldsymbol{x} \in \mathcal{C}$ into a nested-optimization–based penalty function similar to equation (3.15).

**Lemma 3.23** (Conic Lagrangian multipliers)**.** Let $\mathcal{C} \subset \mathbb{R}^n$ be a pointed convex cone, $\boldsymbol{c} \in \mathbb{R}^n$, $A \in \mathbb{R}^{m \times n}$, and $\boldsymbol{b} \in \mathbb{R}^m$. Then,

$$\inf_{\boldsymbol{x} \in \mathcal{C}} \{\langle \boldsymbol{c}, \boldsymbol{x} \rangle : A\boldsymbol{x} = \boldsymbol{b}\} = \inf_{\boldsymbol{x} \in \mathbb{R}^n} \Big[\langle \boldsymbol{c}, \boldsymbol{x} \rangle + \sup_{\boldsymbol{\lambda} \in -\mathcal{C}^*} \langle \boldsymbol{\lambda}, \boldsymbol{x} \rangle + \sup_{\boldsymbol{\mu} \in \mathbb{R}^m} \langle \boldsymbol{\mu}, \boldsymbol{b} - A\boldsymbol{x} \rangle\Big]. \tag{3.32}$$

*Proof.* The linear part related to $\boldsymbol{\mu}$ is as before; only consider the $\boldsymbol{\lambda}$ part.

1. Let $\boldsymbol{x} \in \mathcal{C}$. Then, for any $\boldsymbol{\lambda} \in -\mathcal{C}^*$, $\langle \boldsymbol{\lambda}, \boldsymbol{x} \rangle \leq 0$. The supremum is attained when $\boldsymbol{\lambda} = \mathbf{0} \in -\mathcal{C}^*$; the inequality is then satisfied and the objective is left unchanged.
2. Let $\boldsymbol{x} \notin \mathcal{C}$. Then, by Corollary 3.22, there exists a $\boldsymbol{\lambda} \in -\mathcal{C}^*$ such that $\langle \boldsymbol{\lambda}, \boldsymbol{x} \rangle > 0$. Any positive multiple of $\boldsymbol{\lambda}$ still is in $-\mathcal{C}^*$; hence, the supremum is infinite, so that this region will not participate in the outer minimization. □

Of course, the general statement of weak duality in Lemma 3.10 still holds, even if the domain of $\boldsymbol{\lambda}$ is replaced by the dual cone; it was a consequence of the far more general min–max inequality (3.18). Explicitly,

**Corollary 3.24** (Dual of a conic program). The dual problem for a conic program in the form of equation (3.29) is given by

$$\sup_{\substack{\boldsymbol{y}\in\mathbb{R}^p\\ \boldsymbol{s}\in\mathcal{C}^*}} \{\langle \boldsymbol{b}, \boldsymbol{y}\rangle : A^\top \boldsymbol{y} + \boldsymbol{s} = \boldsymbol{c}\}. \tag{3.33}$$

**Example 3.25** (Important cones and their duals). The following cones are self-dual, i.e., their duals are identical to the cones themselves:

$$\mathcal{C}_{\text{pos}} = [0,\infty)^n \qquad \text{(nonnegative cone)} \tag{3.34a}$$

$$\mathcal{C}_{\text{quad}} = \left\{\boldsymbol{x}\in\mathbb{R}^n : x_1^2 \geq \sum_{i=2}^n x_i^2, x_1 \geq 0\right\} \qquad \text{(quadratic cone)} \tag{3.34b}$$

$$\mathcal{C}_{\text{psd}} = \{X \in \mathbb{S}^s : \langle \boldsymbol{y}, X\boldsymbol{y}\rangle \geq 0 \ \forall \boldsymbol{y}\in\mathbb{R}^s\} \qquad \text{(positive semidefinite cone)} \tag{3.34c}$$

Note that the quadratic cone is also known as *second-order cone*, *norm-cone*, *Lorentz cone*, or *ice-cream cone*. For the semidefinite cone, the Hilbert–Schmidt/Frobenius inner product is used in the definition of the dual cone.

**Definition 3.26** (Conic KKT conditions). The *Karush–Kuhn–Tucker conditions* for a point $(\boldsymbol{x}, \boldsymbol{\lambda}, \boldsymbol{\mu})$ of a convex conic program read

$$A^\top \boldsymbol{\mu} + \boldsymbol{\lambda} = \boldsymbol{c}; \quad \boldsymbol{x}\in\mathcal{C}; \quad A\boldsymbol{x} = \boldsymbol{b}; \quad \boldsymbol{\lambda}\in\mathcal{C}^*; \quad \langle\boldsymbol{\lambda}, \boldsymbol{x}\rangle = 0. \tag{3.35}$$

**Remark 3.27** (Conic vs. linear). Note that the only noteworthy formal difference to the linear case is that the complementary slackness condition is weaker: it now only holds for the scalar product instead of every component.

**Lemma 3.28** (KKT sufficiency). The KKT conditions of a primal or dual convex conic program are sufficient for global optimality.

*Proof.* The proof of Lemma 3.15 directly carries over. All equalities are unchanged; the only inequalities were $\langle\boldsymbol{\lambda}^\star, \bar{\boldsymbol{x}}\rangle \geq 0$ for any feasible point $\bar{\boldsymbol{x}}$ and any optimal Lagrange multiplier $\boldsymbol{\lambda}^\star$ of the primal problem; and $\langle\tilde{\boldsymbol{\lambda}}^\star, \bar{\boldsymbol{s}}\rangle \geq 0$ for any feasible point $\bar{\boldsymbol{s}}$ and any optimal Lagrange multiplier $\tilde{\boldsymbol{\lambda}}^\star$ of the dual problem. As the respective Lagrange multipliers are members of the dual cones, these inequalities still hold. □

However, Farkas's lemma for linear programs, Lemma 3.17, needs a modification.

**Lemma 3.29** (Farkas [CK77; MOS24a]). For a convex conic program in standard (primal) form, exactly one of the following statements is true:

1. The problem is feasible. Additionally, the KKT conditions are necessary for optimality, and primal–dual correspondence in the sense of Lemma 3.14 and strong duality hold.
2. (*Limit-feasibility*) The problem is infeasible, but there is a sequence $(\boldsymbol{x}_n) \subset \mathcal{C}$ such that
$$\lim_{n\to\infty} \|A\boldsymbol{x}_n - \boldsymbol{b}\| = 0. \tag{3.36}$$
3. The dual problem is unbounded.

Similar alternatives hold true for the dual.

The limit-feasible case did not exist in Lemma 3.17, and it can be excluded if the problem is strictly feasible, i.e., if the feasible set has nonempty interior; additionally, generic problems are not limit-feasible, so that a random perturbation will transform a limit-feasible problem to one of the other cases.

Checking for limit-feasible problems is usually done *a posteriori* (and automatically by any solver) simply by comparing whether primal and dual problems yielded finite, but numerically different solutions.

Finally, note that solvers usually do not operate on the original problem directly. Instead, they will often first form an augmented model, the so-called *homogeneous model* that includes both the primal and dual problem.

**Definition 3.30** (Homogeneous primal-dual model [GT57; ART03]). Given a pair of primal and dual convex conic optimization problems in standard forms as defined in equations (3.29) and (3.33), its *homogeneous model* is given by

$$\operatorname*{find}_{\substack{\boldsymbol{x}\in\mathcal{C}\\ \boldsymbol{y}\in\mathbb{R}^p\\ \boldsymbol{s}\in\mathcal{C}^*\\ \tau,\kappa\geq 0}} \text{ such that } \begin{cases} A\boldsymbol{x} - \tau\boldsymbol{b} = \boldsymbol{0}; \\ A^\top\boldsymbol{y} + \boldsymbol{s} - \tau\boldsymbol{c} = \boldsymbol{0} \\ -\langle\boldsymbol{c},\boldsymbol{x}\rangle + \langle\boldsymbol{b},\boldsymbol{y}\rangle - \kappa = 0. \end{cases} \tag{3.37}$$

This feasibility problem has the advantage that it always has a solution, regardless of whether the primal or dual problems are feasible, bounded, or have a finite solution—and based on the solution of the homogeneous model, in particular the value of the optimal $\tau$ and $\kappa$, it can be precisely determined which of these cases apply.

### 3.5.3 Interior-point methods

Nowadays, interior-point methods are probably the most important class of algorithms used to solve convex conic problems. There are many implementations using various algorithms.

Most of them either work on just a few cones—for example, there are many solvers that can deal with second-order cones only; and some that are restricted to positive semidefinite matrices. Others work on a combination of them, and only in the last few years, there have been results that allowed to design practical algorithms working on cones that are not self-dual (see Remark 3.42).

The modern theory of interior-point methods goes back to a seminal monograph by Nesterov and Nemirovskii [NN94]; again, they use a penalty reformulation with a more intricate penalty function than just the linear one. Their insight starts from considering the classical Newton method (i.e., minimization with the Newton direction (3.11) and fixed unit step length), which is quadratically convergent in the Euclidean norm:

**Theorem 3.31** (Newton method)**.** Let $f\colon \mathbb{R}^n \to \mathbb{R}$ be strongly convex with constant $m$ and $\boldsymbol{\nabla}^2 f$ be Lipschitz continuous with constant $L$, i.e.,

$$\langle \boldsymbol{h}, \boldsymbol{\nabla}^2 f(\boldsymbol{x})\boldsymbol{h}\rangle \geq m\|\boldsymbol{h}\|^2 \ \forall \boldsymbol{x}, \boldsymbol{h} \in \mathbb{R}^n \tag{3.38a}$$

$$\left\|(\boldsymbol{\nabla}^2 f(\boldsymbol{x}) - \boldsymbol{\nabla}^2 f(\boldsymbol{y}))\boldsymbol{h}\right\| \leq L\|\boldsymbol{x}-\boldsymbol{y}\|\|\boldsymbol{h}\| \ \forall \boldsymbol{x}, \boldsymbol{y}, \boldsymbol{h} \in \mathbb{R}^n. \tag{3.38b}$$

Then,

$$\left\|\boldsymbol{\nabla} f(\boldsymbol{x}_{i+1})\right\| \leq \frac{L}{2m^2}\left\|\boldsymbol{\nabla} f(\boldsymbol{x}_i)\right\|^2, \tag{3.39}$$

so that minimizing $f$ iteratively using the Newton direction (3.11) leads to quadratic convergence close to the minimum.

Such a Euclidean description is not necessarily a good choice any more in other geometries naturally given by the problem structure; taking this into account, a better way is to look at the Euclidean metric that the second-order differential of $f$ induces *infinitesimally*. This, together with a further convergence assumption, leads to the following definitions.

**Definition 3.32** (Directional derivative)**.** For a function $f\colon \mathcal{X} \to \mathbb{R}$ and $k \in \mathbb{N}$, define the $k^{\text{th}}$ *differential* of $f$ at $\boldsymbol{x} \in \mathring{\mathcal{X}}$ *in the directions* $\{\boldsymbol{h}_k\}_k \subset \mathbb{R}^n \supset \mathcal{X}$,

$$D^k f(\boldsymbol{x})[\boldsymbol{h}_1, \dots, \boldsymbol{h}_k] := \frac{\partial^k}{\partial t_1 \cdots \partial t_k} f(\boldsymbol{x} + t_1\boldsymbol{h}_1 + \cdots + t_k\boldsymbol{h}_k)\Big|_{t_1=\cdots=t_k=0}. \tag{3.40}$$

Note that in particular, $Df(\boldsymbol{x})[\boldsymbol{h}] = \langle \boldsymbol{\nabla} f(\boldsymbol{x}), \boldsymbol{h}\rangle$ and $D^2 f(\boldsymbol{x})[\boldsymbol{h}, \boldsymbol{h}] = \langle \boldsymbol{h}, \boldsymbol{\nabla}^2 f(\boldsymbol{x})\boldsymbol{h}\rangle$.

**Definition 3.33** (Local norm)**.** Let $\mathcal{C} \subset \mathbb{R}^n$ be a closed convex set with nonempty interior and $f\colon \mathcal{C} \to \mathbb{R}$ a convex function with—for simplicity—nonsingular Hessian in $\mathring{\mathcal{C}}$. The *local norm* of $\boldsymbol{h} \in \mathbb{R}^n$ with respect to $f$ at $\boldsymbol{x} \in \mathcal{C}$ is defined by

$$\|\boldsymbol{h}\|_{\boldsymbol{x}}^2 := D^2 f(\boldsymbol{x})[\boldsymbol{h}, \boldsymbol{h}]. \tag{3.41}$$

**Definition 3.34** (LHSC barrier functions). Let $\mathcal{C} \subset \mathbb{R}^n$ be a closed convex cone with nonempty interior. Let $f\colon \mathcal{C} \to \mathbb{R}$ be a convex function. $f$ is called a *logarithmically homogeneous self-concordant (LHSC) barrier function* on $\mathcal{C}$ if

1. $f$ is a barrier function of $\mathcal{C}$, i.e., for every sequence $(\boldsymbol{x}_n)_{n=1}^{\infty} \subset \mathring{\mathcal{C}}$,
$$\lim_{n\to\infty} \boldsymbol{x}_n \in \partial\mathcal{C} \Rightarrow \lim_{n\to\infty} f(\boldsymbol{x}_n) = \infty, \tag{3.42a}$$
so $f$ diverges on the boundary of $\mathcal{C}$.
2. $f$ is a strongly 1-self-concordant function, i.e., for every $\boldsymbol{x} \in \mathring{\mathcal{C}}$ and $\boldsymbol{h} \in \mathbb{R}^n$,
$$\left|D^3 f(\boldsymbol{x})[\boldsymbol{h}, \boldsymbol{h}, \boldsymbol{h}]\right| \leq 2\|\boldsymbol{h}\|_{\boldsymbol{x}}^3, \tag{3.42b}$$
so the Hessian does not change very rapidly in the local Euclidean metric induced by the Hessian itself. Furthermore, the level sets $\{\boldsymbol{x} \in \mathring{\mathcal{C}} : f(\boldsymbol{x}) \leq t\}$ must be closed for every $t \in \mathbb{R}$.
3. $f$ is logarithmically homogeneous, i.e., there exists a $\vartheta > 0$ such that for each $\boldsymbol{x} \in \mathring{\mathcal{C}}$ and every $t > 0$,
$$f(t\boldsymbol{x}) = f(\boldsymbol{x}) - \vartheta \log t. \tag{3.42c}$$
The value $\vartheta$ is called the *parameter* of the barrier.
4. $f$ is a $\vartheta$-self-concordant barrier, i.e.,
$$\left|Df(\boldsymbol{x})[\boldsymbol{h}]\right| \leq \sqrt{\vartheta}\|\boldsymbol{h}\|_{\boldsymbol{x}}; \tag{3.42d}$$
this means that the Newton decrement $\|\boldsymbol{p}_\mathrm{N}\|_{\boldsymbol{x}}$ that measures the local norm of the Newton direction $\boldsymbol{p}_\mathrm{N}$ and indicates the distance to the solution is bounded from above.

**Remark 3.35.** Note that while listed separately, the first three properties imply the last one. As already suggested by the notation, the parameter for the logarithmic homogeneity is then the same as for the self-concordant barrier [NN94, corollary 2.3.2].
Only the third property requires $\mathcal{C}$ to be a cone.

The procedure of any interior-point method thus consists of moving the description of the feasible set into the objective by means of an LHSC barrier function that describes the set. Associated with this barrier is a nonnegative multiplier which controls the weight of the barrier with respect to the objective. Clearly, when the multiplier goes to zero, the barrier function approaches the infinite step function in equation (3.14), which is the exact equivalent description of the original optimization problem. However, this also means that finding a solution numerically is very hard, as the function resembles a discontinuous one more and more.

Therefore, a *path-following interior point method* starts with some finite value for the multiplier for which the penalized objective is well-behaved and solves it using a variant of the Newton method. Next, the multiplier is reduced according to some updating strategy; i.e., the problem becomes harder to solve numerically, but more accurate. Again, the updated problem is solved, with the important difference that an initial point is now known from the last iteration; and provided that the reduction of the multiplier was not too strong, this initial point will be fairly close to the optimal point of the updated problem. Note that when the iterate is close to the optimal point, Newton's method is guaranteed to converge quadratically for suitably well-behaved functions—and given that the objective was linear and the penalty term an LHSC barrier function, this is indeed the case. Therefore, quick convergence for the slightly more difficult problem is assured; in fact, the number of steps required in the Newton iteration is upper bounded by some constant that depends on the path tolerance (a parameter that upper bounds the Newton decrement) and update strategy alone. By repeating this process multiple times, the method follows a path through a family of problems and their solutions parameterized by just a single value: the multiplier of the penalty function. Once this value is suitably small, the procedure can be terminated, giving the actual optimum with high accuracy.

**Lemma 3.36** (Convergence of the path-following method [NS96, theorem F.12]). The total number of Newton steps required to find a solution that is $\varepsilon$-close to the true objective is bounded from above by

$$c\sqrt{\vartheta}\log\left(\frac{2\vartheta\lambda_0}{\varepsilon}\right), \tag{3.43}$$

where the constant $c$ depends on the path tolerance and update strategy alone and $\lambda_0$ is the initial value of the multiplier.

**Remark 3.37** (Updating the problem). After every iteration, the problem is updated; as usual, there are two competing strategies: short steps ensure a very quick convergence of the subsequent Newton method—typically within one or two iterations. However, they require many outer iterations to arrive at the actual problem. Long-step path-following methods in turn make a lot of progress from iteration to iteration, at the expense of the per-iteration runtime.

Most algorithms make use of predictor–corrector ideas collected by Mehrotra [Meh92], requiring the homogeneous form introduced in equation (3.37). The algorithm is an adaptive heuristic that aims at improving the "fairly close" search direction from the previous iteration before starting the next; while its execution requires more computational effort, the improvements are worth it. "Heuristic" might imply that the Mehrotra algorithm can fail. This is indeed true for simple implementations, where it might break convergence proofs or for which there are even counterexamples that lead to a diverging algorithm [NW06]. However, safeguards are available that can prevent such a behavior. In practice, the number of Newton steps in a good implementation of an interior point method is then not just bounded by $\mathcal{O}(\sqrt{\vartheta})$, but almost constant—between ten and 30 iterations are usually sufficient for any problem size.

Therefore, solving a convex conic optimization problem using interior point methods is no more difficult than a sequence of $\mathcal{O}(\sqrt{\vartheta})$ unconstrained problems using Newton's method. Of course, this statement hides some important aspects: First, the per-iteration cost. Newton's method requires to solve a linear system with the coefficient matrix given by the Hessian and the right-hand side by the gradient of the barrier function. Second, it is assumed that an LHSC barrier function is known and can be computed efficiently. In principle, there is a guarantee that an LHSC barrier exists:

**Lemma 3.38** (Universal barrier [NN94, theorem 2.5.1])**.** There exists an absolute constant $c$ such that each closed convex cone $\mathcal{C} \subset \mathbb{R}^n$ admits an LHSC barrier with parameter $cn$. If $\mathcal{C}$ does not contain any one-dimensional affine subspace of $\mathbb{R}^n$, then a possible choice is

$$f(\boldsymbol{x}) = \mathcal{O}(1) \log\left|\mathcal{C}^\circ(\boldsymbol{x})\right|, \tag{3.44a}$$

where

$$\mathcal{C}^\circ(\boldsymbol{x}) := \left\{\phi \in \mathbb{R}^n : \langle \phi, \boldsymbol{y} - \boldsymbol{x} \rangle \leq 1 \ \forall \boldsymbol{y} \in \mathcal{C}\right\} \tag{3.44b}$$

is the polar of $\mathcal{C}$ at $\boldsymbol{x}$, $\mathcal{O}(1)$ is some absolute constant, and $|\bullet|$ the Lebesgue $n$-dimensional measure.

However, the universal barrier is usually too complicated to be used in practice, therefore, obtaining a barrier function for a convex cone is a prerequisite for actually using the cone.

**Remark 3.39** (Complexity)**.** There are convex problems that are nevertheless hard; for example, the set of separable states in quantum information is a convex set, therefore it is *in principle* amenable to an optimization via interior-point methods. However, this would require finding a suitable LHSC barrier function; and even if one was found, it would have to be computable—including derivatives—efficiently. Given that the problem of classifying a quantum state as separable is an $\mathcal{NP}$-hard problem [Gha10] that could then be solved in polynomial time via a convex optimization over this set, such an efficient barrier cannot exist unless $\mathcal{P} = \mathcal{NP}$.

**Example 3.40** (Standard barriers). For the cones defined in Example 3.25, barrier functions are well-known:

$$f_{\text{pos}}(\boldsymbol{x}) = -\sum_{i=1}^{n} \log(x_i); \qquad \vartheta_{\text{pos}} = 1 \quad \text{(nonnegative cone)} \tag{3.45a}$$

$$f_{\text{quad}}(\boldsymbol{x}) = -\log\left(x_1^2 - \sum_{i=2}^{n} x_i^2\right); \qquad \vartheta_{\text{quad}} = 2 \quad \text{(quadratic cone)} \tag{3.45b}$$

$$f_{\text{psd}}(X) = -\log\det X; \qquad \vartheta_{\text{psd}} = s \quad \text{(positive semidefinite cone)} \tag{3.45c}$$

A wealth of other cones together with their barrier functions and derivatives can be found in [CKV23; CKV24].

**Remark 3.41** (Efficiency of semidefinite programs). To estimate[7] the efficiency of interior-point algorithms for semidefinite programs, it must first be made clear how the matrix variables are parameterized. This is usually done by a linear combination of fixed symmetric coefficient matrices, a *linear matrix inequality*, where the coefficients are the decision variables: $\sum_{i=1}^{n} x_i M_i \succeq 0$. Therefore, there are two parameters: the side dimension $s$ of the matrices $M_i$, and the number $n$ of decision variables required for its construction.

Assume for the moment that the $M_i$ are elementary matrices, i.e., $i$ is a double index $(i_1, i_2)$ and $M_i$ is zero except for a one in the entry $(i_1, i_2)$ and $(i_2, i_1)$; let the parameterization be dense, $n = \frac{s(s+1)}{2}$. In this case, depending on the matrix format, there is no to little extra cost involved with constructing the concrete matrix. This matrix, at least in a feasible solver, is already positive semidefinite, and a Cholesky decomposition can be done in $\mathcal{O}(\frac{1}{3}s^3)$, from which the determinant can be calculated by multiplying the squared diagonal.

The gradient has $n$ elements, and it is given by a vectorization of the inverse of the semidefinite matrix. Given the factorization, this is an $\mathcal{O}(\frac{2}{3}s^3)$ operation.

The calculation of the $n \times n$ Hessian from the gradient is simple, as it is the outer product of the gradient with itself; but despite the simple operation, these are $\mathcal{O}(n^2) = \mathcal{O}(s^4)$ elements that have to be calculated. In particular in this case, it is not required to actually store the Hessian, as its individual elements can be calculated so easily—provided a matrix-free linear system solver is implemented to perform the Newton step.

This step then finally has to solve $(\boldsymbol{\nabla}^2 f)^{-1} \boldsymbol{\nabla} f$—a square linear system with $n = s^2$

7 Here, naive implementations are assumed; by considering practically efficient algorithms that rely on fast matrix multiplication, the cost may be dropped—see [Jia+20] for an example of a solver that exploits this. Even better, sparsity patterns in the matrices might lead to a significant reduction—but, when implemented despite the problem being dense, could actually be worse.

unknowns. This is an $\mathcal{O}(n^3) = \mathcal{O}(s^6)$ operation, which ultimately leads to the conclusion that interior-point solvers with a semidefinite constraint of side dimension $s$ scale as $\mathcal{O}(s^{6.5})$.
Now, assuming that there are fewer, but more complicated $M_i$, this will have a negligible effect on the actual construction of the matrix (unless it was a no-operation before due to packed storage). Calculating the gradient and Hessian will still require inverses, but now they are interspersed with matrix multiplications that involve the coefficient matrices. This will yield $\mathcal{O}(n^2 s^2 + ns^3)$ (see [Nem96, section 11.3] for a more detailed description) as the cost for getting these two, which is no more than $\mathcal{O}(s^6)$ unless the matrix is overparameterized; then, the system has to be solved as before; hence, the total scaling is still $\mathcal{O}(s^{6.5})$, but matrix-free methods for solving the system become much harder, therefore also requiring $\mathcal{O}(n^2) = \mathcal{O}(s^4)$ in storage.

**Remark 3.42** (Symmetric and nonsymmetric cones)**.** Practical interior-point algorithms usually solve the homogeneous model in equation (3.37) and go back to work by Nesterov and Todd [NT97]. Importantly, they require the convex cone $\mathcal{C}$ to be a *symmetric cone* [Gül96]. The Koecher–Vinberg theorem [Koe57; Vin63] relates these special types of cones to formal Jordan algebras; and these in turn were completely characterized by Jordan, von Neumann, and Wigner in their seminal paper [JNW34]. This characterization shows that there are four families and an exceptional symmetric cone; therefore, solvers that require symmetric cones can never support more than a combination of the following constraint types:

0. nonnegative real numbers (as a special case of the next items),
1. quadratic cones, equation (3.34b),
2. cones of positive semidefinite symmetric matrices, equation (3.34c)
3. cones of positive semidefinite Hermitian matrices,
4. cones of positive semidefinite quaternionic matrices, and
5. an exceptional 27-dimensional cone, the Albert algebra.

Only items 0 to 2 are usually implemented; item 3 would be useful, but is often seen as redundant, as it can also be reproduced by representing the Hermitian matrix by a symmetric one of doubled size in terms of real and imaginary parts (and, if care is taken, with no extra equality constraints [Wan23]). In particular for quantum information problems where complex numbers prevail, this is an undesirable approach, as the associated real-valued problem may be equivalent in terms of *what* can be modeled, but the underlying operations would be much more efficient if they were able to exploit the Hermitian structure.

While nonnegative, quadratic and positive semidefinite cones together already allow for an impressive range of problems, there are still lots of problems that are outside of their reach but nevertheless (contrary to the example in Remark 3.39) *could* be implemented efficiently. Most prominently, these are the exponential cone

$$\mathcal{C}_{\text{exp}} = \{\boldsymbol{x} \in \mathbb{R}^3 : x_1 \geq x_2 \mathrm{e}^{x_2/x_3}, x_1, x_2 \geq 0\} \tag{3.46a}$$

and the power cones for $\alpha \in (0, 1)$

$$\mathcal{C}_{\text{pow},\alpha} = \{\boldsymbol{x} \in \mathbb{R}^3 : x_1^\alpha x_2^{1-\alpha} \geq |x_3|, x_1, x_2 \geq 0\}, \tag{3.46b}$$

and their duals. The cones are nonsymmetric, and indeed in recent years, some solvers have been generalized to handle some nonsymmetric cones. This is not straightforward; for example, the Mehrotra predictor only works on symmetric cones in its original formulation [DA22].

Given the form of the homogeneous model in equation (3.37), it seems only natural that implementations for non-self-dual cones require LHSC barrier functions for both the primal and the dual cone. However, this also limits the range of what can be done: knowing the barrier of either cone in general does not help in finding the barrier of the other; and indeed, there are some interesting cones for which only one of the two is known. The Skajaa–Ye algorithm [SY15], despite using the homogeneous model, is able to work with one of the two barriers only. While its performance is poor, improvements have been made [CKV23] and led to the development of the solvers Alfonso [PY22] and in particular Hypatia [CKV22] that can—with good performance, though subpar compared to symmetric solvers—handle an impressive number of cones directly. Most relevant for QI problems, this enables optimizations over entropies not only of scalars (by the exponential cone), but also of matrices directly. Without such powerful native cone support, only bounds by means of semidefinite relaxations can be modeled [FF18].

A different route to nonsymmetric optimization was developed and implemented in the `DDS` solver [KT20; KT25], which does not use an explicit homogenization. While the algorithm in [KT20] requires the availability of the convex conjugate of the primal LHSC barrier function (which is a possible LHSC barrier function for the dual cone, and is known for a number of cones [KAV24]), the closed-source implementation appears not to strictly depend on it. Sharing the same roots [Tun01] as `DDS`, a nonsymmetric algorithm has also been implemented in `Mosek` [DA22], restricted to the exponential, power, and more recently geometric mean cone.

### 3.5.4 Augmented Lagrangian first-order methods

The large demands of interior-point methods stem from the second-order information that is required to perform accurate steps in the iterations. To reduce these requirements, algorithms must rely on less information—which opens the avenue for first-order methods [Bec17] (for a review, see [DSS21]). This usually leads to far more required outer iterations than the ten to 30 in interior-point methods, easily ranging in the thousands—in particular, the "tail" of the optimization, i.e., the last tiny steps to increase precision to a desirable degree, can become the domineering part [HY12]. However, the individual iterations are much cheaper, which in principle allows first-order methods to scale to larger problem sizes, if one is able to accept less precise results. To give some numbers, symmetric interior-point methods come with a default precision of $10^{-7}$ to $10^{-8}$; first-order methods are only viable with a precision of $10^{-3}$ to $10^{-4}$. For special problem classes, there might be techniques to improve upon the final solution; see for example the "solution polishing" implemented by OSQP [Ste+20].

#### 3.5.4.1 ADMM-based

The majority of first-order solvers relies on the *alternating direction method of multipliers* (ADMM) [GM75; GM76; Gab83; Eck89] to optimize the homogeneous self-dual model introduced in Definition 3.30. The "alternating directions" come into play because the problem of, say, minimizing an objective of the form $f_1(\boldsymbol{x}) + f_2(\boldsymbol{x})$ can equivalently be written as $f_1(\boldsymbol{x}) + f_2(\boldsymbol{z})$ subject to the constraint $\boldsymbol{x} = \boldsymbol{z}$. The constraint then gives rise to a set of Lagrangian multipliers $\boldsymbol{\mu}$ so that there are now three kinds of variables: $\boldsymbol{x}$, $\boldsymbol{z}$, and $\boldsymbol{\mu}$. An ADMM iteration successively finds new values for these variables by fixing the others, thus alternating in which direction it is currently optimizing.

Note that the homogeneous self-dual model in Definition 3.30 is a feasibility problem; there is no objective. Therefore, it can equivalently be written as a maximization of an indicator-like function over the feasible set. But this is straightforward to split in two parts, as there are conic constraints giving rise to one indicator function and linear equality constraints giving rise to a second one. This establishes the basic form required for ADMM, and the relevant theory has been worked out in [ODo+16] and implemented in the solver `SCS`.

In order to perform the iterations on one variable, the algorithm has to project the variables in question onto the cone, i.e., it solves the problem

$$\mathbb{P}_{\mathcal{C}}\boldsymbol{x} := \operatorname*{arg\,min}_{\boldsymbol{y}\in\mathcal{C}} \|\boldsymbol{x} - \boldsymbol{y}\|. \tag{3.47}$$

Despite being formulated as an optimization problem itself, this projection can be done analytically for a lot of cones.

**Example 3.43** (Projection onto $\mathcal{C}_{\text{psd}}$). The most relevant case for QI problems is the positive semidefinite cone. In this case, the projection is achieved by performing a spectral decomposition of the input matrix, cutting off all negative eigenvalues, and re-multiplying to obtain the positive part of the matrix. While this is a costly procedure, scaling as $\mathcal{O}(s^3)$ with the side dimension $s$ of the matrix, it is still quadratically better than the per-iteration cost in interior-point algorithms, which was $\mathcal{O}(s^6)$ according to Remark 3.41; and it will for sure require no more than $\mathcal{O}(s^2)$ memory compared to potentially $\mathcal{O}(s^4)$ in interior-point methods.

#### 3.5.4.2 Burer–Monteiro factorization and manifold optimization

There are other first-order methods that, for the special case of the positive semidefinite cone, rely on a low-rank Burer–Monteiro factorization [BM03]: $X \in \mathcal{C}_{\text{psd}} \Leftrightarrow X = RR^\top$ with some rectangular $R$ whose number of columns may be reduced if the optimal solution has low rank. This ansatz is problematic, as it removes the convexity from the problem—leading to a general nonlinear program without any global guarantees. However, in certain cases [BM05; BVB16; BVB20; Cif21; Let+23], it can be guaranteed that the reformulation does not introduce any new local minima. If good bounds on the rank are known[8], this can, for generic but not arbitrary objectives [OSV22], lead to a significant reduction in the resource consumption and yield a polynomial-time algorithm [CM22; HG24].

Furthermore, the Burer–Monteiro factorization is a particular way to represent elements from the Riemannian manifold of symmetric positive semidefinite matrices with fixed rank; it therefore stands to reason that Riemannian optimizers can be particularly well-suited for those convex conic problems that can be rewritten in a factorized way without compromising global optimality. Indeed, this was first exploited in [Jou+10] and recently extended to include equality constraints [Wan+23; WH25]; the published benchmarks seem very promising.

#### 3.5.4.3 Matrix sketching

Another approach tailored to the positive semidefinite cone is given by matrix sketching techniques. Similar to the Burer–Monteiro factorization, the matrices are not stored in its full form, but instead, it is assumed that the most relevant information is captured in a much smaller subspace. However, what is stored is not an eigendecomposition, but a (Nyström) sketch of the matrix [HMT11; Git13; Li+17; Tro+17]. Such an approach is reported to scale extremely well and work with moderate precision; for details, see [Yur+21].

8 For a problem featuring only linear constraints and a positive semidefinite cone, $\lfloor \frac{\sqrt{8m+1}-1}{2} \rfloor$, where $m$ is the number of linear constraints, is such a bound [Bar95; Pat98].

### 3.5.5 Cutting planes and proximal bundle methods

The Burer–Monteiro factorization and matrix sketching are methods specially developed for semidefinite optimization; they reduce the complexity of the problem description by making assumptions about the (approximate) rank of the optimal solution. A more general way to reduce complexity is—reminiscent of the trust region method introduced in section 3.3.2—to never actually work on the true problem, but instead to iteratively build up a model that resembles the function.

The cutting planes method [Rus06, chapter 7.2] can be applied to convex programs with a compact feasible set. In fact, it even allows for nondifferentiable objectives as long as they are convex. The linear equality constraints that appear explicitly in the standard form in equation (3.29) are moved to the feasible set, which no longer is a cone, but convex and compact. The problem therefore now reads

$$\inf_{\boldsymbol{x} \in \mathcal{C}} f(\boldsymbol{x}) \tag{3.48}$$

with $\mathcal{C}$ convex and compact and $f$ convex. A linear solver on $\mathcal{C}$ must be available.

The objective function is then replaced by a lower linear approximation, a *cutting plane*, based on any subgradient $\boldsymbol{g} \in \partial f(\boldsymbol{x}_0)$ at some point $\boldsymbol{x}_0$: due to convexity,

$$f(\boldsymbol{x}) \geq f(\boldsymbol{x}_0) + \langle \boldsymbol{g}, \boldsymbol{x} - \boldsymbol{x}_0 \rangle. \tag{3.49}$$

By minimizing the linearized version, a minimum is obtained due to the compactness of $\mathcal{C}$. The subgradient at this minimum gives rise to a new cutting plane, and the piecewise maximum of both defines the next approximation. This is iterated, and more and more linear functions become part of the piecewise maximum, so that it resembles the original objective better and better. When at a minimal point, the actual objective coincides with the approximation, the true global minimum has been found.

While this method is quite simple, does not even require differentiability, and will converge to the true minimum at least asymptotically, the steadily growing number of cutting planes that have to be stored makes it rather impractical. It is therefore important to be able to delete cutting planes again as soon as they are found to no longer be of use; but this is not a simple problem, as even cutting planes that have no effect in one iteration might become relevant later on.

The *proximal bundle method* [Rus06, chapter 7.3] resolves the uncontrolled growth of the cutting planes method. Now $\mathcal{C}$ no longer has to be bounded; only a closed convex set is required. However, a *quadratic* solver on the search space is necessary, and the objective needs to be lower semicontinuous.

The objective for iteration $k$ is regularized into a master problem with a penalty term based on proximity to the previous point:

$$\tilde{\boldsymbol{x}}_k := \arg\min_{\boldsymbol{x}\in\mathcal{C}} \left[ f_k(\boldsymbol{x}) + \frac{\alpha}{2}\|\boldsymbol{x} - \boldsymbol{x}_{k-1}\|^2 \right], \tag{3.50}$$

where $\alpha > 0$ and as in the cutting planes method, the objective was replaced by a piecewise maximum of previously accumulated cutting planes:

$$f_k(\boldsymbol{x}) := \max_i [f(\boldsymbol{x}_i) + \langle \boldsymbol{g}_i, \boldsymbol{x} - \tilde{\boldsymbol{x}}_i \rangle]. \tag{3.51}$$

$\{\tilde{\boldsymbol{x}}_i\}_i$ are the points of the minima obtained in previous iterations, $\boldsymbol{g}_i \in \partial f(\boldsymbol{x}_i)$, and $i$ runs over some indices smaller than the current iteration.

The point $\tilde{\boldsymbol{x}}_k$ obtained by equation (3.50) may not necessarily lead to a sufficient descent (in fact, it might not descend at all, as it only involves the model function $f_k$). The latter, reminiscent, of the Armijo condition in equation (3.12a), is defined as

$$f(\tilde{\boldsymbol{x}}_k) \leq f(\boldsymbol{x}_{k-1}) - \beta \underbrace{[f(\boldsymbol{x}_{k-1}) - f_k(\tilde{\boldsymbol{x}}_k)]}_{\text{predicted decrease at new point}} \qquad \text{with } \beta \in (0,1). \tag{3.52}$$

This means that the actual function value at the new point must be smaller than the old value by at least a certain fraction of what the model predicts it *should* be. If this is fulfilled, the point $\tilde{\boldsymbol{x}}_k$ was a good choice and will become the output of the current iteration, $\boldsymbol{x}_k := \tilde{\boldsymbol{x}}_k$ (*descent step*); if it is not, the current point will not be changed, $\boldsymbol{x}_k := \boldsymbol{x}_{k-1}$ (*null step*); however, the cutting plane model is still updated because the $\tilde{\boldsymbol{x}}_i$ feature in equation (3.51).

Since now, the master problem is a quadratic problem, for which many specialized algorithms exist—though they are not as efficient as linear ones—the problem is essentially solvable; indeed, global optimality of the master problem can be verified:

**Lemma 3.44** (Strong duality of the master problem)**.** Equation (3.50) satisfies Slater's constraint qualification in Theorem 3.11 and therefore features strong duality.

*Proof.* Write the problem explicitly as

$$\arg\min_{\kappa\in\mathbb{R},\boldsymbol{x}\in\mathcal{C}} \left\{ \kappa + \frac{\alpha}{2}\|\boldsymbol{x} - \boldsymbol{x}_{k-1}\|^2 : f(\boldsymbol{x}_i) + \langle \boldsymbol{g}_i, \boldsymbol{x} - \tilde{\boldsymbol{x}}_i \rangle \leq \kappa \ \forall i \right\}, \tag{3.53}$$

then choose any interior point $\boldsymbol{x} \in \mathring{\mathcal{C}}$ and set $\kappa$ to be larger than the evaluation of any linear constraint at this point. Automatically, $(\kappa, \boldsymbol{x})$ is interior feasible. □

Similar to the cutting planes method, the proximal bundle method converges to the true minimum at most with an infinite number of steps; the proximal parameter $\alpha$ can additionally be tuned to improve efficiency by analyzing whether the step lengths are appropriately sized. If they are found to be too long (e.g., by checking whether sufficient decrease is violated a lot), $\alpha$ should be increased, if they are too short, say, due to $f(\boldsymbol{x}_k) = f_{k-1}(\boldsymbol{x}_k)$, $\alpha$ should be decreased.

Contrary to the cutting planes method, however, the proximal bundle method does not require to keep all previous cuts in memory; instead, they can be aggregated.

**Lemma 3.45** (Cut aggregation)**.** No more than two cuts need to be maintained in the proximal bundle method.

*Proof.* The proof works by induction. The first iteration has no cuts; the second iteration has one cut from the first. In the third iteration, there are two cuts, and the master problem that has to be solved is explicitly given by equation (3.53) with $i \in \{1, 2\}$. Its Lagrangian is

$$\mathscr{L}((\kappa, \boldsymbol{x}), \boldsymbol{\lambda}) = \kappa + \frac{\alpha}{2}\|\boldsymbol{x} - \boldsymbol{x}_{k-1}\|^2 + \sum_{i=1}^{2} \lambda_i (f(\boldsymbol{x}_i) + \langle \boldsymbol{g}_i, \boldsymbol{x} - \tilde{\boldsymbol{x}}_i \rangle - \kappa); \tag{3.54}$$

after solving the problem, there is an optimal tuple $(\kappa^\star, \boldsymbol{x}^\star, \boldsymbol{\lambda}^\star)$ that satisfies the KKT conditions in equation (3.17).

Note that knowing the optimal values $\boldsymbol{\lambda}^\star$, the Lagrangian could have been written in the following way:

$$\mathscr{L}((\kappa, \boldsymbol{x}), \tilde{\lambda}) = \kappa + \frac{\alpha}{2}\|\boldsymbol{x} - \boldsymbol{x}_{k-1}\|^2 + \tilde{\lambda}\left[\sum_{i=1}^{2} \lambda_i^\star (f(\boldsymbol{x}) + \langle \boldsymbol{g}_i, \boldsymbol{x} - \tilde{\boldsymbol{x}}_i \rangle) - \kappa\right]; \tag{3.55}$$

this is due to the first KKT condition $\frac{\partial \mathscr{L}}{\partial \kappa} = 0 \Leftrightarrow \sum_i \lambda_i^\star = 1$. The original master problem is therefore equivalent to solving

$$\begin{aligned}
&\operatorname*{arg\,min}_{\kappa \in \mathbb{R}, \boldsymbol{x} \in \mathcal{C}} \left\{ \kappa + \frac{\alpha}{2}\|\boldsymbol{x} - \boldsymbol{x}_{k-1}\|^2 : \sum_{i=1}^{2} \lambda_i^\star [f(\boldsymbol{x}_i) + \langle \boldsymbol{g}_i, \boldsymbol{x} - \tilde{\boldsymbol{x}}_i \rangle] \le \kappa \right\} \qquad (3.56)\\
&\quad = \operatorname*{arg\,min}_{\kappa \in \mathbb{R}, \boldsymbol{x} \in \mathcal{C}} \left\{ \kappa + \frac{\alpha}{2}\|\boldsymbol{x} - \boldsymbol{x}_{k-1}\|^2 : \sum_{i=1}^{2} \lambda_i^\star [f(\boldsymbol{x}_i) - \langle \boldsymbol{g}_i, \tilde{\boldsymbol{x}}_i \rangle] + \left\langle \sum_{i=1}^{2} \lambda_i^\star \boldsymbol{g}_i, \boldsymbol{x} \right\rangle \le \kappa \right\},
\end{aligned}$$

which has the form of a single cut.

Clearly, knowing the Lagrangian multipliers requires solving the problem, so the third iteration requires solving a problem with two cuts. However, after the iteration, those two cuts are compacted into a single one; therefore, the fourth iteration starts with the same form as the third. □

**Remark 3.46** (Convergence rate of proximal bundle methods)**.** If the difference between $f(\boldsymbol{x}_k)$ and the actual minimum should be less than $\varepsilon$, then the number of required iterations scales as $\mathcal{O}(\varepsilon^{-3})$. If additional criteria hold such as quadratic growth, or if the proximal parameter $\alpha$ is tuned or even adaptive, this rate of convergence can be improved up to $\mathcal{O}(\log \varepsilon^{-1})$. For a detailed table and proofs, see [DG23a].

## 3.6 Generic global optimization

While the previous section dealt with problems in a particular form—namely linear or convex conic—many interesting problems do not have such a structure. In this case, the general framework of duality (Lemma 3.10) might still allow to verify the quality of a solution candidate or give (potentially infinitely large) bounds on the distance to the optimal solution, but does not provide a way to obtain one. There are various approaches to finding global solutions; Neumaier groups them in incomplete (heuristic), asymptotically complete, complete, and rigorous [Neu04]. Only the last three will find a true global minimum, only the last two can know when the minimum is reached up to a certain tolerance, and only rigorous methods also work with inexact arithmetic. While naturally, the latter are the most interesting ones, even incomplete algorithms are not to be dismissed as useless; and whether a complete algorithm is rigorous or not is often not considered in convergence proofs. The following sections will only give a brief overview over incomplete global optimization methods that are relevant for this thesis, and mention one of the most famous other approaches to give an example of a complete procedure.

### 3.6.1 Constraints and global optimization

Generic global optimizers typically do not optimize the problem themselves, but they employ other solvers to do this as a subproblem. They then use the results to modify the problem or parameters for the subsolver and iterate. The fully general form of an optimization problem in equation (3.1) allows for more than any such solver can handle. Arbitrary constraints in the form of the functions $g_i$ and $h_j$ usually even complicate the selection of an initial point too much; however, no constraints at all means that the search space is unbounded, which is a severe problem when making any kind of global statement. Therefore, equation (3.1) will now be restricted to problems where the constraints are described by the set $\mathcal{X}$—at least from the point of view of the global optimizer, it does not matter what the subsolver does with the set. It is assumed that the subsolver can handle $\mathcal{X}$ in some way, though of course no global requirements need to be fulfilled by its solutions.

### 3.6.2 Heuristics

*For heuristic methods, $\mathcal{X}$ must be such that it is easily possible to find points in the set, possibly close to other points.*

A local optimizer that uses any of the methods presented in section 3.3 still requires an initial point to start its process. If the objective has more than just one stationary point, starting from different initial points will potentially lead to different local minima. Therefore, running the local optimizer multiple times with various initial points increases the chances of actually finding the global minimum.

This simple train of thoughts is quite successful if the shapes of the objective and feasible region are not too contrived and if the local minima (that are not also global) are few or at least in few well-localized basins. However, they are still “incomplete” methods. The only concern of a global optimizer heuristic is the selection of initial points; it does not have to know anything about the problem and it can work with numerous local optimizers that it simply calls.

There is a multitude of heuristics to choose initial points; the simplest one probably is to divide the (bounded) search space in an equidistant grid and start from each grid point. This method is in principle easy to turn into a “complete” method that is guaranteed to find, in infinite time, the global optimum; for this, the grid just has to be refined after each iteration, and the best local solution is used as a starting point on the next subgrid. However, the scaling in the number of decision variables is exponential, so that such an algorithm is viable only for extremely small problems.

Most other global heuristics are based on stochastic methods.

Basinhopping [WD97], for example, randomly perturbs the previous initial point and then accepts or rejects the new point depending on the local minimization result. While it appears to be quite successful in a lot of physical and chemical problems, I found its performance for the QI problems considered in this thesis to be rather modest; improvements in individual runs of the local optimizer were small, which, given the high cost of one local run, results in a disappointing performance.

A survey on a number of global optimizer heuristics can be found in [AGK19]; there, the TikTak optimizer was reported to repeatedly surpass all others. Indeed, I could observe that TikTak also seems to be quite promising for small QI problems (where “small” dimensions of the involved quantum systems can still mean fifty decision variables), at least when compared to other global heuristics.

### 3.6.3 Relaxations

*For relaxation methods, the requirements on $\mathcal{X}$ vary greatly.*

Relaxation-based solvers take the original problem and convert it to some form that can be optimized more easily and whose optimal solution has a well-defined relationship with the original solution—typically, they are underestimators of the true minimum. Often, the relaxations are iterative and can be improved by increasing the size of the relaxed problem. In a way, the cutting planes algorithm introduced in section 3.5.5 can be seen as some kind of relaxation, as here, instead of the true objective, a model that gives a guaranteed lower bound is optimized, and by adding more and more cuts, the relaxation becomes better and better. In this case, the relaxation and the original problem share the same decision variables and their graphs are directly related. However, relaxations may also transform the problem into something completely different. An example for this is polynomial optimization, which is a central topic of this thesis and will be the focus of chapter 4.

### 3.6.4 Branch and bound

*For branch and bound methods, $\mathcal{X}$ must be such that it is possible to subdivide the set into disjoint subsets that have comparable volumes.*

The branching principle underlies many complete global optimizers. In each iteration, the problem is split into multiple subproblems, which together cover the full problem. After branching, the algorithm tries to find bounds on the individual subproblems that hopefully enable the elimination of a large subset. In the worst case, no subproblem gives a bound that allows its removal, so that all of them have to be carried to the next iteration and be split again.

Note that branch and bound particularly suggests itself in the case of discrete or even binary decision variables, where an immediate way of splitting a problem lies in fixing one discrete variable to all of its values separately, or at least, subdividing the set of possible values.

Branch and bound works particularly well if the description of $\mathcal{X}$ can be made explicit by a number of separable constraints, i.e., constraints that depend only on the sum of expressions that depend on a smaller subset of variables [DNS97; Bli+01]. This is called *constraint propagation*, and it allows to avoid a lot of branching steps.

# 4 Polynomial optimization

Polynomial optimization problems (POPs) are a subset of the general optimization problem defined in equation (3.1): they are restricted to functions $f$, $g_i$, $h_j$ that are polynomials; likewise, $\mathcal{X} \subset \mathbb{F}^n$ must be a basic semialgebraic set, i.e., a set where membership is defined by the satisfaction of a finite number of polynomial equality and inequality constraints.

**Remark 4.1** (Complex numbers)**.** POPs are well-defined under the aforementioned restriction if $\mathbb{F} = \mathbb{R}$. For $\mathbb{F} = \mathbb{C}$, some additional conditions must be imposed, as neither the objective nor the inequality constraints are allowed to yield numbers with nonvanishing imaginary part. Either the complex numbers have to parameterized in terms of real and imaginary parts; then, by suitably enforcing conditions on the coefficients of the polynomials, it can be guaranteed that they are real-valued at all times. Alternatively, the conjugate variables have to be explicitly included in the list of variables for the polynomials [JM18]; then, it is also possible to enforce real-valuedness by constraints on the coefficients.

**Proposition 4.2** (Capability of POPs)**.** Any of the symmetric cones (see Remark 3.42) can be rewritten in terms of a set of polynomial constraints—in particular, polynomial optimization problems are at least as descriptive as semidefinite programs.

*Proof.* Clearly, the nonnegativity constraint and the definition of a quadratic cone, equation (3.34b), are polynomials themselves. There are various equivalences that characterize the symmetric positive semidefinite cone; for example, $M \succeq 0 \Leftrightarrow M = RR^\top$, which puts a quadratic polynomial in every component of a positive semidefinite matrix. For an $m \times m$ matrix, these are $\frac{m(m+1)}{2}$ constraints and new variables; an approach that requires significantly less constraints and no auxilliary variables is to demand the coefficients of the characteristic polynomial of $M$ strictly alternate in sign [HJ12, corollary 7.2.4]. As there are only $m$ nonconstant coefficients, this gives rise to $m$ polynomial constraints of increasing degree, which might be a better or worse description depending on the rest of the problem. Extending the field to the complexes or quaternions, which can be parameterized by the reals, still allows to apply both criteria. Finally, the Albert algebra is also fully parameterized by a $3 \times 3$ octonionic matrix with linear parameters. $\square$

**Remark 4.3** (Noncommutative POPs). Instead of $\mathbb{F} = \mathbb{R}$ or $\mathbb{F} = \mathbb{C}$, the decision variables that enter the polynomials may also be noncommutative—in particular for QI problems, where operators are the central objects, this becomes relevant. Indeed, there is a separate theory of noncommutative polynomial optimization (going back to [Hel02; HM04]) which was developed for QI problems [NPA07; NPA08] and formalized in [PNA10].
In the following, only the commutative variant will be presented because it originated much earlier and a lot more results are known; for a textbook on noncommutative POPs, see [BKP16] and [GPR12, chapter 8]; and for recent extensions [LGG23; Kle$^+$24; PLG25].
The decision variables in a noncommutative polynomial optimization problem are not only the operators in a Hilbert space, but also the space itself. This implies that the dimension of the operators is not fixed explicitly; rather, it is a part of the optimization process. The maximal size of the operators that can be extracted from noncommutative polynomial optimization problems depends on the relaxation level.
By adding commutativity constraints explicitly, a noncommutative POP can effectively be turned into a commutative POP. In turn, commutative polynomial optimization also allows to address noncommutative operators—by parameterizing them explicitly in terms of their matrix components, therefore fixing the size inherently already in the problem description.

## 4.1 Univariate unconstrained polynomial optimization over $\mathbb{R}$

There are various ways to globally solve POPs; the focus of this thesis is on the sums-of-squares or moment hierarchy, which is a relaxation technique as defined in section 3.6.3. To explain the concept, assume for the moment that $\mathbb{F} = \mathbb{R}$, the original optimization problem is univariate, i.e., $n = 1$, and that no constraints are present.

**Definition 4.4** (Degree-bound polynomials). The convex cone of (univariate) polynomials up to degree $d$ in $x$ over $\mathbb{R}$ is defined by

$$\mathbb{R}_d[x] := \left\{ p \colon \mathbb{R} \to \mathbb{R}, p(x) = \sum_{i=0}^{d} \alpha_i x^i : \alpha_i \in \mathbb{R} \right\}. \tag{4.1}$$

If any degree is permitted, the index $d$ is omitted: $\mathbb{R}[x] = \mathbb{R}_\infty[x]$; note that $\mathbb{R}[x]$ is a commutative ring with unity.

**Definition 4.5** (Nonnegative polynomials). The pointed convex cone of nonnegative polynomials up to degree $2d$ is defined by

$$P_{2d}[x] := \{ p \in \mathbb{R}_{2d}[x] : p(x) \geq 0 \ \forall x \in \mathbb{R} \}. \tag{4.2}$$

Likewise, the convex set of strictly positive polynomials up to degree $2d$ is given by

$$P_{2d}^{+}[x] := \{p \in \mathbb{R}_{2d}[x] : p(x) > 0 \ \forall x \in \mathbb{R}\}. \tag{4.3}$$

Again, the index $2d$ may be omitted if no degree bound is imposed.

Using these two definitions, the optimization problem over the decision variable $x$ can be rewritten into an optimization over the cone of nonnegative polynomials:

**Lemma 4.6.** As the infimum is the greatest lower bound,

$$\inf_{x \in \mathbb{R}} p(x) = \sup_{\ell \in \mathbb{R}} \{\ell : p - \ell \in P_{2d}[x]\}. \tag{4.4}$$

Now, $P_{2d}[x]$ is certainly a much more complicated optimization region than $\mathbb{R}$, so this reformulation might seem like a bad idea. However, in the light of section 3.5.2, a reformulation of a nonlinear program into a convex conic program is quite reasonable, as it enables the use of all the helpful theory. In particular, if it were possible to find a computable LHSC barrier function that describes the cone, an interior-point algorithm would be able to provide a globally optimal answer to the problem; or if there were a way to project onto this cone, the task could be delegated to a first-order solver. Indeed both are possible due to the following correspondence.

**Theorem 4.7.** Every nonnegative univariate polynomial can be written as the sum of the squares of two polynomials.

*Proof.* Let $p \in P_{2d}[x]$. Note that this already implies that $p$ is of even degree with a positive highest coefficient $\alpha^2$. Using the fundamental theorem of algebra and the fact that the roots $z_k$ of $p$ must occur in conjugate pairs, $p$ can be represented as

$$\begin{aligned} p(x) &= \alpha^2 \prod_{k=1}^{d} (x - z_k)(x - \bar{z}_k) = \alpha^2 \prod_{k=1}^{d} |x - z_k|^2 = \alpha^2 \left| \prod_{k=1}^{d} (x - z_k) \right|^2 \\ &= \alpha^2 \operatorname{Re}^2 \left[ \prod_{k=1}^{d} (x - z_k) \right] + \alpha^2 \operatorname{Im}^2 \left[ \prod_{k=1}^{d} (x - z_k) \right]. \end{aligned}$$

Here, complex-valued roots must occur in conjugate pairs as a consequence of the real-valued coefficients of $p$; for real-valued roots (where the conjugation is unnecessary, but also not harmful), this is due to their required even degree, as an odd degree would imply a change of sign of the polynomial at this position. □

This leads to the definition of another set of polynomials, namely those that can be represented as the sums of squares of other polynomials. Conceptually, this is different from the set of nonnegative polynomials, although those two coincide in the univariate case.

**Definition 4.8** (Sums-of-squares polynomials)**.** The pointed convex cone of (univariate) *sums-of-squares (SOS) polynomials* up to degree $2d$ is defined by

$$\Sigma_{2d}[x] := \left\{ \sum_{i=1}^{m} q_i^2(x) : m \in \mathbb{N}, q_i \in \mathbb{R}_d[x] \right\}. \tag{4.5}$$

The index $2d$ may be omitted if no degree bound is imposed.

The concept and representation of SOS polynomials is much easier to characterize than positive polynomials; in fact, this is a semidefinite program [Nes00].

**Theorem 4.9** (Semidefinite characterization of SOS membership)**.** A semidefinite program can be used to check whether $p \in \Sigma_{2d}[x]$.

*Proof.* Let $p(x) = \sum_{i=0}^{2d} \alpha_i x^i$. Write $\tilde{\boldsymbol{x}} := (x^0, x^1, \dots, x^d)^\top$, then

$$p(x) = \langle \tilde{\boldsymbol{x}}, G\tilde{\boldsymbol{x}} \rangle$$

for some non-unique $G \in \mathbb{R}^{d\times d}$ (Gram matrix) that depends on the $\alpha_i$. Note that it is always possible to choose $G \in \mathbb{S}^d$.

$\Leftarrow$: Let $G \succeq 0$. $p$ is SOS since $G = \sum_i g_i \boldsymbol{g}_i \boldsymbol{g}_i^\top$ with $g_i \geq 0$, so

$$\langle \tilde{\boldsymbol{x}}, G\tilde{\boldsymbol{x}} \rangle = \sum_i g_i \langle \tilde{\boldsymbol{x}}, \boldsymbol{g}_i \rangle \langle \boldsymbol{g}_i, \tilde{\boldsymbol{x}} \rangle = \sum_i \left( \sqrt{g_i} \langle \tilde{\boldsymbol{x}}, \boldsymbol{g}_i \rangle \right)^2. \tag{4.6}$$

$\Rightarrow$: Let $p$ be SOS, i.e.,

$$p(x) = \sum_{i=1}^{m} \left( \sum_{j=1}^{d} g_{i,j} x^j \right)^2 = \sum_{i=1}^{m} \langle \tilde{\boldsymbol{x}}, \boldsymbol{g}_i \rangle^2 = \sum_{i=1}^{m} \langle \tilde{\boldsymbol{x}}, \boldsymbol{g}_i \rangle \langle \boldsymbol{g}_i, \tilde{\boldsymbol{x}} \rangle = \langle \tilde{\boldsymbol{x}}, G\tilde{\boldsymbol{x}} \rangle \tag{4.7}$$

with $G = \sum_i \boldsymbol{g}_i \boldsymbol{g}_i^\top$. Although this is not necessarily a spectral decomposition (the $\boldsymbol{g}_i$ need not form an orthogonal basis, they may not even span the whole space), still $G$ is positive semidefinite: $\langle \boldsymbol{y}, G\boldsymbol{y} \rangle = \sum_i \langle \boldsymbol{y}, \boldsymbol{g}_i \rangle^2 \geq 0 \ \forall \boldsymbol{y}$. □

**Remark 4.10** (SOS cone)**.** This shows that the cone of sums-of-squares polynomials is a *projected spectrahedron*: it is a spectrahedron—i.e., something that can be described by a linear matrix inequality—but in a higher dimension. The matrix has $\frac{d(d+1)}{2}$ distinct entries, but only $2d$ are required for the description of a polynomial in $\Sigma_{2d}[x]$. This “overspecification” is projected out by the linear map $\langle \tilde{\boldsymbol{x}}, G\tilde{\boldsymbol{x}} \rangle$ that combines all the entries that give rise to the same power.

**Corollary 4.11** (Solution of a univariate, unconstrained POP over $\mathbb{R}$—primal problem).

$$\inf_{x\in\mathbb{R}} p(x) = \sup_{\substack{\ell\in\mathbb{R}\\ G\in\mathbb{S}^{d+1}}} \left\{\ell : G \succeq 0, G_{0,0} = \alpha_0 - \ell, \sum_{\substack{i,j=0\\ i+j=k}}^{d} G_{i,j} = \alpha_k \ \forall k = 1,\dots,2d\right\} \tag{4.8}$$

Note that when a solver is invoked with the problem in Corollary 4.11, it will give a solution for *its* decision variables, which are $\ell$ and the matrix $G$ in Corollary 4.11. While this matrix can be used to write down a sums-of-squares representation of the original polynomial minus its global lower bound, this is not the quantity of interest in the original problem: $x$ is not readily available. However, $x$ can be recovered from the *moment representation* of the POP. For this, consider substituting the nonconvex powers of $x$ by linear variables, and moving the exponentiation to constraints. Then,

$$\inf_{x\in\mathbb{R}} p(x) = \inf_{\substack{x\in\mathbb{R}\\ \boldsymbol{y}\in\mathbb{R}^{2d+1}}} \left\{\sum_{i=0}^{2d} \alpha_i y_i : y_i = x^i \ \forall i\right\}. \tag{4.9}$$

Next, observe that with $\tilde{\boldsymbol{x}} := (x^0, x_1, \dots, x^d)^\top$ as before,

$$\tilde{\boldsymbol{x}}\tilde{\boldsymbol{x}}^\top = \begin{pmatrix} x^0 & \cdots & x^d \\ \vdots & \ddots & \vdots \\ x^d & \cdots & x^{2d}\end{pmatrix} = \left(x^{i+j}\right)_{i=0,\, j=0}^{d \quad\ d} \overset{!}{=} \left(y_{i+j}\right)_{i=0,\, j=0}^{d\quad\ d} =: H(\boldsymbol{y}) \tag{4.10}$$

must hold, where the letter $H$ was chosen due to the fact that so-defined matrix is of Hankel form [TA99]. Therefore, $y_i = x^i \ \forall i \Leftrightarrow H(\boldsymbol{y}) = \tilde{\boldsymbol{x}}\tilde{\boldsymbol{x}}^\top$.

So far, these rewritings were exact, but they just shuffled around the nonconvexity. In the next step, the nonconvex constraint on $H(\boldsymbol{y})$ is relaxed—for note that $\tilde{\boldsymbol{x}}\tilde{\boldsymbol{x}}^\top$ is a positive semidefinite matrix. Since $H(\boldsymbol{y}) \succeq 0$ is a necessary, but usually by no means sufficient criterion for $H(\boldsymbol{y}) = \tilde{\boldsymbol{x}}\tilde{\boldsymbol{x}}^\top$, the solution set will only be enlarged by this replacement, making the infimum smaller. However, for the univariate case, this is actually exact:

**Theorem 4.12** (Solution of a univariate, unconstrained POP over $\mathbb{R}$—dual problem). Let $p(x) = \sum_{i=0}^{2d} \alpha_i x^i$, then

$$\inf_{x\in\mathbb{R}} p(x) = \inf\left\{\sum_{i=0}^{2d} \alpha_i y_i : \boldsymbol{y} \in \mathbb{R}^{2d+1}, y_0 = 1, H(\boldsymbol{y}) \succeq 0\right\}. \tag{4.11}$$

*Proof.*

$$\begin{aligned}\inf_{x\in\mathbb{R}} p(x) &\geq \inf\left\{\sum_{i=0}^{2d} \alpha_i y_i : \boldsymbol{y} \in \mathbb{R}^{2d+1}, y_0 = 1, H(\boldsymbol{y}) \succeq 0\right\}\\ &= \alpha_0 + \inf_{\boldsymbol{y}\in\mathbb{R}^{2d}}\left\{\sum_{i=1}^{2d} \alpha_i y_i : y_1 \begin{pmatrix} 0 & 1 & 0 & \cdots \\ 1 & 0 & 0 & \cdots \\ 0 & 0 & 0 & \cdots \\ \vdots & \vdots & \vdots & \ddots\end{pmatrix} + \cdots + \begin{pmatrix} 1 & 0 & 0 & \cdots \\ 0 & 0 & 0 & \cdots \\ 0 & 0 & 0 & \cdots \\ \vdots & \vdots & \vdots & \ddots\end{pmatrix} \succeq 0\right\}\end{aligned}$$

Moving to the dual problem (the procedure for an SDP is explicitly described in [BV04, example 5.11]), a lower bound is derived:

$$\begin{aligned}
&\overset{\text{duality}}{\geq} \alpha_0 + \sup_{G\succeq 0}\left\{\left\langle -\begin{pmatrix} 1 & 0 & 0 & \cdots \\ 0 & 0 & 0 & \cdots \\ 0 & 0 & 0 & \cdots \\ \vdots & \vdots & \vdots & \ddots \end{pmatrix}, G\right\rangle : \left\langle \begin{pmatrix} 0 & 1 & 0 & \cdots \\ 1 & 0 & 0 & \cdots \\ 0 & 0 & 0 & \cdots \\ \vdots & \vdots & \vdots & \ddots \end{pmatrix}, G\right\rangle = \alpha_1, \dots\right\} \\
&= \sup_{G\succeq 0}\left\{\alpha_0 - G_{0,0} : \sum_{\substack{i,j=0\\ i+j=k}}^{d} G_{i,j} = \alpha_k \ \forall k = 1, \dots, 2d\right\}
\end{aligned}$$

and with the definition $\ell := \alpha_0 - G_{0,0} \Leftrightarrow G_{0,0} = \alpha_0 - \ell$,

$$\begin{aligned}
&= \sup_{\substack{\ell\in\mathbb{R}\\ G\succeq 0}}\left\{\ell : G_{0,0} = \alpha_0 - \ell, \sum_{\substack{i,j=0\\ i+j=k}}^{d} G_{i,j} = \alpha_k \ \forall k = 1, \dots, 2d\right\} \\
&= \inf_{x\in\mathbb{R}} p(x)
\end{aligned}$$

by Corollary 4.11. □

Note that if the solver returns a matrix $H(\boldsymbol{y})$ that has rank one for equation (4.11), then the correspondence with the powers of $x$ was realized perfectly and $y_1$ is $x$. In this case, the minimizer can be extracted from the dual problem with ease. However, equality of the optimal values of the relaxed and exact problem does not imply that $\boldsymbol{y}$ must necessarily coincide with the powers of $x$; for if $\{x_k\}_k$ are all optimal global solutions to the problem, every $H(\boldsymbol{y}) = \sum_k \mu_k \tilde{\boldsymbol{x}}_k \tilde{\boldsymbol{x}}_k^\top$ with $\mu_k \geq 0$, $\sum_k \mu_k = 1$ still solves the relaxed problem. Note that the $\tilde{\boldsymbol{x}}_k$ most likely are not orthogonal, therefore a simple eigendecomposition of $H(\boldsymbol{y})$ is insufficient for the reconstruction of the original solutions. Instead, a Hankel decomposition must be done, which is not a standard task in linear algebra. However, for univariate problems, it is still well-known as Hamburger's truncated moment problem [Ham20], and a way to extract the solutions has already been described more than 200 years ago [Pro95]. This method is not numerically stable—it requires the solution of two linear systems, one with a Hankel, the second with a Vandermonde coefficient matrix, and finding the roots of a univariate polynomial. In particular the solution of the Vandermonde system is very challenging [Gau20].

For the multivariate case that will be discussed subsequently, further algorithms have been developed which do not face these numerical difficulties to such an extreme degree; they of course also work in the univariate case. A popular algorithm was described in [HL05] and has been successfully employed by software packages for many years. It requires a Cholesky decomposition of $H(\boldsymbol{y})$, its reduction to column echelon form, and a Schur decomposition of a random matrix with the size of $\operatorname{rk} H(\boldsymbol{y})$ constructed out of the reduced matrix. An alternative algorithm was proposed more recently [HKM18], but to my knowledge, my package `PolynomialOptimization.jl` (see section 4.10.3) is the only implementation that uses it. This algorithm computes a singular value decomposition of a rectangular part of $H(\boldsymbol{y})$, followed by an eigendecomposition of a random matrix constructed out of the SVD.

**Remark 4.13** (Barrier for the SOS cone). In the light of sections 3.5.2 and 3.5.3, it might be tempting to try to derive an efficiently computable LHSC barrier function for the primal problem in Corollary 4.11 or the dual problem in Theorem 4.12 *directly* without taking the SDP detour.
No positive result has been obtained for the former formulation so far. Note that knowing the barrier function for the semidefinite cone does not help in this case, as the process of intersecting the $\frac{d(d+1)}{2}$ variables of the semidefinite matrix with the linear space defined by the SOS constraints cannot be easily inverted (or at all, given that the matrix representation is not even unique). Therefore, no clear way is seen that avoids the introduction of $\mathcal{O}(d^2)$ auxiliary variables.
The situation looks much better for the dual problem, where only $\mathcal{O}(d)$ variables enter the semidefinite cone. Indeed, in Corollary 4.68, it will be shown that the *dual* cone of $\Sigma_{2d}[x]$ is a simple spectrahedron, and therefore its barrier function *can* be derived using the semidefinite characterization [Nes00], and this has also been implemented [PY19; KCV23] using the Skajaa–Ye algorithm [SY15] (which does not require both primal and dual barrier, see Remark 3.42).

## 4.2 Multivariate unconstrained polynomial optimization over $\mathbb{R}$

In a next step, the polynomial optimization problem is extended to the multivariate case: $\boldsymbol{x} \in \mathbb{R}^n$. The definitions of nonnegative and SOS polynomials trivially carry over, and it is equally simple to show membership in the cone of multivariate SOS polynomials by means of an SDP similar to Theorem 4.9, now using a multivariate monomial basis; however, there is no analog to Theorem 4.7. This goes back to a theorem by David Hilbert (which led to the formulation of his 17th problem).

**Theorem 4.14** (Hilbert [Hil88]). A nonnegative polynomial $p \in P_{2d}[\boldsymbol{x}]$ can be written as the sum of squares of other polynomials precisely in the following cases:

1. $\boldsymbol{x} \equiv x \in \mathbb{R}$,
2. $2d = 2$, $\boldsymbol{x} \in \mathbb{R}^n$, or
3. $2d = 4$, $\boldsymbol{x} \in \mathbb{R}^2$.

**Remark 4.15** (SOS vs. nonnegative polynomials). A lower bound on the volume of nonnegative polynomials in $n$ variables of degree $2d$ scales as $\mathcal{O}(n^{-1/2})$, while an upper bound on the volume of the corresponding sums-of-squares polynomials has scaling $\mathcal{O}(n^{-d/2})$ (and these bounds are asymptotically tight for fixed degree [GPR12, chapter 4]); therefore, as Blekherman puts it, "there are significantly more nonnegative polynomials than sums of squares" [Ble06].

The first explicit example of a nonnegative polynomial that is not a sum of squares was given more than 70 years[1] after Hilbert's proof; it is the famous Motzkin form [Mot83, page xvi]

$$m(x, y, z) = x^4y^2 + x^2y^4 - 3x^2y^2z^2 + z^6, \tag{4.12}$$

whose membership in $P[x, y, z]$ can easily be shown by applying the arithmetic–geometric mean inequality on $\{x^4y^2, x^2y^4, z^6\}$.

Meanwhile, several important results showed that there is still hope; the first being Artin's affirmative answer to Hilbert's 17th problem.

**Theorem 4.16** (Hilbert's 17th problem/Artin [Art27])**.** Any nonnegative multivariate polynomial can be written as a sum of squares of rational functions.

**Corollary 4.17.** For every $p \in P_{2d}[\boldsymbol{x}]$, there is a nonzero prefactor $q \in \Sigma_{2d'}[\boldsymbol{x}]$ such that $qp \in \Sigma_{2(d+d')}[\boldsymbol{x}]$.

**Remark 4.18** (Practical application)**.** The theorem does not contain any statement about the degree of the SOS polynomials involved, and is therefore not directly applicable in numerical optimization. If the only question is to verify membership in the cone of nonnegative polynomials, introducing a degree truncation on $q$ yields a finite-dimensional convex optimization problem that can, e.g., be modeled via semidefinite programs. Clearly, this can then only be sufficient conditions for membership, as the degree cutoff is arbitrarily chosen. The decision variables reflect the coefficients in $q$ and the SOS decomposition.
However, if the question is about minimizing a function, note that the constant term in $p$ contains the *unknown* lower bound; therefore, $q(\boldsymbol{x})p(\boldsymbol{x})$ again involves a multiplication of decision variables, which once more is a nonconvex problem.

Bisecting on the lower bound allows to still use Corollary 4.17 for optimization problems, leading to convex subproblems. Another way is to exploit an explicit form of the prefactor $q$ due to Pólya [Pól28] and later generalized by Reznick [Rez95; MNR23]:

**Lemma 4.19** (Explicit prefactor)**.** In Corollary 4.17, the prefactor $q$ can be chosen as $q(\boldsymbol{x}) := \|\boldsymbol{x}\|^{2r}$ for sufficiently large $r$ provided that $p$ is a strictly positive homogeneous polynomial.

1 Reznick writes that the reasons for such a long period of no counterexample "may be more psychological than mathematical." [Rez00]

This leaves open the question of how large to choose $r$ and what consequence the "strictly positive" part has on the numerics. Regarding the former, Reznick also provides an answer:

**Lemma 4.20** (Degree bound [Rez95]). Let $p \in P_{2d}^{+}[\boldsymbol{x}]$ be a strictly positive homogeneous polynomial. If

$$r \geq \frac{n2d(2d-1)}{4\log 2}\frac{\sup\{p(\boldsymbol{x}) : \|\boldsymbol{x}\| = 1\}}{\inf\{p(\boldsymbol{x}) : \|\boldsymbol{x}\| = 1\}} - \frac{n+2d}{2}, \tag{4.13a}$$

then there exist homogeneous *linear* polynomials $\{\sigma_j\}_j \subset \mathbb{R}_1[\boldsymbol{x}]$ such that

$$\|\boldsymbol{x}\|^{2r} p(\boldsymbol{x}) = \sum_{j=1}^{N} \sigma_j^{2(d+r)}. \tag{4.13b}$$

with $N = \binom{n+2d+r-1}{n-1}$.

This is one of the few existing results on degree bounds, and often, the resulting bounds are very high if they can even be calculated. At the same time, the statement dictates the precise form of all polynomials involved, which allows to significantly reduce the complexity of the optimization problem to be solved. Given that the prefactor is now known, it is no problem at all to put an unknown lower bound into the constant of $p$ without losing convexity (with the caveat that the value of the constant term influences the degree bound).

A different result that does not work with a Hilbert–Artin prefactor instead *adds* a small perturbation.

**Theorem 4.21** (Approximating SOS polynomials [Las06]). Every nonnegative polynomial $p$ is almost a sum of squares in the following sense:

$$\lim_{\varepsilon \to 0^+} \|p - p_\varepsilon\|_1 = 0 \quad \text{where} \quad p_\varepsilon(\boldsymbol{x}) = p(\boldsymbol{x}) + \varepsilon \sum_{k=0}^{r_\varepsilon} \sum_{j=1}^{n} \frac{x_j^{2k}}{k!} \in \Sigma[\boldsymbol{x}], r_\varepsilon \in \mathbb{N} \tag{4.14}$$

and the norm is the $\ell_1$-norm of the coefficients.

This implies that SOS polynomials are dense, with respect to this norm, in the set of nonnegative polynomials, an observation that was first made by Berg for polynomials on the unit cube [BCR76]. Note how this is still compatible with Blekherman's statement: there are significantly more nonnegative polynomials than sums of squares—when fixing the degree. But when allowing the SOS representation to have *any* degree (Theorem 4.21 does not put any bound on $r_\varepsilon$), then it is always possible to find an arbitrarily good approximation.

**Example 4.22.** The previous "workarounds" to make SOS optimization succeed can be applied to the Motzkin form. Its minima are on the vertices of any scaled unit cube $(x, y = \pm z)$ and also the points for which $z$ and either $x$ or $y$ are zero. The minimum value is exactly zero (so the Motzkin form is *not* a strictly positive polynomial).

1. Hilbert–Artin (Corollary 4.17)
   To represent this as an SDP, note that the nonzero condition on $q$ has to be enforced; else, the solver will just always pick $q \equiv 0$. This can be done by simply imposing an arbitrary positive value as a lower bound on the trace of the semidefinite matrix representing $q$.

   As expected, with $q \equiv 1$, no result is found, verifying that the Motzkin polynomial is not SOS. Already allowing for a quadratic $q$, however, leads to a success; the semidefinite program has $166$ linear constraints, one semidefinite matrix for $q$ with side dimension $4$ and one for the SOS polynomial $\sigma$ representing the product $qp$ with side dimension $35$. Note that many solvers struggle with this problem due to numerical issues.

   The returned solution is of course numerical, but it already allows to guess a possible exact form by rounding the coefficients. The ansatz

   $$\begin{aligned}(x^2 + y^2 + \alpha z^2)m(x, y, z) &= a(x^3 y - xy^3)^2 + b(xz^3 - xy^2 z)^2 \\ &\quad + b(yz^3 - x^2 yz)^2 + c(x^2 y^2 - z^4)^2 + d(x^3 y + xy^3 + 2xyz^2)^2\end{aligned}$$

   can be obtained from an eigendecomposition of the returned matrices and liberally chopping smaller values to zero. Comparing coefficients leads to a very simple system of equations and then to the exact equality

   $$\begin{aligned}(x^2 + y^2 + 4z^2)m(x, y, z) &= (x^3 y - xy^3)^2 + (xz^3 - xy^2 z)^2 \\ &\quad + (yz^3 - x^2 yz)^2 + 4(x^2 y^2 - z^4)^2.\end{aligned}$$

   Sometimes, it is indeed possible to guess exact representations based on the numerical solution; and for small numbers of variables, there exist algorithms that can obtain these representations systematically [MD21].

2. Pólya–Reznick (Lemma 4.19)
   The Motzkin form is already homogeneous, so that no further homogenization reformulations have to be done. While Lemma 4.19 holds for strictly positive polynomials, speculating that numerically, this difference will play no role, it can be applied anyway to the Motzkin form.

   Indeed, already $r = 1$ yields a successful result, and $(x^2 + y^2 + z^2)m(x, y, z)$ can be represented as a sum of five squared polynomials. The semidefinite program has 165 linear constraints and one semidefinite matrix with side dimension 35. Again the problem is numerically challenging and will fail with a lot of solvers.

Once more, the numerics allows to represent the decomposition:

$$4(x^2+y^2+z^2)m(x,y,z) = (x^3y - xy^3)^2 + 3(x^3y + xy^3 - 2xyz^2)^2$$
$$+ 2(xz^3 - yz^3 + x^2yz - xy^2z)^2 + 2(xz^3 + yz^3 - x^2yz - xy^2z)^2 + 4(x^2y^2 - z^4)^2.$$

The simplified form of the prefactor had a more complicated form of the sums of squares as a consequence, although the degree in this case did not increase—which is in general not guaranteed. Also note that the interior-point solver found *a* solution, which typically is a solution of high rank (low rank solutions are on the boundary of the feasible set, where the barrier diverges, and are therefore much harder to reach). Just because a five-component decomposition was found here does not imply that there is no decomposition with fewer terms.

3. Reznick (Lemma 4.20)
   To obtain a value of $r$, the Motzkin form has to be minimized over the unit sphere. However, any vertex of the scaled unit cube with edge length $\frac{2}{\sqrt{3}}$, as well as any combination of $x = \pm 1$, $y = \pm 1$, $z = 0$ is a root of the Motzkin form over the unit sphere; therefore, the degree bound is infinite. Even without the bound, it is still possible to try to find a representation in the form of Lemma 4.20, where only homogeneous polynomials occur on both sides. The size of the basis to represent the SOS polynomial can therefore be reduced by only considering homogeneous polynomials; this is as effective as dropping one variable from the problem.

   The result is the exact same as before, as already the previously given decomposition only involved homogeneous polynomials. However, this time, the semidefinite program only had 45 linear constraints and a semidefinite matrix with side dimension 15, but still, the numerical issues prevail.

4. Lasserre (Theorem 4.21)
   To exploit the denseness result, a perturbation term has to be added to the Motzkin form, but then, no prefactor is required any more. Finding the minimal $\varepsilon$ that is possible for a given $r_\varepsilon$ is a semidefinite program. While in principle, $r_\varepsilon$ could be chosen smaller than half of the total degree, here, this will not give rise to an SOS polynomial for any $\varepsilon$. However, increasing $r_\varepsilon$ leads to smaller and smaller bounds, although this becomes more and more costly—see table 4.1. The minimal $r_\varepsilon = 3$ is more efficient than Hilbert–Artin and Pólya–Reznick, but less efficient than Reznick; already $r_\varepsilon = 4$ is as expensive as the others while still only giving a modest $\varepsilon$.

| $r_\varepsilon$ | $\varepsilon$ | side dimension |
|---|---|---|
| 2 | $\infty$ | 20 |
| 3 | $6.1 \cdot 10^{-2}$ | 20 |
| 4 | $1.9 \cdot 10^{-2}$ | 35 |
| 5 | $8.9 \cdot 10^{-3}$ | 56 |
| 6 | $4.3 \cdot 10^{-3}$ | 84 |
| 7 | $2.0 \cdot 10^{-3}$ | 120 |
| 8 | $1.1 \cdot 10^{-3}$ | 165 |
| 9 | $5.4 \cdot 10^{-4}$ | 220 |

**Table 4.1.** Minimal required perturbation $\varepsilon$ for the Motzkin form to "certify" membership in the SOS cone according to Theorem 4.21.

## 4.3 Multivariate constrained polynomial optimization over $\mathbb{R}$

### 4.3.1 Positivstellensätze

Adding constraints to a POP is conceptually very clear and follows exactly the spirit of the Lagrangian multipliers introduced in section 3.4 and generalized to the conic framework in section 3.5.2. All that remains is to specify from which set the multipliers are taken such that they are able to take the role of the duals of the constraints.

Finding these multipliers is quite intuitive. In the following,

$$\mathcal{S} := \{\boldsymbol{x} \in \mathbb{R}^n : g_i(\boldsymbol{x}) \geq 0 \ \forall i \in \mathcal{I}, h_j(\boldsymbol{x}) = 0 \ \forall j \in \mathcal{E}\} \tag{4.15}$$

will always be a basic semialgebraic set[2]. Given any POP in the form

$$f^\star := \inf_{\boldsymbol{x} \in \mathcal{S}} f(\boldsymbol{x}), \tag{4.16}$$

without the constraints, this was reformulated as finding the largest $\ell$ such that $f(\boldsymbol{x}) - \ell$ is a nonnegative polynomial—for all $\boldsymbol{x} \in \mathbb{R}^n$. While in the original formulation, adding the constraints makes the solution set *smaller*, when going to the equivalent formulation on polynomials, they make the solution space *larger*: for now, it is sufficient if $f(\boldsymbol{x}) - \ell$ is a nonnegative polynomial *on the basic semialgebraic set defined by the constraints only*!

Denote this set as $P_{2d}[\boldsymbol{x} \in \mathcal{S}]$. Clearly, $p \in P_{2d}[\boldsymbol{x}] \Rightarrow P_{2d}[\boldsymbol{x} \in \mathcal{S}]$, as the nonnegative criterion is checked on less points in the right-hand set.

2 Note that here, in accordance with how the theory of polynomial optimization is usually phrased, the inequality condition of equation (3.1) was changed from $g_i \leq 0$ to $g_i \geq 0$.

Now, let $r \in P_{2d-\deg g_1}[\boldsymbol{x}]$. Then, $g_1 r \in \mathbb{R}_{2d}[\boldsymbol{x}]$; but since $g_1$ is nonnegative on the feasible set, $g_1 r$ is also nonnegative there: $g_1 r \in P_{2d}[\boldsymbol{x} \in \mathcal{S}]$. This holds true for the multiplication with any of the inequality constraint polynomials $g_i$, and also for the multiplication with arbitrary products of them.

Finally, let $q \in \mathbb{R}_{2d-\deg h_1}[\boldsymbol{x}]$. Then, $h_1 q \in \mathbb{R}_{2d}[\boldsymbol{x}]$; but even more, since $h_1$ vanishes on the feasible set, $h_1 q$ also vanishes there: $h_1 q = 0 \ \forall \boldsymbol{x} \in \mathcal{S}$. Again, this holds true for the multiplication with any of the equality constraint polynomials $h_j$, and also for the multiplication with arbitrary products of them.

All the conditions given so far are necessary for a polynomial to be a member of $P_{2d}[\boldsymbol{x} \in \mathcal{S}]$. The question of whether they are, in their entirety, also sufficient, is answered by a so-called positivstellensatz[3], which first requires several definitions.

**Definition 4.23** (Cone[4])**.** The *cone* generated by the polynomials $\{g_i\}_{i\in\mathcal{I}}$ is the set

$$\operatorname{cone}(\boldsymbol{g}) := \left\{ \sum_{\mathcal{S}\in\mathcal{P}(\mathcal{I})} \sigma_{\mathcal{S}} \prod_{i\in\mathcal{S}} g_i : \sigma_{\dots} \in \Sigma[\boldsymbol{x}] \right\}, \tag{4.17}$$

where $\mathcal{P}(\bullet)$ denotes the power set.

**Remark 4.24.** The definition of a cone is a formalization of the idea from above: it contains a linear combination of all possible combinations of *distinct* products of elements from $\{g_i\}_{i\in\mathcal{I}}$ (including none), where the prefactor is given by an SOS polynomial. These combinations are not continued *ad infinitum*, as duplicates are not allowed; so the sum has exactly $2^{|\mathcal{I}|}$ elements. However, *no degree bound* is imposed on the prefactors!

**Definition 4.25** (Ideal)**.** Let $(\mathcal{R}, +, \cdot)$ be a commutative ring. A subset $\mathcal{T} \subset \mathcal{R}$ is an *ideal* if and only if it is

1. closed with respect to addition, i.e., $(\mathcal{T}, +)$ is a subgroup of $(\mathcal{R}, +)$, and
2. absorbing elements by multiplication, i.e., $rx \in \mathcal{T}$ for every $r \in \mathcal{R}$ and $x \in \mathcal{T}$.

**Example 4.26** (Polynomial ideal)**.** All polynomials that vanish at least on the same set of points as the $\{h_j\}_{j\in\mathcal{E}}$ form an ideal. It is *generated* by the $h_j$, which is written as

$$\langle \boldsymbol{h} \rangle := \left\{ \sum_{j\in\mathcal{E}} h_j p_j : p_j \in \mathbb{R}[\boldsymbol{x}] \right\}. \tag{4.18}$$

Note that no degree bound can be involved here.

3 Literal translation from German: theorem on positive points. Strictly speaking, the answer is given by a nichtnegativstellensatz, i.e., theorems on nonnegative points, which is usually not spelled out.

4 Not to be confused with the cones from convex conic programming in Definition 3.18. Also note that some authors name Definition 4.23 *preorder* and don't have a separate name for Definition 4.27.

**Definition 4.27** (Preorder)**.** The *preorder* of the basic semialgebraic set $\mathcal{S}$ is given by the Minkowski sum of its cone and ideal:

$$\operatorname{preorder}(\mathcal{S}) := \operatorname{cone}(\boldsymbol{g}) + \langle \boldsymbol{h} \rangle = \{u + v : u \in \operatorname{cone}(\boldsymbol{g}), v \in \langle \boldsymbol{h} \rangle\}. \tag{4.19}$$

**Theorem 4.28** (Krivine–Stengle positivstellensatz [Ste74; Lau08])**.** Let $\mathcal{S}$ be a basic semialgebraic set. Then,

$$p \in P(\boldsymbol{x} \in \mathcal{S}) \Leftrightarrow \exists q, r \in \operatorname{preorder}(\mathcal{S}), k \in \mathbb{N}_0 \text{ such that } qp = p^{2k} + r, \tag{4.20a}$$

$$p \in P^+(\boldsymbol{x} \in \mathcal{S}) \Leftrightarrow \exists q, r \in \operatorname{preorder}(\mathcal{S}) \text{ such that } qp = 1 + r, \tag{4.20b}$$

$$p(\boldsymbol{x}) = 0 \ \forall \boldsymbol{x} \in \mathcal{S} \Leftrightarrow \exists k \in \mathbb{N}_0 \text{ such that } -p^{2k} \in \operatorname{preorder}(\mathcal{S}), \text{ and} \tag{4.20c}$$

$$\mathcal{S} = \emptyset \Leftrightarrow -1 \in \operatorname{preorder}(\mathcal{S}). \tag{4.20d}$$

**Remark 4.29** (Practical application)**.** While the Krivine–Stengle positivstellensatz provides an exact characterization of positive polynomials on a basic semialgebraic set in terms of SOS polynomials, it is not very practical:

1. The SOS polynomials involved may have arbitrary degree; this is the same problem as in the unconstrained case and can be approached numerically only with a—hopefully low—degree truncation.
2. Finding a suitable value for $k$ is also an inconvenience; but, numerically, $P$ and $P^+$ are hardly distinguishable, so that the strict form with $k = 0$ is always applicable, although it may run into numerical difficulties.
3. The cone contains $2^{|\mathcal{I}|}$ SOS polynomials, where each corresponds to a semidefinite variable. While in a scheme with consistent degree truncation, the one of highest degree will be $\sigma_\emptyset$ associated with the empty product in Definition 4.23, their exponential number is still a significant burden. This is a new problem introduced by the optimization on the semialgebraic set.
4. The positivstellensatz can be used for *certification*, but not for *optimization*: for then, the constant term in $p$ is a decision variable itself, so that $qp$ is a product of unknowns. This is also not a new problem and was already discussed for the unconstrained case in Remark 4.18.

While for some cases there are degree bounds that in principle allow to cope with the first point—and one such bound will be presented subsequently—they are usually far from practical; still, experience tells that the first issue might often not be too dramatic. To solve the last two points, there are weaker, but more helpful, variants of the positivstellensatz.

**Lemma 4.30** (Schmüdgen's positivstellensatz [Sch91]).
Let $\mathcal{S} = \{\boldsymbol{x} \in \mathbb{R}^n : g_i(\boldsymbol{x}) \geq 0 \ \forall i \in \mathcal{I}\}$ be a compact basic semialgebraic set. Then,

$$p \in P^+[\boldsymbol{x} \in \mathcal{S}] \Rightarrow p \in \operatorname{cone}(\mathcal{S}). \tag{4.21}$$

Schmüdgen's variant removes the prefactor $q$ at the expense of demanding compactness and is therefore amenable to direct optimization. It feels quite natural provided a denseness result similar to Theorem 4.21 holds on basic semialgebraic sets—which indeed is the case [LN07]. However, it still contains exponentially many SOS polynomials due to the cone. This is finally simplified by Putinar.

**Definition 4.31** (Quadratic module)**.** The *quadratic module* generated by the polynomials $\{g_i\}_{i \in \mathcal{I}}$ is the set

$$\operatorname{qmodule}(\boldsymbol{g}) := \left\{\sigma_0 + \sum_{i \in \mathcal{I}} \sigma_i g_i : \sigma_0, \sigma_i \in \Sigma[\boldsymbol{x}] \ \forall i \in \mathcal{I}\right\}. \tag{4.22}$$

**Theorem 4.32** (Putinar's positivstellensatz [Put93])**.** Let $\mathcal{S}$ be an *archimedean* basic semialgebraic set, i.e., $\mathcal{S}$ must be contained in the norm ball $\|\boldsymbol{x}\|^2 \leq R \in \mathbb{R}$. Then,

$$p \in P^+[\boldsymbol{x} \in \mathcal{S}] \Rightarrow p \in \operatorname{qmodule}(\boldsymbol{g}) + \langle \boldsymbol{f} \rangle. \tag{4.23}$$

**Remark 4.33.** Schmüdgen's and Putinar's positivstellensatz were formulated using a one-sided implication "$\Rightarrow$" only. This is in fact the hard part; for if any representation of $p$ in terms of the cone or quadratic module/ideal was known, then for sure $p \in P[\boldsymbol{x} \in \mathcal{S}]$ (but without strict positivity).

As with the different choices of prefactors in the unconstrained case, choosing between Krivine–Stengle, Schmüdgen, and Putinar will give rise to relaxations of different quality at the same level of degree truncation; a recent overview can be found in [LS25]. In practice, Putinar's formulation is by far the most common one. The best currently known bound on the maximal degree that is required in Putinar's positivstellensatz is given in the following.

**Lemma 4.34** (General Putinar degree bound [BMP25]).
Let $\mathcal{S} := \{\boldsymbol{x} \in \mathbb{R}^n : g_i(\boldsymbol{x}) \geq 0 \ \forall i \in \mathcal{I}\} \subset (-1, 1)^n$ be a nonempty basic semialgebraic set. Define $\varepsilon := \min_{\boldsymbol{x} \in \mathcal{S}} p(\boldsymbol{x}) > 0$.
In order to certify $p \in P_d^+[\boldsymbol{x} \in \mathcal{S}]$ using Putinar's positivstellensatz[5], the total degree of all polynomials involved in the quadratic module need not be larger than

$$c n^2 |\mathcal{I}| \left(\max_{i \in \mathcal{I}} \deg g_i\right)^6 d^{2L} \left(\frac{\|f\|}{\varepsilon}\right)^{7L+3}, \tag{4.24}$$

5 Note that on semialgebraic sets, $\deg p$ may be odd.

where $c$ and $L$ are some positive constants and $\|\bullet\|$ is the maximum norm of coefficients in the Bernstein basis; for details, see [BMP25]. While $L$ is large in general, if for every $\boldsymbol{x} \in \mathcal{S}$, the gradients of active constraints $\{\boldsymbol{\nabla} g_i(\boldsymbol{x}) : i \in \mathcal{I}, g_i(\boldsymbol{x}) = 0\}$ are linearly independent, then $L = 1$.

While this degree bound applies to all inequality-bound basic semialgebraic sets that are contained in the interior of the unit cube and has “only” a polynomial scaling (which is already a large improvement compared to the best previously known exponential bound in [NS07]), the exponents are still large. From practical experience, exact cutoffs can often be found at very low degrees.

This is quite relevant for polynomial optimization, where the polynomial under study is $p - \ell$, i.e., the difference between $p$ and the (unknown) lower bound $\ell$ that is to be found; so for Lemma 4.34, $\varepsilon$ defines the accuracy with which the lower bound is obtained. Naturally, the best possible accuracy is desired here, which for interior-point solvers means $\varepsilon \sim 10^{-8}$. Were the upper bound in Lemma 4.34 even close to tight, an $\varepsilon^{-10}$ dependence in the best possible case would be fatal.

Indeed, Putinar’s positivstellensatz (with finite truncation) holds true even when the polynomial has exact zeros—not only in special cases [Mar08; Sch05; Lau07; Nie13], but also generically under the archimedean condition [Nie14]—but not universally, and deciding if it holds true outside of a special case is in general $\mathcal{NP}$-hard [Var26]. Furthermore, there is a (quite isolated) result on the *minimal* degree required [BS24]: when $\mathcal{S} := [-1, 1]^n$ and $p = \prod_{i=1}^n (1 - x_i^2) + \varepsilon$ with $n \geq 2$, the total degree of all polynomials in the quadratic module has to be in $\Omega(1 / \sqrt[8]{\varepsilon})$. This weak dependence—while showing that indeed $\varepsilon = 0$ might not be possible in all cases—is more in accord with what practical experience tells.

### 4.3.2 Matrix constraints

While the generic form of the basic semialgebraic set given in equation (4.15) already covers this special case, it is particularly relevant for problems in quantum information to consider constraints of the form $G(\boldsymbol{x}) \succeq 0$; i.e., imposing a membership in the positive semidefinite cone on a matrix whose entries are polynomials in $\boldsymbol{x}$. By definition,

$$G(\boldsymbol{x}) \succeq 0 \Leftrightarrow \boldsymbol{y}^\top G(\boldsymbol{x}) \boldsymbol{y} \geq 0 \ \forall \boldsymbol{y} \Leftrightarrow \boldsymbol{y}^\top G(\boldsymbol{x}) \boldsymbol{y} \in P[\boldsymbol{x}, \boldsymbol{y}] \tag{4.25}$$

directly translates into finding out whether a polynomial is nonnegative in an augmented set of variables. Naturally, relaxing $P$ to $\Sigma$ then leads to the definition of sums-of-squares matrices:

**Definition 4.35** (Sums-of-squares matrices, scalarized). The set of $m \times m$ *sums-of squares matrices* up to degree $2d$ is defined by

$$\Sigma_{2d}^m[\boldsymbol{x}] := \{S \in \mathbb{R}_{2d}^{m\times m}[\boldsymbol{x}] : \boldsymbol{y}^\top S(\boldsymbol{x})\boldsymbol{y} \in \Sigma[\boldsymbol{x}, \boldsymbol{y}], \boldsymbol{y} \in \mathbb{R}^m\}. \tag{4.26}$$

In the proof to Proposition 4.2, several methods were already mentioned how membership in the positive semidefinite cone can be rewritten in terms of multiple scalar polynomial constraints. One of them was the characterization of any positive semidefinite matrix by the existence of a Cholesky factorization: $G \succeq 0 \Leftrightarrow G = M^\top M$, where $M$ might be rectangular. Once again, this is a "square" certificate for nonnegativity; in fact, if no requirements on $M$ are made, it is an exact equivalence. If one demands that $M$ itself be a matrix whose entries are polynomials, it is indeed only a certificate (except for the univariate case [CLR80]), and an alternative definition of SOS matrices.

**Definition 4.36** (Sums-of-squares matrices, factorized). The set of $m \times m$ *sums-of-squares matrices* up to degree $2d$ is defined by

$$\Sigma_{2d}^m[\boldsymbol{x}] := \{M^\top M \in \mathbb{R}_{2d}^{m\times m}[\boldsymbol{x}] : M \in \mathbb{R}_d^{t\times m}[\boldsymbol{x}], t \in \mathbb{N}\}. \tag{4.27}$$

Both definitions are in fact equivalent; however, while Definition 4.35 increases the variable number by $m$ and the total degree by 2, there is a more efficient characterization of Definition 4.36 without the detour to the scalar case.

**Lemma 4.37** (Scherer, Hol [SH06]). Let $S \in \mathbb{R}^{m\times m}[\boldsymbol{x}]$ be a polynomial matrix and let $\boldsymbol{u}(\boldsymbol{x})$ be a monomial basis with monomials up to the maximal degree in $S$.

$$S \in \Sigma^m[\boldsymbol{x}] \Leftrightarrow \exists Z \succeq 0 : S(\boldsymbol{x}) = [\boldsymbol{u}(\boldsymbol{x}) \otimes \mathbb{1}_m]^\top Z[\boldsymbol{u}(\boldsymbol{x}) \otimes \mathbb{1}_m] \tag{4.28}$$

Let the maximal degree in $S$ be $d$, then this characterization requires a semidefinite matrix $Z$ with side dimension $\binom{n+d}{n}m$, while the expanded scalar version would need side dimension $\binom{(n+m)+(d+1)}{n+m}$; the former is much more efficient than the latter.

Using the characterization of SOS matrices, a matrix-valued "Putinar"-certificate can be derived[6].

**Definition 4.38** (Quadratic module, matrix version). The *quadratic module* generated by the polynomials $\{g_i\}_{i\in\mathcal{I}}$ and the polynomial-valued matrices $\{G_k\}_{k\in\mathcal{M}}$ is the set

$$\operatorname{qmodule}(\boldsymbol{g}, \boldsymbol{G}) := \left\{\sigma_0 + \sum_{i\in\mathcal{I}} \sigma_i g_i + \sum_{k\in\mathcal{M}} \langle S_k(\boldsymbol{x}), G_k(\boldsymbol{x})\rangle : \begin{array}{r} \sigma_0, \sigma_i \in \Sigma[\boldsymbol{x}]\ \forall i \in \mathcal{I}, \\ S_k \in \Sigma^{\dim G_k}[\boldsymbol{x}]\ \forall k \in \mathcal{M} \end{array}\right\}. \tag{4.29}$$

6 The nomenclature used here is meant to make the similarities between the matrix-valued positivstellensatz and Putinar's scalar version obvious; however, the matrix version is due to Scherer and Hol [SH06]. In fact, they derived a more general version for robust polynomial optimization.

**Theorem 4.39** (Generalized Putinar's positivstellensatz)**.** Let

$$\mathcal{S} := \{\boldsymbol{x} \in \mathbb{R}^n : g_i(\boldsymbol{x}) \geq 0 \ \forall i \in \mathcal{I}, h_j(\boldsymbol{x}) = 0 \ \forall j \in \mathcal{E}, G_k(\boldsymbol{x}) \succeq 0 \ \forall k \in \mathcal{M}\} \tag{4.30}$$

be an archimedean basic semialgebraic set with explicit PSD constraints. Then,

$$p \in P^+[\boldsymbol{x} \in \mathcal{S}] \Rightarrow p \in \text{qmodule}(\boldsymbol{g}, \boldsymbol{G}) + \langle \boldsymbol{f} \rangle. \tag{4.31}$$

A degree bound analogous to Lemma 4.34, but adapted to the case of semidefinite constraints, has also been derived recently [Hua25].

## 4.4 Sparsity methods

The resulting semidefinite programs for polynomial optimization problems can be huge. While one reason is that the theoretically known upper bounds on degree cutoffs are very loose, the main issue is the scaling of the size of the monomial basis with respect to variable number $n$ and degree $d$, which is $\binom{n+d}{n}$. Given the $\mathcal{O}(s^{6.5})$ scaling of inner point algorithms with the side dimension $s = \binom{n+d}{n}$ of the semidefinite constraint matrix, this is prohibitive. However, there are methods that allow to reduce the size of the monomial basis; some of them are *exact*, i.e., they will only eliminate monomials whose coefficients in the solution of the full problem are guaranteed to be zero anyway, so that the reported optimal value is indeed as good as possible for the given degree bound. A second set of methods are instead approximations that lack these guarantees except for special cases [Nie+25], but will typically not hamper optimality too much.

### 4.4.1 Exact sparsity methods

#### 4.4.1.1 Newton polytope

**Definition 4.40** (Support and Newton polytope)**.** Let $p(\boldsymbol{x}) = \sum_{\boldsymbol{i}} \alpha_{\boldsymbol{i}} \boldsymbol{x}^{\boldsymbol{i}}$, then the *support* of $p$ is given by

$$\text{supp}(p) := \{\boldsymbol{i} : \alpha_{\boldsymbol{i}} \neq 0\} \tag{4.32}$$

and the associated *Newton polytope* is its convex hull,

$$\mathcal{N}(p) := \text{conv}(\text{supp}(p)). \tag{4.33}$$

**Theorem 4.41** (Reznick [Rez78])**.** Let $p$ be an SOS polynomial:

$$p(\boldsymbol{x}) = \sum_i q_i^2(\boldsymbol{x}) \Rightarrow \mathcal{N}(q_i) \subset \frac{1}{2}\mathcal{N}(p). \tag{4.34}$$

The Newton polytope method is very powerful in reducing the dimension of the semidefinite matrices required in the optimization.

**Example 4.42** (Continuation of Example 4.22)**.** When applying the various theorems to certify that the Motzkin form is nonnegative, now the Newton polytope is employed to reduce the dimensionality of the optimization problem. Whether a polynomial is SOS or not must not change under this reduction, while the solvers might give different numerical solutions for the decomposition. It is of particular note that the numerical difficulties reported before are completely removed by using the Newton polytope here. This is not unexpected: interior-point algorithms are by design very good in finding solutions in the interior of the cones; but the previous decompositions had just a few components, so that the optimal semidefinite matrices are quite rank-deficient. As the barrier function diverges when approaching the boundary of the cone—here, the set of noninvertible matrices—this is very problematic. Naturally, by moving to the Newton polytope, the matrices are much smaller, so the number of vanishing eigenvalues is reduced by a lot—or even nonexistent.

1. Hilbert–Artin (Corollary 4.17)
   In this case, the polynomial $q(x, y, z)m(x, y, z)$ is investigated, where $q$ may potentially contain all monomials up to degree 2. Previously, this led to the side dimension 35 for $\sigma$; after applying the Newton polytope, only 13 elements are needed, namely

| | | | | |
|---|---|---|---|---|
| $x^3y$ | $xy^3$ | $xz^3$ | $x^2y$ | $z^3$ |
| $x^2y^2$ | $xy^2z$ | $yz^3$ | $xy^2$ | |
| $x^2yz$ | $xyz^2$ | $z^4$ | $xyz$ | |

   Not only is the smaller SDP much more accessible numerically, but the returned solutions are simpler—basically, the exact solution that was given previously can be read off directly without solving any system of equations.
2. Pólya–Reznick (Lemma 4.19)
   Here, $q(x, y, z) = x^2 + y^2 + z^2$ is already fixed, so that it contains fewer powers than before. The side dimension after applying the Newton polytope now has only nine elements:

| | | | | |
|---|---|---|---|---|
| $x^3y$ | $x^2yz$ | $xy^2z$ | $xz^3$ | $z^4$ |
| $x^2y^2$ | $xy^3$ | $xyz^2$ | $yz^3$ | |

   Again, this small problem is simple to solve and the solution has the same five-squares structure as before.
3. Reznick (Lemma 4.20)
   This is unchanged with respect to Example 4.22, as every monomial listed in the basis given by the Newton polytope already has the same degree; therefore, taking the intersection with a homogeneous basis will not change the problem.

4. Lasserre (Theorem 4.21)
   The Newton polytope does not change the scaling in table 4.1 (unless for the infeasible $r_\varepsilon = 2$ case); the added terms require a fully dense basis. This makes Theorem 4.21 rather unattractive compared to the other methods.

#### 4.4.1.2 Further methods

The Newton polytope is the most popular way of reducing the number of possible basis monomials, but it is not the only exact one; various other results have been found over the years which can be employed on their own or in combination with others.

**Theorem 4.43** (Kojima, Kim, Waki [KKW05b]). Let $p(\boldsymbol{x}) = \sum_i q_i^2(\boldsymbol{x})$ be an SOS polynomial and $\mathcal{Q} = \cup_i \operatorname{supp}(q_i)$ the common support of all the decomposition polynomials. Partition $\mathcal{Q} = \mathcal{M} \cup \mathcal{N}$ into two disjoint sets and let $\mathcal{N}' := \mathcal{N} + \mathcal{N} \equiv \{\boldsymbol{x} + \boldsymbol{y} : \boldsymbol{x}, \boldsymbol{y} \in \mathcal{N}\}$. Then,

$$\left[\mathcal{N}' \cap \operatorname{supp}(p) = \emptyset \wedge \mathcal{N}' \cap (\mathcal{Q} + \mathcal{M}) = \emptyset\right] \Rightarrow \mathcal{Q} \subset \mathcal{M}. \tag{4.35}$$

An iterative way to build the decomposition $\mathcal{M} \cup \mathcal{N}$ and put as many supports as possible into $\mathcal{N}$ so that they can be discarded is described in [KKW05b].

**Theorem 4.44** (Diagonal consistency [Löf09]/special case in [CLR95, Proposition 3.7]). Let $p(\boldsymbol{x}) = \sum_i q_i^2(\boldsymbol{x})$ be an SOS polynomial. Then,

$$\{\boldsymbol{j} : 2\boldsymbol{j} \notin \operatorname{supp} p, \nexists \boldsymbol{k} \neq \boldsymbol{\ell} \text{ s.t. } 2\boldsymbol{j} = \boldsymbol{k} + \boldsymbol{\ell}\} \cap \bigcup_i \operatorname{supp}(q_i) = \emptyset, \tag{4.36}$$

*Proof.* Let a squared monomial $\boldsymbol{x}^{2\boldsymbol{j}}$ not appear as a term in $p$ and additionally, it may not be generated by the multiplication of two *distinct* other monomials $\boldsymbol{x}^{\boldsymbol{k}}$ and $\boldsymbol{x}^{\boldsymbol{\ell}}$. Therefore, the diagonal element in the row and column associated with $\boldsymbol{j}$ in the positive semidefinite matrix that represents the coefficients of the $q_i$ must be zero—which is the coefficient of $\boldsymbol{x}^{2\boldsymbol{j}}$ in $p$. Hence, the whole row and column must be zero; so they can just be dropped from the matrix and $\boldsymbol{x}^{\boldsymbol{j}}$ removed from the basis. □

Finally, the Newton polytope and diagonal consistency can be generalized into an algorithm that is based on facial reduction [PP14]. This originates in the simple observation that according to Definition 4.8, SOS polynomials form a convex cone; the optimization can thus be formally written as in Definition 3.19, with no need to invoke the semidefinite representation. Whenever a convex conic program is feasible, but not strictly feasible—i.e., its relative interior is empty while the feasible set is nonempty—the dimension of the cone can be reduced without changing the original problem [BW81; PP18]. For this, a reduction certificate must be found, which is another conic feasibility problem—not necessarily of a lesser dimension than the original problem, although it might be highly structured, which can lead to a smaller cost.

However, the cone may be simplified by an outer approximation—an approach that will also be used later in section 4.7 to completely depart from the semidefinite representation. In facial reduction, such an approximation will never be erroneous, i.e., change the result of the original optimization, but it may not lead to the smallest possible equivalent description of the problem. On the positive side, the approximation may be much easier to solve—for example, facial reduction on semidefinite programs can be approximated using linear programs [PP18], which is the same level of complexity that was necessary for the Newton polytope method.

Applied to SOS programs, [PP14] introduces an algorithm that is able to iteratively reduce the dimension of the SOS cone, parameterized by a set of monomials. It is based on a polyhedral approximation of the cone, leading to a sequence of linear programs. The quality of the reduction of course depends on the approximation, for which various choices are possible. The authors show two choices that exactly correspond to the Newton polytope and the diagonal consistency criterion, respectively, and they propose a simple new approximation that is never worse than either of the other two.

#### 4.4.1.3 Constrained methods

The previous methods were given for membership in the cone of SOS polynomials; by using the explicit form dictated by a positivstellensatz, they can also be applied to SOS polynomials on basic semialgebraic sets. This will be exemplified using Putinar's positivstellensatz.

Assume an SOS certificate of $p(\boldsymbol{x})$ on the basic semialgebraic set $\mathcal{S}$ defined as in equation (4.15) is to be found. Equation (4.23) can then be written as

$$p - \sum_{i \in \mathcal{I}} \sigma_i g_i - \sum_{j \in \mathcal{E}} p_j h_j \in \Sigma[\boldsymbol{x}] \tag{4.37}$$

where $\sigma_i \in \Sigma[\boldsymbol{x}]$ and $p_j \in \mathbb{R}[\boldsymbol{x}]$. By moving the constraint terms to the other side, the "overall" procedure is just finding a single sums-of-squares polynomial, although when looking in more detail, this contains several more terms that are to be found. Note that while $\sigma_i$ and $p_j$ are unknown, by introducing a global degree truncation, the maximal possible support of them is easily determined. Multiplying out the terms will then give a superset of what was $\operatorname{supp}(p)$ in the unconstrained case. All the techniques from Theorems 4.41, 4.43, and 4.44 are therefore still applicable, although they now start with a larger support of the polynomial.

### 4.4.2 Inexact sparsity methods

#### 4.4.2.1 Correlative sparsity

The first inexact sparsity method was described by Waki et al. in 2006 [Wak+06]. It analyzes the polynomial optimization or SOS membership problem with graph-theoretic means, which has been the blueprint for other inexact sparsity methods subsequently developed.

**Definition 4.45** (CSP graph). Let $p(\boldsymbol{x}) = \sum_{\boldsymbol{i}} \alpha_{\boldsymbol{i}} \boldsymbol{x}^{\boldsymbol{i}}$ with $\boldsymbol{x} \in \mathbb{R}^n$, then the *correlative sparsity pattern graph* is an undirected graph with nodes $\{1, \dots, n\}$ that has an edge $(k, \ell)$ if and only if the variables $x_k$ and $x_\ell$ appear simultaneously in a term in $p$, i.e.,

$$\exists \boldsymbol{i} \in \operatorname{supp} p : i_k > 0 \wedge i_\ell > 0. \tag{4.38}$$

From the CSP graph, all maximal cliques are extracted (often from a chordal extension of the CSP graph, as finding maximal cliques of a non-chordal graph is $\mathcal{NP}$-hard). Each maximal clique $\mathcal{C}_i$ is a subset of the variables (or rather, their indices). The “full” condition $\sigma \in \Sigma[\boldsymbol{x}]$ is now replaced by $\sigma = \sum_i \sigma_i$ where $\sigma_i \in \Sigma[x_{\mathcal{C}_{i,1}}, x_{\mathcal{C}_{i,2}}, \dots]$; effectively, this corresponds to partitioning the semidefinite matrix that represents $\sigma$ into a *block diagonal* matrix (with potentially overlapping blocks)—although there is no guarantee that the off-diagonal blocks would in fact be zero were the full problem to be solved. As conceptually, this corresponds to manually enforcing zero-constraints on the off-diagonal blocks of the large matrix, the feasible set is tightened and the supremum in the process of finding a nonnegative polynomial only becomes smaller. The lower bound to the infimum of the original problem can at worst be lower than necessary, but it will still be a lower bound. In practice, instead of putting additional constraints on the same large matrix, which would be detrimental and not helpful, the given sparsity pattern allows to decompose the large matrix into several smaller ones [Agl+88], improving runtime and memory consumption.

Incorporating constraints into the correlative sparsity by proceeding as in section 4.4.1.3 is possible, but impractical: By rewriting the constrained problem according to equation (4.37), several different cases have to be considered that crucially depend on the globally chosen degree cutoff:

1. If the global degree is such that a prefactor $\sigma_i$ is of degree zero (i.e., the degree of the constraint matches the global degree), then every term in this constraint will contribute an edge in the CSP graph, as if it were the objective.
2. If the degree of the prefactor is one for an equality constraint[7], then every variable will be multiplied with every term in the constraint. Therefore, every variable that appears *anywhere* in the constraint will have an edge to every other vertex in the CSP graph.

7 As the prefactors of inequality constraints are sums of squares, they cannot have odd degree.

3. If the degree of the prefactor is larger than one, then every variable will already be multiplied with all other variables in the prefactor alone. The CSP graph is therefore a full graph; it is its own single maximal clique and the CSP procedure will not yield any useful results.

Given that the last case is very likely to be the relevant one, the approach has to be modified slightly. The main issue was that the prefactor itself already mixes all variables and therefore leads to full graph; to avoid this, the form of the prefactor has to be modified. As suggested in [KKW05a], for correlative sparsity this is done such that a prefactor will be restricted to contain only the variables that appear in the associated constraint. The last case is now excluded; however, still every pair of variables that occur anywhere in a constraint polynomial (not just inside a term) gives rise to an edge in the CSP graph—unless the prefactor is of degree zero. This last observation was the motivation in [JM18] to consider "important" constraints for which the prefactor is chosen as usual, inducing lots of edges, and "less important" ones for which the prefactor is manually set to be just a nonnegative number, leading to fewer new terms in the CSP graph. The authors also give a heuristic that may be used to find out which constraints might be worth promoting in importance, leading to an iterative version of correlative sparsity.

Note that this kind of sparsity scheme may as well be employed in section 4.4.1.3 as a preprocessing step on which to apply the exact methods (which in total are then no longer guaranteed to give the same results as the dense relaxation).

#### 4.4.2.2 Term sparsity

While correlative sparsity can be helpful for reducing the problem size, it might also lead to worse bounds. Continuing along this road, *term sparsity* [WML21a; WML21b] will reduce the problem size even further, possibly lowering the bounds more. Its main use is in the presence of constraints, where it tries to bring the term sparsity pattern that was previously exploited for the objective to the constraints as well[8]. It is an iterative procedure that converges to a fixed point; each iteration will produce more dense relaxations, potentially coming closer to the fully dense case.

The algorithm starts by first considering all monomials that are involved at any place in the problem—they must for sure occur in the relaxation.[9] Let $p \in \mathbb{R}[\boldsymbol{x}]$ be a polynomial and

8 The schemes for term sparsity are different from the iteration heuristic in [JM18] for promoting the importance of constraints. Term sparsity works with graphs based on the problem description alone; importance promotion instead requires solving the relaxation. Additionally, in term sparsity, the contributions of the constraints can be of "intermediate" levels instead of the binary choice "important" or "unimportant."

9 Note that the description here looks quite different from the ones in the original papers on term sparsity (which however are also inconsistent among each other). This is only a superficial difference; in practice, the algorithm does the same as what the original ones seem to intend. The presentation here is streamlined to work with as few nested definitions as possible.

$\mathcal{S}$ a basic semialgebraic set according to equation (4.15). Then define the (initial) *support union* as

$$\mathcal{U}^{(k)} := \operatorname{supp}(p) \cup \bigcup_{i\in\mathcal{I}} \operatorname{supp}(g_i) \cup \bigcup_{j\in\mathcal{E}} \operatorname{supp}(h_j) \cup 2\mathcal{B}_p \tag{4.39}$$

with $k := 0$. Here, $\mathcal{B}_p$ is the monomial basis associated with the SOS membership test according to the Putinar certificate (see Theorem 4.9). In general, define $\mathcal{B}_q$ for any constraint $q = g_i$ or $q = h_j$ as well to be the monomial basis associated with the SOS prefactor for this constraint. The appearance of the term $2\mathcal{B}_p$ in the support union is heuristic and serves to improve the convergence [WML21a].

From the support union, a term sparsity pattern (TSP) graph $G_q^{(k)}$ can be constructed for every polynomial $q$ that appears in the problem. This is an undirected graph with vertices given by $\mathcal{B}_q$ that has an edge between the monomials $\boldsymbol{x}^{\boldsymbol{i}}$ and $\boldsymbol{x}^{\boldsymbol{j}}$ if and only if

$$\exists \boldsymbol{\alpha} \in \operatorname{supp}(\tilde{q}) : \boldsymbol{i} + \boldsymbol{j} + \boldsymbol{\alpha} \in \mathcal{U}^{(0)}. \tag{4.40}$$

Here, $\tilde{p} := 1$, $\tilde{g}_i := g_i$, and $\tilde{h}_j := h_j$ (so $\tilde{q}$ is the polynomial that is multiplied with the SOS prefactor associated to $q$).

Now each TSP graph is chordally extended to $\overline{G_q^{(k)}}$—this might for example be the maximal chordal extension that completes connected components ("block extension," giving rise to the TSSOS hierarchy [WML21a]) or the smallest chordal extension (which is the CS-TSSOS hierarchy [WML21b]). Each extension will group certain vertices in the graphs together; the bases of the SOS polynomials can now be partitioned into several smaller ones according to the grouping.

When the next iteration $k + 1$ of term sparsity is to be computed, the new support union $\mathcal{U}^{(k+1)}$ is defined similar to before; only now, the supports of the polynomials are replaced by the supports of the extended graphs $\overline{G_q^{(k)}}$ from the last iteration. To formalize this, the notion of a graph support is required.

**Definition 4.46** (Graph support)**.** The *p-support* of an undirected graph $G$ with edge set $\mathcal{E}(G)$ with respect to a polynomial $p$ is defined as

$$\operatorname{supp}_p(G) := \{\boldsymbol{i} + \boldsymbol{j} + \boldsymbol{\alpha} : \{\boldsymbol{i}, \boldsymbol{j}\} \in \mathcal{E}(G), \boldsymbol{\alpha} \in \operatorname{supp} p\}. \tag{4.41}$$

If $p$ is omitted, it is implicitly understood as the constant polynomial $1$.

With this definition, the support union at iteration $k + 1$ is given by

$$\mathcal{U}^{(k+1)} := \operatorname{supp}(\overline{G_p^{(k)}}) \cup \bigcup_{i\in\mathcal{I}} \operatorname{supp}_{g_i}(\overline{G_{g_i}^{(k)}}) \cup \bigcup_{j\in\mathcal{E}} \operatorname{supp}_{h_j}(\overline{G_{h_j}^{(k)}}). \tag{4.42}$$

**Remark 4.47.** When the chordal extension is the block extension, the fixed point is guaranteed to give the same optimal value as the dense relaxation (it might well be a dense pattern at the end).

**Remark 4.48** (Correlative term sparsity)**.** Correlative and term sparsity may also be considered together, leading to even smaller semidefinite matrices [Wan⁺22].

Again, in the presence of constraints it is possible to interleave the iterations in (correlative) term sparsity schemes with exact methods. While the use of such methods for the choice of the initial basis was already suggested in [WML21b], an exact method such as the Newton polytope could also be applied to reduce the size of a basis due to inexact sparsity further[10], which has not been suggested in the literature yet.

**Remark 4.49** (Matrix constraints)**.** Inexact sparsity methods can be extended to the case of semidefinite constraints [MWG24], but then Remark 4.47 no longer holds.

## 4.5 Symmetry methods

Sparsity methods remove elements from the monomial basis in order to reduce the size of the semidefinite matrices. Symmetry methods also perform modifications to the basis, and they are also exact methods. However, instead of simply removing individual monomials, the choice of monomials as basis elements itself is adapted. Using the monomial basis in the proof to Theorem 4.9 was merely a convenient and simple choice, but any kind of basis for the space of degree-bound polynomials could have been used instead to expand $p$ in (for the explicit description of how to use other bases, see Lemma 4.67 on page 107). If the optimization problem happens to be invariant under the action of a certain finite group $\mathcal{G}$—i.e., for all $\gamma \in \mathcal{G}$, $q \circ \gamma = q$ holds for all polynomials $q$ involved in the problem—then a *symmetry-adapted* basis can be constructed to represent the polynomials. The large semidefinite matrices associated with the SOS prefactors then factor according to the irreducible structure of the group. For details, see [GP04] and the extension to constrained problems in [Rie⁺13].

Note that combining symmetry and sparsity methods—in particular, exact sparsity methods—may not immediately work out. These methods were traditionally developed for the monomial basis, i.e., they can make use of the multiplicative identity $\boldsymbol{x}^{\boldsymbol{i}}\boldsymbol{x}^{\boldsymbol{j}} = \boldsymbol{x}^{\boldsymbol{i}+\boldsymbol{j}}$. As soon as the basis is changed, the multiplication law between two different basis elements has to be adjusted. However, for example the principle of diagonal consistency in Theorem 4.44 can be easily modified to work in a particular symmetry-adapted basis.

10 This may be useful only if constraints are present, for then the input to the exact methods is determined by the support of the objective, which will not be changed by sparsity methods. However, it is still possible to take the intersection of an exact with an inexact sparsity method *once* in order to reduce the size of the basis.

## 4.6 Multivariate constrained polynomial optimization over $\mathbb{C}$

Allowing the decision variables in polynomial optimization problems to become complex numbers is a crucial requirement for quantum information problems. In order to still make a minimization possible—which requires an order relation—the conjugates of the decision variables must be introduced as explicit variables; then, as already mentioned in Remark 4.1, by enforcing suitable conditions on the coefficients of the objective and the inequality constraints, these can be made real-valued.

In principle, the previous theory is already sufficient to completely cover the complex case. Once the problem was formulated in a complex manner, real and imaginary parts can be taken everywhere and everything is mapped back to the real-valued picture.

However, in [JM18], it was shown that by moving to an intrinsically complex-valued representation of SOS polynomials, a more refined control over the degree cutoff is possible. In this representation, only the non-conjugated variables are taken as part of the monomial basis; but when constructing the polynomial from its Hermitian matrix representation, this basis is multiplied in its original form from one side and in its conjugate form from the other side. This still allows to reconstruct the polynomial, but with differently chosen degree cutoffs. If the solver supports the Hermitian positive semidefinite cone natively, the presence of explicit semidefinite constraints as introduced in section 4.3.2 no longer implies a doubling of the side dimension; the structure can be used directly.

Different from the real-valued case, every element in the moment matrix now represents a *unique* monomial—no other moment matrix element will yield the same monomial. Therefore, in the unconstrained case, the moment matrix is basically (apart from the lower bound, which appears once on the diagonal) fixed by the problem and there is not much left for the optimizer to do. Thus, the complex-valued hierarchy of unconstrained problems is trivial; only when constraints are involved, an actual optimization starts to happen.

Regarding sparsity, using the intrinsically complex-valued formulation will not be an issue for inexact methods. Slight alterations have to be made when defining the support union [WM22], and for the purpose of correlative sparsity, $z$ and $\bar{z}$ should be considered as the same variable. The situation is different when looking at exact sparsity methods, for which no results have been published so far—for a good reason. Due to the aforementioned trivialization of the unconstrained case, every monomial that occurs in the objective (potentially after adding constraints as in section 4.4.1.3) in a conjugated or non-conjugated manner must be present in the basis, and nothing more. The latter directly follows from diagonal consistency: if a monomial were added which is not present in the objective, the coefficient of in front of its absolute value—which is zero—would occur on a diagonal of the semidefinite matrix; hence, the whole row and column must be zero, allowing to remove monomial from the basis again.

## 4.7 Modifying the relaxation: diagonal dominance

Sparsity and symmetry methods allow to reduce the size of the basis involved in the SOS relaxation of the polynomial optimization problem. While this is already desirable from the numerical point of view by making problems less degenerate, a main motivation comes from the issue of optimizing over huge semidefinite matrices—and therefore, ultimately, the expensive scaling of the PSD cone. If it were a linear cone or even a second-order cone instead, the mark of what constituted "huge" would shift by orders of magnitude; and indeed, there is a very simple relationship between the semidefinite and the linear cone.

**Theorem 4.50** (Geršgorin [Ger31]). Let $S := (s_{i,j})_{i=1,\,j=1}^{n \quad n} \in \mathbb{C}^{n\times n}$, then

$$\operatorname{spec} S \subset \bigcup_{i=1}^{n} \bar{U}_{r_i}(s_{i,i}) \quad \text{where} \quad r_i := \sum_{\substack{j=1\\ j\neq i}}^{n} |s_{i,j}| \tag{4.43}$$

and $\bar{U}_r(z_0) := \{z \in \mathbb{C} : |z - z_0| \leq r\}$ is the closed disk at $z_0$ with radius $r$.

This is a simple consequence of applying the triangular inequality to the eigenvalue equation. In the case of symmetric matrices, the eigenvalues will be real; therefore, the disk simply becomes a closed interval.

An immediate consequence of the Theorem is that if every diagonal element $s_{i,i}$ is larger or equal to the corresponding $r_i$, $S$ must be positive semidefinite; these sufficient conditions, which can easily be checked using $2n$ linear inequalities, define the *diagonally dominant* (DD) cone—an inner approximation of the PSD cone. Recall that a polynomial optimization problem, expressed via the SOS relaxation, is essentially the maximization in equation (4.4), where membership in the cone of nonnegative polynomials is replaced by membership in the quadratic module. This is in turn modeled via semidefinite constraints; and now, these constraints are tightened using the DD cone instead. The feasible set is smaller than before, so the supremum will not increase; the DD-SOS formulation still gives a global lower bound. While none of the usual statements about asymptotic convergence hold any more for this hierarchy, the DD-SOS hierarchy might still converge (though sometimes, an individual level in the hierarchy can even be infeasible); and even if not, its lower bounds can be meaningful to estimate the quality of a solution that was found by other means.

This idea was introduced in [AM14; AH15] and found greater recognition through [AM19]. A rather unexplored possibility is the change of basis that is already mentioned in the latter papers. This is based on the simple observation that a similarity transform $U^{-1}SU$ with any regular matrix $U$ does not change the spectrum; so checking $U^{-1}SU \in \mathrm{DD}$ is a different sufficient condition for $S \succeq 0$. In fact, if the "correct" $U$ is chosen—its columns forming an eigenbasis of the optimal $S^\star$—the corresponding DD optimization is guaranteed to deliver the optimal result. Since this correct $U$ cannot be known beforehand, an iterative scheme starting from $U = \mathbb{1}$ and using the Cholesky factor of the last solution as input to the next can easily be proven not to decrease from iteration to iteration, and will likely yield better results than the original optimization. However, note that such a rotation is computationally demanding, as it turns the coefficient matrix in the linear program from a sparse (only a few matrix entries are relevant in each linear constraint) into a dense one (all matrix entries feature in each constraint). This usually entails severe performance penalties; a possible, but not yet explored way to circumvent this is to implement the rotated DD as a native cone in a nonsymmetric solver, e.g., by computing its LHSC barrier function explicitly.

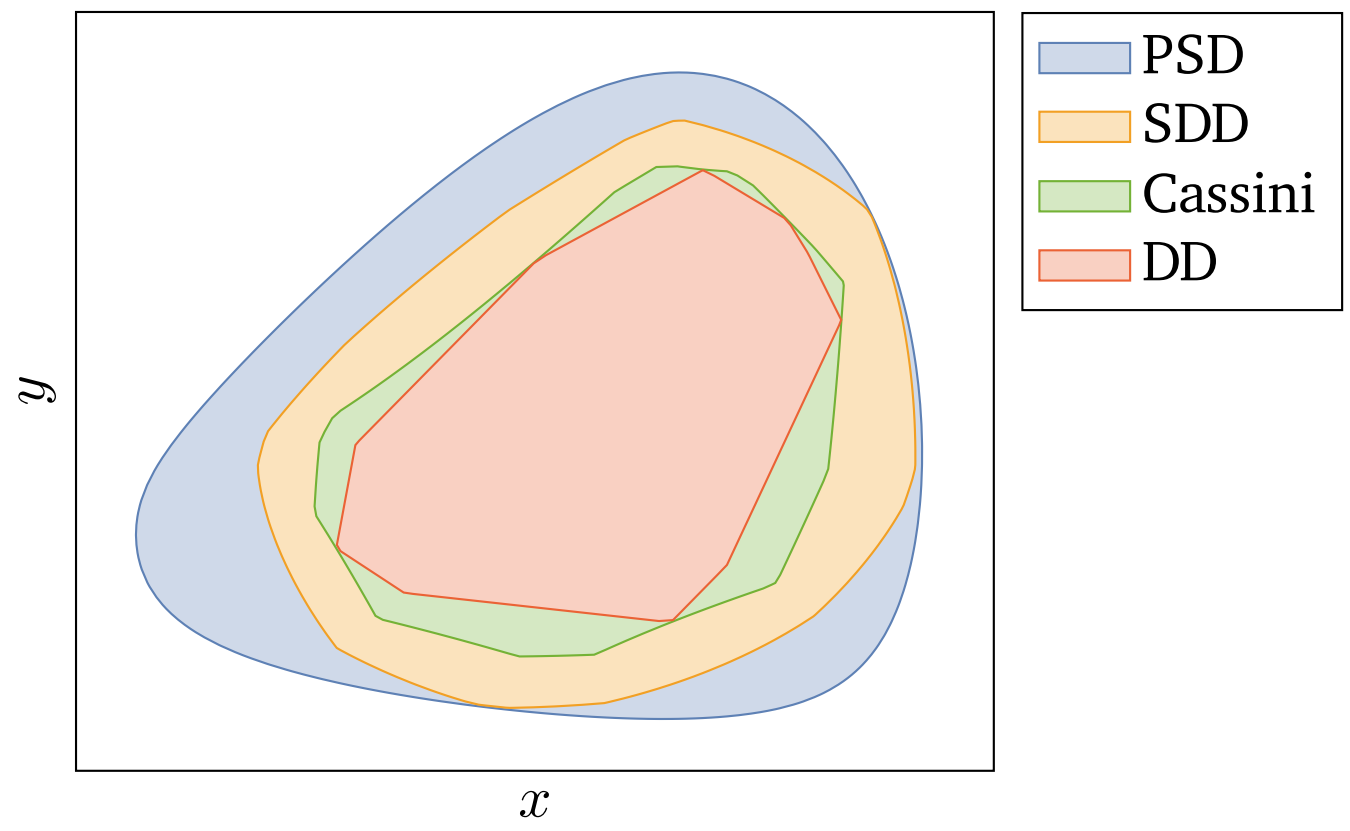

**Figure 4.1.** Comparison between the positive semidefinite cone (PSD) and its inner approximation through Geršgorin's (DD), Brauer's (Cassini) and Boman's (SDD) theorems.
The region plot was generated by checking whether $\mathbb{1}+xA+yB$ with two random symmetric $5 \times 5$ matrices $A$ and $B$ was a member of the respective cone.

A less restrictive inner approximation of the PSD cone is given by the *scaled diagonally dominant* (SDD) cone. Obviously, $S \succeq 0 \Leftrightarrow DSD \succeq 0$ with a diagonal and positive $D$ holds as well. The set of all such matrices comprises the SDD cone, and due to a theorem by Boman [Bom+05], an $n \times n$-dimensional SDD matrix can be described via $\frac{n(n-1)}{2}$ second-order cones. It is therefore representable in a convex optimization problem with $\mathcal{O}(n^2)$ fast and small cones instead of a single large one. Of course, iterative similarity transforms can lead to improvements here as well.

Figure 4.1 visualizes the inclusion relations of the various cones. For completeness, the criterion of Cassini's ovals due to Brauer [Bra47] is also included, which is an extension of Geršgorin's theorem to the second-order case: Multiplying pairs of DD conditions, the theorem yields $s_{i,i}s_{j,j} \geq r_i r_j \ \forall i > j \Rightarrow S \succeq 0$, which again is a set of $\frac{n(n-1)}{2}$ quadratic inequalities. While every DD matrix must also satisfy Brauer's check, the connection to SDD matrices is not as evident and seems not to be explored. The numerical example in figure 4.1 suggests that the Cassini criterion is stronger than SDD matrices; and given that the multiplicative condition is not convex, SDD seems to be the better choice in all respects.

## 4.8 The dual problem

So far, polynomial optimization was described in terms of sums-of-squares: a polynomial had to be representable in terms of an SOS polynomial, potentially with additional certificates of nonnegativity or ideal membership with respect to a basic semialgebraic set. The resulting optimization problem after a degree cutoff is a semidefinite program whose dual problem can be easily formulated. Primal and dual problem are related through Farkas's Lemma 3.29; and Lasserre showed [Las01] that the limit-feasible case can be excluded. In many situations, it can be useful to check whether the dual formulation is more suited to the input format that a certain solver expects. Many solvers only support an interface in the conic form, i.e., a condition "let $A(\boldsymbol{x}) \succeq 0$", instead of "let $X \succeq 0$, work with $X_{i,j}$;" so mimicking the latter with the former is naturally more inefficient.

For the univariate unconstrained case, the dual problem was explicitly formulated in Theorem 4.12. In the following, this will be generalized to the multivariate and constrained case. While this is straightforward, a few definitions are required; the real-valued case is considered first.

**Definition 4.51** (Canonical basis)**.** The canonical basis of monomials in $n$ variables up to degree $d$ is indexed by the set of exponents

$$\mathcal{B}_{n,d} := \{\boldsymbol{i} \in \mathbb{N}_0^d : |\boldsymbol{i}| \leq d\}. \tag{4.44}$$

**Definition 4.52** (Riesz functional)**.** Let $p(\boldsymbol{x}) := \sum_{\boldsymbol{i}\in\mathcal{B}_{n,d}} \alpha_{\boldsymbol{i}} \boldsymbol{x}^{\boldsymbol{i}}$ be a polynomial and $(y_{\boldsymbol{i}})_{\boldsymbol{i}\in\mathcal{B}_{n,d}} \subset \mathbb{R}$ be a (multi-)indexed sequence with at least the same index set as the coefficients of $p$. The *Riesz functional* is defined as $L_{\boldsymbol{y}}(p) := \sum_{\boldsymbol{i}\in\mathcal{B}_{n,d}} y_{\boldsymbol{i}}\alpha_{\boldsymbol{i}}$.

**Definition 4.53** (Truncated moment matrix)**.** Let $\boldsymbol{y} \equiv (y_{\boldsymbol{i}})_{\boldsymbol{i}\in\mathcal{B}_{n,d}} \subset \mathbb{R}$ be a (multi-)indexed sequence. Define the *truncated moment matrix* $M_d(\boldsymbol{y})$ with rows and columns indexed by $\mathcal{B}_{n,d}$ such that for $\boldsymbol{i}, \boldsymbol{j} \in \mathcal{B}_{n,d}$,

$$M_d(\boldsymbol{y})_{\boldsymbol{i},\boldsymbol{j}} := y_{\boldsymbol{i}+\boldsymbol{j}}, \tag{4.45}$$

i.e., $M_d(\boldsymbol{y})$ is a multivariate Hankel matrix.

**Definition 4.54** (Localizing matrix)**.** Let $g(\boldsymbol{x}) := \sum_{\boldsymbol{k}\in\mathcal{B}_{n,d'}} \gamma_{\boldsymbol{k}} \boldsymbol{x}^{\boldsymbol{k}}$ where $d' := \deg g$. Let $\boldsymbol{y} \equiv (y_{\boldsymbol{i}})_{\boldsymbol{i}\in\mathcal{B}_{n,d}} \subset \mathbb{R}$ be a (multi-)indexed sequence.
Then, the *truncated localizing matrix* $M_d(g\boldsymbol{y})$ is defined for $\boldsymbol{i}, \boldsymbol{j} \in \mathcal{B}_{n,d}$ as

$$M_d(g\boldsymbol{y})_{\boldsymbol{i},\boldsymbol{j}} := \sum_{\boldsymbol{k}\in\mathcal{B}_{n,d'}} \gamma_{\boldsymbol{k}} y_{\boldsymbol{i}+\boldsymbol{j}+\boldsymbol{k}}. \tag{4.46}$$

**Example 4.55.** Let $n = 2$ and $g(\boldsymbol{x}) = a - bx_1^2 - cx_2^2$.

$$L_{\boldsymbol{y}}(g) = ay_{0,0} - by_{2,0} - cy_{0,2}$$

$$M_2(\boldsymbol{y}) = \left(\begin{array}{c|cc|ccc} y_{0,0} & y_{1,0} & y_{0,1} & y_{2,0} & y_{1,1} & y_{0,2} \\ \hline y_{1,0} & y_{2,0} & y_{1,1} & y_{3,0} & y_{2,1} & y_{1,2} \\ y_{0,1} & y_{1,1} & y_{0,2} & y_{2,1} & y_{1,2} & y_{0,3} \\ \hline y_{2,0} & y_{3,0} & y_{2,1} & y_{4,0} & y_{3,1} & y_{2,2} \\ y_{1,1} & y_{2,1} & y_{1,2} & y_{3,1} & y_{2,2} & y_{1,3} \\ y_{0,2} & y_{1,2} & y_{0,3} & y_{2,2} & y_{1,3} & y_{0,4} \end{array}\right)$$

$$M_1(g\boldsymbol{y}) = a\left(\begin{array}{c|cc} y_{0,0} & y_{1,0} & y_{0,1} \\ \hline y_{1,0} & y_{2,0} & y_{1,1} \\ y_{0,1} & y_{1,1} & y_{0,2} \end{array}\right) - b\left(\begin{array}{c|cc} y_{2,0} & y_{3,0} & y_{2,1} \\ \hline y_{3,0} & y_{4,0} & y_{3,1} \\ y_{2,1} & y_{3,1} & y_{2,2} \end{array}\right) - c\left(\begin{array}{c|cc} y_{0,2} & y_{1,2} & y_{0,3} \\ \hline y_{1,2} & y_{2,2} & y_{1,3} \\ y_{0,3} & y_{1,3} & y_{0,4} \end{array}\right)$$

**Remark 4.56** (Origin of the names)**.** The problem of finding the (unconstrained) minimum of a function $p$ alternatively can be seen as finding a nonnegative, normalized *measure* $\mu$ that minimizes the integral $\int p \,\mathrm{d}\mu$; naturally, this will be satisfied by any convex combination of Dirac measures at the minima of $p$. This is illustrated in figure 4.2. Restricting the minimization to a basic semialgebraic set $\mathcal{S}$ is easily done by formally demanding that $\mu$ be part of the set $\mathcal{M}(\mathcal{S})$ of nonnegative and normalized Borel measures on $\mathcal{S}$; "nonnegative" means that for all continuous test functions $\phi\colon \mathcal{S} \to \mathbb{R}$ that are nonnegative on $\mathcal{S}$, $\int \phi \,\mathrm{d}\mu \geq 0$ must also hold.
Formally,

$$p^\star = \inf_{x\in\mathbb{R}} \sum_{\boldsymbol{k}\in\mathcal{B}_{n,2d}} \alpha_{\boldsymbol{k}} \boldsymbol{x}^{\boldsymbol{k}} = \inf_{\mu\in\mathcal{M}(\mathbb{R}^n)} \left\{ \int \sum_{\boldsymbol{k}\in\mathcal{B}_{n,2d}} \alpha_{\boldsymbol{k}} \boldsymbol{x}^{\boldsymbol{k}} \,\mathrm{d}\mu : \int \mathrm{d}\mu = 1, \mu \geq 0 \right\}.$$

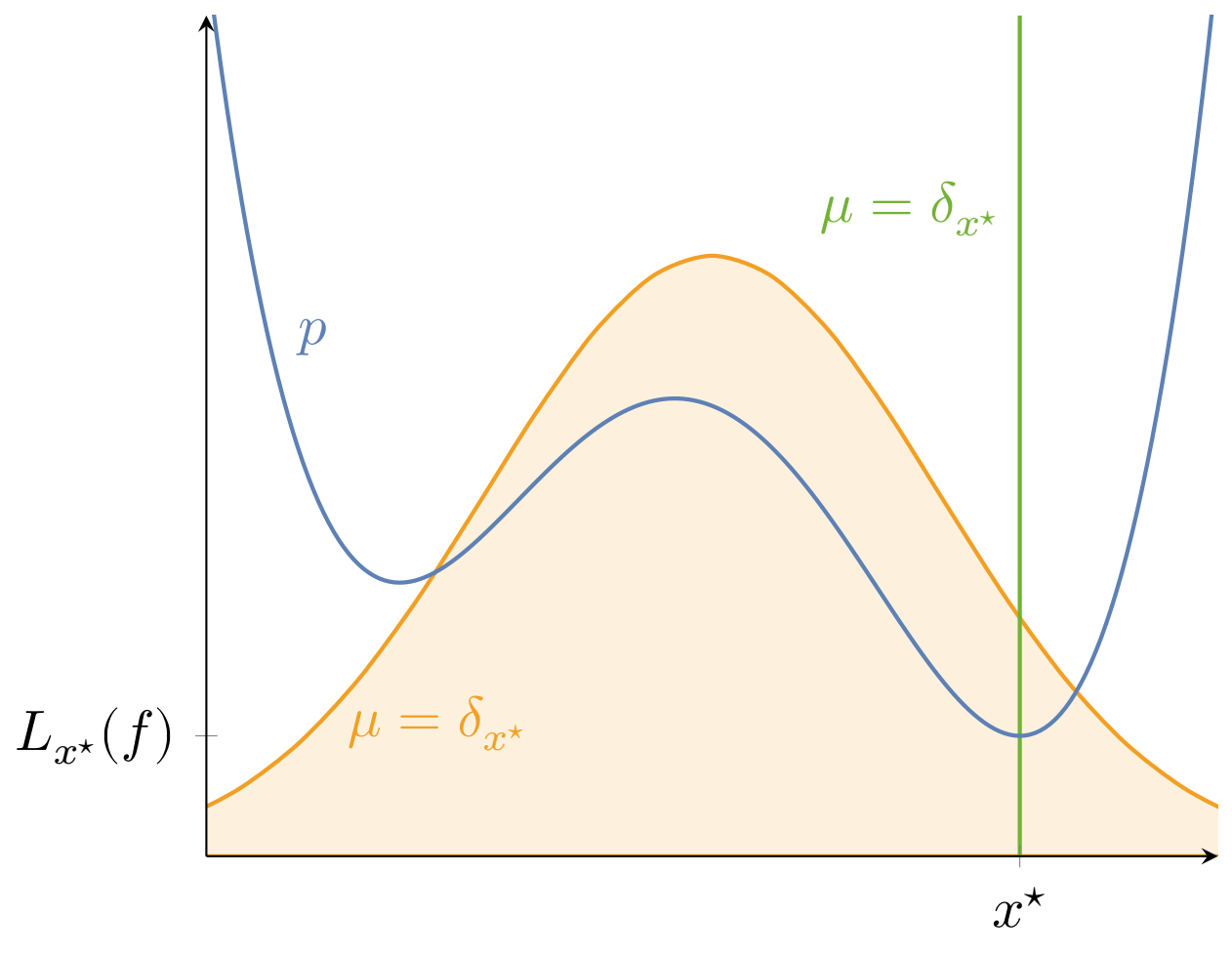

**Figure 4.2.** Minimizing a function vs. finding an appropriate measure
Instead of minimizing the function $p$ (blue), an integral $\int p \,\mathrm{d}\mu$ can be considered. Conceptually, this will sum up each value of $p$ with a certain weight given by the normalized measure $\mu$ (shaded in yellow for an arbitrary measure); when $\mu$ is concentrated at the true minimum (green), the integral is minimized.

The above definition of nonnegativity of the measure is inserted.

$$= \inf_{\mu\in\mathcal{M}(\mathbb{R}^n)} \left\{ \int \sum_{\boldsymbol{k}\in\mathcal{B}_{n,2d}} \alpha_{\boldsymbol{k}} \boldsymbol{x}^{\boldsymbol{k}} \,\mathrm{d}\mu \;\middle|\; \int \mathrm{d}\mu = 1, \forall \phi\colon \mathbb{R}^n \to \mathbb{R} \text{ continuous} : \phi \geq 0 \Rightarrow \int \phi \,\mathrm{d}\mu \geq 0 \right\}$$

The set of all nonnegative functions is too difficult to describe; the problem is relaxed to the set of all sums-of-squares polynomials, and a degree truncation is introduced for practical purposes.

$$\geq \inf_{\mu\in\mathcal{M}(\mathbb{R}^n)} \left\{ \int \sum_{\boldsymbol{k}\in\mathcal{B}_{n,2d}} \alpha_{\boldsymbol{k}} \boldsymbol{x}^{\boldsymbol{k}} \,\mathrm{d}\mu \;\middle|\; \int \mathrm{d}\mu = 1, \forall \sigma \in \mathbb{R}_d[\boldsymbol{x}] : \int \sigma^2 \,\mathrm{d}\mu \geq 0 \right\}$$

Now $\sigma$ is easily parameterized.

$$= \inf_{\mu\in\mathcal{M}(\mathbb{R}^n)} \left\{ \int \sum_{\boldsymbol{k}\in\mathcal{B}_{n,2d}} \alpha_{\boldsymbol{k}} \boldsymbol{x}^{\boldsymbol{k}} \,\mathrm{d}\mu \;\middle|\; \int \mathrm{d}\mu = 1, \forall \boldsymbol{\beta} \in \mathbb{R}^{|\mathcal{B}_{n,d}|} : \int \sum_{\boldsymbol{i},\boldsymbol{j}\in\mathcal{B}_{n,d}} \beta_{\boldsymbol{i}} \beta_{\boldsymbol{j}} \boldsymbol{x}^{\boldsymbol{i}+\boldsymbol{j}} \,\mathrm{d}\mu \geq 0 \right\}$$

Moving all coefficients in front of the integrals and arranging the integrals of the powers in a matrix, this is the definition of the matrix being positive semidefinite.

$$= \inf_{\mu\in\mathcal{M}(\mathbb{R}^n)} \left\{ \sum_{\boldsymbol{k}\in\mathcal{B}_{n,2d}} \alpha_{\boldsymbol{k}} \int \boldsymbol{x}^{\boldsymbol{k}} \,\mathrm{d}\mu : \int \mathrm{d}\mu = 1, \left( \int \boldsymbol{x}^{\boldsymbol{i}+\boldsymbol{j}} \,\mathrm{d}\mu \right)_{\boldsymbol{i},\boldsymbol{j}\in\mathcal{B}_{n,d}} \succeq 0 \right\}$$

Finally, recognizing that since the measure $\mu$ now only enters by its moments $\int \boldsymbol{x}^{\boldsymbol{k}} \,\mathrm{d}\mu$ for $\boldsymbol{k} \in \mathcal{B}_{n,2d}$, the measure itself can be replaced by its vector of moments.

$$p^\star = \inf_{\boldsymbol{y}\in\mathbb{R}^{|\mathcal{B}_{n,2d}|}} \left\{ \sum_{\boldsymbol{k}\in\mathcal{B}_{n,2d}} \alpha_{\boldsymbol{k}} y_{\boldsymbol{k}} : y_{\boldsymbol{0}} = 1, M_d(\boldsymbol{y}) \succeq 0, \exists \mu \in \mathcal{M}(\mathbb{R}^n), \mu \geq 0, y_{\boldsymbol{k}} = \int \boldsymbol{x}^{\boldsymbol{k}} \,\mathrm{d}\mu \;\forall \boldsymbol{k} \in \mathcal{B}_{n,2d} \right\}$$

Note that here, the left-hand side is indeed $p^\star$ again. The detour with the restriction to verify nonnegativity of $\mu$ for degree-bound SOS polynomials was helpful for understanding how this expression arises, but it is undone by again demanding $\mu \geq 0$. While the measure is no longer part of the explicit decision variables, it is still implicitly present in the problem, which is a complication that can now be relaxed in a different manner.

$$\geq \inf_{\boldsymbol{y} \in \mathbb{R}^{|\mathcal{B}_{n,2d}|}} \left\{ \sum_{\boldsymbol{k} \in \mathcal{B}_{n,2d}} \alpha_{\boldsymbol{k}} y_{\boldsymbol{k}} : y_{\boldsymbol{0}} = 1, M_d(\boldsymbol{y}) \succeq 0 \right\}$$

In summary, the assumption underlying the dual perspective is that there exists a nonnegative measure (on the semialgebraic set defined by the constraints) that is compatible with the moments given in the moment matrix. The measure itself is never used in the optimization problem itself, only the functional $\int \bullet \, \mathrm{d}\mu$ by means of its Riesz representation, which is stored in $\boldsymbol{y}$.

Applying Putinar's positivstellensatz (Theorem 4.32), truncated at order $d$, to the polynomial optimization problem $\inf_{\boldsymbol{x} \in \mathcal{S}} p(\boldsymbol{x})$ with $\mathcal{S}$ defined according to equation (4.15) yields

$$\sup\Bigl\{ \ell : \ell \in \mathbb{R}, \sigma_0 \in \Sigma_d[\boldsymbol{x}], \sigma_i \in \Sigma_{d - \lceil \deg g_i/2 \rceil}[\boldsymbol{x}] \ \forall i \in \mathcal{I}, p_j \in \mathbb{R}_{2d - \deg h_i}[\boldsymbol{x}] \ \forall j \in \mathcal{E}, \\ p - \ell = \sigma_0 + \sum_{i \in \mathcal{I}} \sigma_i g_i + \sum_{j \in \mathcal{E}} p_j h_j \Bigr\}, \quad (4.47)$$

Following the lines of the derivation in Remark 4.56, its dual so-called *Lasserre relaxation* of order $d$ is thus given by

$$\inf\Bigl\{ L_{\boldsymbol{y}}(p) : \boldsymbol{y} \in \mathbb{R}^{\binom{n+2d}{n}}, y_{\boldsymbol{0}} = 1, M_d(\boldsymbol{y}) \succeq 0, \\ M_{d - \lceil \deg g_i/2 \rceil}(g_i \boldsymbol{y}) \succeq 0 \ \forall i \in \mathcal{I}, M_{d - \lceil \deg h_i/2 \rceil}(h_i \boldsymbol{y}) = 0 \ \forall j \in \mathcal{E} \Bigr\}. \quad (4.48)$$

Note that equality constraints were written by setting the corresponding localizing matrix to zero; in this way, no new definitions had to be introduced. However, this matrix will have a lot of duplicates, leading to the same equality constraints. For a numerical solver, having many linearly dependent constraints can be a severe problem[11], so care should be taken to use every unique equality only once.

In a manner of speaking, the SOS formulation "extracts" entries from a semidefinite variable and puts constraints on them; the moment formulation instead "puts" entries into a semidefinite variable, which is the native format for many solvers as described before.

11 Solvers often contain a preprocessing stage to remove linear dependencies. However, this can introduce new problems, for in floating point arithmetic, the question of linear dependence cannot be answered unambiguously in most cases. It is therefore much better to avoid generating linearly dependent constraints instead of relying on a presolving phase to remove them, if possible.

**Remark 4.57** (Reformulation–Linearization). The approach to polynomial optimization by means of SOS or moment relaxations was predated by the reformulation–linearization technique (RLT) [ST92]. Using the RLT, the original set of constraints is first artificially inflated by introducing a lot of redundant constraints. This is in close analogy with the lines that led to Definitions 4.23, 4.25, and 4.27 of the cone, ideal, and preorder; but this time, the prefactors are not arbitrary (nonnegative) polynomials, but instead other constraints: if $p_1(\boldsymbol{x}) \geq 0$ and $p_2(\boldsymbol{x}) \geq 0$ for every $\boldsymbol{x} \in \mathcal{X}$, then for sure also $p_1(\boldsymbol{x})p_2(\boldsymbol{x}) \geq 0 \; \forall x \in \mathcal{X}$. After this *reformulation* of the problem, it is *linearized*: a monomial $\boldsymbol{x}^{\boldsymbol{i}}$ is replaced by a variable $y_{\boldsymbol{i}}$, exactly as it is done in the moment formulation. The emerging linear program can then be solved and might even lead to a tight relaxation of the original problem through the correlations induced by the redundant constraints. While the RLT shares many ideas with the moment hierarchy, the latter is in general superior in terms of tightness [Las02], although the former does not require the costly semidefinite constraints, which is a distinct advantage. Still, the RLT also converges asymptotically (but not necessarily finitely) [Las05], and interesting ideas have been developed [SF02] to modify the cuts in the RLT to be more “inspired” by the semidefinite geometry—though still linear.
In a recent result, the RLT has been extended to a reformulation–perspectification technique [MHZ21]: if the functions in the original problem were not just polynomials, but also contained convex conic parts—say, exponential cones for entropic optimization—then the relaxed reformulated problem is not a linear program anymore. Instead, perspective functions are introduced to preserve the convexity, leading to a relaxation of the polynomial parts together with the original convex parts. While this connection has not been explored yet, it certainly seems viable to use this technique to extend the moment relaxation approach of polynomial optimization problems to allow for convex non-polynomial functions.

The interpretation of the objective is clear: by strong duality, it is again a global lower bound to the original problem. However, the decision variables $\boldsymbol{y}$ now directly contain the moments of the original decision variables, and therefore possibly information about their optimal value. Ways to extract this data were already mentioned on page 76.

Instead of extracting the data, it might be sufficient to certify that a given bound is tight, which is often possible and easier than applying a solution extraction algorithm:

**Lemma 4.58** (Curto–Fialkow [CF00]). Let $\boldsymbol{y}^\star$ be the solution of a Lasserre relaxation of order $d$. Define $g_0 := p$. The relaxation was globally optimal if

$$\operatorname{rk} M_d(\boldsymbol{y}^\star) = \operatorname{rk} M_{d - \max\limits_{i \in \{0\} \cup \mathcal{I}} \lceil \deg g_i / 2 \rceil}(\boldsymbol{y}^\star).$$

This so-called *flat truncation criterion* actually verifies the existence of a $\operatorname{rk} M_d(\boldsymbol{y}^\star)$ atomic measure—i.e., a measure that is made up of $\operatorname{rk} M_d(\boldsymbol{y}^\star)$ Dirac measures (point evaluations) with positive weights. This implies that the last inequality in Remark 4.56 was in fact an equality—so the actual minimum must necessarily coincide with the bound.

A result by Blekherman [Ble14], recently extended for constrained problems, compares the rank of the moment matrix to a function of the number of variables only:

**Lemma 4.59** (Lasserre [Las25])**.** Let $\boldsymbol{y}^\star$ be the solution of a Lasserre relaxation of order $d$ in $n$ variables. The relaxation was globally optimal if

$$\operatorname{rk} M_d(\boldsymbol{y}^\star) \leq n - \max_{i \in \mathcal{J}} \lceil \deg g_i \,/\, 2 \rceil + 1.$$

**Remark 4.60** (PSD constraints [HL06])**.** The correspondence between Putinar's positivstellensatz and the Lasserre relaxation can be extended to explicitly cover positive semidefinite polynomial constraints. Previously, they were tackled by means of SOS matrices and Lemma 4.37. In the dual approach, write the matrix $G \in \mathbb{R}^{m \times m}[\boldsymbol{x}]$ in the matrix constraint $G \succeq 0$ in terms of a finite family $\{\mathit{\Gamma}_{\boldsymbol{k}}\}_{\boldsymbol{k}} \subset \mathbb{R}^{m \times m}$ of symmetric matrices: $G = \sum_{\boldsymbol{k}} \mathit{\Gamma}_{\boldsymbol{k}} \boldsymbol{x}^{\boldsymbol{k}}$. The corresponding localizing (block) matrix $M_d(G\boldsymbol{y})$ is then

$$M_d(G\boldsymbol{y})_{\boldsymbol{i},\boldsymbol{j}} := \sum_{\boldsymbol{k}} \mathit{\Gamma}_{\boldsymbol{k}} y_{\boldsymbol{i}+\boldsymbol{j}+\boldsymbol{k}}. \tag{4.49}$$

**Example 4.61.** Let $n = 2$, $G(\boldsymbol{x}) = \begin{pmatrix} 1-4x_1x_2 & x_1 \\ x_1 & 4-x_1^2-x_2^2 \end{pmatrix}$.

$$M_1(G\boldsymbol{y}) = \left(\begin{array}{c|cc} M_{0,0}(G\boldsymbol{y}) & M_{1,0}(G\boldsymbol{y}) & M_{0,1}(G\boldsymbol{y}) \\ \hline M_{1,0}(G\boldsymbol{y}) & M_{2,0}(G\boldsymbol{y}) & M_{1,1}(G\boldsymbol{y}) \\ M_{0,1}(G\boldsymbol{y}) & M_{1,1}(G\boldsymbol{y}) & M_{0,2}(G\boldsymbol{y}) \end{array}\right)$$

where

$$M_{0,0}(G\boldsymbol{y}) = \begin{pmatrix} y_{0,0}-4y_{1,1} & y_{1,0} \\ y_{1,0} & 4-y_{2,0}-y_{0,2} \end{pmatrix} \qquad M_{2,0}(G\boldsymbol{y}) = \begin{pmatrix} y_{2,0}-4y_{3,1} & y_{3,0} \\ y_{3,0} & 4y_{2,0}-y_{4,0}-y_{2,2} \end{pmatrix}$$

$$M_{1,0}(G\boldsymbol{y}) = \begin{pmatrix} y_{1,0}-4y_{2,1} & y_{2,0} \\ y_{2,0} & 4y_{1,0}-y_{3,0}-y_{1,2} \end{pmatrix} \qquad M_{1,1}(G\boldsymbol{y}) = \begin{pmatrix} y_{1,1}-4y_{2,2} & y_{2,1} \\ y_{2,1} & 4y_{1,1}-y_{3,1}-y_{1,3} \end{pmatrix}$$

$$M_{0,1}(G\boldsymbol{y}) = \begin{pmatrix} y_{0,1}-4y_{1,2} & y_{1,1} \\ y_{1,1} & 4y_{0,1}-y_{2,1}-y_{0,3} \end{pmatrix} \qquad M_{0,2}(G\boldsymbol{y}) = \begin{pmatrix} y_{0,2}-4y_{1,3} & y_{1,2} \\ y_{1,2} & 4y_{0,2}-y_{2,2}-y_{0,4} \end{pmatrix}$$

**Remark 4.62** (Complex-valued problems [JM18])**.** In the complex-valued case, the "normal" and conjugated variables must be treated as distinct. Consequently, the vector of moments now has two multi-indices: one for the normal complex variables and another for their conjugated version. The moment matrix element $(\boldsymbol{i}, \boldsymbol{j})$ is no longer formed by the addition of indices $y_{\boldsymbol{i}+\boldsymbol{j}}$; instead, it is simply $y_{\boldsymbol{i},\boldsymbol{j}}$, and similarly for the localizing matrices. Consequently, the moment matrix is no longer a Hankel matrix. While the machinery introduced before will work with these new definitions, the flat truncation criterion has to be modified.

**Lemma 4.63** (Flat truncation replacement for complex-valued problems [JM18]). Assume that one of the constraints of the multivariate optimization problem is a ball, $\|\boldsymbol{z}\| \leq r$ for some $r > 0$. Consider the optimal moment vector $\boldsymbol{y}^\star$ for some relaxation level $d$. If there exists a $t \in \mathbb{Z}$ with $\max_{i \in \{0\} \cup \mathcal{I}} \lceil \deg g_i \,/\, 2 \rceil \leq t \leq d$ such that one of the following two points is true, global optimality is attained.

- $\operatorname{rk} M_t(\boldsymbol{y}^\star) = 1$
- $t \geq 2$, $\operatorname{rk} M_t(\boldsymbol{y}^\star) = \operatorname{rk} M_{t-d_K}(\boldsymbol{y}^\star)$ and for all $1 \leq i < j \leq n$,

$$\begin{pmatrix} M_{t-d_K}(\boldsymbol{y}^\star) & M_{t-d_K}(z_i \boldsymbol{y}^\star) & M_{t-d_K}(z_j \boldsymbol{y}^\star) \\ M_{t-d_K}(\bar{z}_i \boldsymbol{y}^\star) & M_{t-d_K}(|z_i|^2 \boldsymbol{y}^\star) & M_{t-d_K}(z_j \bar{z}_i \boldsymbol{y}^\star) \\ M_{t-d_K}(\bar{z}_j \boldsymbol{y}^\star) & M_{t-d_K}(z_i \bar{z}_j \boldsymbol{y}^\star) & M_{t-d_K}(|z_j|^2 \boldsymbol{y}^\star) \end{pmatrix} \succeq 0,$$

where $d_K \coloneqq \max\{2, d_1, d_2, \dots\}$.

## 4.9 Numerical considerations

### 4.9.1 Monomial basis

Both the SOS and the moment approach made extensive use of the monomial basis. In section 4.5, it was already mentioned that the choice of a different basis might be preferable if the problem gives rise to a certain structure. However, even for completely unstructured problems, the choice of the monomial basis has drawbacks. A very common issue in numerical optimization is *scaling*: floating-point numbers only have a limited precision, so that the result of an arithmetic operation applied on numbers with vastly different orders of magnitude may not be accurately representable. However, the monomial basis naturally contains variables with a multitude of different powers. Therefore, even if the original decision variables all have the same order of magnitude, high-and low-degree monomials will not be as well-behaved. Scaling the basis—i.e., introducing a degree-dependent scaling prefactor on the monomials—might help against this issue, unless the required inverse scaling in the polynomials reintroduces the problem at a different place.

The operations that a solver has to perform are of course dependent on the type of solver itself. For interior-point methods, the LHSC barrier of the semidefinite cone has to be calculated and gradients and Hessians taken, which involves the determinant and inverse of the moment matrix. For univariate problems, this means that a classic Hankel matrix has to be inverted—if a solver is aware of this fact, it can employ fast or superfast algorithms [TA99].

While this might seem unnecessary in the univariate case, where the largest matrix side dimension is just $\deg p + 1$, Hankel matrices also quickly become ill-conditioned [Bec00]; using a structure-exploiting algorithm therefore is still advisable, as they do not seem to be as susceptible to bad conditioning [Xi+14; XX16].

For multivariate problems, preliminary numerical results and [HKM18] suggest that the condition number of multivariate Hankel matrices for a fixed number of variables is still exponential in the degree; however, for a fixed degree, it can be affine in the number of variables[12]. In practice, problems usually feature rather low degrees, but a high number of variables, so that the bad conditioning is not necessarily an issue—but the matrix sizes grow, and no fast algorithms for multivariate Hankel matrices are currently available.

In general, if conditioning turns out to lead to numerical issues, the choice of basis can be modified in a more radical manner than just scaling: by using another set of orthogonal polynomials. A basis with good numerical properties is given by the Chebyshev polynomials $\{T_i(x)\}_{i\in\mathbb{N}_0}$ of first kind. They obey the multiplicative identity $2T_iT_j = T_{i+j} + T_{|i-j|}$ [Han02], giving rise to a Toeplitz + Hankel structure. Fast algorithms also exist for such a structure; how to employ these has been spelled out in [Gen+03]. The extension to the multivariate case can be done similarly to the monomial case simply by taking products of Chebyshev polynomials with different variables; again, fast methods then no longer exist.

### 4.9.2 SOS cone and dual, basis-independent version

So far, changing the basis was only mentioned as a possibility; now the SOS cone and its dual will be explicitly described without reference to the monomial basis. To do this in most generality, the case of a field $\mathbb{F} \in \{\mathbb{R}, \mathbb{C}\}$ will be considered.

**Notation** (Real and complex case). There are subtle differences between the real and complex case. In the real case, squaring a polynomial in $\mathbb{R}_d[\boldsymbol{x}]$ yields a polynomial in $\mathbb{R}_{2d}[\boldsymbol{x}]$. In the complex case, it is assumed that space $\mathbb{C}_d[\boldsymbol{z}]$ contains only the "unconjugated" variables $\boldsymbol{z}$; the square of the *absolute value* (as an order relation needs to be available) then lies in $\mathbb{C}_d[\boldsymbol{z}, \bar{\boldsymbol{z}}]$ if the (complex) degree is defined to be the maximum of the degree in "unconjugated" and conjugated variables, as done in [JM18].
This leads to notational difficulties when trying to handle both cases at the same time. Therefore, the complex-valued notation will also be enforced for the real-valued case with the understanding that then, $\bar{\boldsymbol{z}} = \boldsymbol{z}$ and $\mathbb{R}_d[\boldsymbol{z}, \bar{\boldsymbol{z}}] \equiv \mathbb{R}_{2d}[\boldsymbol{z}]$.

With this convention, some concepts that were only discussed informally or used implicitly in sections 4.6 and 4.8 can now be defined formally.

12 These results were obtained by maximizing the smallest eigenvalue of a matrix with the imposed structure under a shared fixed trace. Randomization of the matrices was also considered. The resulting condition numbers are indications on how good the conditioning can get. It should also be noted that minimizations of the eigenvalue spread can yield condition numbers that are orders of magnitude worse.

**Definition 4.64** (SOS cone, general case)**.** The pointed convex cone of *Hermitian sums-of-squares*[13] *(HSOS) polynomials* up to complex degree $d$ is defined by

$$\Sigma_d[\boldsymbol{z}, \bar{\boldsymbol{z}}] := \left\{ \sum_{i=1}^{m} |q_i|^2 : m \in \mathbb{N}, q_i \in \mathbb{F}_d[\boldsymbol{z}] \right\}. \tag{4.50}$$

The SOS cone can be neatly identified with a vector of coefficients with respect to a functional basis.

**Notation** (Polynomial bases)**.** Let $\boldsymbol{w} := (w_\ell)_{\ell=1,\dots,L}$ be any polynomial basis of the space of polynomials $\mathbb{F}_d[\boldsymbol{z}]$. Let $\boldsymbol{v} := (v_u)_{u=1,\dots,U}$ be any polynomial basis of the space of polynomials $\mathbb{F}_d[\boldsymbol{z}, \bar{\boldsymbol{z}}]$.

**Definition 4.65** (Hermitian inner product and part)**.** The *Hermitian inner product* of two vectors is defined as

$$\langle \boldsymbol{z}, \boldsymbol{z}' \rangle_{\mathrm{H}} := \langle \operatorname{Re} \boldsymbol{z}, \operatorname{Re} \boldsymbol{z}' \rangle + \langle \operatorname{Im} \boldsymbol{z}, \operatorname{Im} \boldsymbol{z}' \rangle \equiv \operatorname{Re} \langle \boldsymbol{z}, \boldsymbol{z}' \rangle \tag{4.51}$$

and the *Hermitian part* of a matrix as

$$M_{\mathrm{H}} := \frac{M + M^\dagger}{2}. \tag{4.52}$$

Naturally, the functions $q_i$ in equation (4.50) can be expanded in $\boldsymbol{w}$, while their absolute squares have a coefficient set with respect to $\boldsymbol{v}$. Therefore, every element in $\mathbb{F}_d[\boldsymbol{z}]$ or $\mathbb{F}_d[\boldsymbol{z}, \bar{\boldsymbol{z}}]$ is isomorphic to a vector in $\mathbb{F}^L$ or $\mathbb{F}^U$, respectively. With this isomorphism, the dual cone of $\Sigma_d[\boldsymbol{z}, \bar{\boldsymbol{z}}]$ according to Definition 3.20 can be written down (where the Hermitian inner product has to be used to get an order relation).

**Definition 4.66** (SOS-dual cone, complex case)**.** The dual cone $\Sigma_d[\boldsymbol{z}, \bar{\boldsymbol{z}}]^*$ to the cone of Hermitian sums-of-squares up to complex degree $d$ is

$$\Sigma_d[\boldsymbol{z}, \bar{\boldsymbol{z}}]^* \simeq \left\{ \boldsymbol{\lambda} \in \mathbb{F}^U \;\middle|\; \langle \boldsymbol{\lambda}, \boldsymbol{\sigma} \rangle_{\mathrm{H}} \geq 0 \; \forall \boldsymbol{\sigma} \in \mathbb{F}^U : \sum_{u=1}^{U} \sigma_u w_u \in \Sigma_d[\boldsymbol{z}, \bar{\boldsymbol{z}}] \right\}. \tag{4.53}$$

**Lemma 4.67** (Nesterov [Nes00])**.** Define the linear operator $\Lambda \colon \mathbb{F}^U \to \mathbb{S}^L$ such that

$$\Lambda_{\mathrm{H}}(\boldsymbol{v}(\boldsymbol{z}, \bar{\boldsymbol{z}})) := \boldsymbol{w}(\boldsymbol{z}) \boldsymbol{w}(\boldsymbol{z})^\dagger. \tag{4.54}$$

Note that this can be satisfied by demanding $\Lambda(\boldsymbol{v}(\boldsymbol{z}, \bar{\boldsymbol{z}})) = \boldsymbol{w}(\boldsymbol{z}) \boldsymbol{w}(\boldsymbol{z})^\dagger$; but only $\Lambda_{\mathrm{H}}$ will be used in the following.

13 Not to be confused with sums of Hermitian squares (SOHS), which is a common term in noncommutative polynomial optimization [KP10] and refers to when the decision variables in the original problem (which are then operators) obey $X = X^\dagger$—for scalar $X$ the equivalent to the *real-valued* case.

Then, for a real-valued polynomial $p = \langle \boldsymbol{p}, \boldsymbol{v} \rangle_{\mathrm{H}} \in \mathbb{F}_d[\boldsymbol{z}, \bar{\boldsymbol{z}}]$ identified by its vector $\boldsymbol{p} \in \mathbb{F}^U$,

$$p \in \Sigma_d[\boldsymbol{z}, \bar{\boldsymbol{z}}] \Leftrightarrow \exists G \succeq 0 : \boldsymbol{p} = \Lambda_{\mathrm{H}}^{\dagger}(G) \tag{4.55a}$$

$$\text{and } p \in \Sigma_d^*[\boldsymbol{z}, \bar{\boldsymbol{z}}] \Leftrightarrow \Lambda_{\mathrm{H}}(\boldsymbol{p}) \succeq 0. \tag{4.55b}$$

Here, $\Lambda_{\mathrm{H}}^{\dagger} : \mathbb{S}^L \to \mathbb{F}^U$ is the adjoint operator of $\Lambda_{\mathrm{H}}$ with respect to the Hermitian inner product,

$$\langle G, \Lambda_{\mathrm{H}}(\boldsymbol{p}) \rangle_{\mathrm{H}} = \langle \Lambda_{\mathrm{H}}^{\dagger}(G), \boldsymbol{p} \rangle_{\mathrm{H}} \ \forall G \in \mathbb{S}^L, \boldsymbol{p} \in \mathbb{F}^U. \tag{4.56a}$$

If $\Lambda(\boldsymbol{p}) = \sum_{u=1}^{U} \Lambda_u p_u$ with coefficient matrices $\Lambda_u \in \mathbb{S}^L$, the adjoint is given by

$$\left[\Lambda_{\mathrm{H}}^{\dagger}(G)\right]_u = \langle \Lambda_u, G \rangle. \tag{4.56b}$$

**Corollary 4.68.** The dual cone $\Sigma_d[\boldsymbol{z}, \bar{\boldsymbol{z}}]^*$ is a spectrahedron, i.e., an intersection of the positive semidefinite cone with an affine space.

An LHSC barrier function for the dual cone $\Sigma_d[\boldsymbol{z}, \bar{\boldsymbol{z}}]^*$ can now be written down immediately: again identifying $p$ with $\boldsymbol{p}$, the barrier is $-\log\det \Lambda_{\mathrm{H}}(\boldsymbol{p})$.

## 4.9.3 Interpolant basis

### 4.9.3.1 Barrier oracles

Having this explicit description, a more unconventional choice of basis can be made: the interpolant basis [LP04; PY19]. A multivariate polynomial in $n$ real variables is completely specified by its value at $\binom{n+d}{n}$ *unisolvent* points—these are points for which the vectors formed by an arbitrary polynomial basis of degree up to $d$ evaluated at the points are all linearly independent. Choosing the monomial basis, this can also be paraphrased as the multivariate Vandermonde matrix of these points being nonsingular.

In general, the number of unisolvent points is the same as the length of the basis of the space. Therefore, interpolating a polynomial in $\mathbb{R}_d[\boldsymbol{x}, \bar{\boldsymbol{x}}] \equiv \mathbb{R}_{2d}[\boldsymbol{x}]$ requires $\binom{n+2d}{n}$ points, while $\mathbb{C}_d[\boldsymbol{z}, \bar{\boldsymbol{z}}]$ needs $\binom{n+d}{n}^2$.

**Proposition 4.69** (Barrier for $\Sigma_d[\boldsymbol{z}, \bar{\boldsymbol{z}}]^*$ in the interpolant basis)**.** Let the basis $\boldsymbol{w}$ of $\mathbb{F}_d[\boldsymbol{z}]$ still be arbitrary, but now choose a set of unisolvent points $\{\boldsymbol{t}_u\}_{u=1,\ldots,U}$ and use the Lagrange basis $v_u(\boldsymbol{t}_{u'}, \bar{\boldsymbol{t}}_{u'}) := \delta_{u,u'}$ for $\mathbb{F}_d[\boldsymbol{z}, \bar{\boldsymbol{z}}]$. With the matrix

$$W := \Big(w_\ell(\boldsymbol{t}_u)\Big)_{\ell=1,\, u=1}^{L \quad U} \in \mathbb{F}^{L \times U} \tag{4.57}$$

and a polynomial $p \in \mathbb{F}_d[\boldsymbol{z}, \bar{\boldsymbol{z}}]$ described by its representation $\boldsymbol{p}$ with respect to $\boldsymbol{v}$ (i.e., by its evaluation at the points $\boldsymbol{t}_u$), the following relations hold for $\Lambda_{\mathrm{H}}(\boldsymbol{p})$ and derivatives of the barrier function $f(\boldsymbol{p}) = -\log\det \mathit{\Lambda}_{\mathrm{H}}(\boldsymbol{p})$:

$$\mathit{\Lambda}_{\mathrm{H}}(\boldsymbol{p}) = W \operatorname{Diag}(\boldsymbol{p}) W^\dagger \tag{4.58a}$$

$$\frac{\partial f(\boldsymbol{p})}{\partial p_u} = -\boldsymbol{w}^\dagger(\boldsymbol{t}_u)\mathit{\Lambda}_{\mathrm{H}}(\boldsymbol{p})^{-1}\boldsymbol{w}(\boldsymbol{t}_u) \quad \Leftrightarrow \quad \boldsymbol{\nabla} f(\boldsymbol{p}) = -\operatorname{diag}(W^\dagger \mathit{\Lambda}_{\mathrm{H}}(\boldsymbol{p})^{-1} W) \tag{4.58b}$$

$$\frac{\partial^2 f(\boldsymbol{p})}{\partial p_u p_{u'}} = \left|\boldsymbol{w}^\dagger(\boldsymbol{t}_u)\mathit{\Lambda}_{\mathrm{H}}(\boldsymbol{p})^{-1}\boldsymbol{w}(\boldsymbol{t}_{u'})\right|^2 \;\Leftrightarrow\; \boldsymbol{\nabla}^2 f(\boldsymbol{p}) = (W^\dagger \mathit{\Lambda}_{\mathrm{H}}(\boldsymbol{p})^{-1} W)^{|\circ 2|}, \tag{4.58c}$$

where $\operatorname{Diag}(\bullet)$ constructs a diagonal matrix from the given vector while $\operatorname{diag}(\bullet)$ extracts the diagonal, and $\bullet^{|\circ 2|}$ denote the elementwise absolute value squared (Hadamard product of the matrix with its conjugate).

*Proof.* Note $w_\ell(\boldsymbol{z})\overline{w_{\ell'}(\boldsymbol{z})} \in \mathbb{C}_d[\boldsymbol{z}, \bar{\boldsymbol{z}}]$ can be expanded with respect to $\boldsymbol{v}$:

$$\begin{aligned} w_\ell(\boldsymbol{z})\overline{w_{\ell'}(\boldsymbol{z})} &= \sum_{u=1}^{U} w_\ell(\boldsymbol{t}_u)\overline{w_{\ell'}(\boldsymbol{t}_u)} v_u(\boldsymbol{z}, \bar{\boldsymbol{z}}) \\ \Leftrightarrow \left(\boldsymbol{w}(\boldsymbol{z})\boldsymbol{w}(\boldsymbol{z})^\dagger\right)_{\ell,\ell'} &= \left(W \operatorname{Diag}(\boldsymbol{v}(\boldsymbol{z}, \bar{\boldsymbol{z}})) W^\dagger\right)_{\ell,\ell'} \end{aligned}$$

This uses the fact that by the definition of the Lagrange basis, the polynomial coefficients are given by the values at the interpolation points, and the product of two polynomials is pointwise in any interpolant basis. By equation (4.54), this already shows the form for $\mathit{\Lambda}_{\mathrm{H}}(\boldsymbol{p})$ with some vector $\boldsymbol{p}$ with respect to the basis $\boldsymbol{v}$. Note that $\boldsymbol{v}(\boldsymbol{z}, \bar{\boldsymbol{z}}) \in \mathbb{R}^U$ for all values of $\boldsymbol{z}$, even if they are complex-valued.

For the derivatives, set $\mathit{\Lambda}_{\mathrm{H}}(\boldsymbol{p}) = \sum_u \mathit{\Lambda}_{\mathrm{H},u} p_u$, where $\mathit{\Lambda}_{\mathrm{H},u} = \boldsymbol{w}(\boldsymbol{t}_u)\boldsymbol{w}^\dagger(\boldsymbol{t}_u)$.

$$\begin{aligned} \frac{\partial f(\boldsymbol{p})}{\partial p_u} &= -\frac{\partial \ln\det \mathit{\Lambda}_{\mathrm{H}}(\boldsymbol{p})}{\partial p_u} = -\frac{1}{\det \mathit{\Lambda}_{\mathrm{H}}(\boldsymbol{p})} \frac{\partial \det \mathit{\Lambda}_{\mathrm{H}}(\boldsymbol{p})}{\partial p_u} \\ &= -\frac{1}{\det \mathit{\Lambda}_{\mathrm{H}}(\boldsymbol{p})} \det \mathit{\Lambda}_{\mathrm{H}}(\boldsymbol{p}) \operatorname{tr}\left(\mathit{\Lambda}_{\mathrm{H}}(\boldsymbol{p})^{-1} \frac{\partial \mathit{\Lambda}_{\mathrm{H}}(\boldsymbol{p})}{\partial p_u}\right) \\ &= -\operatorname{tr}(\mathit{\Lambda}_{\mathrm{H}}(\boldsymbol{p})^{-1}\mathit{\Lambda}_{\mathrm{H},u}) = -\boldsymbol{w}^\dagger(\boldsymbol{t}_u)\mathit{\Lambda}_{\mathrm{H}}(\boldsymbol{p})^{-1}\boldsymbol{w}(\boldsymbol{t}_u) \end{aligned}$$

using Jacobi's formula for the derivative of a determinant [HJ12]. Going on,

$$\begin{aligned} \frac{\partial^2 f(\boldsymbol{p})}{\partial p_u \partial p_{u'}} &= \boldsymbol{w}^\dagger(\boldsymbol{t}_u)\mathit{\Lambda}_{\mathrm{H}}(\boldsymbol{p})^{-1} \frac{\partial \mathit{\Lambda}_{\mathrm{H}}(\boldsymbol{p})}{\partial p_{u'}} \mathit{\Lambda}_{\mathrm{H}}(\boldsymbol{p})^{-1}\boldsymbol{w}(\boldsymbol{t}_u) \\ &= \boldsymbol{w}^\dagger(\boldsymbol{t}_u)\mathit{\Lambda}_{\mathrm{H}}(\boldsymbol{p})^{-1}\boldsymbol{w}(\boldsymbol{t}_{u'})\boldsymbol{w}^\dagger(\boldsymbol{t}_{u'})\mathit{\Lambda}_{\mathrm{H}}(\boldsymbol{p})^{-1}\boldsymbol{w}(\boldsymbol{t}_u) \\ &= \left|\boldsymbol{w}^\dagger(\boldsymbol{t}_u)\mathit{\Lambda}_{\mathrm{H}}(\boldsymbol{p})^{-1}\boldsymbol{w}(\boldsymbol{t}_{u'})\right|^2, \end{aligned}$$

and the expressions in terms of the matrix product follow by inspection. □

**Remark 4.70.** In [PY19], the matrix $W$ was defined as the transpose of what is used here. This is due to the fact that the paper only considers the real-valued case, so the transposition will need to occur at some place in the equations anyway. However, when moving to the complex-valued case, the definition here allows to combine all transpositions and conjugations into adjoints, while the other would require the transposition of one matrix and the conjugation of another.

#### 4.9.3.2 Performance

The reason for using an interpolation basis instead of one based on orthogonal polynomials is found in potential computational benefits.

**Proposition 4.71** (Computational requirements [PY19])**.** An interior-point solver that directly works with the barrier function of $\Sigma_d[\boldsymbol{z}, \bar{\boldsymbol{z}}]^*$ only uses

- $\mathcal{O}(LU^2)$ time for the computation of barrier gradient and Hessian,
- $\mathcal{O}(LU)$ working memory,
- $\mathcal{O}(U^2)$ space to store the Hessian.

A "non-informed" semidefinite solver has a matrix side dimension $L$ and $\mathcal{O}(L^2)$ independent variables, yielding $\mathcal{O}(L^4)$ operations to arrive at the Hessian (see Remark 3.41); for real-valued problems, this can be better than the interpolant version in some parameter regimes (high $d$, low $n$). However, then the linear system $(\boldsymbol{\nabla}^2 f)^{-1}\boldsymbol{\nabla} f$ has to be solved, which leads to the $\mathcal{O}(L^6)$ runtime. In the interpolant version, the solver already knowing that the dimension of the cone is $U$ instead of $L^2$, solving the linear system only takes $\mathcal{O}(U^3)$ time. While this is exactly the same run time behavior as for the interpolant version in the complex-valued case where $U = L^2$, it is considerable less in the real-valued case with $U = \binom{n+2d}{n} \ll \binom{n+d}{n}^2 = L^2$. Even more important are the gains in memory in the real-valued case, as usually this is a more critical bottleneck than runtime.

#### 4.9.3.3 Choosing interpolation points

In [PY19], the authors argue that not every interpolation basis is a good choice and they demonstrate that some bases give rise to a more stable numerical behavior than others. While they can explicitly construct these for the univariate and bivariate case using Chebyshev and Padua points, for more variables, the theoretically best-conditioned choices arising from Fekete points cannot be obtained directly.

I will follow a completely different approach in choosing a good interpolation basis; instead of looking at conditioning, my choices will allow to speed up the operations that are involved in the various calculations. Given that the main benefit of an interpolation basis is evident in the real-valued case, this will be the focus. The final result is summarized in the following Theorem.

**Theorem 4.72** (Improved computational requirements)**.** An interior-point solver that directly works with the barrier function of $\Sigma_d[\boldsymbol{x}]^*$ can use only

- $\mathcal{O}(L^\omega + UN \log N)$ time for the computation of barrier gradient and Hessian,
- $\mathcal{O}(LU)$ working memory,
- $\mathcal{O}(U^2)$ space to store the Hessian,

where $\omega$ is the exponent of matrix multiplication, and heuristically, $N \sim U$.
Alternatively, the Hessian need not be calculated explicitly, reducing the time for the computation of barrier gradient and product of the Hessian with a vector to $\mathcal{O}(L^\omega + LN \log N + LN_\mathrm{r}U)$ with only $\mathcal{O}(L^2)$ working memory in total, where $N_\mathrm{r} = o(L)$.

**Remark 4.73.** Note that in the interesting regimes of large $n$ and small $d$, the $UN \log N$ term dominates over $L^\omega$. Considering that the Hessian has $\mathcal{O}(U^2)$ entries, the resulting $\mathcal{O}(UN \log N)$ timing is close to optimal.

*Proof.* The proof will be divided in several parts.

The first part considers the explicit calculations of gradient and Hessian. I will show the complexities, tacitly postulating the existence of certain matrix decompositions that would be helpful for efficient computations. Based on these desiderata, in the remaining parts of the proof, I will turn to the construction of the matrices, which first requires a proper definition of the interpolation points. While it will turn out that the described procedure always works, the possibility of finding *efficient* parameters—in particular, $N \sim U$—is algorithmic, though backed with substantial numerical results.

**Part 1: Operations** I will assume that the matrix $W$ defined in equation (4.57) can be decomposed as

$$W = ST, \tag{4.59}$$

where $S \in \{0,1\}^{L \times N_\mathrm{c}}$ with $L \leq N_\mathrm{c} < U$ is a sparse binary matrix with no more than $N_\mathrm{r} := 2^{\min(n,d)-1} \ll L$ nonzero entries per row, and $T \in \mathbb{R}^{N_\mathrm{c} \times U}$ is a submatrix of the Discrete Cosine Transform (DCT) matrix of size $N \sim U$.

Therefore, the product $W\boldsymbol{x}$ can be calculated by filling an $N$-dimensional vector with entries from $\boldsymbol{x}$, performing a DCT in $\mathcal{O}(N \log N)$ time, finally followed by the $S$-multiplication in $\mathcal{O}(LN_\mathrm{r})$ time. Conversely, the product $W^\top \boldsymbol{y}$ consists in filling an $N$-dimensional vector with entries from $S^\top \boldsymbol{y}$ in $\mathcal{O}(LN_\mathrm{r})$ time, performing a related DCT in $\mathcal{O}(N \log N)$ time, finally followed by the extraction of $U$ entries from the vector. Since $LN_\mathrm{r} < N \log N$, the overall scaling for these matrix–vector products is $\mathcal{O}(N \log N)$; the space requirements are $\mathcal{O}(N)$ for the temporary vector and $\mathcal{O}(LN_\mathrm{r})$ for $S$.

After these preliminaries, look at the following matrix product, which according to Proposition 4.69 allows to extract both the gradient and the Hessian of the barrier function:

$$W^\top \big[\underbrace{W \operatorname{Diag}(\boldsymbol{p}) W^\top}_{\Lambda(\boldsymbol{p})}\big]^{-1} W. \tag{4.60}$$

1. $\Lambda(\boldsymbol{p})$ can be obtained in $\mathcal{O}\big(L(UN_\mathrm{r}+N \log N)\big)$ time by explicitly calculating every column of $\operatorname{Diag}(\boldsymbol{p})W^\top$ in $\mathcal{O}(UN_\mathrm{r})$ time followed by the left-multiplication by $W$.
2. A Cholesky decomposition, which immediately tells about the feasibility of the point $\boldsymbol{p}$, factorizes $\Lambda(\boldsymbol{p}) = C^\top C$ in $\mathcal{O}(L^\omega)$ time. By inverting the upper triangular matrix $C^{-1} =: \tilde{C}$ in $\mathcal{O}(L^\omega)$ time, the inverse is given in a factorized form as $\Lambda(\boldsymbol{p})^{-1} = \tilde{C}\tilde{C}^\top$.
3. The $u^\text{th}$ element of the gradient of the barrier function is
$$-(W^\top \tilde{C}\tilde{C}^\top W)_{u,u} = -\sum_\ell (W^\top \tilde{C})_{u,\ell}{}^2. \tag{4.61}$$
The complete column $(W^\top \tilde{C})_{:,\ell}$ can be calculated in $\mathcal{O}(N \log N)$ time, so that the full gradient is obtained in $\mathcal{O}(LN \log N)$ time with no further auxilliary memory.
4. The Hessian could be obtained by computing $W^\top \tilde{C}$ similarly (along with the gradient in no extra time) and storing it as a temporary. Then performing the $(U \times L)(L \times U)$ matrix product with its transpose, this would require at most $\mathcal{O}(LU^2)$ time.

   A faster way is to multiply the inverted Cholesky factors in $\mathcal{O}(L^\omega)$ time to obtain the actual inverse $\Lambda(\boldsymbol{p})^{-1}$. The Hessian is then given by the elementwise square of $W^\top \Lambda(\boldsymbol{p})^{-1} W$. Proceeding from left to right, $W^\top \Lambda(\boldsymbol{p})^{-1}$ can be assembled columnwise in $\mathcal{O}(LN \log N)$ time and stored to a temporary $U \times L$ matrix. Then, the product $(W^\top \Lambda(\boldsymbol{p})^{-1})W$ can similarly be done columnwise in $\mathcal{O}(UN \log N)$ time, which is the dominant contribution. The remaining entrywise square trivially requires $\mathcal{O}(U^2)$ time.
5. The product of the Hessian with a vector $\boldsymbol{x}$ (without calculating the Hessian explicitly) can be obtained from the identity
$$\begin{aligned}\sum_{u'} \frac{\partial^2 f(\boldsymbol{p})}{\partial p_u \partial p_{u'}} x_{u'} &= \sum_{u'} \boldsymbol{w}^\top(\boldsymbol{t}_u)\Lambda(\boldsymbol{p})^{-1}\boldsymbol{w}(\boldsymbol{t}_{u'})\boldsymbol{w}^\top(\boldsymbol{t}_{u'})\Lambda(\boldsymbol{p})^{-1}\boldsymbol{w}(\boldsymbol{t}_u)x_{u'} \\ &= \boldsymbol{w}^\top(\boldsymbol{t}_u)\Lambda(\boldsymbol{p})^{-1}\Lambda(\boldsymbol{x})\Lambda(\boldsymbol{p})^{-1}\boldsymbol{w}(\boldsymbol{t}_u).\end{aligned} \tag{4.62}$$
The $L \times L$ matrix $\Lambda(\boldsymbol{x})$ can be calculated as before in $\mathcal{O}\big(L(UN_\mathrm{r} + N \log N)\big)$ time and the product $P := \Lambda(\boldsymbol{p})^{-1}\Lambda(\boldsymbol{x})\Lambda(\boldsymbol{p})^{-1}$, using either backsubstitution or the actual inverse, in $\mathcal{O}(L^\omega)$. An eigendecomposition $P = VDV^\top$, taking another $\mathcal{O}(L^\omega)$ time (as $P$ may well be indefinite), then allows to use the same strategy as for the gradient: the $u^{\text{th}}$ element of the Hessian product is $\sum_\ell D_{\ell,\ell}(W^\top V)_{u,\ell}{}^2$. Therefore, the Hessian product requires the same time as the gradient and $\mathcal{O}(L^2)$ extra memory for the eigendecomposition.

Note that nonsymmetric interior-point solvers can exploit efficient implementations of further operations. In particular, the third-order directional derivative $D^3 f(\boldsymbol{p})[\boldsymbol{h}, \boldsymbol{h}, \boldsymbol{h}]$, see Definition 3.32, may also be calculated with the same time complexity as the gradient or Hessian product and $\mathcal{O}(L^2)$ extra memory, which may be the same memory already used for the Hessian product. Details can be found in appendix A.

**Part 2: Interpolation points** Recall that the matrix $W$ contains the evaluation of a polynomial basis for $\mathbb{R}_d[\boldsymbol{x}]$ at the interpolation points. The second part of the proof therefore focuses on the construction of interpolation points. For convenience, in the remainder of this proof only, all indices will start with zero. Define the shorthand $[\![n]\!] := \{0, 1, \dots, n-1\}$.

Let the interpolation points $\{\boldsymbol{t}_u\}_{u \in [\![U]\!]}$ be given by

$$t_{u,i} := D^{\text{II}}(N)_{r(i),c(u)}. \tag{4.63}$$

Here, $D^{\text{II}}(N)$ is an $N$-dimensional Discrete Cosine Transform-II matrix,

$$D^{\text{II}}(N)_{r,c} := \cos\left[\frac{\pi}{N}\left(c + \frac{1}{2}\right)r\right]. \tag{4.64}$$

The injections $r \colon [\![n]\!] \to [\![N]\!]$ and $c \colon [\![U]\!] \to [\![N]\!]$ as well as $N$ itself (clearly, $N \geq U$, so that enough columns are available) must now be chosen carefully to satisfy the requirement that the points $\{\boldsymbol{t}_u\}_{u \in [\![U]\!]}$ form a unisolvent set for $\mathbb{R}_{2d}[\boldsymbol{x}]$. This means that a polynomial basis for degree $2d$ evaluated at these points must yield $U$ linearly independent vectors. Without loss of generality, $r$ is in the following assumed to be strictly monotonically increasing.

The univariate Chebyshev polynomials $\{T_j(x)\}_{j \in \mathbb{N}_0}$ form a polynomial basis [Han02], and by multiplication (analogous to the monomials, where now the exponent of the variable becomes the index of the corresponding Chebyshev polynomial), a multivariate Chebyshev basis can be constructed. Define the multivariate Chebyshev matrix as

$$\begin{aligned} V_{2d} &:= \left(\prod_{i=0}^{n-1} T_{j_i}(t_{u,i})\right)_{\boldsymbol{j} \in \mathcal{B}_{n,2d}, u \in [\![U]\!]} \\ &= \left[\prod_{i=0}^{n-1} T_{j_i}\left(\cos\left[\frac{\pi}{N}\left(c(u) + \frac{1}{2}\right)r(i)\right]\right)\right]_{\boldsymbol{j} \in \mathcal{B}_{n,2d}, u \in [\![U]\!]}. \end{aligned} \tag{4.65}$$

Note that $r(i) = 0$ can be immediately excluded: Assume that $r(i) = 0$ for some $i$. Pick all indices in $\mathcal{B}_{n,2d}$ that vanish everywhere except for the position $i$ (of which there are $2d > 1$). Each factor in the product in the definition of $V_{2d}$ apart from this $i$ then collapses to $T_0(\bullet) \equiv 1$, while the contribution from $i$ is always $T_{j_i}(0)$, regardless of $u$. Therefore, all rows corresponding to the monomials $x_i^{j_i}$, with any power, are proportional to the all-ones row—which is also explicitly present at index $\boldsymbol{j} = \mathbf{0}$—and thus linearly dependent. The Chebyshev matrix is singular, contradicting unisolvence.

Exploiting the definition of the Chebyshev polynomials, $T_j(\cos\theta) = \cos(j\theta)$, and the cosine product formula $2\cos\theta_1\cos\theta_2 = \cos(\theta_1+\theta_2) + \cos(\theta_1-\theta_2)$, $V_{2d}$ can be written as

$$V_{2d} = \left[\prod_{i=0}^{n-1}\cos\left[\frac{\pi}{N}\left(c(u)+\frac{1}{2}\right)r(i)j_i\right]\right]_{\boldsymbol{j}\in\mathcal{B}_{n,2d},u\in[\![U]\!]} \tag{4.66}$$

$$= \left[\frac{1}{2^n}\sum_{\boldsymbol{\sigma}\in\{-1,1\}^n}\cos\left[\frac{\pi}{N}\left(c(u)+\frac{1}{2}\right)\sum_{i=0}^{n-1}\sigma_i r(i)j_i\right]\right]_{\boldsymbol{j}\in\mathcal{B}_{n,2d},u\in[\![U]\!]}. \tag{4.67}$$

Consider the expression $\cos[\frac{\pi}{N}(c+\frac{1}{2})x]$ with $x \geq 0$. While $x = N + \delta$ with $\delta > 0$, keep replacing $x$ by $|N - \delta|$, until $x \in [0, N]$, leaving the cosine unchanged, and denote by # how often the replacement was done. If now $x = N$, then the expression vanishes; else, it is given by $(-1)^{\#}D^{\mathrm{II}}(N)_{x,c}$. The repeated replacement can be explicitly described by $x \mapsto |\mathrm{wrap}_N(x)| := |(N + x) \bmod 2N - N|$, where # is odd if and only if the result was negative before taking the absolute value.

Therefore, equation (4.67) can effectively be spelled out as a map

$$V_{2d} = \frac{1}{2^n}AD^{\mathrm{II}}(N)\big|_{:,c([\![U]\!])}, \tag{4.68}$$

where the notation $\bullet|_{:,\mathcal{S}}$ indicates only the columns with indices in $\mathcal{S}$ are retained, deleting the rest. The sparse matrix $A \in \mathbb{Z}^{U\times N}$ can be built by iterating through all exponents $\boldsymbol{j} \in \mathcal{B}_{n,2d}$ and all $\boldsymbol{\sigma} \in \{-1,1\}^n$ and assigning the value 1 if $\mathrm{wrap}_N(x := \sum_{i=0}^{n-1}\sigma_i r(i)j_i) \geq 0$ and the value $-1$ for $-N < \mathrm{wrap}_N(x) < 0$ at the row indexed by $\boldsymbol{j}$ and the column with index $|\mathrm{wrap}_N(x)|$; duplicate positions are summed over.

Clearly, a prerequisite for $V_{2d}$ to be invertible is $A$ having full row rank. In fact, for every such $A$, there is a $c$ such that $V_{2d}$ is regular: since $D^{\mathrm{II}}$ itself is regular, $AD^{\mathrm{II}}(N)$ has rank $U$, its columns form a generating system and therefore, a basis with $U$ elements can be chosen. Finding $c$ will thus not be discussed here, as it is automatic.

The nontrivial part is finding $N$ and $r$. While checking through all possible combinations at the moment seems to be the only feasible way, performing several possible optimizations along the way can possibly speed up the process greatly. The following algorithm is essentially only a description of successive applications of Gaussian elimination with shortcuts to exit early in case of failure and to then use the failure information for skipping potentially many of the $\binom{N}{n}$ candidates for $r$. It can be parallelized easily by dividing the search space with respect to $r$; due to the early exits, dynamic rescheduling is essential for high performance.

- Iterate through all $N \geq U$ and $r$ (monotonically increasing).
- Gradually build the matrix $A$, starting with rows which are indexed only by the first variable, then the second two variables...
- When adding a row, immediately try to make it conformant to row echelon form; by identifying a new pivot element, if possible, or applying row operations until a pivot element can be found.
- If after row operations, a row is empty, the corresponding exponent was invalid, yielding a linear dependency for the given choice of $N$ and $r$. Since this issue was revealed by considering only the first $i$ variables, it cannot be fixed by changing $r$ at indices larger than $i$; so all these combinations can be skipped.
- If all rows could be constructed, then $(N, r)$ is a valid choice; terminate.

The whole process can be carried out in exact rational arithmetics. While numerical instabilities in Gaussian elimination that are common in floating-point arithmetics are excluded by this choice, care must still be exercised, as the native (64-bit) register size may soon be insufficient to hold the data. In floating-point arithmetics, numerical issues are usually circumvented using a pivoting strategy—the most common being partial pivoting. This requires to scan the whole column at the next position for its largest absolute value, then permute the rows such that this largest element is the next pivot. Given the nature of the on-demand construction of $A$, this pivoting strategy is not available, as the complete column will not be known until the end. A more complicated pivoting strategy that completely constructs all rows corresponding to a given number of variables, then selects the most suitable element, might be applicable, still retaining the skipping shortcut. Instead, to generate the numerical data presented here, a switch to arbitrary precision numbers is performed when necessary, not performing any kind of pivoting.

*A Graph theory perspective.* A prerequisite for $A$ to have full rank is that every exponent $\boldsymbol{j} \in \mathcal{B}_{n,2d}$ can be matched with a unique index $x \in [\![N]\!]$. While classical graph matching algorithms [HK73; Kar73] have recently seen large improvements [Mad13; Bra+20; Che+22] to work in almost linear time, they need the full graph as an input, which is basically given by the complete sparsity pattern of $A$. As this precludes both the early exit as well as the skipping strategy in the above algorithm, an improvement due to advances in graph theory is not expected.

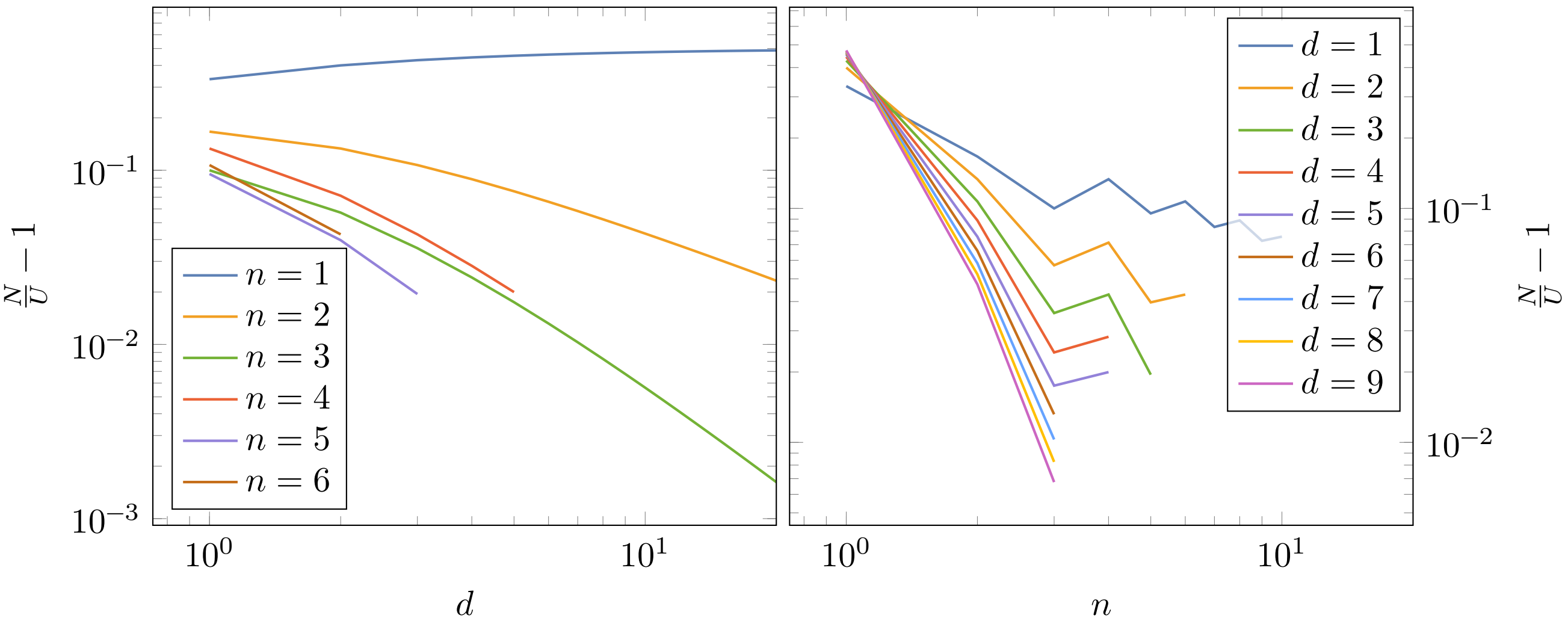

**a)** for various variable numbers **b)** for various relaxation degrees

**Figure 4.3.** Scaling of $N$ with respect to $U$ for various configurations

*Termination.* Note that the algorithm described above will eventually terminate, for choosing $r(i) = \frac{(2d)^i - 1}{2d-1}$ and $N = \frac{(2d)^{n+1}-1}{2d-1}$ is always a permitted—yet very uneconomic—choice. This corresponds to associating powers of the first variable alone with the indices 1 to $2d$, of the second variable with $2d + 1$ to $2d(2d + 1)$, ..., leaving enough space that no allowed products can ever hit another index. While there may still be duplicates from the differences, since $N$ is chosen such that no wrapping will happen, the all-sums term is always strictly monotonically increasing and gives the largest index; it will therefore ensure linear independence.

*Scaling.* The data shown in figure 4.3 is fully consistent[14] with the conjecture

$$N - U = \binom{x+d}{x+1} \frac{x+1}{d} - 1, \tag{4.69}$$

with $x := \lfloor \frac{n}{2} \rfloor$, except for the $n = 1$ case where $N - U = d$. Considering the scaling in $d$, with the exception of $d = 1$ which has $N - U \to \frac{L}{2}$ as $n \to \infty$, all higher degrees yield $\frac{N-U}{L} \to 2^{-d}$. Considering the scaling in $n$, with the exception of $n = 1$ which has $N - U \to L$ as $d \to \infty$, all higher variable numbers yield $\frac{N-U}{L} \to 0$. The initial statement $N \sim U$ therefore seems to be well-motivated.

*Connections to additive combinatorics.* It should be mentioned that the problem has some similarity to the subject of generalized Sidon or $B_h[g]$-sets [OBr04; TV06]. A Sidon set

14 About 650 data points were obtained starting from $N = U$ as the minimum; after this, $N = U + d$ was used as the starting point to save simulation time.

is a set of distinct integers such that all pairwise sums are unique; likewise, a $B_h[g]$-set requires that sums of $h$ elements have at most $g$ solutions. Due to the sum over $\boldsymbol{\sigma}$, the problem here is not completely equivalent to constructing a $B_{2d}[1]$-set of length $n$ over an as small as possible finite ring; the previous argument for termination no longer holds exactly, as the wrapping at $N$ invalidates the assumption that the all-sums term is always strictly monotonically increasing. Therefore, any candidate for $(N, r)$ constructed using a $B_{2d}[1]$ sequence must still be subjected to the final check of linear independence; and existing bounds, e.g., [JTT22], cannot be transferred immediately.

Finally note that the calculation of the points is an offline task that is completely independent from the actual polynomial optimization problem; it depends on $n$ and $d$ only and can therefore be done once in advance and re-used whenever such an optimization is required.

**Part 3: Basis polynomials and $W$** So far, only the interpolation points were defined. The barrier function requires the matrix $W$ that contains the evaluation of a polynomial basis for $\mathbb{R}_d[\boldsymbol{x}]$ at the interpolation points.

Recall that the matrix $V_{2d}$ was defined as the evaluation of the multivariate Chebyshev polynomials up to degree $2d$ at the interpolation points. Obviously, the multivariate Chebyshev polynomials up to degree $d$ are a valid choice of basis for $\mathbb{R}_d[\boldsymbol{x}]$; and $V_d$ is a submatrix of $L$ rows of $V_{2d}$ (the top rows in a degree-order, which is assumed in the following). The relationship of $V_{2d}$ with the DFT-II matrix in equation (4.68) implies that

$$V_d = \frac{1}{2^n} A\big|_{[\![L]\!],:} D^{\mathrm{II}}(N)\big|_{:,c([\![U]\!])}. \tag{4.70}$$

Since $V_d$ is the evaluation of a polynomial basis for $\mathbb{R}_d[\boldsymbol{x}]$ at the interpolation points and is already defined in product form, it is quite natural to turn this definition into one that is compatible with $W = ST$, where $S \in \{0,1\}^{L\times N_\mathrm{c}}$ is very sparse with at most $N_\mathrm{r} \ll L$ nonzeros per row and $T$ is related to the DCT transform.

Note that by definition, $A$ has no more than

$$N_\mathrm{r} = 2^{\min(n,d)-1} \ll L \tag{4.71}$$

entries per row[15]: The rows in this matrix are indexed by $\boldsymbol{j} \in \mathcal{B}_{n,d}$, while the columns stem from the possible values due to different $\boldsymbol{\sigma} \in \{-1,1\}^n$. Given the even symmetry of the cosine, the entry $\sigma_i$ corresponding to the first nonzero $j_i$ can effectively be fixed to $+1$; and there cannot be more than $d$ nonzero entries in $\boldsymbol{j}$, leading to the upper bound of $2^{\min(n,d)-1}$ entries per row. However, $A$ has $N > N_\mathrm{c}$ columns. Assuming no duplicates arise, an upper

15 If only one of $n$ or $d$ scale, $N_\mathrm{r}$ is effectively $\mathcal{O}(1)$; and even for $n = d$, $\frac{N_\mathrm{r}}{L} = \mathcal{O}(2^{-n}\sqrt{n})$, i.e., $N_\mathrm{r}$ is still exponentially smaller than $L$.

bound for the number of nonzero columns in $A$ can be derived; it is given by

$$N_{\mathrm{c}} \leq N_{\mathrm{c}}^{\sharp} := 1 + \sum_{d'=1}^{d} \sum_{n'=1}^{n} \binom{n}{n'} \binom{d'-1}{n'-1} 2^{\min(n',d')-1}. \tag{4.72}$$

Here, after treating the $\boldsymbol{j} = \mathbf{0}$ case separately, for every $|\boldsymbol{j}|$ given by $d'$, the number of exponents of degree exactly $d'$ that make use of all $n'$ variables is determined[16] and then multiplied by the number of distinct entries such an exponent can have at most. $\frac{N_{\mathrm{c}}^{\sharp}}{U}$ is maximal for $n = d = 1$ with value $\frac{2}{3}$; therefore $N_{\mathrm{c}} < U$. Consequently, there is a set $\mathcal{Z}$ of size $N_{\mathrm{c}}$ such that $A|_{[\![L]\!],\mathcal{Z}} \in \mathbb{Z}^{L \times N_{\mathrm{c}}}$ has no vanishing columns.

While the derived matrix already has the correct dimensionality, $\frac{1}{2^n} A|_{[\![L]\!],\mathcal{Z}}$ is not necessarily a binary matrix. Although not a strict requirement for the scaling improvement, it would benefit an actual computation, saving both storage and multiplication operations. Note that a row $\boldsymbol{j}$ in $S$ corresponds exactly to a basis polynomial $w_{\boldsymbol{j}}$. Since no normalization is required, the basis polynomials can be scaled arbitrarily without changing the basis property. Therefore, if all nonzero values in a row of $S$ are equal, this can effectively be treated as if they were one. To ensure that this is indeed possible, the previous algorithm to find $(N, r)$ should check whether all entries are the same before trying to remodel a row to echelon form; else, the choice of $r$ is invalid. In practice, this further condition turned out to be quite helpful, as for all considered examples, it led to a speedup due to early discardings without increasing the value of $N$, lending an additional justification why a binary matrix is even demanded. Table 4.2 shows the smallest possible $(N, r)$-choices for small values of $n$ and $d$.

Now $S$ has been defined; the missing piece $T$ is then trivially $T := D^{\mathrm{II}}(N)|_{\mathcal{Z},c([\![U]\!])}$. □

**Table 4.2.** Smallest possible choices for $N$ and assignments for $r(i)$ with small variable numbers and degrees. $r(0) = 1$ always holds.

**a)** $n = 1$

| $d$ | $N$ |
|---|---|
| 1 | 4 |
| 2 | 7 |
| 3 | 10 |
| 4 | 13 |
| 5 | 16 |

**b)** $n = 2$

| $d$ | $N$ | $r(1)$ |
|---|---|---|
| 1 | 7 | 4 |
| 2 | 17 | 5 |
| 3 | 31 | 7 |
| 4 | 49 | 9 |
| 5 | 71 | 11 |

**c)** $n = 3$

| $d$ | $N$ | $r(1)$ | $r(2)$ |
|---|---|---|---|
| 1 | 11 | 4 | 5 |
| 2 | 37 | 8 | 23 |
| 3 | 87 | 14 | 23 |
| 4 | 169 | 13 | 46 |
| 5 | 291 | 22 | 57 |

**d)** $n = 4$

| $d$ | $N$ | $r(1)$ | $r(2)$ | $r(3)$ |
|---|---|---|---|---|
| 1 | 17 | 3 | 8 | 11 |
| 2 | 75 | 7 | 23 | 36 |
| 3 | 219 | 7 | 34 | 131 |
| 4 | 509 | 9 | 55 | 140 |
| 5 | 1021 | 11 | 76 | 275 |

16 In order to consider only exponents that use all variables, let $\boldsymbol{j} := \tilde{\boldsymbol{j}} + \mathbf{1}$, where $\mathbf{1}$ is the all-1 vector. If $\boldsymbol{j}$ has degree $d'$, $\tilde{\boldsymbol{j}}$ has degree $d' - n'$. The number of exponents for $n'$ variables of degree exactly $d' - n'$ is given by $\binom{n'+(d'-n')-1}{n'-1} = \binom{d'-1}{n'-1}$, which therefore is the number of exponents that use all variables. These $n'$ variables then still have to be distributed over the whole set of $n$ variables, for which there are $\binom{n}{n'}$ combinations. In total, this yields the binomial identity $\binom{n+d'-1}{n-1} = \sum_{n'=1}^{n} \binom{n}{n'}\binom{d'-1}{n'-1}$.

### 4.9.4 Beyond standard interior-point solvers

Despite many improvements over the recent years, including the approach outlined in the previous section, interior-point solvers still struggle with the large relaxations arising from polynomial optimization problems. So although they are the gold standard for convex optimization, it makes sense to look at other types of solvers that might perform well.

#### 4.9.4.1 Accuracy concerns

First-order solvers do not require the Hessian and therefore scale much better; however, they often come with an accuracy penalty. For polynomial optimization problems, this does not mean that the solution is "just" less accurate—instead, the solution of the *relaxed* problem looses its accuracy. The methods to reconstruct the solution to the original problem, working with rank criteria, necessarily require a good precision in the returned moment matrix[17]. This means that the lost accuracy leads to a complete inability to reconstruct the original solution or certify optimality of the bound. Therefore, solvers that do not allow to go to high accuracies are less suitable for polynomial optimization problems if one is interested in more than bounds.

#### 4.9.4.2 Exploiting low ranks

A promising approach for a solver is to exploit low ranks. The justification for this comes from the fact that for every globally optimal solution to the original problem, the corresponding moment matrix can immediately be written down as a rank-one matrix. By construction, it is a feasible point in the moment relaxation with the same objective value as the original problem. If the relaxation order (the degree cutoff) was high enough, the optimal objective of the relaxation and the problem coincide, so that the rank-one matrix is indeed also a global solution of the moment relaxation. This suggests to look for solver algorithms that try to exploit the low-rankedness of the optimal moment matrix—although, unless the problem has a unique solution, even having a sufficient relaxation order does not guarantee that every solution will necessarily be of low rank.

At the optimum of the semidefinite program—assuming the limit-feasible case can be excluded—the complementary slackness condition, equation (3.17e), will translate into a relationship between the rank of the primal and associated dual optimal variable: their sum is the side dimension of the matrix [AHO97]. The format in which a solver accepts the

17 In [KPV18], it was shown that the extraction of minimizers is robust; this justifies why the algorithms work for numerical solvers at all. However, practical experience reveals that an accuracy of $10^{-3}$ instead of $10^{-7}$ can often lead to a complete failure in the solution extraction; though note that this experience was gained using the algorithm in [HKM18], not the Gelfand–Naimark–Segal-based one in [KPV18], as the latter is usually not used in commutative polynomial optimization.

data defines what this solver considers the "primal" and what the "dual" form; sometimes, Definition 3.19 and Corollary 3.24 are swapped. It is important to be aware of the form for which the solver expects a low-rank structure—the problem must be formulated such that the matrix variable in *this* form corresponds to the *moment matrix*, while the other form will then automatically be of high rank and reflect the matrix for the SOS representation.

The vast majority of, and perhaps even all, low-rank solvers assumes that the problem in the primal form stated in Definition 3.19, where $\mathcal{C}$ is a product of positive semidefinite cones and nonnegative variables, defines the low rank variable; explicitly:

$$\inf_{\substack{\boldsymbol{x},\{X_i\}_i\\ x_j\geq 0\ \forall j\\ X_i\succeq 0\ \forall i}} \left\{\langle \boldsymbol{c},\boldsymbol{x}\rangle + \sum_i \langle C_i, X_i\rangle : \langle \boldsymbol{a}_k,\boldsymbol{x}\rangle + \sum_i \langle A_{k,i}, X_i\rangle = b_k\ \forall k\right\}. \tag{4.73}$$

While this is the natural form of an SOS problem, the opposite is needed. Constructing a moment problem in this primal form thus requires to cast the operation of "putting a scalar variable at some position in a semidefinite matrix," which would be the dual formulation natural for the moment matrix, into the operation "enforcing equality constraints between matrix entries." Given that the moment matrix contains each variable—and only this variable—at isolated positions, this is not a particularly difficult process[18].

#### 4.9.4.3 Comparing specialized solvers

In the following, I will compare various solvers that exploit the low-rank structure of moment matrices or which otherwise might seem particularly suitable for polynomial optimization problems. All solvers mentioned here are the result of (sometimes very) recent research papers. Usually, these papers do not consider polynomial optimization problems in their applications; therefore, the assessment made here is not based on the results reported by the authors. Instead, I implemented an interface to all the solvers for polynomial optimization problems and tested them on large-scale quantum information problems (here, "large-scale" means that the side dimension of the largest semidefinite variable of the relaxation exceeds 1 000—which is still tiny in the original form).

**SketchyCGAL** The seminal paper [Yur⁺21] introducing the first-order primal–dual low-rank solver SketchyCGAL has spurred interest in low-rank solvers beyond the long-known Burer–Monteiro factorization-based methods (see section 3.5.4.2). The solver maintains only a small (Nyström) sketch of the matrix, which is where the low-rank assumption enters; despite this, it offers global optimality guarantees. However, while this might look ideal,

18 The situation can change when sparsity methods are employed, as the moment matrices now no longer necessarily cover all variables that might appear in the localizing matrices.

the accuracy of `SketchyCGAL` is quite low. This goes beyond the slow tail convergence of first-order solvers; experiments show that `SketchyCGAL` also severely struggles to even converge to the correct order of magnitude within an acceptable time for large problems, which makes it unsuitable for polynomial optimization. In addition, `SketchyCGAL` requires a bound on the trace of the involved moment matrices; a bad bound can be very detrimental for convergence. Often, new solvers are benchmarked only against problems such as MaxCut that naturally have the exact trace easily available, but this is not the case for generic polynomial optimization problems[19].

**`ProxSDP`** The first-order primal–dual solver `ProxSDP` [MV22] chooses a different way to mitigate the costs of the projection onto the semidefinite cone usually required for such solvers (see section 3.5.4.1). While it *does* perform the eigendecomposition necessary for the projection, it uses an iterative solver that calculates only the first $r$ dominant eigenvectors, where $r$ is the unknown target rank that is adaptively increased. Even if the decomposition is not exact, it still leads to a valid point in the semidefinite cone, which, however, is no longer the closest point; but an error bound is readily available. This simple idea works remarkably well and makes `ProxSDP` very suitable for use in large polynomial optimization problems if bounds are desired. The slow convergence due to the first-order nature remains an obstacle if accurate moment matrices for solution extraction are required.

**`Loraine`** The concept of interior-point solvers seems to be in opposition to exploiting low-rankedness: by construction, an interior-point solver will converge to a solution in the interior of the semidefinite cone—i.e., a full-rank solution. Nevertheless, the solver `Loraine` [SS24] is an interior-point solver that can exploit low ranks. If a polynomial optimization problem has a *unique* solution and the relaxation order is sufficient, there is also only one optimal moment matrix of rank one, so even interior-point methods will converge to it. `Loraine` now replaces the direct solvers that are usually employed to solve the linear system in Definition 3.30 in each interior-point iteration with an iterative Krylov solver together with a preconditioning that can profit from low-rank iterates.

19 This statement is not completely accurate for polynomial optimization problems where at least some bounds on the variables are known—or, mathematically speaking, their solution space is a compact set. Very often, this is the case; for quantum information problems, it usually follows from the trace normalization condition on the semidefinite variables. In this case, plugging in the bounds for all variables gives the largest possible trace in the moment matrix. Of course, this is an extremely loose bound, which is usually inadequate for solvers that require such a bound.
Mai, Lasserre, and Magron show in [MLM23] that it is possible to *rewrite* a bounded polynomial optimization problem, introducing new squared variables in the original problem that convert inequality into equality constraints. Depending on whether a sphere constraint is present or not, another variable needs to be introduced to explicitly constrain all previous variables to a ball that covers the bounded region. The moment matrix can then be transformed via similarity, preserving the sign of all eigenvalues, and this transformed matrix has a constant trace.

Loraine has two issues: one is a very inefficient implementation, the other that non-unique solutions can invalidate the low-rank assumption. The latter is quickly confirmed in experiments with an improved implementation: non-unique solutions lead to more and more iterations of the inner Krylov solver and preconditioner (though it is actually advantageous to turn the preconditioner off in such cases), increasing the runtime disproportionately.

**LoRADS** The solver LoRADS [Han+25] combines a Burer–Monteiro-based initial solution stage with an ADMM optimizer that can exploit low ranks and is warm-started by the solution from the first stage. Experiments show that for polynomial optimization problems, the first stage is actually quite promising; but when the solver enters the second stage, ADMM first looses quite a bit of the progress that was already made and then takes rather long—often terminating due to a maximum iteration number—to achieve convergence, which is expected to be sublinear. Before the recent second release, the dual solutions were very unreliable, but now an optimality criterion based on the dual infeasibility is included in the termination criteria. Turning off the second stage is an option that should always be tried. While the LoRADS solver is still in an early stage, it is rather promising for extremely large polynomial relaxations. Note that a successor, cuLoRADS [Han+24] claims superior performance through the use of a GPU; however, as no source code was made available, I was unable to test this on polynomial optimization problems.

**STRIDE** An interesting new perspective was introduced in [Yan+23]. The STRIDE algorithm works on the relaxed problem as well as the original problem together. After obtaining a preliminary low-precision solution of the relaxed moment problem using a first-order solver, the corresponding candidates of the original problem are extracted from the moment matrix ("rounding"). Then they are passed on to a generic nonlinear solver for the original problem, which improves on the solution candidate using only a local search. If the nonlinear solver terminates with a better result than was previously found, the results are lifted to a rank-one moment matrix, which is then the input for the next iteration—hopefully accelerating the slow tail convergence.

While the idea itself sounds striking, actually performing these steps is a very delicate task: How can a candidate for the original problem even be extracted if the relaxed solution is of low quality (or not even close to a solution, if the algorithm prematurely terminated because of a maximum iteration or time constraint)? For this reason, the way to extract an initial point for the nonlinear solver deserves a lot of attention; the authors propose to use a problem-specific function that operates on the most relevant eigenvectors of the current "moment matrix." Particularly for constrained problems, such a point might not even be feasible, which either requires an infeasible-start nonlinear solver or a projection onto the feasible set of the original problem, if possible.

What qualifies a local solution found by the nonlinear optimizer to be “better” than a preliminary solution of the relaxed problem? For a minimization, an immediate criterion seems to be: if its optimal value is lower. Note that while the relaxation is a lower bound to the original problem, this is only true at optimality, and the moment relaxation approaches this lower bound as a minimization—so it is indeed possible that such a relation between the two values holds. However, neither the relaxed solution (in the case of premature termination) nor the local solution (in the case of an infeasible initial point) have to be feasible to a satisfactory tolerance. But is an invalid local solution better than an invalid relaxed solution if it is lower, and should the infeasibilities (and, in the case of the relaxed problem, the duality gap) also be considered in the decision? The authors of `STRIDE` only look at problems for which the projection onto the feasible set is so simple to perform that the nonlinear solver always starts with a feasible point.

Even assuming that the nonlinear solver converged to some valid local minimum which is better than the current value of the relaxed problem, the difficulties continue. It is easy to generate a moment matrix based on the local solution, which, as for low-rank solvers, corresponds to the primal variable. However, `STRIDE` employs a primal–dual solver and is now missing the appropriate slack matrix for the dual problem[20]. As this corresponds to the SOS decomposition of the original problem, it is not readily available from the nonlinear solver. Using a meanwhile overwritten primal variable with an old slack variable leads to unpredictable results. Therefore, even with a safeguarding step that accepts a new primal point from the local optimization only if it yields a better objective, the next few steps might very well be worse than they would have been had the nonlinear optimization been skipped.

While the just-mentioned safeguarding step still allows to prove global convergence for `STRIDE`, in the light of these difficulties it is not surprising that the authors “are not able to establish conditions under which the local search algorithms can provide provably better rank-one candidates than the iterates generated” by their inexact projected gradient method used to tackle the semidefinite relaxation [Yan+23]. Indeed, when I implemented `STRIDE`, I found the local optimization step to be detrimental; most of the time, its results were discarded anyway by the safeguarding, but when they were accepted, the next couple of iterations were much worse than before.

`STRIDE` is composed of several steps: finding an initial point using a semiproximal alternating direction method of multipliers (sPADMM); projecting onto the semidefinite feasible set using a symmetric block Gauss–Seidel-based accelerated proximal gradient method (sGS-APG) and accelerated by a limited-memory BFGS; and finally passing over to the nonlinear solver. In my implementation of the algorithms lined out in [Yan+23], the sPADMM and sGS-APG steps worked well to converge to a moderately precise solution; however,

20 The vector of duals to the constraints will be freshly calculated anyway from its relation to the slack variables; its absence from the original optimization problem is therefore irrelevant.

both L-BFGS and the nonlinear solver were not helpful. Unfortunately, the closed-source implementation of the authors does not allow a direct comparison; they might have implemented additional steps omitted in the paper that improve their algorithms. However, as the algorithms are described, I found them not to be particularly helpful for solving difficult polynomial optimization problems.

**`ConicBundle` and `SpecBM Primal`** The proximal bundle method introduced in section 3.5.5 is a memory-efficient first-order method. When the semidefinite constraints are made explicit, it is called the *spectral bundle* method [HR00]. Note that the spectral bundle method requires knowledge of a (good) upper bound to the trace of the semidefinite variable. If the trace is not known (see also footnote 19), a method for adaptively adjusting the trace can easily be devised, which in principle allows for unknown traces. However, the success of such a heuristic is not at all guaranteed; indeed, Helmberg warns in the documentation for `ConicBundle` [Hel20]: “If the primal problem has a bounded optimal solution, this will end at some point, but $\gamma$ might get too large for keeping numerical stability. In general this option is still a bit hazardous, the other two variants [constant or bounded trace] are clearly to be preferred.” In my experiments using `ConicBundle`, I found that the library works rather well, keeping the low-rank structure of the primal form, but only if the trace of the optimal moment matrix is specified beforehand. Bounded-trace and adaptive options failed to produce any meaningful results.

Traditional spectral bundle methods are efficient only if the *primal* problem has low-rank solutions [DG23b]. However, very recently, the spectral bundle algorithm `SpecBM Primal` was introduced in [LDZ25], which is efficient if the *dual* solution has low rank—which allows to use the much more natural problem formulation. I implemented this algorithm, also incorporating a heuristic for an unspecified trace, which appears to work better than the one of `ConicBundle` applied on the rewritten problem. Nevertheless, convergence is rather slow, and the algorithm requires hyperparameters that control the number of stored past and current eigenvectors. These critically control the convergence behavior and currently, there is no proper way to find a good set during the problem analysis[21].

The algorithm still stores the matrices completely and needs to perform an eigendecomposition (albeit with few eigenvalues only) in every iteration. The matrix sketching techniques mentioned for `SketchyCGAL` may be used instead; this has been proposed first in [DG23b] and implemented and open-sourced with more sophisticated constraints in [AM24]. However, both papers assume low-rankedness in the primal variable, and the latter solver currently only supports a single semidefinite variable, which makes it less suitable for general polynomial optimization problems.

21 In [DG23b], linear convergence is shown if the number of current eigenvectors exceeds the largest rank of all optimal moment matrices—which, if finite, is the number of solutions.

#### 4.9.4.4 Other recent developments

The last months have seen a large increase in the literature concerning huge semidefinite programs; this immense influx of new ideas is both desirable and also a large burden. The following is only an incomplete list of potentially relevant algorithms. When a solver is said to work with the low-rank primal form, this always means that—as before—the solver allows to extract items from a low-rank semidefinite variable, i.e., the moment-based approach has to be rewritten introducing lots of equality constraints.

- [BGP21] demonstrates an interior-point inspired algorithm in primal form that converges to low-rank solutions. No implementation is provided.
- [LL24] is a low-rank interior-point solver for polynomial optimization problems that uses arbitrary precision arithmetic and an interpolant basis approach. An open-source Julia package is provided.
- [Asp+23] switches between first-order relaxations and second-order Newton iterations in the original problem, simplified by identifying which inequalities are strict and which can be dropped in the vicinity of the optimum. It appears to work well for unique solutions; but the generalization to non-unique problems with solution extraction techniques might suffer from the precision problems mentioned before. When a solution is obtained by this algorithm, some test of global optimality has to be implemented. No implementation is provided.
- [Wan+23] contains the low-rank Burer–Monteiro-based Augmented Lagrangian method `SDPDAL` for the primal form with global guarantees. No implementation is provided.
- [TT24] presents a manifold-based feasible low-rank solver for the primal form with global guarantees using a rank-adaptive Burer–Monteiro-like factorization. Its discussion on how to obtain a dual variable from the primal one might be of independent interest. No implementation is provided.
- [MSC24] introduces the low-rank first-order solver `HALLaR` for the primal form with global guarantees. By now, a closed-source solver with file-based input is provided.
- [WH25] provides a manifold-based Augmented Lagrangian method for the primal form with global guarantees using a rank-adaptive Burer–Monteiro-like factorization. An open-source MATLAB implementation is provided.
- [Xu+21] gives a way to *verifying* global optimality of a solution instead of *finding* this solution. Again, the problem is that only the moment matrix can be constructed from the supposedly optimal solution, but not the corresponding SOS matrix. The authors decompose the positive semidefinite constraints on the moment and localizing matrices in terms of principle minors (though a more efficient way would be to use the weakly alternating sign of the characteristic polynomial coefficients), then they write down the KKT conditions for the reformulated problem, plug in the supposedly optimal solution (this will reduce all nonlinearities to numbers) and solve the remaining linear program for the Lagrange multipliers. If a solution exists, the point was indeed optimal. No implementation is provided.

#### 4.9.4.5 Working with research software

In the previous subsections, a large—but still incomplete—list of algorithms was given, which might be beneficial for polynomial optimization problems. While some of the mentioned literature also contains benchmarks for polynomial optimization problems, many do not: their focus is on other types of problems. To consistently check whether the algorithms introduced in the papers are actually useful in the polynomial case is a laborious endeavor: most often, no implementation is made public; even if code is accessible, the programming language varies and usually, the algorithms are only proofs-of-concept without much optimization, sometimes containing even blatant inefficiencies. Documentation and comments are notoriously absent.

A fair comparison is therefore only possible if one were to implement optimized versions of all algorithms consistently in one programming language and benchmark them among a variety of different polynomial optimization problems, including all the special cases mentioned before (equality, inequality, and PSD constraints, real and complex problems, fully dense and at least with exact sparsity methods).

In order to judge the different solvers as in section 4.9.4.3, I took the chore of implementing efficient, extended or from-scratch versions where necessary:

- `SketchyCGAL`
  The official implementation[22] by the author is in MATLAB. There are a few more available in Julia and C++, with varying levels of sophistication and efficiency. None of them allow for more than a single semidefinite variable.

  My implementation in Julia supports an arbitrary number of semidefinite variables as well as various methods to calculate the minimal eigenvector.

- `ProxSDP`
  The official solver[23] is implemented in Julia and rather well-written.

  The only critique is that the rather brief high-level documentation does not explain how to call the solver directly, avoiding the `JuMP` interface. However, the conversion routine, together with the paper, allows to infer this.

22 `https://github.com/alpyurtsever/SketchyCGAL`
23 `https://github.com/mariohsouto/ProxSDP.jl`

- `Loraine`

  The official solver[24] is a rather literal translation of the original MATLAB implementation to Julia. It is penetrated with unnecessary allocations that will not only reduce the efficiency, but lead to crashes for large problems. Fields are often left untyped, which removes all the advantage of the compiled language.

  My implementation follows the logic of the original code, eliminating most of these deficiencies; but I dropped the direct solver as well as the handling of dense data matrices, as both are not relevant for the large-scale problems the solver ought to handle.

- `LoRADS`

  The official solver[25], implemented in C, internally stores the matrices in packed format, but needs to convert them to full storage in the iteration.

  This back-and-forth conversion takes a significant amount of time; with my changes, full storage is always maintained, leading to noticeable speedups. The first release contained lots of dead code and came without any explanation of how to use the source code; it also included a bug that would eventually lead to heap corruption. Shortly before finishing this thesis, a second release was made public, cleaning up many of these issues.

- `STRIDE`

  I implemented this solver from scratch, as the official repository[26] only contains read-protected MATLAB source code.

- `ConicBundle`

  The solver [Hel20] is implemented in C++, with a very limited interface in C.

  I wrote a script to generate wrappers of the C++-only functionality (e.g., everything regarding semidefinite programming), although this, given the size of the 20-years-old library, is only an initial starting point which leaves many things to be improved. I also wrote a Julia wrapper of the package[27], which is available via the Julia package manager.

- `SpecBM Primal`

  The official solver[28] is implemented in MATLAB; for the quadratic subproblem, it creates its data in `SeDuMi` format, which is then converted to the format expected by `Mosek`. Only a single semidefinite variable is supported.

  My implementation in Julia avoids unnecessary allocations and directly produces the data in the form required for the subsolver; currently, `Mosek` and `Hypatia` are supported. Additionally, multiple semidefinite variables are supported.

24 https://github.com/kocvara/Loraine.jl
25 https://github.com/COPT-Public/LoRADS
26 https://github.com/MIT-SPARK/STRIDE
27 https://github.com/projekter/ConicBundle.jl
28 https://github.com/soc-ucsd/SpecBM

This list, while representing months of work, is still only a beginning and reveals some of the many problems in scientific programming. Reproducibility is one of the core tenets of (data-generating) science, which is undermined by refraining from publishing the source code that is used to produce the results in their papers. While this might still be understandable, if regrettable, for code that is developed by companies which eventually want to sell their software, for publicly funded research, it is rather mysterious.

As source code and algorithm are not the same, it is always a danger to judge one by the other: a bad or faulty implementation can shed doubts on a promising algorithm; and a finely tuned implementation with many optimizations that go beyond the core algorithm might no longer resemble it well enough for the proofs to even apply, or become a crucial part for the original algorithm to actually work well. This makes it all the more important that the research community is able to check whether the results shown in papers are actually representative of the algorithms described therein.

## 4.10 Software frameworks

In order to solve polynomial optimization problems in practice, it is not sufficient to have good numerical solvers for semidefinite programs (or the dual SOS cone) available. The solvers must also be called with the proper data resulting from the original optimization problem. While this might seem a trivial observation, it is nevertheless crucial and a stumbling block whose resolution can be more resource-hungry than the actual call to the solver. Formulating the relaxation "by hand" and passing the data on manually is out of the question for all but the most trivial problems. A software framework is therefore needed that takes as input the polynomials, entered in a way that is simple to do for a human, builds the relaxation according to the specifications given by the user (e.g., regarding sparsity/symmetry), sets up a solver of the user's choice, calls the solver, parses the returned moment data, and extracts solutions. Section 4.10.1 gives an overview of existing frameworks; rather in-line with the trend currently observable in mathematical programming, MATLAB (mainly for older software) and Julia are the prevalent languages.

### 4.10.1 Existing frameworks

`GloptiPoly` [HL02; DL09], a MATLAB toolbox for dense relaxations, has been able to do some of the aforementioned tasks from the early days of polynomial optimization. It was initially designed to use `SeDuMi` [Stu99] as a semidefinite solver, although since version 3, other solvers can also be employed via the modeling framework `YALMIP` [Löf04]. With only little development over the years, `GloptiPoly` is still notoriously inefficient and cannot be used for practically relevant problems. Nevertheless, it implements an interesting feature: moment substitution. Instead of adding equality constraints in polynomials, when these are in a particularly simple form, they may be used as substitution rules in the moment matrices. This is a simplified version of applying the machinery of Gröbner bases [CLO15; Buc65] to the problem, which is often mentioned as another way of working with equality constraints. However, Gröbner bases are extremely challenging to compute for generic problems, so that having lots of linear constraints and potentially redundant monomials in the solver might still be much more efficient overall than their elimination. In simple cases, moment substitution can thus be a useful intermediate step between Gröbner bases and not exploiting the constraints at all. The most widely used solution extraction algorithm [HL05] was originally developed for `GloptiPoly`.

`SOSTools` [Pap+21] is another MATLAB package that has been available for quite some time. Only the most recent update to version 4 has improved the performance to such a degree that the solver becomes the bottleneck in the problem formulation. Scalar and matrix-valued constraints are supported; exact sparsity methods (Newton polytope, diagonal consistency, and facial reduction) are implemented. Complex-valued problems and inexact sparsity methods are not supported; neither are DD/SDD formulations. A number of solvers can be used; `SOSTools` generates its data in the format native for `SeDuMi` and converts it for other solvers, if necessary; this can be quite inefficient.

`YALMIP` [Löf09] itself also offers a module for SOS problems and moment relaxations, which supports scalar and matrix-valued constraints, exact sparsity methods (Newton polytope, diagonal consistency), sign symmetry, correlative sparsity and DD/SDD formulations, but many of these features are undocumented. Complex-valued problems are not supported.

`SparsePOP` [WM11; MT16] is a MATLAB package that implements a couple of algorithms for sparsity developed by its authors that are not found elsewhere, though none of them is documented. Scalar and matrix-valued constraints are supported; complex-valued problems and DD/SDD formulations are unsupported. A number of solvers can be used; `SparsePOP` generates its data in the format native for `SeDuMi` and converts it for other solvers, if necessary.

`Moment` [GA24] is the most recent framework for MATLAB (with main parts in C++), published during the writing of this thesis; it is mainly intended for noncommutative polynomial optimization. Recognizing the problem of formulating the moment relaxation as a crucial bottleneck, this software tries to do it in the most efficient way possible. Curiously, still, `Moment` relies on `CVX` or `YALMIP` to then pass the semidefinite formulation on to the solvers (which the authors see as a feature, providing solver-agnosticism). `Moment` supports manual moment substitution, but no sparsity or a DD/SDD formulation. Complex-valued problems are in a way supported, as they usually are for noncommutative optimization, but this is different from the notion of a complex-valued problem used here.

`SumOfSquares.jl` (and the extension `MomentOpt.jl`) [Leg+17; Wei+19] is a more recent Julia package with extensive features. It supports scalar and to some degree matrix-valued constraints, bases different from the monomial one, exact and approximative sparsity, arbitrary symmetries, and the DD/SDD formulation (though without rotation). Complex-valued problems (in the sense of section 4.6) are not supported. Interestingly, the package offers different choices of bases—not only the monomials, but also arbitrary polynomial bases such as the Chebyshev one, improving numerical stability. Most recently, changes to support the interpolant-based SOS cone were made. `SumOfSquares.jl` is built on top of the modeling framework `JuMP` [Lub+23], which makes a large set of solvers available; however, this also means that the modeling layer of `JuMP` lies between the solver and the semidefinite relaxation. As `JuMP` is certainly one of the most efficient modeling frameworks available, the overhead can be negligible—or significant, depending on how well the conversion between the internal format of `JuMP` (or, rather, `MathOptInterface`) and the solver is implemented or even possible. While `SumOfSquares.jl` offers many features, it is often not easy to see how to combine these together, which poses a significant barrier to overcome when first using the package. This becomes particular cumbersome when optimization with multiple matrix constraints and sparsity methods is involved, which surpasses the simple examples in the documentation.

`MomentPolynomialOpt.jl` [MB20] is another Julia package that relies on `JuMP` to formulate the moment problem. It is rather restricted in its functionality; no complex-valued problems, matrix-valued constraints, sparsity, or DD/SDD are supported. However, it implements some interesting algebraic algorithms such as the extraction of exact (rational) SOS certifiers [BKM24].

`TSSOS.jl` [MW21] is a Julia package that is specifically designed for inexact sparsity methods. It supports the Newton polytope in the unconstrained case, complex-valued problems, and recently also matrix-valued constraints and symmetry methods. However, many of these features do not combine together. For some reason, only a small subset of solvers is supported although `TSSOS` relies on `JuMP` as well to construct the problem data, with all drawbacks previously mentioned.

In Macaulay2, the package `SumsOfSquares.m2` can be used [CKP20]; it is mainly intended to provide rational certificates for sums-of-squares. It is very limited in its functionality and constructing the semidefinite model seems to be more resource-demanding than solving it.

In Python, `Ncpol2sdpa` is a package that is based on the `Picos` modeling system and `Sympy` to enter the expressions with all the performance drawbacks. It offers scalar equality and inequality constraints, where for the former it uses monomial substitution as mentioned in `GloptiPoly`. The only type of supported sparsity is correlative sparsity. Complex-valued problems, matrix constraints, or DD/SDD are not supported.

Finally, `SumOfSquares.py` [Yua20] is another Python-based framework that supports scalar equality and inequality constraints. The only supported type of sparsity is the Newton polytope. Complex-valued problems, matrix constraints, or DD/SDD are not supported.

### 4.10.2 Requirements and critique

The previous section listed the current state-of-the-art for polynomial optimization frameworks. Feature support varies a lot, and many of the frameworks are not very suitable for larger problems. However, this is not the only problem: code quality is often lacking. For one, this relates to the notoriously underdeveloped documentation typical for research-level software [HF22]: the packages are often only explained by short examples for the end user; the code itself is far from self-explanatory, missing structure and comments. This is particularly bad when it turns out that the code deviates from the papers that laid out the theoretical foundation without any mention or explanation—is this an error or simply an alternative or more efficient way to do a task in practice? For all but the most trivial functions, only a test suite can establish confidence in the correctness, and not all of the frameworks take this task seriously. Good tests are comprised of "global" tests that verify the overall functionality, e.g., by using lots of examples from the literature to verify that they give the same result (or reveal errors in the literature, which is unfortunately also common). However, there must also be tests that are specifically designed for individual methods, calling them with parameters that invoke special cases or could lead to problems. It is in particular the latter category of tests that is absent far too often.

These requirements concerning code quality are absolutely indispensable when a software framework is to be used publicly. Other items are subject to design decisions. As laid out before, most solver frameworks rewrite the polynomial optimization problems into semidefinite programs by using another modeling framework in turn. This has a huge advantage, as it immediately brings a lot of flexibility: solvers can be switched easily, and the often difficult question of how a certain solver actually requires the data has already been addressed elsewhere—eliminating a possible source of error (or deferring it to a separate, hopefully well-tested, piece of software). However, it is also a disadvantage: any modeling framework necessarily introduces further overhead, if only because it has to store the data in its own internal format. In the best case, the internal format of the framework and the solver coincide *and* the framework can access the internals of the solver[29]. In the worst case, complicated rewritings are required that might not only take time, but also consume more memory than what is available. It is therefore desirable from a performance point of view that the polynomial optimization framework avoid any intermediate layer and access the solver in the most low-level way possible. The data for the SDP should be constructed from the polynomial optimization problem directly in the way required by the solver.

29 This is usually a problem for commercial solvers. Whatever internal and undocumented format they use, it is not exposed; instead an API has to be used that builds up the problem incrementally. So even if the internal formats of the modeling framework and the solver were the same, the data could not necessarily just be referenced or at least copied verbatim.

### 4.10.3 New framework: `PolynomialOptimization.jl`

The issues mentioned previously were the reason for the development of a new framework for polynomial optimization during this thesis, `PolynomialOptimization.jl`. Given the focus on quantum information problems, the framework had to support both complex-valued data as well as polynomial matrix constraints in addition to the normal scalar equality and inequality constraints.

#### 4.10.3.1 The choice of language

`PolynomialOptimization.jl` is written in Julia [Bez+17]; its development began with Julia 1.9 and the current version is based on Julia 1.10, which is the latest long-term support version of Julia. Compatibility with Julia 1.11 has not been tested yet. Julia is a programming language whose development started in 2009. It is based on just-in-time (JIT) compilation to machine code through a LLVM backend [LA04]. The JIT compiler allows a great flexibility in method[30] signatures: it is not required to specify the types of variables beforehand; the compilation for the actual types happens in the moment the method is invoked. This implies a lack of some of the checks that can be done by a fully static compilation process[31], but it does not compromise performance in the slightest—apart from the compilation cost, which can now no longer be delegated to the developer, although precompilation can defer it at least partly to the time of package installation.

Furthermore, Julia seems to establish itself more and more as the language of choice for reference implementations of new numerical solvers, which makes it easy to interface them. Solvers written with a well-defined API and ABI can be addressed directly using native system calls (if an interface does not already exist, which is rare); interoperability with other interpreted languages is also high through `JuliaInterop` packages[32].

#### 4.10.3.2 Creating a problem

In the following, short code snippets serve as a step-to-step guide on how to start using the framework. While the snippets mainly come from the guided tour of the documentation, I will also explain design decisions and implementation details that proved crucial for performance and scalability. The package has been in development for over two years and seen several redesigns of internal concepts when it became evident that the first direct

30 In Julia nomenclature, a *function* is just an access point for something that can be called; the *method* is an actual instance of the function with a certain tuple of parameter types—a function can have many methods. This is not related to object-oriented programming in which methods are functions/procedures/subroutines that *belong* to a class.

31 To do the very active Julia community justice, it should be pointed out that there are static analysis packages that can step in to fill the gap.

32 `https://github.com/JuliaInterop`

approach is not scalable or flexible enough. The knowledge gained from these attempts is later summarized in section 4.10.3.7.

To start using the framework, it has to be imported, together with an implementation of the `MultivariatePolynomials` interface [Leg22] to formulate the problems in the first place.

```
julia> using PolynomialOptimization, DynamicPolynomials

julia> @polyvar x[1:3];

julia> prob = poly_problem(1 + x[1]^4 + x[2]^4 + x[3]^4 + x[1]^2*x[2]^2 +
                           x[1]^2*x[3]^2 + x[2]^2*x[3]^2 + x[2]*x[3])
Real-valued polynomial optimization problem in 3 variables
Objective: 1.0 + x₂x₃ + x₃⁴ + x₂²x₃² + x₂⁴ + x₁²x₃² + x₁²x₂² + x₁⁴
```

This constructs a polynomial optimization problem with three real-valued variables and a fourth-order polynomial as its objective that is to be minimized.

The `poly_problem` constructor is the entry point for every polynomial optimization problem; it takes various optional keyword arguments: `zero` and `nonneg` each expect a vector of polynomials that are constrained to be zero or nonnegative, while `psd` needs a vector of matrices of polynomials that must be positive semidefinite.

Multiple options exist to modify the problem in order to help convergence; for example, `tighter=true` will invoke Nie's scheme of adding the necessary optimality conditions from the KKT conditions in Theorem 3.9 as explicit constraints to the problem [Nie19]. Nie showed that the unknown Lagrange multipliers can be replaced by polynomials—this is quite intuitive when comparing the generic Lagrangian with a positivstellensatz. Convergence of the problem with added KKT conditions is better than without or even possible for the first time; however, note that the required degrees of the polynomials are not known beforehand. They must obey a certain form which can be derived by automatically differentiating the original problem. This form can lead to contradictions if the chosen degree was not large enough; in fact, it might happen that no degree is ever sufficient. This is not discussed in [Nie19], although Example 5.6 in the paper exhibits exactly this behavior; the given solution there is erroneous, which was found when using the results of the paper in the test suite for `PolynomialOptimization.jl`.

#### 4.10.3.3 Creating a relaxation

Unless some modifying options were specified, `prob` is just a wrapper for the original polynomial optimization problem, which is pretty-printed when inspected; no relaxation has been created so far. This can now be done by constructing one of the relaxations defined in the corresponding submodule:

- `Relaxation.Dense` is the starting point for all further relaxations; it corresponds to the dense monomial basis of a specified or automatically chosen degree. As detailed below, internally, a monomial is stored as an integer. Therefore, a dense basis is in fact only a unit range—it can be constructed instantaneously and requires constant memory.
- `Relaxation.Custom` allows to specify a user-defined basis; it is the only relaxation different from the dense one that can be constructed directly from a problem.
- `Relaxation.Newton` is a relaxation based on the Newton polytope. It requires another relaxation as an input—typically a dense one—and checks whether each monomial is actually in the half Newton polytope of the Putinar representation of the input relaxation. The relaxation also works with constraints using the technique described in section 4.4.1.3, and may provide a further filter of a sparse relaxation.
  Internally, the function requires a linear solver, which, following the common paradigm of the framework, is addressed directly. Importantly, the linear program that has to be solved for each candidate monomial always has the same structure; only some coefficients change. For best efficiency, the linear program is therefore only constructed once with small modifications from monomial to monomial. This kind of parametric programming is often not or not very well-supported by modeling frameworks[33].

  The Newton polytope implementation offers various preprocessing steps that aim at reducing the number of variables in the linear program by first eliminating unnecessary points in the convex hull. The Akl-Toussaint heuristic [AT78] is enabled by default, as it is rather cheap to compute; both a randomized as well as a full search can be requested manually. The `verbose` option controls progress monitoring. Note that while there are more sophisticated algorithms for convex hulls than the application of successive linear programs, these all work in low dimensions (often at most 3D); and here, the dimension corresponds to the number of variables in the problem.

  After determining the minimally and maximally possible exponents for each variable, the main algorithm iterates through all candidate monomials and checks membership in the convex hull. It allows for multi-threading, and if necessary even distributed computing. The resulting implementation is extremely efficient—see appendix B for examples.
- `Relaxation.SparsityCorrelative` applies the correlative sparsity from section 4.4.2.1 to an already existing relaxation. Various keyword arguments allow to define high-order and low-order constraints as suggested in [JM18]. Chordal completion is enabled by default and can be disabled.

33 To support this claim, note that during the development of the framework solver, I found a number of critical bugs in `Mosek`, which have by now been fixed. Some of them were related to the modification of solved problems and would have been exposed for a long time had they been used by modeling frameworks. To quote the `Mosek` support, I have apparently been doing these task modifications “as literally the first user ever.”

- `Relaxation.SparsityTermBlock` and `Relaxation.SparsityTermCliques` are the term sparsity relaxations from section 4.4.2.2 using connected components or maximal cliques as graph extension. These relaxations support the `iterate!` method; in each iteration, the choice between connected components and maximal cliques can be overwritten; and some constraints or the objective may also be excluded from iteration. Chordal completion for clique determination can be turned on or off.
- `Relaxation.SparsityCorrelativeTerm` is a convenient shorthand. Thanks to the modular interface, it only instantiates a `SparsityCorrelative` relaxation, which is then passed on to one of the term relaxations depending on the keyword arguments.

For this very small example, none of the sparsity methods make much sense; therefore, a dense relaxation is used. It can be constructed explicitly:

```
julia> rel = Relaxation.Dense(prob)
[ Info: Automatically selecting minimal degree cutoff 2
Relaxation.Dense of a polynomial optimization problem
Variable cliques:
  x[1], x[2], x[3]
PSD block sizes:
  [10 => 1]
Relaxation degree: 2
```

Hence, the optimization will involve a PSD variable of side dimension 10. The optional second parameter to `Relaxation.Dense` allows to specify the relaxation degree manually.

More information about the relaxation can be obtained by inspecting the groupings of the relaxation: this corresponds to the separate bases whose sizes determine the side dimensions of the variables. In the dense case, the objective and all constraints have one basis each, but in the sparse case, multiple bases can be present.

```
julia> Relaxation.groupings(rel)
Groupings for the relaxation of a polynomial optimization problem
Variable cliques
================
[x₁, x₂, x₃]

Block groupings
===============
Objective: 1 block
  10 [1, x₃, x₂, x₁, x₃², x₂x₃, x₂², x₁x₃, x₁x₂, x₁²]
```

#### 4.10.3.4 Optimizing the relaxation

To actually run the optimization, one of the supported solvers has to be loaded. The solvers that were mentioned in section 4.9.4.3 as having been implemented by me are available automatically when loading `PolynomialOptimization.jl`, though this may change in the future should they be externalized into standalone packages. Table 4.3 lists all currently supported solvers.

**Table 4.3.** Overview of the solvers that are implemented in `PolynomialOptimization.jl`

| Solver | Package | License | Method |
|---|---|---|---|
| `Clarabel` | `Clarabel.jl` | Apache | moment |
| `ConicBundle`[a] | `ConicBundle.jl` | GPL3 | primal moment |
| `COPT` | `COPT.jl` | commercial | moment |
| `Hypatia` | `Hypatia.jl` | MIT | moment |
| `LANCELOT` | `GALAHAD.jl` | BSD | nonlinear |
| `Loraine` | included | MIT | primal moment |
| `LoRADS` | external[b] | MIT | primal moment |
| `Mosek` | `Mosek.jl` | commercial | SOS, moment |
| `ProxSDP` | `ProxSDP.jl` | MIT | primal moment |
| `SCS` | `SCS.jl` | MIT | moment |
| `SketchyCGAL` | included | | primal moment |
| `SpecBM` | included | | SOS |
| `STRIDE`[a] | included | | moment and nonlinear |

*a* While `ConicBundle` and `STRIDE` are supported, they turned out not to be very successful, so that the interface is available only in a separate branch. This still provides the means for experimenting and perhaps improving performance.

*b* `LoRADS` is a C library that must be compiled separately; the interface is already included.

The last column in the table lists the method that is used for the translating the relaxation into an actual optimization problem.

"Moment" is based on the Lasserre formulation, where the (scalar) primal variables are moments that are assembled into a semidefinite cone—this is the most common formulation for modern general-purpose semidefinite solvers.

"SOS" corresponds to the case in which the solver has monolithic semidefinite primal variables and their entries are extracted to be put into constraints—the PSD variables then are the coefficient matrices for the SOS representation in Putinar's positivstellensatz.

Finally, "primal moment" corresponds to the case in which the interface of the solver is as with "SOS"—the primal variables are again matrices, conforming with the definition of the primal form in Definition 3.19. However, since the solver exploits low-rank structures on these variables, they must correspond to the moment matrices as explained in section 4.9.4.2. *Constructing* this formulation is less efficient compared to the other cases, as the format cannot be built directly from the problem data; it has to be converted. Additionally, the required equality constraints increase the problem size, but still, the low-rank structure can make it worthwhile. This interface is experimental; for efficiency reasons, it makes certain assumptions about the relaxation—in particular, that each moment that occurs *somewhere* in the problem also occurs in a moment matrix (not just in a localizing matrix). This is always the case for dense relaxations, but sparsity might break the assumption.

Note that `LANCELOT` is special: It is a nonlinear solver that works on the problem directly instead of on a relaxation. Consequently, it does not provide any global guarantees, but the problem also does not grow before solving. The interface was mainly provided because `STRIDE` requires a nonlinear solver, but it can also be used on its own.

Now is the time to perform the optimization; this is demonstrated with each method.

```
julia> using Clarabel, Mosek

julia> res1 = poly_optimize(:Clarabel, rel)
Polynomial optimization result
Relaxation method: Dense
Used optimization method: ClarabelMoment
Status of the solver: SOLVED
Lower bound to optimum (in case of good status): 0.9166666672624658
Time required for optimization: 0.0017566 seconds

julia> res2 = poly_optimize(:MosekSOS, rel)
Polynomial optimization result
Relaxation method: Dense
Used optimization method: MosekSOS
Status of the solver: Mosek.MSK_SOL_STA_OPTIMAL
Lower bound to optimum (in case of good status): 0.916666648370901
Time required for optimization: 0.0028159 seconds

julia> res3 = poly_optimize(:Loraine, rel)
Polynomial optimization result
Relaxation method: Dense
Used optimization method: LoraineMoment
Status of the solver: STATUS_OPTIMAL
Lower bound to optimum (in case of good status): 0.9166666664309234
Time required for optimization: 0.006898 seconds
```

In all cases, the optimization was successful and delivered a lower bound to the global minimum of the function. Note that the timings shown here exclude compilation. The first argument to `poly_optimize` is always a symbol consisting of the name of the solver, optionally followed by `Moment` or `SOS` to disambiguate if multiple methods are supported. Various options can be supplied—in particular, verbose output during the construction of the problem and from the solver can be enabled using the `verbose=true` keyword.

#### 4.10.3.5 DD and SDD optimization

Instead of relying on semidefinite matrices, the (scaled) diagonally dominant form can also be employed, potentially with rotations. For this, a keyword argument in the optimization method is available:

```
julia> res_dd = poly_optimize(:Clarabel, rel,
                              representation=RepresentationDD())
Polynomial optimization result
Relaxation method: Dense
Used optimization method: ClarabelMoment
Status of the solver: SOLVED
Lower bound to optimum (in case of good status): 0.5000000105896211
Time required for optimization: 0.0009393 seconds
```

Rewriting the cone using the DD representation has decreased the lower bound; it is now far from optimal. Re-optimizing the problem, automatically using a Cholesky factor of the previous SOS matrix as rotation matrix will improve the bound:

```
julia> getproperty.(((global res_dd; res_dd = poly_optimize(res_dd))
                      for _ in 1:5), :objective)
5-element Vector{Float64}:
 0.7882579368640298
 0.8744697855281065
 0.9122828985708029
 0.9160797358944991
 0.916563314002245
```

So indeed, by a succession of rotations, the objective again approaches the previous value. Similarly, `RepresentationSDD` can be used for the scaled diagonally dominant cone, and it is possible to choose different representations for each grouping in the relaxation. The translation to DD or SDD will use all capabilities of the solver: for complex-valued problems, the DD cone will use quadratic constraints or double the matrix size; in any case, the $\ell_1$- or $\ell_\infty$-norm cone, depending on the problem formulation, is preferred if available.

Note that in the previous code listing, warnings were omitted. They state that the chosen solver does not support re-optimization, but that the optimization task has to be reconstructed from scratch. In fact, while the framework provides enough information to modify the previous task, currently, none of the solver implementations supports this, which was a deliberate design decision based on experiments. It turned out that using rotations on large-scale problems raises the issue of constructing the constraint matrix in the first place, which is now fully dense as already mentioned in section 4.7. Therefore, nondiagonal DD/SDD rotations should be restricted to small-scale cases—where the PSD form would be perfectly feasible anyway. This does not mean that the rotated DD/SDD form has no future: a native implementation of the cones in a solver, parameterized by the rotation matrix, would keep the coefficient matrix sparse. The corresponding interface in `PolynomialOptimization.jl` already exists; once a solver declares support for the cone, it will immediately be used.

#### 4.10.3.6 Optimality certificate, solution extraction, and SOS decomposition

After optimizing the problem with various solvers and representations, finally the question is: was the bound optimal? Any of the previous optimization results can be used:

```
julia> optimality_certificate(res1)
:Optimal

julia> optimality_certificate(res_dd)
:Unknown
```

This optimality certificate is based on the flat truncation criterion in Lemma 4.58.

If a certificate can be obtained, it should be possible to extract solutions from the optimization.

```
julia> poly_all_solutions(res1)
2-element Vector{Tuple{Vector{Float64}, Float64}}:
 ([-1.873515685740035e-20, -0.4082426616610129, 0.4082426611580507],
 ↪  5.324274354734371e-10)
 ([-3.4809857219002294e-19, 0.40824266166101386, -0.4082426611580515],
  ↪  5.324275464957395e-10)

julia> poly_all_solutions(res_dd)
2-element Vector{Tuple{Vector{Float64}, Float64}}:
 ([2.3657129145500863e-7, -0.40953165432967403, 0.4175499396753497],
  ↪  0.00020366431026463963)
 ([1.5566047247690354e-6, 0.41070617776349994, -0.41874739291313934],
  ↪  0.00023260190058127517)
```

The solutions are presented as a vector of potential candidates: the first item in every vector is the supposed optimal point, the second item indicates the quality of the solution. This is given by the maximum of all constraint violations (if there are any), or the difference between the predicted lower bound and the actual evaluation of the objective at the given point. If this value is numerically zero, as for the first case that used the PSD cone (line 69 in the above listing), the lower bound is known to be exact.

Note that a solution such as the one from `res_dd` with a mediocre, but probably still useful quality, can be taken as the starting point for a refinement using a nonlinear solver. `LANCELOT` can be used for this; calling `poly_optimize` on the *problem* (not the relaxation) will return a function that expects an initial starting point with which to invoke `LANCELOT`.

```
75 julia> sol = first(poly_solutions(res_dd));

76 julia> nlopt = poly_optimize(:LANCELOT, prob);

77 julia> nlopt(sol)
78 (0.9166666666666979, [5.742736005918116e-12, -0.4082483684519115,
   ↪  0.40824844585271064])
```

Just to demonstrate the possibility, `poly_solutions` was used, which provides a lazy unordered iterator through all the solutions that are collected by `poly_all_solutions`; this may be helpful for sparse problems where stitching together solutions from various cliques can lead to a great many results. As demonstrated, the initial point close to the optimum allowed `LANCELOT` to refine the result. Instead taking the origin as a random guess would have resulted in a local minimum:

```
79 julia> nlopt(zeros(3))
80 (1.0, [0.0, 0.0, 0.0])
```

Finally, most of the solvers are primal–dual solvers: they reveal the solution for both the moments and the SOS certificate. The latter can be extracted explicitly:

```
81 julia> SOSCertificate(res1)
82 Sum-of-squares certificate for polynomial optimization problem
83 1.0 + x₂x₃ + x₃⁴ + x₂²x₃² + x₂⁴ + x₁²x₃² + x₁²x₂² + x₁⁴ - 0.9166666672624658
84 = (-0.2492464642498061 + 0.0 + 0.0 + 0.0 + 0.6811832979888123x₃² -
   ↪  0.13316036665354078x₂x₃ + 0.6812065898234094x₂² + 0.0 + 0.0 +
   ↪  0.7152479291533441x₁²)²
85 + (-0.00015840159683466377 + 0.0 + 0.0 + 0.0 + 0.6360091467299869x₃² +
   ↪  0.0011902230323755974x₂x₃ - 0.6338687294141168x₂² + 0.0 + 0.0 -
   ↪  0.0018514462038562963x₁²)²
```

```
+ (0.050599062148224884 + 0.0 + 0.0 + 0.0 - 0.35322230868481075x₃² -
↪  0.40669620906370346x₂x₃ - 0.3569973936050725x₂² + 0.0 + 0.0 +
↪  0.6183225665112776x₁²)²
+ (0.0 - 0.6307886589586179x₃ - 0.6307886578526084x₂ + 0.0 + 0.0 + 0.0 + 0.0 +
↪  0.0 + 0.0 + 0.0)²
+ (0.0 + 0.0 + 0.0 + 0.0 + 0.0 + 0.0 + 0.0 - 0.5707311041008931x₁x₃ -
↪  0.5706316416095126x₁x₂ + 0.0)²
+ (0.1365622366906534 + 0.0 + 0.0 + 0.0 - 0.08194926810609472x₃² +
↪  0.6553389692882523x₂x₃ - 0.08198096663343958x₂² + 0.0 + 0.0 +
↪  0.32572101226607164x₁²)²
+ (0.0 + 0.0 + 0.0 + 0.0 + 0.0 + 0.0 + 0.0 - 0.43861947273884583x₁x₃ +
↪  0.4386959251861786x₁x₂ + 0.0)²
+ (0.0 + 0.0 + 0.0 + 0.4527802849918676x₁ + 0.0 + 0.0 + 0.0 + 0.0 + 0.0 + 0.0)²
```

This decomposes the original objective minus the global lower bound into a sum of squares. Indeed, expanding the right-hand side shows agreement in all coefficients up to a numerical error of less than $10^{-8}$.

#### 4.10.3.7 Observations

While developing the package, profiling various strategies that can be commonly found in the literature or other packages was possible. This has reflected in the reconsideration, removal or redesign of several features.

**Equality constraints and Gröbner bases** As mentioned in section 4.10.1, Gröbner basis methods [CLO15; Buc65] are a way to account for equality constraints. If such a basis for all equality constraints is available, all other polynomials can be reduced with respect to it and the full monomial basis can be replaced with a standard basis[34]. The initial version of `PolynomialOptimization.jl` made use of Gröbner bases when possible; however, it turned out that the savings in the SDP are often marginal for quantum information problems, where, different to the often-cited examples from combinatorics, the Gröbner basis is not known beforehand, while the cost of its calculation can be disproportionately large. Additionally, by taking every polynomial modulo the ideal, the entries in the moment matrix are no longer monomials but polynomials with an unknown number of terms instead. Having to account for this fact prevents some useful optimizations during problem construction. Therefore, in a redesign of the package internals, all Gröbner basis methods were completely removed.

34 A standard basis contains only those monomials that are not divisible by any leading term of the polynomials in the Gröbner basis.

**Polynomial representation** For interfacing with the user, a comfortable way of entering polynomials is required; the `MultivariatePolynomials` interface [Leg22] was chosen for this. Using the most popular `DynamicPolynomials` implementation is often sufficient even in the internal representation. However, for very large problems, the overhead incurred by the rather inefficient multiplications can become quite relevant; for this reason, a new internal representation was developed, which works without allocations. This representation is able to identify a monomial by its position in a degree-lexicographic basis with respect to a multivariate exponent set. While the multiplication of two monomials is now slightly slower from the algorithmic point of view, the fixed size of the result—usually a 64-bit unsigned integer, but this can be generically parameterized—means that no vectors have to be allocated to hold the exponents, giving a net speedup. Since the chosen order is degree-lexicographic (with the possibility to limit the exponent range of individual variables), the storage is efficient: there are no gaps[35].

At the moment, this internal representation does not implement all the features of `MultivariatePolynomials`, as they are not required for the tasks here. However, it would certainly be possible to complete the interface and turn the submodule into a fully-fledged package independent of `PolynomialOptimization.jl`.

**Dense vs. sparse** Inexact sparsity methods seem very promising; however, they require graph-based calculations to derive the sparsity pattern. Similarly to Gröbner bases, this can be a very significant part of the overall computation time; and as the algorithms themselves do not scale very well, there is not much to be gained by more optimization. However, before these algorithms are invoked, sparsity frameworks often require to build the full dense basis, which is then filtered according to some sparsity criterion. For very large problems, already the construction of this dense basis is prohibitive, not only feeding it to a solver.

This is automatically solved by the internal representation of monomials as integers; a dense basis of arbitrary size is completely specified by just two integers: the number of variables and the maximum degree. Since iteration through this basis can be made slightly more efficient than random access, a lazy iterator was additionally implemented.

35 As an example, consider storing monomials in 100 variables, knowing that the degree will never exceed five. A bit-packed storage (like the `SIMDPolynomials` package [Ma23] uses and which is usually quoted as the fastest polynomial implementation available in Julia) that also allows to map the multiplication of monomials into a simple addition of indices would require to reserve $\lceil \log_2 5 \rceil = 3$ bits per variable, giving 300 bits in total. However, even disregarding the excess due to the fact that five is not a power of two, it is possible to represent, e.g., the monomial $x_1^5 \dots x_{100}^5$, although its total degree for sure excludes it from the predetermined range. In fact, $5^{100} - \binom{100+5}{100} \sim 5^{100}$—i.e., almost all—representable monomials, still disregarding the bit alignment, are not in the class of valid monomials. Of course, this is due to the restriction of the *total* degree instead of the individual *variable* degrees—but this is by far the most common situation. It is easy to see that 64 bits cannot represent a lot of variables even with rather low degrees if such an arithmetically favorable representation is chosen. In this example, the degree-ordered variant would require only 27 bits.

**Clique merging** When several sparse bases are found that all differ by only very few elements, they will lead to multiple semidefinite matrices in the solver, as opposed to one only slightly larger matrix without the sparsity analysis. If the size difference is small enough, the sparse version can consume (significantly) more resources than the dense one. To avoid such an issue, the concept of *clique merging* was introduced in the `COSMO` solver [GCG21] and has also been applied to polynomial optimization problems in `TSSOS` [WML21a]; it is a graph-based algorithm that merges such bases together when a heuristic cost estimation says that it is beneficial to do so. While `PolynomialOptimization.jl` also includes an optimized version of a clique merging algorithm, in practice this appeared to be not particularly helpful. The algorithm itself can take up a significant portion of the run-time, where the situations in which it is really helpful are quite rare; therefore, it is disabled by default.

**Modeling middleware** While it was repeatedly claimed that `PolynomialOptimization.jl` uses no modeling framework, it in fact provides one of its own that comes with no performance penalty at all. That this is possible was a result from a first version in which each of the solvers was completely separate and had to work with the groupings and polynomials in the problem directly.

It became evident that solvers can be divided into two groups. The first group are those with a “data-based” API, where the entire structure that makes up the problem is specified beforehand in memory—this is usually a collection of dense vectors and matrices in compressed sparse column or row format—and then a single call to the solver consumes the data. The second group are solvers with an “instruction-based” API, which builds the problem incrementally, for example by appending rows or columns to the constraint matrices. As already mentioned in footnote 29, commercial solvers all fall into the second category, open-source solvers use the first approach.

With this split in mind and, importantly, not having to provide a generic middleware that allows to later read the problem specification again—only the solution is of interest—it is indeed possible to build a zero-cost intermediate layer between the polynomial input and the solver format. This intermediate layer must be able to generate the data in various formats—for example, solvers may expect positive semidefinite matrices by specifying the lower or upper triangle; often, but not always, with non-diagonal entries multiplied by $\sqrt{2}$ to preserve the norm; or they may expect the full matrix with columns or rows stacked. Instead of such a vectorized input format, they may also require the data in coordinate format with three vectors: rows, columns, and values; again for the lower or upper triangle, scaled or unscaled, or the full matrix; with $32$ bit or $64$ bit integer indices.

While this is a plethora of different formats, oftentimes, the algorithms to generate the data already in the desired form differ only in two or three lines, for example when defining the range of a loop and a conditional multiplication of off-diagonals. If designed appropriately, this branching depending on the desired format can be resolved at compile-time without a need for code duplication. Additionally, while the solvers can roughly be put in one or the other group, their interfaces are still so different that no intermediate layer could be able to address them directly. Therefore, the solver implementation *has* to provide a function that, e.g., sets a coefficient in a constraint matrix, and this function will be called in an inner loop. Julia allows to inline these callbacks, completely eliminating the performance penalty.

Obviously, the intermediate layer must be made aware of the features that are supported by the solver, for it will then automatically rewrite every non-supported constraint or variable type, similar to the bridges in other modeling frameworks—if possible, by directly generating the data for the supported cones. This was designed to make the most out of the solvers: the support of `Hypatia` for exotic cones such as the $\ell_\infty$-norm cone or the complex-valued semidefinite cone is honored and these cones are used whenever they are applicable.

Handling complex-valued problems natively turned out to be particularly laborious, and the intermediate layer has to undertake a lot of work in order to support the vast majority of solvers that only know real-valued cones. While the simple equivalence

$$M \succeq 0 \Leftrightarrow \begin{pmatrix} \operatorname{Re} M & -\operatorname{Im} M \\ \operatorname{Im} M & \operatorname{Re} M \end{pmatrix} \succeq 0 \tag{4.74}$$

holds, an *efficient* implementation of this translation in the case where $M$ contains products of a constraint polynomial, monomials from the current grouping, and monomials from its conjugate grouping is not trivial. Coefficients for the same monomial (treating real and imaginary part separately) must be merged, but lookups should be avoided if possible; and coefficients that have to be zero *in total* should never be added, even if for a single term, they are present.

**Solution extraction** For a long time, the method in [HL05] was the only known procedure for extracting a solution from a moment matrix (unless the moment matrix was of rank 1), and it has been implemented in almost all frameworks. However, a more efficient algorithm was developed more recently [HKM18] and this is the one implemented in `PolynomialOptimization.jl`. Both algorithms require dense relaxations, i.e., all monomials up to a certain degree cutoff must be present. This is no longer satisfied when sparsity methods—even exact sparsity methods—are applied. The simplest way to proceed is to introduce a random perturbation linear in all variables to the objective as suggested in [Wak$^+$06]; this way, the perturbed problem will have a unique solution, leading to a rank-one moment matrix. Of course, the resulting solution will now be for a different problem (and sparsity methods might not work as well on the new problem due to the new terms); but provided the perturbation was small, it is close to a solution of the original problem and can successfully be refined using a nonlinear solver.

Another possibility is to apply a heuristic extraction; there are certainly many ways to do this. I devised and implemented an algorithm based on observations on a number of test problems. This heuristic is empirically very successful if there is, apart from possible sign (or phase) differences, mainly one solution present in the moment matrix. Quite often, two solutions differ only in the sign of one or multiple variables[36]. Based on the intuition that the moment matrix is a linear combination of the lifted solutions—the Dirac measures—the absolute value of such a variable can be extracted by looking at even powers of this variable, taking the first such occurrence in the moment matrix; the sign of the variable is marked as unknown. If no nonnegative even power was found, odd powers can be used to extract the value of the variable, including its sign; but in the case of the aforementioned superposition, two values of opposite sign would add up, and, if the weights in the superposition are the same—which is quite common—even cancel; so using odd powers is only a last resort. This assignment is done for all variables that can be found to occur alone somewhere in the moment matrix.

If due to sparsity, no monomial is present that contains a certain variable alone, but it appears together with a variable that already has a known value or magnitude, information can be obtained as well. Special care must be taken with the moment value zero; if all other variables were already inferred to be nonzero, the remaining one must be zero. If the value of the moment is nonzero, sign assignments can also possibly be made if a variable with unknown sign occurs to an odd power, and the rest of the monomial has a known sign.

Even after all these assignments, the signs of some variables might still be unknown. Then, pick one variable and fix its sign; re-run the sign assignments, invert the sign and re-run once more. This can give multiple possible solutions and is the reason why it might be beneficial to use the iterator version `poly_solutions`, which generates the next solution on demand instead of the exhaustive `poly_all_solutions`.

For complex-valued problems where the phases are unknown, the concept is essentially the same, although now more than just two possible assignments must be considered; and where previously, even powers were able to uniquely determine the absolute value, it is no longer clear whether *any* power will eventually remove all the phases.

**Outlook** While `PolynomialOptimization.jl` already implements many state-of-the-art features, allowing to tackle even optimization problems which previously exceeded all reasonable resources, there is still a vast potential for further improvement.

The only exact sparsity method implemented so far is the Newton polytope; but several more were mentioned in section 4.4.1. Given that facial reduction [PP14] encompasses

36 This might be indicative of a sign symmetry present in the problem, i.e., a group transformation that leaves the polynomials in the problem invariant when changing the sign of a subset of variables. If such a sign symmetry is known beforehand, it can be used to reduce the size of the relaxation [Löf09].

both the Newton polytope as well as the diagonal consistency criterion, it is the obvious choice for upcoming versions. It has the further advantage that when constraints are present and handled according to section 4.4.1.3, the coefficients in the SOS polynomials are not arbitrary—the fact that they must give rise to an SOS polynomial can be explicitly modeled in the facial reduction process. An implementation of the facial reduction algorithm is currently under development.

An issue of polynomial optimization problems with matrix constraints is that the size of the semidefinite matrix in the solver corresponds to the product of the basis size with the side dimension of the matrix in the original problem. Given the importance of matrix constraints, the recent idea in [MWG24] to select a different basis for each possible row (or column) index of the constraint matrix can be helpful. While [MWG24] only considers the rather exotic case of a matrix in the objective, the more common case of a matrix constraint can once again be incorporated following the idea in section 4.4.1.3. Merging this with facial reduction—the scalarized Definition 4.35 of SOS matrices may be used to constrain the coefficients—should result in a powerful exact constrained sparsity method even for matrix constraints.

Note that `PolynomialOptimization.jl` currently depends on the monomial basis; it is not possible to use symmetry-adapted bases. This allows for very efficient implementations, but the fact remains that symmetry often results in large savings. A compromise will have to be found to incorporate symmetry methods—and support for arbitrary bases also opens up the possibility to use bases different from the monomial one for reasons of numerical stability, even when no symmetry is known. Finally, the noncommutative realm also relies on an extension of the current basis implementation, keeping the monomial structure, but with an order-preserving internal representation.

# 5 Application: quantum state transmission though a lossy channel

In this chapter, I will demonstrate how the optimization methods introduced in chapters 3 and 4 can be used efficiently to obtain answers to problems in quantum information. As an exemplary application, I will consider lossy transmission: using a fixed quantum channel whose action consists in discarding a subset of its input, how can a maximally entangled state be shared through this channel? Given the importance of this scenario for quantum key distribution, often-neglected issues in scalable quantum communication are discussed first.

## 5.1 Motivation

The question of how reliable quantum information can be transmitted through a quantum channel is very tightly connected to the quantum channel capacity [Llo97]. In section 2.4.2, a very general perspective on this task was given that is driven in particular by experimental needs. However, even when completely disregarding what is experimentally feasible, finding an optimal protocol that saturates the channel capacity for an arbitrary channel is not easy.

In [DP22b], I considered the loss channel, which is particularly relevant for quantum communication. Quantum key distribution, providing guarantees of information-theoretically secure communication [BB14; SP00; Ren05] necessarily requires the exchange of quantum information between the communicating parties. Naturally, they will be separated by a significant distance; due to the exponential attenuation in fiber [Agr10], this is a considerable challenge. Even quantum repeaters [Bri+98] that are meant to provide an effective quantum analogue of classical amplification have a certain "soft" limit on their minimal distance due to practical economic constraints; in fact, if the existing telecom infrastructure is to be used with dark fiber networks that are already installed in the ground, the distance between existing repeater stations ($60 - 80\,\mathrm{km}$ [Agr10]) and the attenuation coefficient of common fibers in the $1550\,\mathrm{nm}$ telecom window ($0.2\,\frac{\mathrm{dB}}{\mathrm{km}} = 0.046\,\frac{1}{\mathrm{km}}$ [Sca+09]) imply that losses of the order of $95\,\%$ have to be considered.

### 5.1.1 Analysis of the classical task

While such extreme losses can always be dealt with by investing into resources such as memory, time, and repetitions, this is often not analyzed carefully enough. Consider a scenario in which Alice wants to communicate with Bob directly over a lossy line—and start with a classical message to clearly show where the quantum scenario begins to play havoc with the available resources. If due to attenuation, the success probability for a single bit to arrive is given by $p_{\text{trans}}$—no errors apart from loss will be considered—then on average, Alice has to repeat the process $p_{\text{trans}}^{-1}$ times. Now, "on average" already implies that there is no guarantee for Alice that Bob will ever receive her bit, regardless of how many repetitions were performed; but Alice can achieve an arbitrary level of confidence by choosing her repetitions appropriately. Since a message will typically not consist of only a single bit but many (call this a *batch*), Alice in fact has to send batches of a lot of bits. Depending on the multiplexing capacities of Alice, Bob, and the communication line, transmitting all these repetitions takes more time than simply sending the message once. Now, Bob has to *store* one bit from every repetition batch to reconstruct the full message; he can immediately discard all the others, should more than just one arrive. Alice, in turn, just keeps sending bits without ever needing to store anything (she might even forget a bit in the original message once its batch was sent, if she does not need it for other purposes).

If a repeater is inserted in the scenario in the middle of the line, a single bit that arrives at the repeater is enough; this bit can be copied as many times as desired and then relayed on to Bob, discarding other potentially arriving copies. The repeater requires no storage capacity to carry out its task[1]. The *latency* of the whole communication line is dictated by the distance between the parties and the speed of light—it is simply a delay until the information from Alice starts to arrive at Bob's site. The *rate* of the process depends on how fast Alice can push her bits out and the others are able to receive them[2]; it is not bounded by the distance between Alice and Bob.

### 5.1.2 Analysis of the quantum task

In a quantum scenario, many of the afore-mentioned features can change fundamentally, as the information that is sent by Alice is often only in the form of *correlations*, i.e., it has not yet manifested as individual bits on Alice's and Bob's side. To establish this then-classical information requires operations on both sites that must be conducted on the correlated quantum objects, and only those. Frequently[3], this does not actually require Alice to keep her part of all the copies alive, as she can measure them at any time, even before sending (or perhaps just prepare in the desired way). Therefore, Alice only has to store classical bits for all of her copies—which is already more as before, where she could just forget everything, but it is still manageable, as classical information storage is available in abundance. To then reduce this classical bit for every copy in a batch to the single message bit, she needs information from Bob. This implies a backwards-communication that can only happen after Bob received the qubits and did his processing. Therefore, the sender Alice only knows what to make of her sent qubits after at least *twice* the latency of the classical communication process. Still, the *rate* is not related to the distance; Alice can push new qubits out as long as she has enough classical storage capacity to keep track of everything she sent.

---

1 As usual, this argument neglects the meta-communication necessary to inform all parties about the current state of the protocol, which then must also be stored at their sites. In this case, the amount of communication and storage required is independent of the message size, so that disregarding it is not an issue for the resource analysis.

2 This does not only refer to their processing time, which can be part of the latency or the rate, but instead to a certain minimal separation of pulses that is dictated by the physical system in order to still be able to distinguish the signals. However, this minimal separation can be mitigated by adding more communication lines.

Now adding a repeater alters the situation drastically. A true quantum repeater must not influence the effectively exchanged states, only the transmission success probability should be increased. Therefore, the quantum correlations that are to be established in the end are correlations between Alice and Bob, but *not* the repeater. As this forbids the repeater from knowing the actual quantum state, it cannot, upon receiving a single qubit, copy it as often as required and relay it to Bob; instead, for every outgoing qubit, there must be an incoming qubit. For this to still have any effect on the success probability, the repeater has to divide the line in two segments, establish a successful transmission in each one individually and then connect them via correlated (Bell-type) measurements—i.e., entanglement swapping [Ben+93]. In order to "establish successful transmission," the repeater has to communicate with the corresponding other party to receive the information about which part of the batch arrived[4]; only once this is known can the swapping occur—else, the two segments are not independent from each other and a failure in one leads to a failure in both, effectively leaving the transmission success probability unchanged. This means that the repeater now has to store *all quantum states* that arrive. If Alice keeps on sending states as fast as possible, they will arrive much faster than the arrival information is available to the repeater, so that it has to have a huge quantum memory available to store all these states until they can be processed. The current technological state puts such a quantum memory in the very far future. The only alternative is for Alice to throttle the sending of her states; she has to accommodate for the communication time, effectively *turning the latency into a rate limitation*.

This is a crucial insight that has to inform the development of all quantum communication

3 Unless the protocol depends on a specific sequence of events conditional on outcomes at the other site.

4 While it is possible to imagine a situation in which both Alice and Bob send *to* the repeater—storing only classical information—so that it automatically knows which parts arrived, this argument breaks down as soon as there is a second repeater which needs to keep alive the quantum state.

lines: Even if attenuation is the only possible source of error and without all considerations of security that may be detrimental to the rate, the necessity of backwards communication will put an upper bound on the rate. This upper bound is determined by the distance between neighboring repeater stations and the speed of light, and there is no technologically feasible way around it except to avoid the backwards communication—i.e., the sender must be able to take the arrival of a suitable amount of information at the other site for granted and cannot query what actually happened. A small proviso can be added: the endpoints in the communication chain *are* allowed to depend on this information provided the end-products of their communication-assisted task are classical bits, as this will only impact the latency, but not the rate. Note that the latter argument applies in particular to quantum key distribution, where the relevant data is classical; the matter is entirely different with distributed quantum computing [Cal+24]. However, there, the focus is usually on a datacenter-like configuration with distances that are much more in control (though not exclusively; this then goes under the name *Quantum Internet*).

Quantum repeaters of the third generation [Mur+16] employ error correcting codes to achieve such an independence from backwards communication; however, "error correction" is a sufficiently broad term that it merely stands for a concept. How to ideally put such a concept into a concrete procedure is not clear, in particular since I found the analysis of a previously suggested scheme to accomplish precisely this [Mun+12] to be faulty and too optimistic by far [DP22b, Appendix C].

## 5.2 Setting

Numerical optimization can now be helpful in constructing protocols that give the best possible rates under reasonable experimental constraints. If a stochastic law of attenuation is given, the only way to transmit more photons is to send considerably more in the first place[5]. Assuming the setting of quantum key distribution, where the actual states that are to be transmitted are rather irrelevant as long as a security proof can be formulated in the end, the quantum state of these photons is *a priori* unknown and the result of an optimization process. The receiver must then be able to reliably identify which of the expected photons has arrived. This marks the difference between an erasure and a deletion [GBP97]; and by investing into obtaining this extra classical information, which is always available regardless of how high the attenuation of the channel, a higher-quality quantum state can be recovered. While "available" does not always equal "experimentally feasible," in this case, the extraction of the information is indeed doable, e.g., via non-destructive photon detectors [Nie+21; Dis+21] or boosted teleportation schemes [EL14].

5 Though an intriguing possibility was pointed at in [MLG22a; MLG22b]: exploiting memory effects in a fiber, it appears to be possible to change the attenuation law for a short time by priming the fiber with a small number of preceding control pulses.

Once the receiver knows exactly which photons arrived (for a continuous message, think of a block-wise transmission), suitable recovery quantum maps can be carried out. Again, these maps are not known, but results of the optimization.

### 5.2.1 Optimization problem

In order to analyze this complicated problem in detail, and with the understanding that these steps may lead to overall inferior results, sender Alice and receiver Bob are thought to communicate point-to-point with no intermediaries. This situation can immediately be applied to the case of a longer chain if Alice and Bob are allowed to be repeaters. In this simplified setting, no assumption about the classicality of the data at the sender's and receiver's site may be made: it must be possible to exchange a most general quantum state. Then, by performing entanglement swapping, entanglement can be generated between the ultimate sender and receiver, which is a necessity for quantum key distribution [CLL04].

The optimization problem for producing a maximally entangled state in this setting now reads

$$\begin{cases} \sup\limits_{\{E_\ell\}_{\ell\in\mathcal{L}},\hat{\varrho}} \ F := \langle\Phi^+|\hat{\varrho}_{\mathrm{f}}|\Phi^+\rangle \\ \quad \text{subject to} \\ \qquad \hat{\varrho} \succeq 0, \quad \operatorname{tr}\hat{\varrho} = 1 \\ \mathscr{C}[E_\ell] \succeq 0, \quad \operatorname{tr}_{\mathrm{B}} \circ E_\ell \preceq \mathbb{1}\ \forall \ell \in \mathcal{L} \\ \qquad \hat{\varrho}_{\mathrm{f}} = \sum\limits_{\ell\in\mathcal{L}} p_{\mathrm{trans}}^{s-|\ell|}(1-p_{\mathrm{trans}})^{|\ell|} \dfrac{E_\ell[\operatorname{tr}_\ell \hat{\varrho}]}{\operatorname{tr} E_\ell[\operatorname{tr}_\ell \hat{\varrho}]} \\ \quad p_{\mathrm{tot}} = \sum\limits_{\ell\in\mathcal{L}} p_{\mathrm{trans}}^{s-|\ell|}(1-p_{\mathrm{trans}})^{|\ell|} \operatorname{tr} E_\ell[\operatorname{tr}_\ell \hat{\varrho}], \end{cases} \tag{5.1}$$

where $s$ is the number of photons sent initially and $\mathcal{L} = \mathcal{P}(\{1, \dots, s\})$ is the set of all possible loss configurations that can occur. Here, it was assumed that the arrival probability of a single photon is always given by $p_{\mathrm{trans}}$. Due to the possibility for global classical postprocessing, the recovery maps $\{E_\ell\}$ need not be successful—failures occurring with the probability $1 - p_{\mathrm{tot}}$ may be filtered out. Equation (5.1) reflects what was described previously: find an initial state $\hat{\varrho}$ defined on $\mathcal{H}_{\mathrm{A}} \otimes \mathcal{H}_1 \otimes \cdots \otimes \mathcal{H}_s$; send the numbered parts, where they undergo all possible loss scenarios; detect which scenario is present and select an appropriate recovery map $E_\ell$ that transforms the systems $\{1, \dots, s\} \setminus \ell$ into B. Overall, maximize the fidelity $F$—the overlap of the resulting state $\hat{\varrho}_{\mathrm{f}}$ with a maximally entangled state $|\Phi^+\rangle$.

Implicit in this description is that the block size $s$ has been fixed in advance, and also that the dimension $d$ of the systems to be sent is predetermined.

To simplify the problem further, now assume that the most significant contributions come from one particular number of losses $|\ell|$ around the expected value $r$ of the binomial

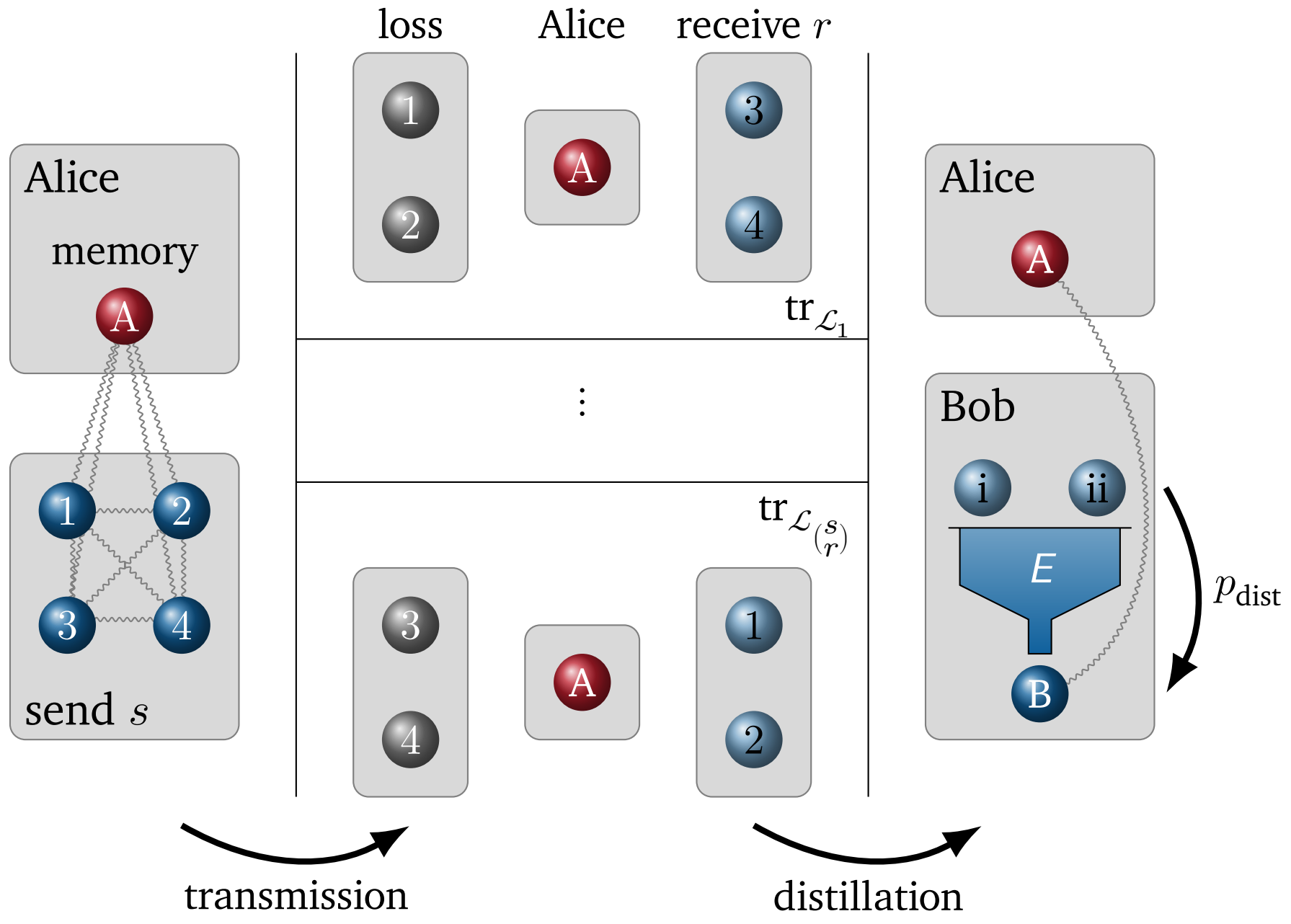

**Figure 5.1.** Visualization of the setting. After preparation at Alice's site, $s$ qubits are sent through the channel, but only $r$ arrive. All possible configurations, corresponding to the loss maps $\{\mathrm{tr}_{\mathcal{L}_i}\}_i$, are independent and equally likely. A local distillation map $E$ then aims at producing a high-quality Bell pair between Alice and Bob with probability $p_{\mathrm{dist}}$; this map does not know which particles were lost. The important initial and final correlations are visualized by wavy lines, but suppressed at the intermediate stages. While the individual parties are named "Alice" and "Bob," they may also be repeaters. (adapted from [DP22b])

distribution of loss configurations. This already drops the number of (matrix-valued) optimization variables significantly from $2^s$ to $\binom{s}{r}$. Nevertheless, realizing such a number of quantum operations, all conditionally on which configuration arrives, is challenging; so now, implement only a single recovery map $E$, regardless of the actual arrival. Still, the arrived photons have to be mapped into the correct input slots of $E$ by means of a quantum switch. What remains is an optimization problem over a single state and quantum map; this is graphically depicted in figure 5.1. Importantly, this problem contains a permutation symmetry and can therefore be described in the fully symmetric subspace alone, which for the state has dimension $s + 1$; furthermore, now the *individual* success probabilities are equal to $p_{\mathrm{tot}}$, not only their average.

### 5.2.2 Solution methods

In order to solve the problem, in [DP22b], I employ multiple different methods.

The *nested optimization scheme* consists of a nonlinear outer optimization as described in sections 3.3 and 3.4 searches over either $\hat{\varrho}$ or $E$; the inner optimizer then sees a semidefi-nite program which can be solved to global optimality using interior-point methods (see section 3.5). Exploiting that the optimal $\hat{\varrho}$ will always be of rank one, fixing $\hat{\varrho}$ by the outer optimization yields better results overall; though for small $r$ and large $s$, taking the nonlinear optimization over $E$ incurs far less variables for which no global guarantees can be made.

Still, the issue of finding a good initial point remains. In [Dat19], *convex iteration*, a heuristic convexification procedure of mainly convex programs with additional nonconvex rank constraints was proposed. All nonlinearities in the optimization program stem from the multiplication of an entry in $\hat{\varrho}$ with an entry in $E$; but a multiplication can be easily written in terms of a rank constraint:

$$x = yz \Leftrightarrow \exists a, b : \operatorname{rk} \begin{pmatrix} a & x & y \\ x & b & z \\ y & z & 1 \end{pmatrix} = 1. \tag{5.2}$$

Therefore, this procedure is applicable here; for details, see [DP22b]. While the convex iteration requires a heuristic to steer out of local optima and does not always converge to high precision, it works well to find approximate good initial points, which can then in turn be fed to the nested optimization scheme.

The importance of efficient implementations cannot be stressed enough, as all convex optimizations (both in the inner part of the nested optimization as well as convex iteration) have a fixed structure; the only change from iteration to iteration is given by specific values of coefficients or right-hand sides. Modeling frameworks such as `Convex.jl` [Ude+14] or `CVXPY` [DB16; Agr+18] allow to use variables in a *parametric* sense—i.e., to assign values to variables before they are passed to the solver, so that they can be used in an arbitrary, not necessarily convex, way. In principle, this makes it possible to construct the underlying solver data only once; however, support for this is extremely restricted. Very often, the solver representation still has to be reconstructed from scratch for every iteration. Therefore, any efficient implementation must necessarily use the solver API directly, constructing the problem manually and changing it from iteration to iteration[6]. In this way, the years-long experience of highly optimized generic interior-point solvers can be combined with new optimization methods.

### 5.2.3 Results

Using the `Mosek` API directly, I was able to find optimal solutions—strictly speaking, only locally optimal due to the overall nonconvexity, but with good confidence that they are

6 Note that for many years, `Mosek` [MOS24b] has built up a reputation as the best semidefinite solver that is available (though in recent years, the Mittelmann benchmarks [Mit03] suggest that `COPT` [Ge+23] should probably be considered as the new champion); but `Mosek`'s API is very peculiar and different from many others, making it particularly difficult for modeling frameworks to exploit reusability of the optimization tasks.

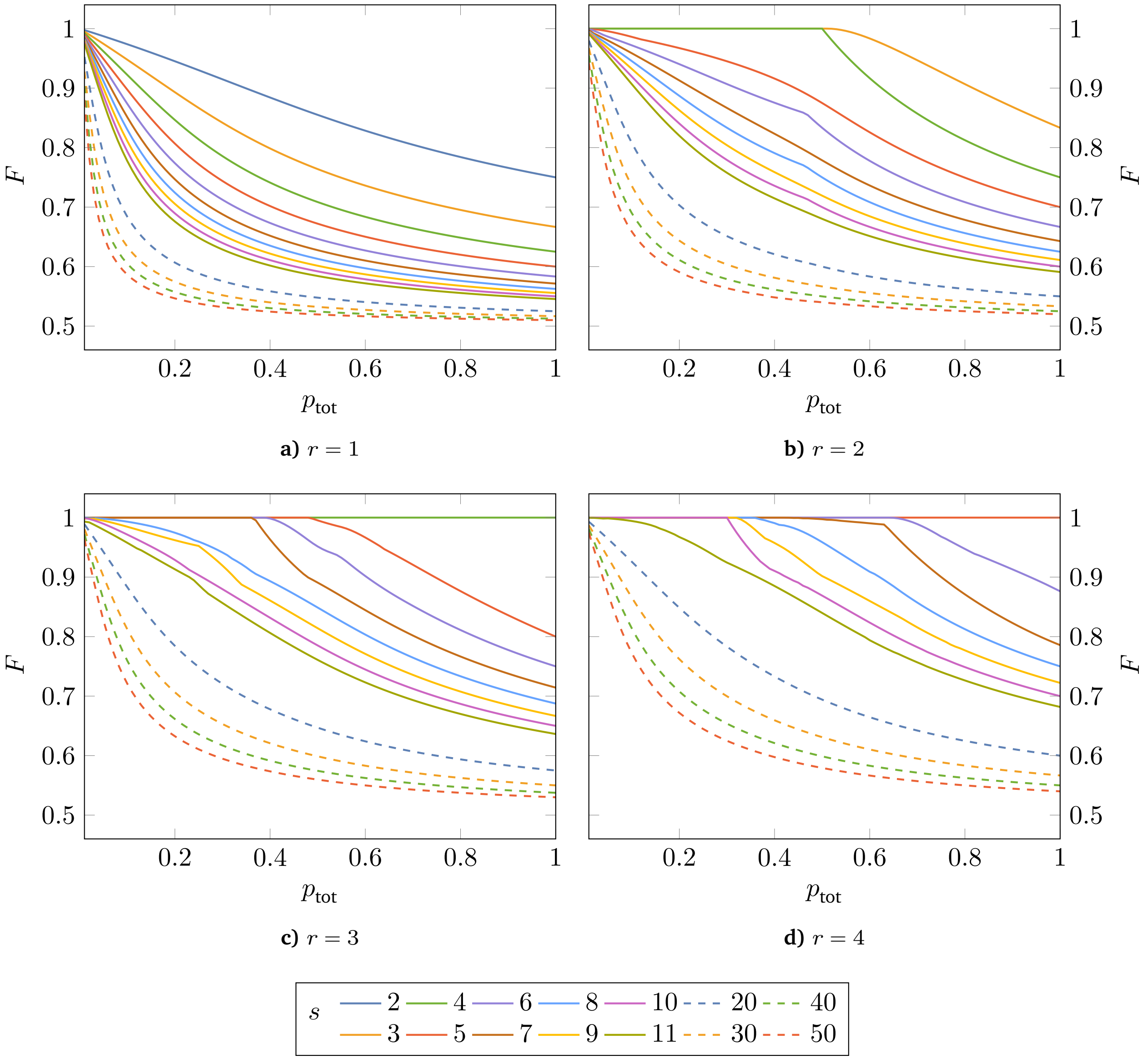

**Figure 5.2.** Lower bounds on the maximally achievable fidelities $F$ for a pure loss channel with $s$ input qubits (given by the color coding) and $r$ output qubits, for various success probabilities $p_{\text{tot}}$. Samples are taken at every integer percent value, but marks were omitted for clarity. Every plot starts with $s = r + 1$; if this line is not visible, it is obstructed by the $F = 1$ line of a higher $s$-value. (taken from [DP22b])

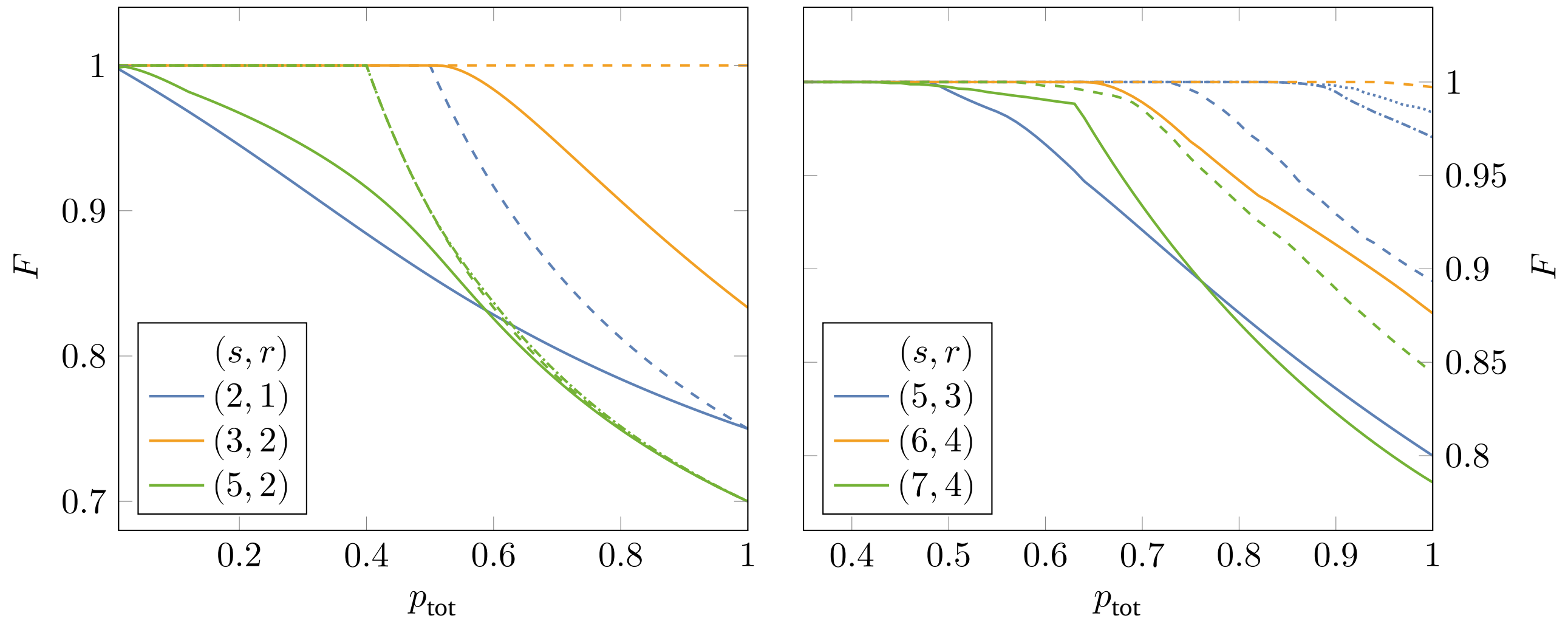

**Figure 5.3.** Comparison between lower bounds on the fidelities for qubit channels (*solid lines*, as in figure 5.2), qutrit channels (*dashed lines*), ququart channels (*dashed–dotted lines*), and ququint channels (*dotted lines*). Various combinations of $s$ and $r$ are color-coded. I only show a channel of higher dimensionality in those cases in which I have been able to find an improved lower bound numerically. (taken from [DP22b])

indeed global—to the permutation-invariant problem. I considered up to $s = 75$ sent qubits; and since a photon can carry more information than just a qubit, I also optimized the task for sending qutrits, ququarts, and ququints. Note that this leads to a massive growth in the number of free variables. Results for qubits are given in figure 5.2 and compared with higher-dimensional carriers in figure 5.3. The effect of allowing for an overall probability of failure can manifest in decidedly nontrivial behavior: as $p_{\text{tot}}$ goes to zero, one or multiple points may be passed at which the fidelity curve is nonsmooth—the optimal initial state then takes a qualitatively different form that would have been suboptimal before. Note that the case $p_{\text{tot}} = 1$ corresponds to the scenario considered for ordinary quantum error correction codes: they must always yield a result.

### 5.2.4 Comparison

Looking at $s = 5$, $r = 3$, and comparing with the known perfect five qubit code [Laf+96], which is able to correct two erasures deterministically, shows the disadvantage introduced by the simplifications before: the five qubit code does not have to operate in the permutationally invariant regime and can therefore protect better than the numerically derived code. However, if the encoding of the five-qubit code is taken and then for the decoding the *same* optimal correction map is applied on the three arrived qubits (which can be found by a semidefinite program), regardless of which these are, results are much worse—so indeed, within the given restriction, the optimization yields excellent encodings and decodings.

Now consider the task of entanglement distillation. In [DP22b], I compare established recurrence schemes for entanglement distillation [Ben+96; Deu+96; Mac98; Deh+03; MO06; FY09; KAJ19], applied once to a state that is optimal for the given scheme, with the procedures derived numerically under the assumption of permutation invariance. In this case, the restriction does not lead to inferior results; in fact, the results obtained by the optimization are always an upper bound to any known entanglement distillation scheme, and usually by a fair margin. This shows that the common setting for entanglement distillation—i.e., to provide a fixed generic procedure that probabilistically increases the entanglement of an arbitrary state—is so far from optimal that even its appealing nature of being independent of whatever channel is given makes its practical use questionable. A thorough understanding of the channel necessitating the distillation, together with control over the initial state, will lead to dramatic improvements in the yield.

## 5.3 Shortcomings

The previous optimizations have two deficiencies: first, the use of generic nonlinear methods does not provide global guarantees; second, the reduction to the permutationally invariant subspace posited a restriction that could exclude potentially beneficial sectors in the search space.

### 5.3.1 Non-global methods

The first issue can be ameliorated or even removed completely by the use of polynomial optimization techniques when sizes of the resulting relaxations are reasonable. Clearly, equation (5.1) is a POP. To start with a nontrivial small-scale example, consider the case of sending three and receiving a single qubit. As mentioned in appendix B, the second level in the relaxation hierarchy will converge to a finite upper bound on the global maximum. This level contains several semidefinite matrices; the largest one, which is the moment matrix, has side dimension 1128, followed by the matrices from the various constraints with side dimensions 376, 188, and 94. Note that all numbers refer to a formulation with real-valued coefficients, which was the assumption in [DP22b] as well. By applying Newton polytope methods, cf. section 4.4.1, the size of the moment matrix can be reduced to side dimension 83. As the Newton polytope does not affect the constraint bases—which would have been minimal here anyway—the moment matrix is now the smallest one present in the problem. While the resulting semidefinite program is still somewhat resource-hungry, it can be solved to excellent accuracy using `Mosek` on a high-end consumer PC.

A comparison of the bound given by `Mosek` and the numerical results found before is made in figure 5.4. Note that here, the upper bound produced by the polynomial optimization routine was postprocessed: values exceeding $1$ were manually replaced by this fundamental bound. This reveals that the polynomial relaxation certifies[7] the optimality of the previously found results with a relative precision of about $5 \cdot 10^{-3}$ or better.

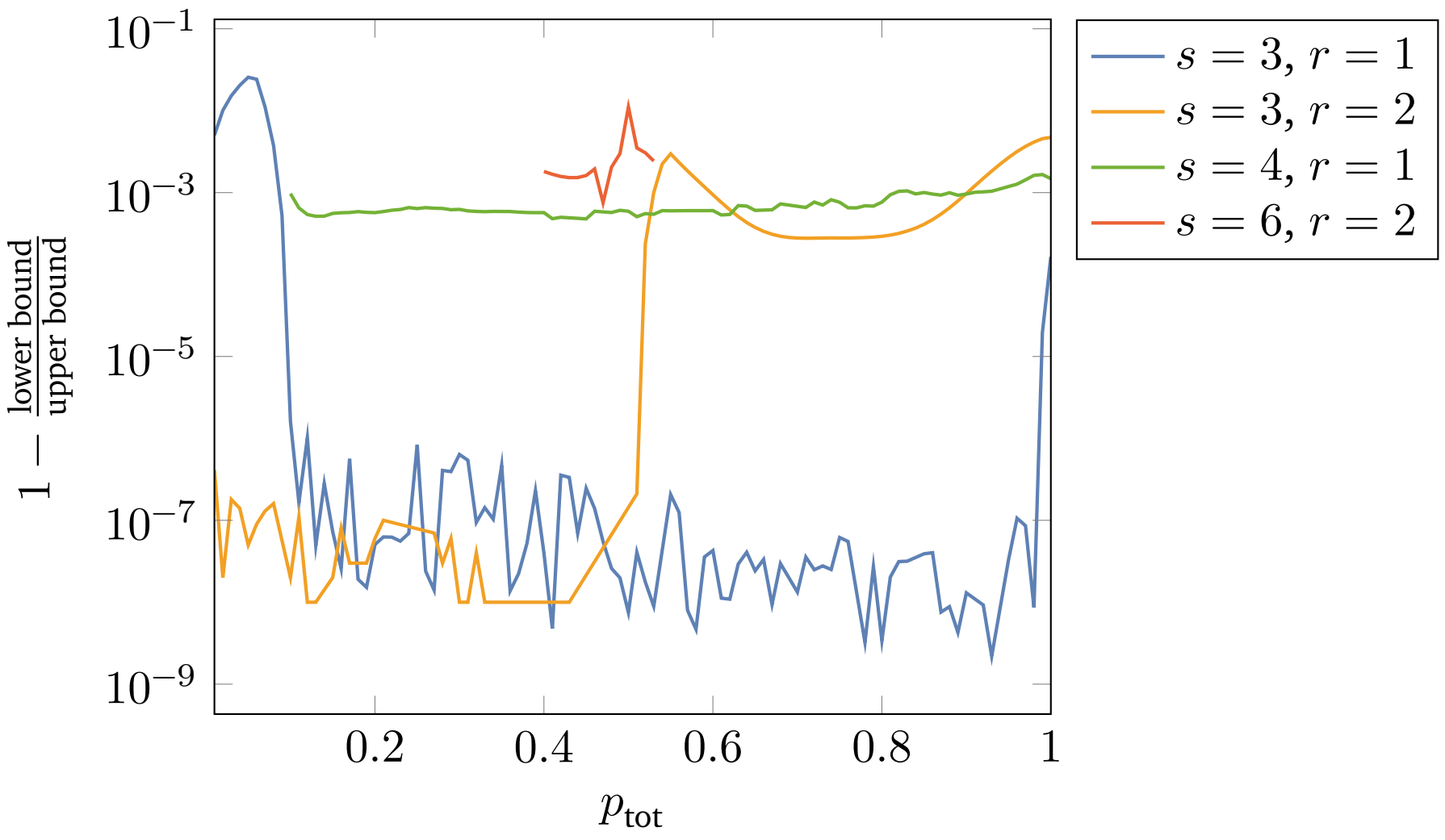

**Figure 5.4.** Comparison of bounds on the maximally achievable fidelity $F$ for a pure loss channel with $s$ input qubits and $r$ output qubit, for various success probabilities $p_{\text{tot}}$. The lower bound is given by the previous nonlinear optimization; the upper bound by a degree-2 polynomial optimization relaxation (clipped to at most $1$). For $s = 3$, the POPs were solved with an interior-point solver with default tolerances of $10^{-8}$; for $s > 3$, a first-order solver with tolerances of $10^{-4}$ was employed.

Note that when employing inexact sparsity methods to reduce the problem size, the results will degrade significantly; for example, the optimal bound $F \leq 0.669128$ for $p_{\text{tot}} = 0.98$ will be increased to $F \leq 0.740980$ using correlative sparsity; block term sparsity will not reduce the problem size; and chordal sparsity gives an obvious bound $F \leq 1.465084$.

The example shows both the promise held by polynomial optimization methods as well as their limitations with challenging semidefinite programs already for small realistic problems.

There are two "neighboring" configurations, where only one parameter is changed: $(s = 3, r = 2)$ and $(s = 4, r = 1)$. In the former, the recovery map now acts on two instead of just one qubit. While again the Newton polytope reduces the size of the moment matrix from $1711$ to $106$, the largest constraint matrix of side dimension $464$ determines the size of

7 Note that "certification" is not used in its strictest sense: as the polynomial program was also only solved up to a certain accuracy, the constraints imposed in the relaxation are typically violated—for example, the smallest eigenvalues of matrices that should be positive semidefinite might be of the order of $-10^{-8}$. While the computationally demanding (and not always feasible [Sch16a]) idea of rational—i.e., exact—certifiers obtained through polynomial optimization has already persued [Pow11; DP22a], for a practical task (as opposed to obtaining fundamental no-go theorems), such a numerical, "weak" certificate is certainly sufficient.

the problem. The memory consumption of interior-point solvers such as `Mosek` is significant[8], but still manageable. Similar to the $r = 1$ case, optimality is certified to a good accuracy, which becomes slightly worse towards the limits $p_{\text{tot}} \to 1$ and $F \to 1$.

The moment matrix in the configuration $s = 4$, $r = 1$ can be reduced using the Newton polytope from its initial side dimension $2211$ to $114$, and again the largest constraint matrix of side dimension $660$ dominates. The first-order solver `SCS` solves this problem to global optimality, obeying the relaxation constraints to an accuracy of about $10^{-4}$, with insignificant memory consumption. For probabilities exceeding $20\,\%$, creating a problem from scratch, determining the Newton polytope and solving it takes about $3$ – $4$ min—solver times increase considerably for lower probabilities. The results, also shown in figure 5.4, are similar to those previously obtained: optimality is certified up to roughly the same accuracy as before. Note that $p_{\text{tot}} < 10\,\%$ was not computed because of the much longer computation times.

To conclude this section, now look at $s = 6$, $r = 2$, which is the smallest configuration with a nonsmooth curve in figure 5.2. The region of interest is the vicinity of $p_{\text{tot}} \approx 0.47$. The moment matrix of side dimension $8128$ is quickly reduced via Newton polytope methods to size $244$; the largest constraint matrix of size $1778$ dominates the problem. Constructing the data for `SCS` takes `PolynomialOptimization.jl` less than $15$ s; and although this semidefinite program is extremely large, the memory consumption is fairly modest with about $7$ GiB in total. The problem was solved in a range around the nonsmooth point; and figure 5.4 certifies that indeed the previously observed behavior appears to be correct—a qualitative change in behavior occurs here. Apart from the results themselves, the solver time is also interesting: with $41$ min, $p_{\text{tot}} = 48\,\%$ is easiest to solve, increasing quickly to $100$ min for $p_{\text{tot}} = 53\,\%$; in the other direction, i.e., for $p_{\text{tot}} < 48\,\%$, the times are about six hours.

### 5.3.2 Permutationally invariant structure

The second deficiency in the previous approach is not due to the numerical method, but lies in the structure imposed on the problem. Motivated both by a simpler physical realization with a single recovery map instead of multiple ones and the fact that the originally exponentially large search space collapses to a linear size, a permutationally invariant basis set was demanded—potentially detrimental to optimality.

While the full set of $2d^s$ basis states would allow for the most freedom, the question can be asked whether an efficiently computable restriction different from the permutationally

8 Note that it may be helpful to turn off or reduce the number of threads: `Mosek` will use multi-threading to accelerate the Cholesky factorization required in every iteration. The redistribution of tasks between the iterations with potential new allocations means that the memory usage may grow until the solver terminates. Solving the problem with $48$ threads requires about $166$ GiB; limited to $16$ threads, approximately $30$ GiB can be saved, and in a single-threaded approach, only half of the memory is required. Of course, the penalty has to be paid by means of a longer solver time.

invariant one still leads to a moderate number of degrees of freedom left to optimize, while at the same time allowing for higher fidelities. In this section, a more intricate way of selecting a restriction will be explored; but while the results will indeed show that the fidelities can sometimes be much better, they will at the same time reveal that the permutationally invariant choice—known to preserve entanglement well under losses [BV12]—is usually superior in terms of numerical accessibility.

The permutationally invariant basis states automatically implied that the success probabilities for every loss configuration were equal; this property is now no longer present. As a consequence, the nonlinear divisions by the individual success probabilities prevent even the nested approach from before. Using the arithmetic–geometric mean inequality,

$$\frac{O_1}{p_1} + \frac{O_2}{p_2} = \frac{O_1 p_2 + O_2 p_1}{p_1 p_2} \geq \frac{O_1 p_2 + O_2 p_1}{\sqrt{p_1 p_2}} \geq \frac{O_1 p_2 + O_2 p_1}{\frac{p_1+p_2}{2}} \geq \frac{O_1 + O_2}{\frac{p_1+p_2}{2}} \tag{5.3}$$

for individual overlaps $O_{1,2} \in [0,1]$ and probabilities $p_{1,2} \in [0,1]$. Therefore, forsaking the individual normalizations of $E_\ell[\operatorname{tr}_\ell \hat{\varrho}]$ in favor of the "global" $\operatorname{tr} \sum_{\ell \in \mathcal{L}} E_\ell[\operatorname{tr}_\ell \hat{\varrho}] = p_{\text{tot}}$ will provide a lower bound to the actual fidelity, restoring convexity.

A very simple approach is to select certain basis states for the input state in the computational basis beforehand, setting all others to zero. This automatically gives rise to a reduced set of basis states for the recovery maps. As it is not clear which basis states should be selected, an exhaustive search over all combinations is a first starting point. However, a careful analysis of symmetries allows to partition this large set into equivalence classes of subbases. This is motivated by the fact that the partial traces of different combinations of basis states may yield the same reduced states—or at least reduced states with the same trace and overlap with the target state. In particular, let a subspace be defined by a basis $\{|\boldsymbol{x}_1\rangle, \dots, |\boldsymbol{x}_k\rangle\}$, with the understanding $\boldsymbol{x}_j = (a_j, s_j^1, \dots, s_j^s) \in \{0,1\} \times \{0, \dots, d-1\}^s$. Now consider two subspaces defined by bases $\mathcal{B}_1$ and $\mathcal{B}_2$, respectively. These subspaces are *simply equivalent* if a permutation of the $s$ sent systems together with an arbitrary renaming of the levels within the systems (or Alice's levels) allows to convert $\mathcal{B}_1$ into $\mathcal{B}_2$. Subspaces that are simply equivalent are clearly in the same equivalence class as previously defined, although a larger choice might be possible.

To underline the importance of recognizing simple equivalence, consider the example $d = 2$, $s = 5$, $k = 3$: There are $465$ bases, but only $10$ of them are not simply equivalent. Increasing to $k = 4$, there are $35\,960$ bases, but only $47$ are not simply equivalent. The software repository developed for [DP22b] has been updated with code that generates all possible basis choices. To identify each equivalence class, a "canonical" representative from the basis is defined by taking the smallest one with respect to a certain ordering. An obvious one is to interpret every computational basis state as a $d$-ary number: $|011\rangle \mathrel{\hat{=}} 3$, $|100\rangle \mathrel{\hat{=}} 4$.

The choice made here instead sorts by position first: it counts how often the individual levels occur separately for each position—for example, in the subbasis $\{|000\rangle, |001\rangle\}$ (considering only the sent part), the first position has level counts $(2, 0)$, the second $(2, 0)$, and the third $(1, 1)$. The order is then induced by the counts in the first position, with tie-breaking in the subsequent ones, favoring higher values in lower levels. While the description of this order seems to be more contrived, it favors more symmetrical basis choices: for example, $\{|000\rangle, |011\rangle, |101\rangle\}$ is a canonical basis in the numerical order, but $\{|001\rangle, |010\rangle, |100\rangle\}$ would be the canonical choice in the second approach.

There is a simple way to put an upper bound on the fidelity that can be produced by a particular basis. Consider an arbitrary initial state expressed in the subspace basis:

$$|\psi\rangle = \sum_i \alpha_i |0\boldsymbol{s}_i\rangle + \sum_j \beta_j |1\boldsymbol{s}_j\rangle. \tag{5.4}$$

The partial traces $\mathrm{tr}_\ell |\psi\rangle\langle\psi|$ can easily be calculated symbolically for any loss configuration $\ell$. Every one of these marginalized states can contain terms with $\alpha^2$, $\alpha\beta$, and $\beta^2$ coefficients. Let the desired maximally entangled state be

$$|\Phi^+\rangle\langle\Phi^+| = \frac{1}{2}(|00\rangle\langle 00| + |00\rangle\langle 11| + |11\rangle\langle 00| + |11\rangle\langle 11|); \tag{5.5}$$

if the marginalized state does not have an $\alpha\beta$ coefficient, it will not be possible for the overlap with $|\Phi^+\rangle$ to exceed $\frac{1}{2}$ for this particular loss configuration. The simple assignment of the output fidelity 1 for all loss configurations for which $\mathrm{tr}_\ell[|\psi\rangle\langle\psi|]$ contains $\alpha\beta$ cross-terms and $\frac{1}{2}$ for all others immediately gives an upper bound for a particular basis choice—and a heuristically motivated order in which the various bases should be investigated.

As a further criterion for the elimination of unnecessary candidates, note that if there is a coefficient $\alpha_j$ ($\beta_j$) which never appears in a product with any $\beta$ ($\alpha$), then this coefficient can for sure be set to zero—i.e., the subspace was too large.

To now apply nonlinear optimization methods on the subspace, an initial vector of coefficients has to be supplied. Assuming again a nested scenario in which the inner optimizer is over the state and the outer one over the set of recovery maps, this means guessing good recovery maps—which is easier than guessing good initial states, as the individual loss configurations can be taken into account separately. The following heuristic is based on a similar argument as the previously derived upper bound. Consider the partial trace for a specific loss configuration; applied on the symbolic initial state, some marginalized received state will arise. Given a term of the form $|a_0\boldsymbol{r}_0\rangle\langle a_1\boldsymbol{r}_1|$, the recovery map has to identify that Alice's part $|a_0\rangle$ corresponds to the output $|\boldsymbol{r}_0\rangle$ and $|a_1\rangle$ to $|\boldsymbol{r}_1\rangle$. Provided that $|\boldsymbol{r}_0\rangle$ never occurs with $|1-a_0\rangle$ on Alice's side (and likewise for $|\boldsymbol{r}_1\rangle$), this is a simple one-to-one mapping and the required entry in the recovery map, $|a_0\rangle\langle\boldsymbol{r}_0|$, can be easily constructed, weighted by a correction factor that accounts for how often the individual basis element occurs. However, if this uniqueness is broken, the best heuristic choice is to pick one assignment over the other, in the hope that the subsequent optimization will then set the other coefficient to zero. Since this might be in conflict with other loss configurations, all possible combinations of singling out one choice should be checked as initial states for the optimization; under the assumption that the number of basis states is low, this is not too much overhead.

Starting from these initial guesses on the maps, the nested optimization is performed; after this, the optimal state is extracted and used as initial state for second nested optimization, where the maps are the convex decision variables. The multi-stage approach begins with two basis states and increases this value until at some point the best possible result is the same as with one basis element less. While this is no guarantee that the larger basis might not yield an improvement or that no further improvement may be found when enlarging the basis once more, it is quite likely that the number of nonlinear optimization variables has grown to such a degree than all further optimizations will also not be able improve.

Looking at results, the configuration $s = 3$, $r = 2$ seems promising: already four basis states improve the deterministic fidelity from $\frac{5}{6} \approx 0.83$ in the permutationally invariant case to $0.87$; and with six basis states, the optimization hits its peak with fidelity $F = \frac{4+\sqrt{2}}{6} \approx 0.9$. A possible initial state is given by

$$|\psi\rangle = \frac{|0\,000\rangle + |0\,001\rangle - |0\,110\rangle + |0\,111\rangle}{2\sqrt{2}} + \frac{|1\,010\rangle + |1\,101\rangle}{2}. \tag{5.6}$$

However, apart from this configuration, improvement is rarely found by the optimization. The $s = 5$, $r = 3$ configuration is slightly better with six basis states and $F = 0.8055 > 0.8$, and can again improve with seven basis states to $F = 0.8333$ with the initial state

$$\begin{aligned} |\psi\rangle = 0.47895(|0\,00000\rangle + |1\,00110\rangle - |1\,11000\rangle) + 0.26683(|0\,00111\rangle - |0\,11001\rangle) \\ - 0.32880|0\,11110\rangle - 0.24762|1\,00001\rangle. \end{aligned} \tag{5.7}$$

These negative results change slightly when lowering the success probability $p_{\text{tot}}$; for example, for $p_{\text{tot}} = 0.8$, improvements can be found in the $(s, r)$ configurations $(2, 1)$, $(3, 1)$, $(3, 2)$, $(4, 1)$, $(5, 1)$, and $(5, 3)$; however, all other configurations, which were computed up to $s = 8$ are inferior compared to the permutationally invariant approach.

It is not clear whether the failures to produce better results actually come from being stuck in a non-global maximum or whether the basis was simply not large enough. Permutationally invariant bases intersect those found here, but unless the number of basis vectors is sufficiently large, they are not contained in the new approach.

Following the lines of section 5.3.1, polynomial optimization methods can also be applied to the extended problem formulation. There are now many more semidefinite constraints due to the different recovery maps, but the general problem structure stays unchanged, as does the fact that the second-order relaxation provides meaningful bounds.

A comparison of the bounds given by a POP with the data from the nested optimization scheme is given in figure 5.5. For the later analysis, it is very important to note that the POP results were obtained using the first-order solver `SCS` with its default tolerances.

Three characteristic configurations are shown in the figure; for the first, the total simulation time was eight hours, for the other two twelve hours. The overall number of basis states depicted is thus an indication of the size and numerical difficulty of the problem—all basis states were solved consecutively until the total time ran out.

For the first configuration, $s = 3$, $r = 2$, $p_{\text{tot}} = 0.8$, almost all bounds coincide rather closely with the value obtained by the nested optimization. Slight possible improvements are indicated; for example, a solution with the basis consisting of

$$\{|0\,001\rangle, |0\,110\rangle, |1\,000\rangle, |1\,111\rangle\} \tag{5.8}$$

was found by the nested optimization scheme, giving $F = 0.928719$; however, the POP gives an upper bound of $F \leq 0.934981$. Re-optimizing this particular case and setting all precision values of `SCS`[9] $\varepsilon_{\text{abs}}$, $\varepsilon_{\text{rel}}$, and $\varepsilon_{\text{infeas}}$ to $10^{-5}$ already reduces the upper bound slightly to $F \leq 0.933737$, at the expense of significantly longer execution times: 25 min vs. 3:50 h. However, it is not possible to certify optimality or extract a solution, so it is not clear whether the bound can actually be attained.

This example already showed that one has to be careful when interpreting the numerical results from POPs solved under the very loose default tolerances of first-order solvers. Consider a second characteristic configuration, $s = 6$, $r = 2$, $p_{\text{tot}} = 1$. The results seem to invalidate the theory of polynomial optimization since the upper bounds are in most cases below concrete instances found by the nonlinear optimizer. While some of these results obviously only deviate very little, the basis choice

$$\{|0\,000001\rangle, |0\,000010\rangle, |1\,000000\rangle\} \tag{5.9}$$

with a result $F = 0.61085$ by the nonlinear optimizer and the bound $F \leq 0.585812$ seems to violate the default tolerances to a significant degree. However, this is not necessarily true; tolerances of the solver are with respect to the relaxed problem; how they are calculated exactly is different from solver to solver and has to be looked up in the corresponding software manual. In this particular case, choosing again the precision $10^{-5}$ from before leads to an upper bound $F \leq 0.65201$, which is perfectly compatible with the existing result from nested optimization. Importantly, in section 4.10.3.7 a heuristic solution extraction

9 See `cvxgrp.org/scs/algorithm/#termination` for a definition.

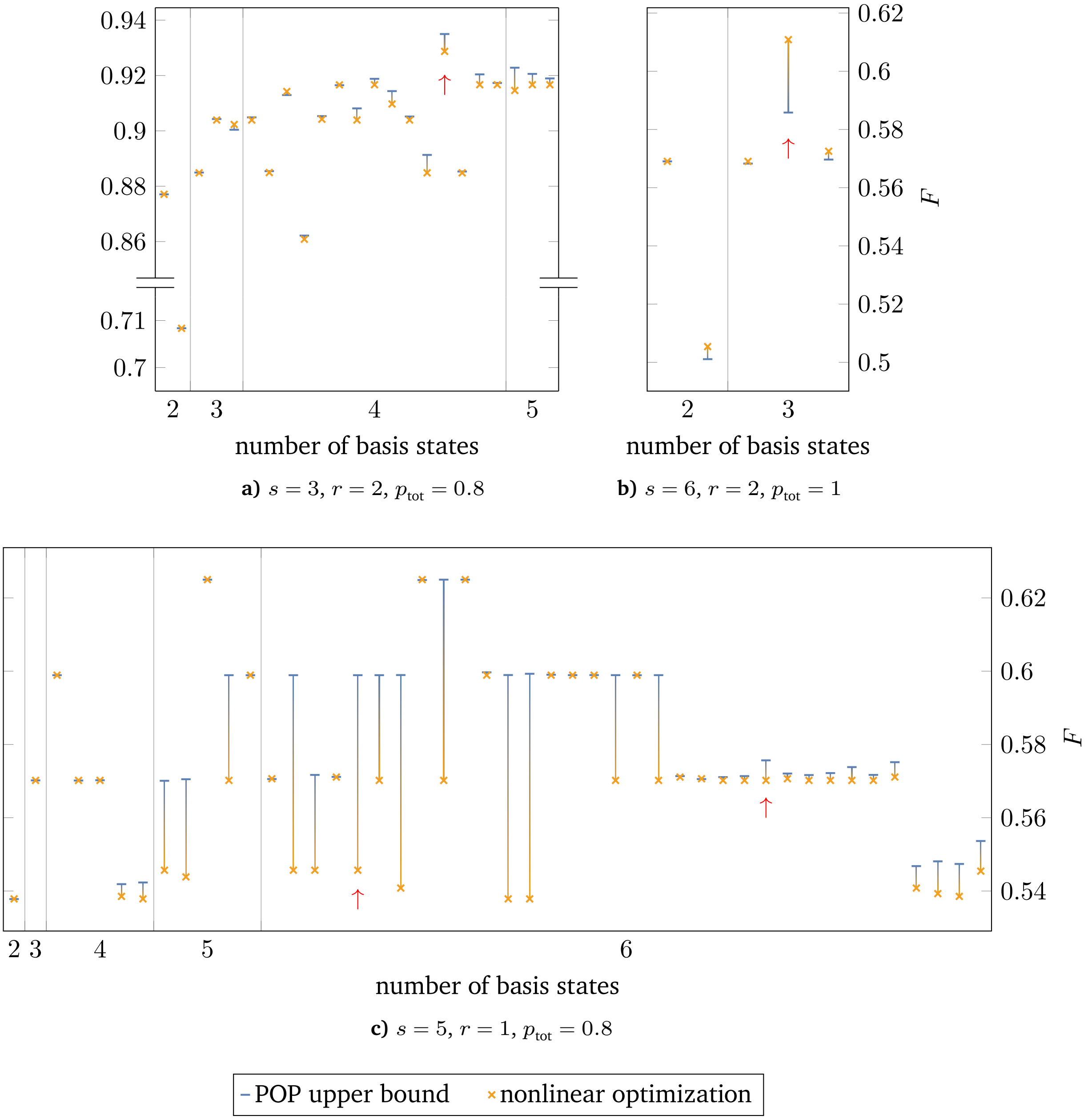

**a)** $s = 3, r = 2, p_{\mathrm{tot}} = 0.8$

**b)** $s = 6, r = 2, p_{\mathrm{tot}} = 1$

**c)** $s = 5, r = 1, p_{\mathrm{tot}} = 0.8$

**Figure 5.5.** Comparison of upper bounds obtained by solving the second-order polynomial relaxation problem with actual values from nonlinear optimization. The configurations marked with red arrows are discussed in the main text.

algorithm was devised which works on sparse problems (recall that the Newton polytope is always applied); it is able to find solution candidates with a total absolute error (as defined in section 4.10.3.6) of only $0.04$—not enough to conclude that the upper bound is attained, but the extracted points for the state are rather close to the ones the nonlinear optimizer found. Indeed, starting the latter at these points will again yield the previously-obtained optimal value $F = 0.61085$.

As a final example, consider the configuration $s = 5$, $r = 1$, $p_{\text{tot}} = 0.8$. Since $r = 1$, the recovery maps are smaller and a large number of basis choices could be optimized in the given time. The plot shows that with growing number of basis states—i.e., with growing number of parameters in the nonlinear solvers—a significant mismatch between the upper bound and the value found by the nested iteration scheme happens more and more often. Consider for example the basis

$$\{|0\,00001\rangle, |0\,00010\rangle, |0\,00011\rangle, |0\,00100\rangle, |1\,00000\rangle, |1\,01011\rangle\} \tag{5.10}$$

with reported value $F = 0.545675$ and bound $F \leq 0.598911$. This bound can indeed also be verified with the higher precision $10^{-5}$, and it is possible to extract an approximative solution using the heuristic algorithm, again with an error of about $0.04$. Starting from the point provided by this solution, the nested optimization scheme immediately finds a valid point with $F = 0.598911$, satisfying the bound. Looking at figure 5.5c, it does not seem surprising that this particular bound can be achieved, as there are a few other basis configurations for which the value was already found—though the minimization also often appears to end up in a local maximum. A more interesting bound is given for the basis

$$\{|0\,00001\rangle, |0\,00010\rangle, |0\,00111\rangle, |1\,00000\rangle, |1\,00101\rangle, |1\,00110\rangle\} \tag{5.11}$$

with the bound $F \leq 0.575663$ compared to the reported value $F = 0.570194$. In higher precision, the bound drops to $F \leq 0.57489$; however, the extracted "solution" with an absolute error of $1.25$ is so imprecise that no starting point can be generated which does not again converge to the same reported value.

# 6 Conclusion

In this thesis, I gave an overview of typical characteristics of optimization problems in quantum information. I introduced the mathematical framework of numerical optimization with a particular emphasis on convex programming, and in addition showed how the theory of polynomial optimization applies these methods to solve—or at least provide bounds—to many problems which lack such a convex structure.

While there are certainly many intrinsically hard optimization problems in quantum information and other subjects, there also have been numerous contributions in the past years trying to exploit special properties and identifying underlying structures that allow to find solutions more reliably or efficiently. This is also true for the field of polynomial optimization itself as well as the convex solvers employed in the process. However, many of these contributions are in the form of complexity results in theoretical papers; and history has shown that practical and predicted performance often do not quite keep up—for better or worse. Even if implementations exist, they naturally lack the many years of ongoing algorithm and hyperparameter tuning that established solvers can bank on; singling out promising candidates is therefore a Sisyphean task. Nevertheless, it is a necessity for sustainable science—good proposals should not effectively be discarded because no-one knows how good they actually are.

In this thesis, I take a first step on this journey by implementing a polynomial optimization framework, `PolynomialOptimization.jl`, with a focus on efficiency. This ranges from the construction of the problem, the application of methods to reduce their sizes, addressing the solver, extracting solutions, and even the solvers themselves by (re)implementing efficient versions of various research-grade software or papers. While the thesis may be completed, the software is most definitely not—there are still many interesting and useful proposals for which support should be added, and with no doubt, the future will bring more. Among them, one such concept was developed in this thesis, exploiting capabilities of nonsymmetric interior-point solvers. The integration of this idea with `PolynomialOptimization.jl` is also left for future work, as it relies on the use of non-monomial bases, which is currently not supported.

Polynomial optimization has been known for many years now, with great promises of global certificates—can it actually deliver? An increasing amount of papers using sums-of-squares method shows that indeed the promise holds—sometimes. But it should not be concealed that when taking a random real-world problem which has polynomial form, for example from quantum information, the straightforward use of polynomial optimization frameworks (if they can be called “straightforward” for people who do not know the theory and “just want to solve” the problem) will usually end up with a memory error. My framework pushes further the boundary of what can be achieved without months of dedicated research just for this problem; but the examples in chapter 5 show that sometimes, this is simply not enough. The experienced practitioner in polynomial optimization might be able to introduce further redundant constraints to improve the results without hampering performance; but the aim of the software framework is to be accessible without such experience. Therefore, the results in chapter 5 are simply an account of solving the problems without problem-specific fine tuning and telling the honest story of success and disappointment which often are close neighbors in polynomial optimization.

But efficiency is not only important for POPs, which is also demonstrated in chapter 5. Efficiently accessing the interface of leading optimization packages allows to incorporate even nonlinear optimization methods which in turn call convex optimizers at superior speed and overall resource consumption. Modeling frameworks are tremendously helpful for *modeling*—for experimenting, quickly designing optimization problems and obtaining solutions in the small-scale case. However, they are then often also employed for large-scale problems or in a parametric way that exceeds their capabilities to reuse previously constructed solver instances. It is important to realize how much potential is wasted for example by this misuse or because the solver ideally suitable for a given problem only exists as a concept paper. This finally allows to conclude that many problems might not be as computationally difficult as they seem.

# A Third-order directional derivative for interpolation basis

Taking up from Proposition 4.69, the third-order derivative is in general, setting $\Lambda \equiv \Lambda_{\mathrm{H}}(\boldsymbol{p})$

$$\begin{aligned}
\frac{\partial^3 f(\boldsymbol{p})}{\partial p_u \partial p_{u'} \partial p_{u''}} &= -\boldsymbol{w}^\dagger(\boldsymbol{t}_u)\Lambda^{-1}\frac{\partial \Lambda}{\partial p_{u''}}\Lambda^{-1}\boldsymbol{w}(\boldsymbol{t}_{u'})\boldsymbol{w}^\dagger(\boldsymbol{t}_{u'})\Lambda^{-1}\boldsymbol{w}(\boldsymbol{t}_u) \\
&\qquad - \boldsymbol{w}^\dagger(\boldsymbol{t}_u)\Lambda^{-1}\boldsymbol{w}(\boldsymbol{t}_{u'})\boldsymbol{w}^\dagger(\boldsymbol{t}_{u'})\Lambda^{-1}\frac{\partial \Lambda}{\partial p_{u''}}\Lambda^{-1}\boldsymbol{w}(\boldsymbol{t}_u) \\
&= -\boldsymbol{w}^\dagger(\boldsymbol{t}_u)\Lambda^{-1}\Big[\boldsymbol{w}(\boldsymbol{t}_{u''})\boldsymbol{w}^\dagger(\boldsymbol{t}_{u''})\Lambda^{-1}\boldsymbol{w}(\boldsymbol{t}_{u'})\boldsymbol{w}^\dagger(\boldsymbol{t}_{u'}) \\
&\qquad + \boldsymbol{w}(\boldsymbol{t}_{u'})\boldsymbol{w}^\dagger(\boldsymbol{t}_{u'})\Lambda^{-1}\boldsymbol{w}(\boldsymbol{t}_{u''})\boldsymbol{w}^\dagger(\boldsymbol{t}_{u''})\Big]\Lambda^{-1}\boldsymbol{w}(\boldsymbol{t}_u) \\
&= -2\,\mathrm{Re}\big(\boldsymbol{w}^\dagger(\boldsymbol{t}_u)\Lambda^{-1}\boldsymbol{w}(\boldsymbol{t}_{u''})\ \boldsymbol{w}^\dagger(\boldsymbol{t}_{u''})\Lambda^{-1}\boldsymbol{w}(\boldsymbol{t}_{u'})\ \boldsymbol{w}^\dagger(\boldsymbol{t}_{u'})\Lambda^{-1}\boldsymbol{w}(\boldsymbol{t}_u)\big). \qquad \text{(A.1)}
\end{aligned}$$

Going back to the real-valued case, the double contraction with an $\boldsymbol{x}$ (due to continuity, it does not matter with which indices) is given by

$$\begin{aligned}
&\sum_{u',u''}\frac{\partial^3 f(\boldsymbol{p})}{\partial p_u \partial p_{u'} \partial p_{u''}} x_{u'} x_{u''} \\
&\quad = -2\sum_{u',u''}\boldsymbol{w}^\top(\boldsymbol{t}_u)\Lambda^{-1}\boldsymbol{w}(\boldsymbol{t}_{u''})x_{u''}\ \boldsymbol{w}^\top(\boldsymbol{t}_{u''})\Lambda^{-1}\boldsymbol{w}(\boldsymbol{t}_{u'})x_{u'}\ \boldsymbol{w}^\top(\boldsymbol{t}_{u'})\Lambda^{-1}\boldsymbol{w}(\boldsymbol{t}_u) \\
&\quad = -2\boldsymbol{w}^\top(\boldsymbol{t}_u)\Lambda(\boldsymbol{p})^{-1}\Lambda(\boldsymbol{x})\Lambda(\boldsymbol{p})^{-1}\Lambda(\boldsymbol{x})\Lambda(\boldsymbol{p})^{-1}\boldsymbol{w}(\boldsymbol{t}_u) \\
&\quad = -2\big(W^\top\Lambda(\boldsymbol{p})^{-1}\Lambda(\boldsymbol{x})\tilde{C}\tilde{C}^\top\Lambda(\boldsymbol{x})\Lambda(\boldsymbol{p})^{-1}W\big)_{u,u} \\
&\quad = -2\sum_{\ell}\big(W^\top\Lambda(\boldsymbol{p})^{-1}\Lambda(\boldsymbol{x})\tilde{C}\big)^2_{u,\ell}. \qquad \text{(A.2)}
\end{aligned}$$

$\Lambda(\boldsymbol{x})$ is again an $\mathcal{O}(L(UN_{\mathrm{r}} + N\log N))$ operation. Investing another $\mathcal{O}(L^2)$ in temporary storage (which may be the same as for the Hessian product), the latter three matrices can be multiplied in $\mathcal{O}(L^\omega)$ time and the full expression for all $u$ in $\mathcal{O}(LN\log N)$.

In total, the third-order directional derivative requires $\mathcal{O}(L^\omega + LN\log N)$ time and $\mathcal{O}(L^2)$ storage, the same as for the gradient.

# B Newton polytope—examples

This appendix is divided in two parts: the first part takes an actual real-world example from quantum information and demonstrates some performance metrics of the Newton polytope implementation in `PolynomialOptimization.jl`.

In the second part, an attempt is made to compare this to other modeling frameworks; this is not as simple as entering the same problem in a different framework and checking the timings. On the one hand, most of the frameworks do not even support a constrained Newton polytope or struggle with semidefinite constraints—comparing realistic quantum information problems is therefore not possible. Unconstrained test problems for numerical optimization, on the other hand, are usually formulated either for few variables—where the calculations should be so quick that they are hard to benchmark—or with a flexible number of variables. I checked numerous problems of the latter type from the classics in [MGH81]; they all share the inconvenient characteristic that their half Newton polytope coincides either fully or almost fully with the dense relaxation. Therefore, an unconstrained problem generated based on the real-world example from the first part is used instead.

All timings were obtained on an Intel i7-11800H CPU with $64\,$GiB physical RAM, a turbo frequency of up to $4.6\,$GHz with single-threaded execution. The linear solver employed by `PolynomialOptimization.jl` was `Mosek` 11.0.12 [MOS24b].

## B.1 A realistic problem

As a real-life example, a simple optimization problem from quantum information is chosen—the transmission through a lossy channel as described in chapter 5, more precisely, equation (14) in [DP22b], for sending three and receiving a single photon. For simplicity, all variables are assumed to be real; the problem has 46 scalar variables, two equality constraints and three semidefinite constraints. The semidefinite constraints and one equality constraint are linear, the objective and the other equality constraint are quadratic. The minimal relaxation order $1$ is insufficient (the solver generates a certificate of primal infeasibility); therefore, order $2$ is chosen. The largest side dimension of a semidefinite matrix in the relaxation is $1128$, and `PolynomialOptimization.jl` requires less than five seconds and insignificant memory before control is passed to the solver `Mosek`, which would then produce an out-of-memory error on a standard computer.

Putting all constraints into the objective according to equation (4.37) leads to 51 537 potential extremal points in the convex hull; the Akl-Toussaint heuristic is able to reduce this number to 33 160 points in just over a second; this heuristic is turned on by default. While a more detailed search would be able to reduce this number to no more than 1455 linearly independent points in roughly three minutes, it is disabled by default.

A quick scan over the remaining points in the convex hull allows to restrict the range of allowed degrees *per variable* that any monomial in the convex hull may have (note that this is the only computation executed by `SumOfSquares.jl` when it approximates the Newton polytope, and even this only in the presence of constraints). This reveals that while the monomials might be quadratic in total, no variable may occur more than linearly, giving rise to 1082 possible members of the dense basis. Running this number of linear programs based on the Akl-Toussaint heuristic in nine seconds gives only 83 monomials at the end, which is the final size of the basis. Of course, the same result is also obtained using the detailed preprocessing, and the search now only takes a third of a second—still, the preprocessing cost in this instance is too large to make it worthwhile. Using all eight physical cores reduces the time after preprocessing to about two seconds.

Due to the semidefinite constraints, the side dimension of the largest matrix in the relaxation is not 83, but 376 instead; however, this is small enough to be solved without requiring a compute cluster. The largest memory consumption of the process during optimization using the `MosekSOS` method was 40 GiB, and in 150 s (including about one second of data parsing by `PolynomialOptimization.jl`), the upper bound to the best possible fidelity 0.6667763 was obtained. Note that the memory consumption can be halved by forcing `Mosek` in a single-threaded mode, at the expense of 505 s spent in the solver. Comparing with the solution obtained via numerical optimization—this is the rightmost point of the orange $s = 3$ line in figure 5.2*a*—which was numerically identically with $\frac{2}{3}$, this is in good agreement. However, it also becomes clear that the second level is not yet sufficient, as the fourth digit deviates; and indeed, no optimality certificate is available and no solution can be extracted.

## B.2 Comparison with other frameworks

As initially mentioned, comparing the Newton polytope implementation of such a realistic problem with semidefinite constraints with other frameworks is not possible due to a lack of their features. The problem is therefore simplified by a great deal: All constraints are dropped; as the problem is now of course unbounded, the objective is squared. The resulting unconstrained optimization problem is much smaller; 463 potential extremal points are reduced to 112 by the Akl-Toussaint heuristic; 596 monomials in the cheap approximation to the Newton polytope are reduced to only 31 in 0.03 s by `PolynomialOptimization.jl`.

Next, a comparison with `TSSOS.jl` is made. For a benchmark to become possible, the code for `tssos_first` is slightly modified to disable all printing and return the Newton polytope immediately. The $31$ monomials are found in $1.38$ s.

`SumOfSquares.jl` usually uses `JuMP` to model the problem and in this process then calls `Certificate.half_newton_polytope` without variable partitioning. While this function is the fastest of all, taking less than a millisecond, it also does not perform any reduction: all $596$ elements remain. However, `SumOfSquares` then also invokes the diagonal consistency filter from Theorem 4.44; this is also able to arrive at $31$ monomials and takes $1.00$ s.

For more distinctive results, it is interesting to consider a larger instance of the problem; so let now $s = 6$, $r = 3$. `PolynomialOptimization.jl` reduces the dense basis of size $7021$ in $8$ s with no significant memory consumption to $145$ elements (single-threaded). `TSSOS.jl` also requires negligible memory, but takes $465$ s to generate the same result. `SumOfSquares` takes $0.03$ s with the filter disabled, but produces a basis of size $6904$. With filter enabled, it did not terminate before exceeding $32$ GiB in memory consumption, which was reached after six minutes.